\documentclass{article}
\usepackage{amssymb}
\usepackage{psfig}
\usepackage{epsfig}

\newtheorem{theom}{Theorem}[section]
\newtheorem{lem}{Lemma}[section]
\newtheorem{defi}{Definition}[section]

\newtheorem{proflem}{Proof}[section]

\makeatletter

\@addtoreset{equation}{section}

\makeatother

\def\squareforqed{\hbox{\rlap{$\sqcap$}$\sqcup$}}
\def\qed{\ifmmode\squareforqed\else{\unskip\nobreak\hfil
\penalty50\hskip1em\null\nobreak\hfil\squareforqed
\parfillskip=0pt\finalhyphendemerits=0\endgraf}\fi \medskip}

\begin{document}

\title{Time Aperiodic Perturbations of  Integrable Hamiltonian Systems\\
\quad\\
\quad\\
}

\author{Alfredo Martinez
\\
Control and Dynamical Systems  107-81 \\ Caltech \\ Pasadena, CA
91125
\hfill \\
\hfill \\
Stephen Wiggins \thanks{This research was supported by
ONR Grant No.  N00014-97-1-0071.} \\
Applied Mechanics and
\\ Control and Dynamical Systems  107-81 \\ Caltech \\ Pasadena, CA
91125
}

\date{\today}

\maketitle

\begin{abstract}
We consider a Hamiltonian $H=H^{0}(p)+\kappa H^{1}(p,q,t)$,
$(p,q)\in {\mathbb{R}}^{n} \times {\mathbb{T}}^n$,
$t\in{\mathbb{R}}$ where $\kappa \in {\mathbb{R}}$ is a small perturbation
parameter and $p$, $q$ are the action and angle variables
respectively. The Hamiltonian generates an autonomous vector
field obtained by extending the phase space making $t$ a
dependent variable and adding its conjugate variable $\tau$.
 In this paper
 we look at
a time aperiodic  perturbation $H^{1}(p,q,t)$ which tends
 as $t\rightarrow \infty$ to either
a time independent perturbation or a time
quasiperiodic perturbation and we prove
a KAM-type theorem.
Extending the phase
space results in the preservation under a small enough perturbation
 of  cylinders of the extended autonomous  system
rather than the usual tori. To prove the theorem we transform the
Hamiltonian $H$ to a normal form which depends on fewer angles,
none if possible. This transformation is done via a near identity
canonical transformation. The canonical transformation is
constructed using the Lie series formalism and by solving for a
generating function. Because of the aperiodic time dependence,
the usual Fourier series methods used to obtain  the generating
function no longer apply. Instead, we use Fourier transform
methods to solve for the generating function and make use of an
isoenergetic non-degeneracy condition which results in a shift of
frequencies associated with each cylinder.
\end{abstract}
\newpage

\tableofcontents
\newpage

\section{Introduction}
\quad\\

Our motivation for this work is the application to the existence of
``flow barriers'' in two dimensional, incompressible,
time-dependent fluid flows.
Over the past 10 years there has been much work in applying the approach
and methods
of dynamical systems theory to the study of transport in fluids from the
Lagrangian point of view.
Suppose one is interested in the
motion of a {\it passive tracer} in a fluid (e.g. dye, temperature, or
any
material that can be considered as having negligible effect on the flow),
then, {\it neglecting molecular diffusion}, the passive tracer follows
fluid
particle trajectories which are solutions of

\begin{eqnarray}\label{eq:u}
  \dot{x}= v (x, t).
 \end{eqnarray}

\noindent
where $v (x, t)$ is the velocity field of the fluid flow, $x \in {\mathbb{R}}^n,
n=2$
or $3$.  When viewed from the point of view of
dynamical systems theory,  the phase space of (\ref{eq:u}) is actually
the
physical space in which the fluid flow takes place.  Evidently,
``structures"
in the phase space of (\ref{eq:u}) should have some influence on the
transport and mixing properties of the fluid.
Babiano $\cite{Babino94}$  and Aref  and
El Naschie $\cite{Aref94}$ provide recent reviews of this approach.

To make the connection with the
large body of literature on dynamical systems theory more concrete
let us consider a less general
fluid mechanical setting.  Suppose
that the fluid is two-dimensional, incompressible, and inviscid. Then we
know
that the velocity field can be obtained from
the derivatives of a scalar valued function, $\psi(x_1,x_2,t)$, known
as
the {\it streamfunction}, as follows

\begin{eqnarray}
\dot{x}_1 &= & {\partial \psi\over\partial x_2} (x_1,x_2,t),\nonumber
\\
\dot{x}_2 &= &-{\partial\psi\over \partial x_1} (x_1,x_2,t),
\hskip 1 truein (x_1,x_2)\in  {\mathbb{R}} ^2.
\label{vel2}
\end{eqnarray}

\noindent
In the context of dynamical systems theory, (\ref{vel2}) is a
time-dependent Hamiltonian  vector field where the streamfunction plays
the
role of the Hamiltonian function.   If the flow is time-periodic then the
study
of (\ref{vel2}) is typically reduced to the study of a two-dimensional
area preserving {\em Poincar\'e map}.
Practically speaking,  the reduction to a Poincar\'e map
means that rather than viewing a particle trajectory as a curve
in continuous time, one views the trajectory  only at  discrete intervals
of time, where the interval of
time is the period of the velocity field.  The value of making this
analogy with Hamiltonian
dynamical systems lies in the fact that a variety of techniques in this
area have immediate
applications to, and implications for, transport and mixing processes in
fluid mechanics.
For example, the persistence of invariant curves in the Poincar\'e map
(KAM curves) gives rise
to barriers to transport, chaos and Smale horseshoes  provide mechanisms
for the
``randomization'' of fluid particle trajectories,  an analytical
technique, Melnikov's method,
allows one to  estimate fluxes as well as describe the
parameter regimes where chaotic fluid particle motions occur, a
relatively new technique,
lobe dynamics,  enables one to  efficiently compute transport between
qualitatively
different flow regimes.

Extending the constructions of Smale horseshoes, Melnikov's method, and
stable and unstable manifolds of hyperbolic trajectories
to vector fields  with aperiodic time-dependence has been done. However,
extensions of KAM theory to vector fields with arbitrary time-dependence
has not been done. This paper represents a first step in that direction.

One of the main stability results for nearly-integrable Hamiltonian
systems is Kolmogorov's theorem  $\cite{Arnold63}$ concerning the preservation
of a set of full-measure-nonresonant  invariant tori.
Such nearly-integrable Hamiltonian
systems are generated by
a real valued Hamiltonian written in action-angle variables
\begin{eqnarray*}
H(p,q)=H^{0}(p)+ \kappa H^{1}(p,q),
\end{eqnarray*}
 where $p=(p_{1},...,p_{n})   \in B \subset
  {\mathbb{R}}^{n}$, $q=(q_{1},...,q_{n})\in {\mathbb{T}}^{n}$ are,
respectively, the action and angle variables, and $\kappa$ is a small perturbation parameter. $\kappa H^{1}(p,q)$ is therefore a small perturbation of the
integrable Hamiltonian $H^{0}(p)$.
For any $p\in B$ the unperturbed angular frequencies are defined by
$\tilde{\lambda} (p)
 =(\lambda _{1},...,\lambda _{n})(p)=(\partial H^{0}/\partial p_{1}(p),
...,\partial H^{0} /\partial p_{n}(p))$.
For the integrable Hamiltonian $H^{0}(p)$ the equations of motion
reduce to $\dot p =-\partial H^{0}/\partial q=0,\quad
\dot q =\partial H^{0} /\partial p =\tilde{\lambda} (p)$ with solutions
$p(t)=p_{0},\quad q(t)=q_{0}+\tilde{\lambda} (p_{0})t\quad ({\mbox{mod}}2\pi)$.
Consequently the phase space is foliated by invariant $n$-tori
 with quasi-periodic motions characterized by the frequency $\tilde{\lambda}
 (p)$.
The result of adding a small
perturbation  to the integrable Hamiltonian $H^{0}(p)$
is the destruction of all tori
except for those whose frequencies satisfy a non-resonance condition.

In the  KAM theorem attention is restricted to tori supporting quasi-periodic motions with
an appropriate nonresonant condition.
Given the set of all frequencies $\Omega \subset {\mathbb{R}}^{n}$,
the frequencies satisfying this nonresonant condition
are called  diaphontine frequencies and as a subset of
$\Omega$ are defined by
\begin{eqnarray*}
\Omega _{\Gamma}=\{\tilde{\lambda} \in \Omega \subset {\mathbb{R}}^{n}|\quad
|\tilde{\lambda} \cdot k |\geq \Gamma \| k \|^{-n},\quad
\forall k \in {\mathbb{Z}}^{n},\quad k\neq 0 \}
\end{eqnarray*}
for some positive constant $\Gamma$.

The method used to prove the  KAM theorem
 in $\cite{Benettin84}$ and
$\cite{Delshams96}$, which this paper follows,
is standard in classical perturbation theory.
The central idea is to construct a suitable canonical transformation
$\psi$ which brings the original Hamiltonian $H$ into a normal form
which depends on fewer angles, none if possible.
The transformation $\psi$ is constructed iteratively as the composition
of successive near-identity canonical transformations
$\phi^{1},\phi^{2},...$ .
Taking the limit as the number of iterative steps
goes to infinity results in the elimination of the
perturbation of the Hamiltonian  leaving
an integrable system that is isomorphic to the original system.
 To construct this canonical transformation
the authors in $\cite{Benettin84}$ and $\cite{Delshams96}$ make use of the Lie series
method which has the advantage of avoiding
 any inversion and thus any reference to the implicit-function
theorem. The Lie series method requires  constructing  a Hamiltonian called
the generating function. At the  $k$th step of the
iterative process
a generating function $\chi_{k}$ is constructed and
the desired near-identity canonical transformation $\phi^{k}$ is obtained
as the flow at time $1$ associated with such  generating function. The
generating function is constructed via the nondegeneracy condition and
by solving a partial differential equation.

The restriction to diophantine tori comes about from the problem of solving
the partial differential equation mentioned above which has the form
\begin{eqnarray}
\label{eqn1111}
\sum _{i=1}^{n}\lambda _{i}\frac{\partial F}{\partial q_{i}}(q)=G(q),
\end{eqnarray}
where the unknown function $F$ and the known function $G$ are defined on the
torus ${\mathbb{T}}^{n}$ and $G$ has zero average.

In this paper we consider two similar KAM theorems where,
in each, there is an aperiodic time dependent term in the perturbation.
In both cases we consider an n-degree of freedom real valued
nearly-integrable Hamiltonian  in action-angle variables
of the form
\begin{eqnarray*}
H(p,q,t)=H^{0}(p)+ \kappa H^{1}(p,q,t),
\end{eqnarray*}
where $p=(p_{1},...,p_{n})\in {\mathbb{R}}^{n}$,
 $q=(q_{1},...,q_{n})\in {\mathbb{T}}^{n}$, $t\in {\mathbb{R}}$ and
  $\kappa \in {\mathbb{R}}$ is a small
perturbation parameter.
 In the first case
the perturbation considered has an exponentially decaying aperiodic
time dependent term and a quasiperiodic term depending only
on the angles $q$
\begin{eqnarray*}
H^{1}(p,q,t)=
\sum _{k\in {\mathbb{Z}}^{n}}g_{k}e^{ik\cdot q}+
\sum _{k\in {\mathbb{Z}}^{n}}
f_{k}(p)e_{k}(t)e^{ik\cdot q},
\end{eqnarray*}
 where $f(p)$ is a bounded function and $e_{k}(t)$
decays exponentially.
The nature of this exponential decay is explained fully in the next section,
 (see definition 1).
In the second case the perturbation
 consists of an exponentially decaying aperiodic
time dependent term and a term depending
in a quasiperiodic manner on the angles $q$ and on time $t$
\begin{eqnarray*}
H^{1}(p,q,t)=
\sum _{k\in {\mathbb{Z}}^{n+m}}g_{k}e^{ik\cdot( q,\theta t)}+
\sum _{k\in {\mathbb{Z}}^{n}}
f_{k}(p)e_{k}(t)e^{ik\cdot q},
\end{eqnarray*}
 where $f(p)$, $e_{k}(t)$ are as above and $\theta=(\theta_{1},...,\theta_{m})$
is the vector of basic frequencies. Although the second case
seems more promising as far as applications is concern, we
first prove the  less technical first case.
In the last section we sketch the proof of the time quasiperiodic
case  and state the theorem.
The only difference in the
assumptions of the time quasiperiodic  theorem involves the diophantine
condition $\lambda \in \Omega _{\Gamma}$ which must hold
for the larger vector $\lambda =(\tilde{\lambda} ,\theta)
\in{\mathbb{Z}}^{n+m}$
rather than just the vector $\tilde{\lambda}\in{\mathbb{Z}}^{n}$.

In both cases we  consider the nearly-integrable  system
generated by the Hamiltonian $H(p,q,t)$ and prove a KAM type theorem
where instead of
tori we show the preservation of cylinders of the form ${\mathbb{T}}^{n}\times {\mathbb {R}}$.
The  proof is similar in method to the one given in $\cite{Benettin84}$.
First, the non-autonomous vector field is made autonomous by
making  time a dependent  variable,
 $t=q_{n+1}$, and  grouping it with the  angles. Similarly a conjugate variable $\tau =p_{n+1}$ is
grouped with the  action variables. We write the Hamiltonian in
a Kolmogorov-type normal form,
which consists of  Taylor expanding the Hamiltonian about
$p=0, \tau =0$ and grouping terms of different orders of magnitude. Note the proof
of this theorem focuses on the preservation of the  cylinder with
$p=0$, $\tau =0$. This can be done without loss of generality since any other
cylinder can be shifted to the ``zero'' cylinder.

Once the Hamiltonian is in Kolmogorov normal form we seek a suitable
canonical transformation which brings the
Hamiltonian into a normal form which depends on fewer angles.
The transformation is constructed iteratively as the product of
near identity canonical transformations $\phi_{k}$; i.e
$\phi =\phi_{1}\circ \phi_{2} \circ \cdot \cdot
\cdot \circ \phi_{n}$. The result is a sequence
of Hamiltonians, $H_{1}=H_{0}\circ \phi_{1},H_{2}=H_{1}\circ \phi_{2}
,...,H_{n}=H_{n-1}\circ \phi_{n}$,
which come closer to the desired normal form. Letting $n\rightarrow \infty$,
we obtain
$H_{\infty}$ which is an integrable Hamiltonian.
The sequence of canonical transformations is obtained using
the Lie series method. This method and other techniques from
the theory of several complex variables will require us to extend the domain of
our real valued functions and consider
analytic functions defined on a complex domain.

To carry out the  Lie series method we introduce a generating function
of the form
\begin{eqnarray*}
\chi(p,\tau,q,t)= \xi\cdot (q,t)+X(q,t)+\sum_{i=1}^{n+1}Y_{i}(q,t)p_{i},
\end{eqnarray*}
where $\xi \in {\mathbb{R}}^{n+1}$. To solve for  $X$ and $Y_{i}$ we  must solve
a partial  differential equation of the form
\begin{eqnarray}
\label{eqn1212}
\sum _{i=1}^{n+1}\lambda_{i} \frac{\partial F}{\partial q_{i}}(q,t) =G(q,t),
\end{eqnarray}
where $\lambda _{i} =\frac{\partial H^{0}}{\partial p_{i}}(0)$ for $i=1,...,n$,
$\lambda _{n+1}=1$, the given function $G(q,t)$ with zero
average  and
the unknown function $F(q,t)$ are defined on ${\mathbb{T}}^{n}\times {\mathbb{R}}$.
The known function $G(q,t)$ will be an expression
related to the perturbation and
the unknown function $F(q,t)$ will be an expression related to $X(q,t)$
or $Y_{i}(q,t)$.

For the time independent case with functions $F(q)$ and
 $G(q)$ defined on ${\mathbb{T}}^{n}$ the partial differential equation
 can be solved in terms of  Fourier coefficients $\cite{Benettin84}$.
Understandably  though, the aperiodic nature of the time dependence
renders the Fourier series procedure inapplicable to the problem at hand.
Instead we must use Fourier transform methods to solve for
$X(q,t)$ and $Y_{i}(q,t)$ and  obtain the appropriate domains of analyticity.
Consequently $F(q,t)$ is written as a Fourier series with time dependent
Fourier coefficients $f_{k}(t)$ and these coefficients are expressed
in terms of the Fourier transform of the time dependent
Fourier coefficients of
$G(q,t)$.
Solving $(\ref{eqn1212})$ for the unknown function $F(q,t)$
restricts the  form of the time dependent perturbation
requiring an exponential
decay in time in order
to obtain convergence of the Fourier coefficients $f_{k}(t)$.

To obtain a value for $\xi$ and completely solve for the generating function
we make use of the nondegeneracy assumption ${\mbox {det}}(\partial \lambda _{i}/\partial p_{i})
\neq 0\quad i=1,...n$.
Note this assumption implies the frequencies are functionally independent
and since $\lambda _{n+1}=1$,  the ratios of the frequencies
$\lambda_{i} \quad i=1,...,n$ to $\lambda _{n+1}$ are functionally
 independent as well.
Thus we obtain what is called an isoenergetic nondegeneracy condition.
This condition is the key feature in solving for $\xi$.

After applying the iterative lemma $n$ times and taking $n\rightarrow
\infty$ we obtain an integrable Hamiltonian $H_{\infty}(\tilde p,
\tilde \tau, \tilde q, \tilde t)$ in the transformed variables
$(\tilde p,
\tilde \tau, \tilde q, \tilde t)$ which
generates an autonomous vector field in ${\mathbb{T}}^{n}\times
{\mathbb{R}}$ with solutions of the form $(\tilde p,\tilde \tau)=
(p_{0},\tau_{0}),
\quad \tilde q =(1+\kappa E\zeta_{A})\tilde{\lambda} t+q_{0}, \quad \tilde t =
(1+\kappa E\zeta_{A})t$. These solutions imply the phase space
${\mathbb{R}}^{n+1}\times {\mathbb{T}}^{n}\times
{\mathbb{R}}$
is foliated by invariant infinite cylinders each sustaining winding
motions identified by the frequency $\tilde{\lambda} (p_{0})$ and
evolving in the $\tilde{t}$ direction.

In Appendix $\ref{Rossby}$ we present the Rossby wave flow
$\cite{Malhotra98}$ with a time
decaying perturbation and show this system satisfies the hypothesis
of the theorem presented in this paper.

\section {Preliminaries}

Although we begin by considering a real Hamiltonian dynamical system,
techniques from several complex variables will be use in the analysis.
This  requires complex extensions of the original functions defined
in
 $\mathbb{R}^{m}$ to functions defined in $\mathbb {C}^{m}$. We begin  describing
 these complex extensions with some notation.
Denote the open
ball of radius  $\rho$ centered at
 $p_{0}$
by
 $B$
so that
 $p_{0}\in  {B} \subset \mathbb {R}^{m}$, and
we assume without loss of generality $\rho<1$.
Also, denote
 $q\equiv (q_{1},...,q_{m})\in \mathbb{T}^{m}$
where we identify functions on
 $\mathbb{T}^{m}$
with functions on
 $\mathbb{R}^{m}$
that are $2\pi$ periodic in each $q_{1},...,q_{m}.$
The complex extension of
 ${B} \times \mathbb{T}^{m}$
is given by
\[
D_{\rho , p_{0}}=\{ (p,q) \in \mathbb {C}^{2m} | 
\quad  \| p-p_{0} \| \leq \rho ,{\mbox{ Re }}q\in {\mathbb{R}}^n {\mbox{ mod }}
2\pi, \| \mbox {Im } q \| \leq \rho \},
\]
where Im $q = (\mbox {Im }q_{1},...,\mbox {Im }q_{m})$
 and $\| v \| =\max _{i}
 |v_{i} |$
for $v\in {\mathbb{C}}^{m}$.
 Define ${\cal{A}}_{\rho ,p_{0}}$
as the set of all complex  continuous functions defined on $D_{\rho ,p_{0}}$,
analytic in the interior of $D_{\rho ,p_{0}}$, $2\pi$ periodic in
the variables $q_{1},...,q_{m}$,
and real for real values of the variables.

Consider the following non-autonomous Hamiltonian
\[H(p ,q,t)=H^{0}(p)+ \kappa H^{1}(p,q,t),\]
 where $p\in \mathbb{R}^{n},
q \in \mathbb{T}^{n}$ are the action and angle variables
respectively, $ t \in \mathbb{R}$
and $\kappa\in {\mathbb{R}}$ is a small perturbation parameter.
The nature of the time dependence is explained at the
end of this section.
Hamilton's equations are given by
\begin{equation}
 \dot{p}=
-\frac {\partial H}{\partial q} =
 -\kappa \frac {\partial H^{1}}
{\partial q}, \quad \quad   \dot{q} =
 \frac {\partial H}{\partial p}= \frac {\partial H^{0}}
{\partial p}+
\kappa  \frac {\partial H^{1}}{\partial p}.
\label{Ham}
\end{equation}

For the system generated by the integrable Hamiltonian
$H^{0}(p)$ we define the angular frequencies $\tilde{\lambda}(p)=
(\frac{\partial H^{0}}{\partial p_{1}},...,
\frac{\partial H^{0}}{\partial p_{n}})=
(\lambda_{1},...,\lambda_{n})$ and  Hamilton's
equations reduce to $\dot p=-\partial H^{0}/\partial q=0,\quad \dot
q=\partial H^{0}/\partial p =\tilde{\lambda}$ with solutions
$p(t)=p_{0},\quad q(t)=q_{0}+\tilde{\lambda}(p_{0})t$. Consequently
the phase space is foliated by invariant n-tori, each of which
sustains  quasi-periodic
motions characterized by the frequency $\tilde{\lambda}(p_{0})$.

To analyze the full Hamiltonian $H(p,q,t)$ with the time dependent
perturbation $H^{1}(p,q,t)$ we  change this non-autonomous
 system into an autonomous system
by making time a dependent variable and effectively extending  
the phase space from ${\mathbb{R}}^n\times {\mathbb{T}}^n$ to
 ${\mathbb{R}}^n\times {\mathbb{R}}
\times {\mathbb {T}}^{n}\times {\mathbb{R}}$.
For notation sake we group $t$ with the angles $q$ and its conjugate
variable
 $\tau$
with the actions  $p$ and write the Hamiltonian
$H(p,\tau,q,t)$ with
$(p,\tau ,q,t)\equiv (p', q')
 \in \mathbb{R}^{n} \times \mathbb{R} \times  \mathbb{T}^{n} \times \mathbb{R} $
\begin{eqnarray*}
H(p',q')\equiv H(p, \tau ,q,t)=H^{0}(p)+ \kappa H^{1}(p,q,t)+\tau \quad \quad
\end{eqnarray*}
or
\begin{eqnarray}
\label{eqnONE}
H(p',q')=\tilde {H}^{0}(p,\tau)+   {H}^{1}(p ,q ,t) =\tilde {H}^{0}(p')
+ \kappa  {H}^{1}(p',q'),
\end{eqnarray}
where
$\tilde {H}^{0}(p')=H^{0}(p)+\tau$.
Given $0<\sigma < \rho <1,  $ we define the complex extension of $ \mathbb {R}^{n} \times
\mathbb {R} \times \mathbb {T}^{n} \times \mathbb {R} $
\begin{eqnarray*}
D_{\rho, \sigma ,p_{0}}= \{ (p,\tau , q,t)\in  \mathbb {C}^{2n+2} |\quad \| p-p_{0} \|
\leq \rho,|\tau | \leq \sigma,{\mbox{ Re }}q\in {\mathbb{R}}^n {\mbox{ mod }}
2\pi , \| \mbox{Im }q \| \leq \rho , |\mbox{Im }t | \leq \sigma \},
\end{eqnarray*}
and define ${\cal{A}}_{\rho , \sigma , p_{0}}$ as the set of all complex continuous functions
defined on $D_{\rho, \sigma ,p_{0}}$, analytic in the interior of $D_{\rho, \sigma ,p_{0}}$,
 $2\pi$ periodic in the variables $q_{1},...q_{n}$, and real for real values of the
variables.
Hamilton's equations for the new autonomous Hamiltonian are  given by
\begin{equation}
\begin{array}{ccccccc}
\quad \quad \quad \quad \quad \quad \dot{p}=-\frac {\partial H}{\partial q} = - \kappa
 \frac {\partial H^{1}}
{\partial q},  \quad \quad   &    \\
  & \\
\quad \quad \quad \quad \quad \quad \dot{\tau}=-\frac {\partial H}{\partial t}=-\kappa
 \frac {\partial H^{1}}
{\partial t}, \quad \quad   &  \\
    &  \quad \quad   \quad \quad \quad \quad \\
\quad \quad \quad \quad \quad \quad \dot{q} = \frac {\partial H}{\partial p}= \frac {\partial H^{0}}
{\partial p}+\kappa \frac {\partial H^{1}}{\partial p},\quad   &  \\
   &  \\
\quad \quad \quad \quad \quad \quad \dot{t}=\frac {\partial H}{\partial \tau}=1 .\quad \quad \quad \quad \quad &  \\
\end{array}
\label{Hamex}
\end{equation}
Of the new  Hamilton's equations, (\ref{Hamex}), the first, third and fourth equations are
equivalent to the original system (\ref{Ham}).
For the system generated by the  Hamiltonian $\tilde{H}^{0}(p')$
Hamilton's equations reduce to $\dot p' =-\partial \tilde{H}^{0}/\partial
q'=0, \quad \dot q'=\partial \tilde{H}^{0}/\partial p'=\lambda(p')$
with solutions $p'(t)=p'_{0},\quad q'(t)=\lambda(p'_{0})t+q_{0}$
where
\begin{eqnarray*}
\lambda (p')=\frac {\partial \tilde {H}^{0}}{\partial p'}(p')=
(\frac { \partial \tilde {H}^{0}}{\partial p_{1}}
,...,\frac { \partial \tilde {H}^{0}}{\partial p_{n}},\frac {\partial \tilde {H}^{0}}{\partial \tau})=(\lambda _{1}
,..., \lambda _{n},1).
\end{eqnarray*}
We will often write
$\lambda =(\tilde{\lambda},1) =(\lambda _{1}
,..., \lambda _{n},1)$. It is instructive to write out the solution
$q'(t)=(q_{1}(t),...q_{n}(t),t)=(\lambda_{1}(p'),...,\lambda_{n}(p'),1)t+
q_{0}$ where $q_{0}\in {\mathbb{T}}^{n}\times {\mathbb{R}}$
and the $n+1$ term of $q_{0}$ is
zero. Consequently the phase space
$(p',q')\in ({\mathbb{R}}^{n+1},{\mathbb{T}}^{n}\times {\mathbb{R}})$ is
foliated by invariant  infinite cylinders each sustaining
winding motions characterized by the
frequency $\tilde{\lambda}(p_{0})$ which evolve along the $t$ direction.

For a vector $v\in {\mathbb {C}}^{n}$ we define the norm
$|v|= \sum_{i}^{n}|v_{i}|$.
For a scalar valued function $f\in {\cal{A}}_{\rho ,\sigma ,p_{0}}$
we use the
norm,
$\| f \|_{\rho ,\sigma ,p_{0}} \equiv \sup _{(p',q')\in D_{\rho ,\sigma
,p_{0}}} |f(p',q')|$.
For vector valued functions $f=(f_{1},...,f_{2n+2})$ with values in
$\mathbb {C}^{2n+2}$ we define $f\in {\cal{A}}_{\rho ,\sigma ,p_{0}}$ if $f_{i}\in
{\cal{A}}_{\rho ,\sigma ,p_{0}}$, $(i=1,...,2n+2)$, and the norm is defined as follows
\[\| f \| _{\rho ,\sigma ,p_{0}} = \max _{i} \|f_{i} \|_{\rho , \sigma ,p_{0}}.\]
For a $(2n+2) \times (2n+2)$ matrix $C$ whose entries $C_{i,j}$ belong to
${\cal{A}}_{\rho ,\sigma ,p_{0}}$ we write the norm of $C$ as
$\| C \|_{\rho ,\sigma , p_{0}} = \max _{i,j} \|C_{i,j} \|_{\rho ,\sigma , p_{0}}$.
If the matrix $C$ is a constant matrix then
$
\| C \| = \max _{i,j}|C_{i,j}|$.

For the integrable Hamiltonian $H^{0}$ recall the frequency map
\[
\tilde{\lambda} : p \rightarrow
\frac{\partial H^{0}}{\partial p} (p) \equiv
\tilde{\lambda} (p).\]
The set of all possible frequencies in ${\mathbb{R}}^{n}$ is denoted by
$
\Omega \equiv \tilde{\lambda} (B).
$
The subset of diophantine frequencies which will be important throughout the
paper is defined for some positive constant $\Gamma$ by
\[
\Omega _{\Gamma} = \{ \tilde{\lambda}
 \in \Omega \subset {\mathbb{R}}^{n} | |\tilde{\lambda} \cdot
k | \geq \Gamma \| k \|^{-n}, \quad \forall k\in {\mathbb{Z}}^{n}, k\neq 0\}.
\]

Given a function $f(p,\tau ,q ,t)=f(p',q')\in {\cal{A}}_{\rho ,\sigma}$ $2\pi $ periodic in
each of the $q_{i}$'s except $q_{n+1} =t$,
 $\overline {f}(p'^{*})$ will denote the average of the function $f$
over the angles $q_{1},...,q_{n}$ and evaluated at $p'^{*}$
\begin{eqnarray*}
\overline {f} (p '^{*},t)= \frac {1}{(2\pi)^{n}}\int ^{2\pi}_{0}\cdot \cdot \cdot \int ^{2\pi}_{0}
f(p'^{*},q,t)dq_{1}\cdot \cdot \cdot dq_{n}.
\end{eqnarray*}

Given a function $f(p,\tau ,q ,t)=f(p',q')\in {\cal{A}}_{\rho ,\sigma}$ $2\pi $
 periodic in
each of the $q_{i}$'s except $q_{n+1} =t$,
we will denote the average of the function $f(p',q')$ over the
angles $q_{1},...,q_{n}$ and over $t$  evaluated at $p'^{*}$ as follows
\begin{equation}
\overline {\overline {f}} (p'^{*})=\lim _{T\rightarrow \infty}
\frac{1}{2T}
 \left( \frac {1}{2\pi } \right) ^{n} \int ^{2\pi}_{0}\cdot \cdot \cdot
\int ^{2\pi}_{0}\int ^{T}_{-T}f(p'^{*},q,t)dq_{1} \cdot \cdot \cdot
dq_{n}dt . \label{aver}
\end{equation}

Note for a given Hamiltonian $H\in {\cal{A}}_{\rho, \sigma ,p_{0}}$,
one can assume $\|H\|_{\rho , \sigma}<1$. This can be obtained
by the change of variables $(p',q')\rightarrow (\alpha p',q')$
with a suitable positive constant $\alpha$.

Now we address the topic of the time dependence of
the perturbation $H^{1}(p,q,t)$. Since we want to consider
a time aperiodic dependence and the perturbation plays a key role
in the solution of the canonical transformation,
the perturbation has to have a particular form
for certain Fourier transforms to converge.
The form of the perturbation consists of an
exponential decay in time of the time dependent
part of the perturbation.
The exponential decay in time of the perturbation
is described by the following definition.
\begin{defi}
\quad\\

We say a complex valued  function $f(z)$, $z=x+iy$ with
 $x,y\in {\mathbb{R}}$,
is of $(C_{1},C_{2},c_{1},c_{2},\nu,\mu)$-exponential
order with respect to $x$ and write
\begin{eqnarray*}
f(z)=\left\{
\begin{array}{cc}
{\cal{O}}\left(e^{-(\nu-\varepsilon)x}\right) & (x \rightarrow \infty)\\
\quad & \quad \\
{\cal{O}}\left(e^{(\mu-\varepsilon)x}\right)\quad & (x \rightarrow -\infty)
\end{array}
\right.,
\end{eqnarray*}
where  $\nu , \mu , \varepsilon >0, \nu> \varepsilon, \mu> \varepsilon$ if there exist constants
$C_{1}, C_{2},c_{1}>0, c_{2}<0$ such that
\begin{eqnarray*}
|f(z)| \leq C_{1} e^{-(\nu-\varepsilon)x}
\quad {\mbox{for}} \quad 0<c_{1}<x<\infty {\mbox{ and}}
\end{eqnarray*}
\begin{eqnarray*}
|f(z)| \leq C_{2} e^{(\mu-\varepsilon)x}
\quad {\mbox{for}} -\infty <x<c_{2} <0.
\end{eqnarray*}
The iterative lemma will require the perturbation to have
a particular form. This form is defined below.
\end{defi}
\begin{defi}
\quad\\

We will refer to a function $F(p',q')$  as having
$(C_{1},C_{2},c_{1},c_{2},\nu,\mu)p',q'$-exponential form
if it can be expressed as follows
\begin{eqnarray*}
&\quad& F(p',q')=f(p')+r(p',q')+ g(p',q'),\\
\quad \\
&\quad& r(p',q')= \sum _{k\in {\mathbb{Z}}^{n}}
s_{k}(p')e^{ik\cdot q},
\quad \\
&\quad& g(p',q')= \sum _{k\in {\mathbb{Z}}^{n}}
h_{k}(p')e_{k}(t)e^{ik\cdot q},
\end{eqnarray*}
where $F(p',q') \in {\cal{A}}_{\rho, \sigma}$ and
 $e_{k}(t),t=t_{R}+it_{I},t_{R},t_{I}\in{\mathbb{R}}$,
is of $C_{1},C_{2},c_{1},c_{2}$-exponential order with respect to $t_{R}$.
When it is not important for
context to specify the constants $C_{1},C_{2},c_{1},c_{2},\nu,\mu$
we will simply refer to functions of exponential order with respect to
$x$ or functions of $p',q'$-exponential form.
\end{defi}
\section{Statement of the Theorem}

We now state the main theorem in terms of the
Hamiltonian $H(p',q')=H^{0}(p)  +
\kappa H^{1}(p,q,t) + \tau$.
\begin{theom}
\label{firststatementofthm}
\quad \\
Consider the Hamiltonian  $H(p',q')=H^{0}(p)  +
\kappa H^{1}(p,q,t) + \tau=\tilde{H} ^{0}(p')+ \kappa  H^{1}(p',q')$
of $(C_{1},C_{2},c_{1},c_{2},\nu,\mu)p',q'$-exponential form
defined on ${{B}}
\times {\mathbb{T}^{n}}$ where
$\tilde{H}^{0}(p')=H^{0}(p)+\tau$
. Fix $p'_{0} \in {{B}}$ and denote
\begin{eqnarray*}
\lambda = \lambda (p'_{0})= \frac{\partial \tilde{H}^{0}}{\partial p'}(p'_{0}),
\quad
\tilde{C}^{*}_{ij}= \frac{\partial ^{2} \tilde{H}^{0}}{\partial p_{i} \partial p_{j}}(p'_{0}),
\quad
C^{*}_{ij}= \frac{\partial ^{2} \tilde{H}^{0}}{\partial p'_{i} \partial p'_{j}}(p'_{0}),
\quad
{\mathbb{I}}= \left(
\begin{array}{cc}
C^{*} & \lambda ^{T} \\
\lambda & 0
\end{array}
\right).
\end{eqnarray*}
Assume there exists positive numbers $\Gamma ,\gamma , \rho , \sigma
,d,\kappa $, all less than one and $L$,  such that $\rho > \sigma $ and
\begin{eqnarray*}
\begin{array}{cc}
 & \tilde{\lambda} \in \Omega _{\Gamma},\quad |\tilde{\lambda} |<L,
\quad \quad \quad \quad \quad \quad \quad  \quad  \\
\quad\\
 & H^{0},H^{1} \in {\cal{A}}_{\rho, \sigma , p_{0}},\quad \quad  \quad  \quad
 \quad \quad  \quad \quad  \\
\quad\\
 & d\| \tilde{v} \| \leq \| \tilde{C} ^{*} \tilde{v} \| \leq
d^{-1} \| \tilde{v} \|, \quad  \forall \tilde{v} \in {\mathbb {C}}^{n},
\end{array}
\end{eqnarray*}
 moreover, $\|H \|_{\rho, \sigma , p_{0}}<1$, (the last condition
will imply the isoenergetic nondegeneracy condition on ${\mathbb{I}}$).
\\
\\
Then there exists  positive numbers $E,\kappa ',E',\zeta ',f,c_{3},\gamma,
\rho'$ 
and $\sigma '$, such that if
$
\|H^{1} \|_{\rho ,\sigma , p_{0}} \leq E
$
one can construct a canonical analytic change of variables
\begin{eqnarray*}
\begin{array}{ccc}
\psi :D_{\rho ', \sigma '} & \rightarrow & D_{\rho ,\sigma ,p_{0}},
\quad \quad  \quad \quad \quad  \\
\quad\\
({\cal{P}}',{\cal{Q}}') & \mapsto & \psi ({\cal{P}}',{\cal{Q}}') = (p',q'),
\end{array}
\end{eqnarray*}
with $\psi \in {\cal{A}}_{\rho ', \sigma '}$, which brings the Hamiltonian
$H$ into the form $H' = H \circ \psi$
given by
\begin{eqnarray}
\label{wantedformofh}
H'({\cal{P}}',{\cal{Q}}') =(H\circ \psi )({\cal{P}}',{\cal{Q}}')= a'
+(1+  \kappa ' E' \zeta ') \lambda
\cdot {\cal{P}}'+{\cal{O}}(\| {\cal{P}}' \|^{2}),
\end{eqnarray}
where $a'\in {\mathbb{R}}$.
The change of variables is near identity in the sense that
\begin{eqnarray*}
\| \psi -identity \|_{\rho ', \sigma} \rightarrow 0 \mbox{ as }
\| H^{1} \|_{\rho ,\sigma , p_{0}} \rightarrow 0.
\end{eqnarray*}
In particular, we can take 
\begin{eqnarray*}
&\quad& \kappa E=\left(\frac{\sigma}{32} \right)^{2\aleph}
\frac{d^4f^4\sigma '}{2^{12}\Lambda ^2},\quad
 \aleph =10n+9,\\
&\quad&
\label{Lambdaexpression}
\Lambda =
\frac{3^{n}2^{2n+12}\varpi ^{3}(6n+6)^{4}(c_{3}+1)^{2}}{\rho '\Gamma ^{3}
\gamma ^{2}}e^{\nu +\mu},\quad
\varpi=2^{4n+1}\left(\frac{n+1}{e}\right)^{n+1}.
\end{eqnarray*}
In the new coordinates Hamilton's Equations are given by
\begin{eqnarray*}
&\quad& \dot{{\cal{Q}}}'= \frac{\partial H'}{\partial {\cal{P}}'}({\cal{P}}',{\cal{Q}}')=
 (1+\kappa ' E' \zeta ')\lambda+
{\cal{O}}(\|{\cal{P}}' \|),\\
\quad\\
&\quad&\dot{{\cal{P}}}'=-\frac{\partial H'}{\partial {\cal{Q}}'}({\cal{P}}',{\cal{Q}}')=
{\cal{O}}(\|{\cal{P}}' \|^{2}).
\end{eqnarray*}
The solutions are given by
\begin{eqnarray*}
&\quad & {\cal{Q}}'(t)=(1+ \kappa ' E' \zeta ') \lambda t   +{\cal{Q}}'_{0}
\quad
\quad  {\cal{P}}'(t)= {\cal{P}}'_{0}=({\cal{P}}_{0},\tilde{\tau}_{0}) = 0.
\end{eqnarray*}
We can write out explicitly  these solutions
${\cal{Q}}'(t)=
({\cal{Q}}_{1}(t),...,{\cal{Q}}_{n}(t),{\cal{T}}(t))=
(1+ \kappa ' E' \zeta ') \lambda t   +{\cal{Q}}'_{0}
=(1+ \kappa ' E' \zeta ')(\tilde{\lambda},1)t  +{\cal{Q}}'_{0}$
which indicates  in the new coordinate system the phase space
${\mathbb{R}}^{n}\times {\mathbb{R}}\times {\mathbb{T}}^{n}\times {\mathbb{R}}$is foliated by
invariant infinite cylinders each sustaining quasi-periodic motions
identified by the frequency $\tilde{\lambda}({\cal{P}}_{0})$ and evolving in
the $t$ direction. Note  under the transformation
the time variable ${\cal{T}}(t)=(1+ \kappa ' E' \zeta ')t$ has shifted
by a constant proportional to the size of the perturbation.
\end{theom}
\section{The Normal Form}

We begin the proof with a trivial rearrangement
of the Hamiltonian that consists of Taylor expanding 
in $p,\tau$ around $p'=0$.
Recall the Hamiltonian
\[ H(p',q')= H^{0}(p)+\tau +\kappa  H^{1}(p,q,t)=\tilde {H}^{0}(p')+
\kappa H^{1}(p',q'). \]
Taylor expanding  the Hamiltonian $H(p',q')$ about $p'=0$ we obtain
\begin{eqnarray*}
\label{taylorform}
 H(p',q')&=&H(0,q')+ \frac {\partial H}{\partial p'}(0,q') \cdot p' +\frac {1}{2}
\frac {\partial ^{2} H}{\partial p'^{2}}(0,q')(p',p')+ {\cal {O}}(\| p' \|^{3})\quad \quad \\
\quad \\
&=& \tilde {H}^{0}(0)+\kappa  H^{1}(0,q') +\lambda \cdot p' +\kappa \partial _{p'}H^{1}(0,q') \cdot p'\\
\quad \\
&\quad& +\frac{1}{2}\partial^{2}_{p'}\tilde {H}^{0}(0)(p',p')+ \kappa
 \frac{1}{2}\partial^{2}_{p'}H^{1}(0,q')(p',p')+
{\cal {O}}(\|p' \|^{3}).
\end{eqnarray*}
The terms can be grouped by order of magnitude in $p'$
\begin{equation}
\label{normalformhamiltonian}
H(p',q')=a+\lambda \cdot p'+  A(q')+ B(q')\cdot  p' +\frac {1}{2} \sum^{n+1}
_{i,j}  C_{i,j}(q')p'_{i}p'_{j} +R(p',q'),
\end{equation}
where
\begin{eqnarray*}
&a& =\quad \overline{\overline{H(0)}}=
 \overline{\overline{\tilde{H}^{0}}}(0)+
\kappa\overline{\overline{H^{1}}}(0)
=\tilde{H}^{0}(0)+
\kappa \overline{\overline{H^{1}}}(0),\\
\quad \\
& \kappa A(q')&=\quad H(0,q')-a= \kappa(  H^{1}(0,q')-
\overline{\overline{H^{1}}}(0)),\\
\quad \\
& \kappa B_{i}(q')&=\quad \frac {\partial H}{\partial p'_{i}}(0,q')-\lambda _{i}=
\kappa  \frac {\partial H^{1}}{\partial p'_{i}}
(0,q'), \\
\quad \\
&C_{ij}(q')&=\quad \frac {\partial ^{2}H}{\partial p'_{i}\partial p'_{j}}(0,q')
=\frac {\partial ^{2} \tilde {H}^{0}}{\partial p'_{i}\partial p'_{j}}(0)+
\kappa \frac {\partial ^{2}  {H}^{1}}{\partial p'_{i}\partial p'_{j}}(0,q'),
\end{eqnarray*}
and
$a\in \mathbb {R},A,B_{i},C_{i,j},R\in A_{\rho , \sigma}, R={\cal{O}}(\| p'^{3}\|)$.
This Hamiltonian has the desired form
 $(\ref{wantedformofh})$ of Theorem $\ref{firststatementofthm}$
 except for the terms $ A(q')$ and
 $ B(q')$ which
constitute the actual perturbation of the Hamiltonian
 $(\ref{normalformhamiltonian})$.
\section{Bounds on the  Normal Form in terms of the Original Hamiltonian}

In what follows we present several results that allow us
to reformulate Theorem $\ref{firststatementofthm}$ for
the Hamiltonian $(\ref{normalformhamiltonian})$ in
normal form. Since we are interested in estimates
for $A,B_{i},C_{i,j}$ and these are defined
in terms of derivatives of the original Hamiltonian,
we often make use of Cauchy's inequality. For scalar valued functions
$f\in {\cal{A}}_{\rho ,\sigma}$, $\delta<\sigma <\rho$ and nonnegative integers
$k_{i},l_{i},i=1,...,n+1$
\begin{eqnarray*}
\left|
\frac{\partial^{k_{1}+...k_{n+1}+l_{1}+...+l_{n+1}}}
{\partial p_{1}^{k_{1}}...\partial p_{n+1}^{k_{n+1}}
\partial q_{1}^{l_{1}}...\partial q_{n+1}^{l_{n+1}}}f(p,q)\right|
\leq
\frac{k_{1}!...k_{n+1}!l_{1}!...l_{n+1}!}
{\delta^{k_{1}+...+k_{n+1}+l_{1}+...+l_{n+1}}}
\|f\|_{\rho, \sigma},\quad \forall(p,q)\in {\cal{D}}_{\rho-\delta,
\sigma-\delta}.
\end{eqnarray*}
\begin{lem}
\label{boundlemmaone}
\quad\\
\\
Recall the Hamiltonian of interest
$H(p',q')=H^{0}(p)+ \kappa  H^{1}(p,q,t)+\tau =\tilde {H}^{0}(p')+
\kappa  \tilde {H}^{1}(p',q')$
where
$\tilde {H}^{0}(p')=H^{0}(p)+\tau$.
Given the $n \times n$ matrix
\\
\[\tilde {C}^{*}_{ij}=\frac {\partial ^{2} \tilde {H}^{0}}{\partial p_{i} \partial p_{j}}(0),\]
assume there is a positive number $d<1$ such that
$d\|\tilde {v} \| \leq \| \tilde {C}^{*}\tilde{v} \| \leq d^{-1} \|\tilde{v} \| \quad
\forall \tilde{v} \in \mathbb {C}^{n}$.
Define the $(n+1) \times (n+1)$ matrix
\[C^{*}_{ij}= \frac {\partial ^{2} \tilde {H}^{0}}{\partial p'_{i} \partial p'_{j}}(0),\]
it follows 
$ \| C^{*}v \| \leq d^{-1} \|v \| \quad \forall v\in
{\mathbb {C}}^{n+1}$.
{Proof}
See appendix $\ref{appbounds}$.
\end{lem}
\begin{lem}(Lemma on bounds)
\label{boundlemmatwo}\\
\\
Given a constant $E$ such that $\|H^{1}\|_{\rho,\sigma}\leq E$
\\
\\
1. $\mbox {max}(\|A \|_{\rho ,\sigma }, \| B \| _{\rho ,\sigma }) < 2 \frac {E}{\sigma}
=E_{1}$.\\
\\
2. There exists a positive number $ m \leq 1$ such that
$ \| Cv \| _{\rho ,\sigma } \leq m^{-1} \| v \| $,
$\forall v \in \mathbb {C}^{n+1}$. In particular  we take $m=\frac {d}{2}$.
{Proof}
See appendix $\ref{appbounds}$.
\end{lem}
\begin{lem}(Isoenergetic Nondegeneracy Condition)
\label{lemmaiso}
\\
\\
Assume $\sum |\lambda _{i}'|<L$. Given
$
\| \tilde{C} ^{*} \tilde{v} \| _{{\rho },{ \sigma }} \geq d \| \tilde{v} \|,
\quad \forall \tilde{v}\in {\mathbb{C}}^{n}$,
$2m>L$.
It follows
$
\| {\mathbb{I}}v \|_{\rho ,\sigma} \geq l\|v \|, \quad \forall
v\in {\mathbb{C}}^{n+2}$
for some $l=\frac{1}{2}|d-L|$.
{Proof}
See appendix $\ref{appbounds}$.
\end{lem}
\begin{lem}
\label{thmisob}
\quad\\
\\
Given
$
\|{\mathbb{I}}v \|_{\rho ,\sigma} \geq l\| v \|_{\rho ,\sigma}
$
with $l=\frac{1}{2}|d-L|$. 
It follows
\begin{eqnarray*}
\left\| \left(
\begin{array}{cc}
\overline{\overline{C}} & \lambda ^{T} \\
\lambda & 0
\end{array}
\right)
v \right\| _{\rho ,\sigma} \geq
f\| v \|_{\rho ,\sigma},
\end{eqnarray*}
where $f=|l-\frac{\kappa E_{1}}{\sigma^2}|$.
\end{lem}
{Proof}
See appendix $\ref{appbounds}$.
\section{Re-statement of the Theorem for the Normal Form}

One can deduce, based on the bounds
on the  normal form in terms of the original Hamiltonian,
theorem $\ref{firststatementofthm}$ from the following theorem.
\begin{theom}{Main Theorem}
\quad \\

For given positive numbers
$
\kappa, \Gamma , \gamma , \rho ,\sigma ,m
$ all less than one and $L$
consider the Hamiltonian $H(p',q')=U(p',q')+P(p',q')$
of $(C_{1},C_{2},c_{1},c_{2},\nu,\mu)p',q'$-exponential form
defined in
$D_{\rho ,\sigma}$ by
\begin{eqnarray*}
&\quad& U(p',q')=a+\lambda \cdot p' + \frac{1}{2}
\sum_{i,j} C_{i,j}p'_{i}p'_{j}+R(p',q'), \\
&\quad& P(p',q')= \kappa\left( A(q')+ \sum_{i} B_{i}(q')p'_{i}\right),
\end{eqnarray*}
with $\| H \|_{\rho ,\sigma }<1$, where $\tilde{\lambda} \in \Omega_{\Gamma}$
and $| \tilde{\lambda } |<L,A,B_{i},C_{i,j},R \in {\cal{A}}_{\rho , \sigma}$
and $R$ is of order $\| p' \|^{3}$.
Assume
\begin{eqnarray*}
&\quad& 2m\|\tilde{v} \| \leq \|\tilde{C}^{*} \tilde{v} \|, \quad \forall
\tilde{v}\in {\mathbb{C}}^{n},\\
\quad\\
&\quad& \| C v \| _{\rho ,\sigma} < m^{-1} \| v\| , \quad \forall v \in
{\mathbb{C}}^{n+1}.
\end{eqnarray*}
\\
Then there exists positive numbers
$
E_{1} ,\kappa _{\infty},E_{1}^{\infty},\zeta_{\infty}, \rho ', \sigma ',m',f$ 
all less than one and $L',c_{3},\gamma$ also
positive
with
\[
\rho '< \rho , \sigma '< \sigma ,m'<m,
\]
\begin{eqnarray*}
f\| v\|\leq
\left \|
\left(
\begin{array}{cc}
\overline{\overline{C}} & \lambda ^T\\
\lambda & 0
\end{array}
\right)
v \right \|
,\quad \forall v\in{\mathbb{C}}^{n+1},
\end{eqnarray*}
such that if
\[
\max ( \| A \| _{\rho ,\sigma } , \| B \| _{\rho ,\sigma  })<E_{1},
\]
we can construct a canonical analytic change of variables
\[
\phi : D_{\rho ', \sigma '} \rightarrow D_{\rho ,\sigma},
\]
with $\phi \in {\cal{A}}_{\rho ' , \sigma '}$, which brings the Hamiltonian
$H$ into the form
\[
H'( {\cal{P}}, {\cal{Q}}) =(H \circ \phi )({\cal{P}}, {\cal{Q}})= a '
+ \lambda ' \cdot {\cal{P}}+ R( {\cal{P}}, {\cal{Q}}),
\]
where $H'$ is of $p',q'$-exponential form,
$\lambda'=(1+\kappa_{\infty} E^{\infty}_{1}\zeta_{\infty})\lambda$,
$a'\in {\mathbb{R}}$ and $R= {\cal{O}}(\| {\cal{P}} \|^{2})
\in {\cal{A}}_{\rho ', \sigma '}$.
\\
\\
The change of variables is near the identity  in the sense that
\[
\| \phi - {\mbox{identity}} \| _{\rho ' ,\sigma '} \rightarrow 0
\quad {\mbox{as}} \quad \| P \| _{\rho ,\sigma }
\rightarrow 0.
\]
In particular, we can take 
\begin{eqnarray*}
&\quad& \kappa E_{1}=\left(\frac{\sigma}{32} \right)^{2\aleph}
\frac{m^4f^4\sigma '}{2^8\Lambda ^2},\quad
\aleph =10n+9,\\
&\quad&
\label{Lambdaexpression}
\Lambda =
\frac{3^{n}2^{2n+12}\varpi ^{3}(6n+6)^{4}(c_{3}+1)^{2}}{\rho '\Gamma ^{3}
\gamma ^{2}}e^{\nu +\mu},\quad
\varpi=2^{4n+1}\left(\frac{n+1}{e}\right)^{n+1}.
\end{eqnarray*}
In the new coordinates Hamilton's Equations are given by
\begin{eqnarray*}
&\quad& \dot{{\cal{Q}}}'= \frac{\partial H'}{\partial {\cal{P}}'}({\cal{P}}',{\cal{Q}}')=
 (1+\kappa_{\infty} E_{1}^{\infty} \zeta_{\infty})\lambda+
{\cal{O}}(\|{\cal{P}}' \|),\\
\quad\\
&\quad&\dot{{\cal{P}}}'=-\frac{\partial H'}{\partial {\cal{Q}}'}({\cal{P}}',{\cal{Q}}')=
{\cal{O}}(\|{\cal{P}}' \|^{2}).
\end{eqnarray*}
The solutions are given by
\begin{eqnarray*}
&\quad & {\cal{Q}}'(t)=(1+ \kappa_{\infty} E_{1}^{\infty} \zeta_{\infty}) \lambda t   +{\cal{Q}}'_{0}
\quad
\quad  {\cal{P}}'(t)={\cal{P}}'_{0}=({\cal{P}}_{0},\tilde{\tau}_{0}) = 0 .
\end{eqnarray*}
We can write out explicitly  these solutions
${\cal{Q}}'(t)=
({\cal{Q}}_{1}(t),...,{\cal{Q}}_{n}(t),{\cal{T}}(t))=
(1+ \kappa_{\infty} E_{1}^{\infty} \zeta_{\infty}) \lambda t   +{\cal{Q}}'_{0}
=(1+ \kappa_{\infty} E_{1}^{\infty} \zeta_{\infty})(\tilde{\lambda},1)t  
+{\cal{Q}}'_{0}$
which indicates  in the new coordinate system the phase space
${\mathbb{R}}^{n}\times {\mathbb{R}}\times 
{\mathbb{T}}^{n}\times {\mathbb{R}}$ is foliated by
invariant infinite cylinders each sustaining quasi-periodic motions
identified by the frequency $\tilde{\lambda}({\cal{P}}_{0})$ and evolving in
the $t$ direction. Note under the transformation,
the time variable ${\cal{T}}(t)=(1+ \kappa_{\infty} E_{1}^{\infty} \zeta_{\infty})t$ 
has shifted
by a constant proportional to the size of the perturbation.
\end{theom}

\section{Perturbations tending to $q-$Periodicity and $t$ Quasiperiodicity}
\label{sectionwithsplit}

We are considering two similar KAM theorems
each involving an  n-degree of freedom real valued
nearly-integrable Hamiltonian  in action-angle variables
of the form
\begin{eqnarray*}
H(p,q,t)=H^{0}(p)+ \kappa H^{1}(p,q,t),
\end{eqnarray*}
where $p=(p_{1},...,p_{n})\in {\mathbb{R}}^{n}$,
 $q=(q_{1},...,q_{n})\in {\mathbb{T}}^{n}$, $t\in {\mathbb{R}}$ and
  $\kappa$ is  small.
After extending $H$ to the complexified domain $D_{\rho, \sigma}$
we have $H\in {\cal{A}}_{\rho, \sigma}$.
In the first case
the perturbation considered has an exponentially decaying aperiodic
time dependent term and a quasiperiodic term depending only
on the angles $q$
\begin{eqnarray*}
H^{1}(p,q,t)=
\sum _{k\in {\mathbb{Z}}^{n}}g_{k}e^{ik\cdot q}+
\sum _{k\in {\mathbb{Z}}^{n}}
f_{k}(p)e_{k}(t)e^{ik\cdot q},
\end{eqnarray*}
 where $f_{k}(p)$ is a bounded function and $e_{k}(t),t=t_{R}+it_{I},$ is
of exponential order with respect to $t_{R}$.
In the second case the perturbation
 consists of an exponentially decaying aperiodic
time dependent term and a term depending
in a quasiperiodic manner on the angles $q$ and on time $t$
\begin{eqnarray*}
H^{1}(p,q,t)=
\sum _{k\in {\mathbb{Z}}^{n+1}}g_{k}e^{ik\cdot( q,\theta t)}+
\sum _{k\in {\mathbb{Z}}^{n}}
f_{k}(p)e_{k}(t)e^{ik\cdot q},
\end{eqnarray*}
 where $f_{k}(p)$, $e_{k}(t)$ are as above and $\theta=(\theta_{1},...,\theta_{m})$
is the vector of basic frequencies.

For each of the cases above,
it is necessary
for the application of the Lie series method
to construct a
generating function.
For this purpose we will formulate
an expression
in terms of the perturbation $H^{1}(p,q,t)$ evaluated at $p=0$
, call it $G(q,t)$,
and solve the partial differential equation
\begin{eqnarray}
\label{eqntwelve}
\sum _{i=1}^{n+1}\lambda_{i} \frac{\partial F}{\partial q_{i}}(q,t) =G(q,t).
\end{eqnarray}
For the first case, the given function
has the form
\begin{eqnarray*}
&\quad&G(q,t)=G_{A}(q,t)+G_{Q}(q)=
\sum _{k\in {\mathbb{Z}}^{n}} g_{k}(t) e^{ik\cdot q}+
\sum_{l\in{\mathbb{Z}}^{n}}\tilde{g} _{l}e^{il\cdot q},
\end{eqnarray*}
and we look for a solution of the form
\begin{eqnarray*}
&\quad&F(q,t)=F_{A}(q,t)+F_{Q}(q)=
\sum _{k\in {\mathbb{Z}}^{n}} f_{k}(t) e^{ik\cdot q}+
\sum_{l\in{\mathbb{Z}}^{n}}\tilde{f} _{l}e^{il\cdot q}.
\end{eqnarray*}
Substituting in $(\ref{eqntwelve})$ we obtain
\begin{eqnarray*}
\sum ^{n+1}_{i=1} \lambda _{i} \frac{\partial F}{\partial q_{i}}=
\sum _{k\in {\mathbb{Z}}^{n}} \left[ i\tilde{\lambda} \cdot k
f_{k}(t)+ \frac{d f_{k}(t)}{dt} \right]e^{ik\cdot q}
+ \sum_{l\in{\mathbb{Z}}^{n}}i(\tilde{\lambda}\cdot l) \tilde{f}_{l}
e^{il\cdot q}.
\end{eqnarray*}
Clearly
$
\sum ^{n+1}_{i=1} \lambda _{i}\frac{\partial F}{\partial q_{i}} -G=0
$
becomes
\begin{eqnarray}
\label{2problemsol}
\sum _{k\in {\mathbb{Z}}^{n}}\left[
 i(\tilde{\lambda} \cdot k)
f_{k}(t)+ \frac{d f_{k}(t)}{dt}-g_{k}(t) \right]e^{ik\cdot q}+
\sum _{l\in {\mathbb{Z}}^{n}}\left[
 i(\tilde{\lambda} \cdot l)\tilde{f}_{l}-\tilde{g}_{l}\right]e^{il\cdot q}=0,
\end{eqnarray}
$(\ref{2problemsol})$ is satisfied if
\begin{eqnarray*}
\left\{
\begin{array}{c}
 i(\tilde{\lambda} \cdot k)
f_{k}(t)+  \frac{d f_{k}(t)}{dt}  =g_{k}(t)\\
\quad \\
\tilde{f}_{l} =  \frac{\tilde{g}_{l}}{ i(\tilde{\lambda} \cdot l)}
\end{array}
\right.
\end{eqnarray*}
and the problem reduces to two problems.
Similarly for the second case, the given function
has the form
\begin{eqnarray*}
G(q,t)=G_{A}(q,t)+G_{Q}(q,t)=
\sum _{k\in {\mathbb{Z}}^{n}} g_{k}(t) e^{ik\cdot q}+
\sum_{l\in{\mathbb{Z}}^{n+m}}\tilde{g} _{l}e^{il\cdot (q,\theta t)},
\end{eqnarray*}
we look for a function of the form
\begin{eqnarray*}
F(q,t)=F_{A}(q,t)+F_{Q}(q,t)=
\sum _{k\in {\mathbb{Z}}^{n}} f_{k}(t) e^{ik\cdot q}+
\sum_{l\in{\mathbb{Z}}^{n+m}}\tilde{f} _{l}e^{il\cdot (q,\theta t)},
\end{eqnarray*}
and the
problems to be solved are
\begin{eqnarray*}
\left\{
\begin{array}{c}
 i(\tilde{\lambda} \cdot k)
f_{k}(t)+  \frac{d f_{k}(t)}{dt}  =g_{k}(t)\\
\quad \\
\tilde{f}_{l} =  \frac{\tilde{g}_{l}}{ i(\tilde{\lambda},\theta) \cdot l}
\end{array}
\right. .
\end{eqnarray*}
Therefore in both KAM theorems solving
$(\ref{eqntwelve})$ is equivalent to solving two p.d.e's;
one p.d.e  involving
quasiperiodic dependent functions, ($G_{Q}(q),F_{Q}(q)$ or
$G_{Q}(q,t),F_{Q}(q,t)$) and a p.d.e
involving  $p',q'$-exponential form functions (
$G_{A}(q,t),F_{A}(q,t)$). In the next section we give
the result for solving the p.d.e involving
quasiperiodic functions.

\section{P.D.E on a Torus}

Consider the real valued functions $F(q)$ and $G(q)$ where
$q=(q_{1},...,q_{n})\in {\mathbb{C}}^{n}$
and ${\mbox{Re }} q=(\mbox{Re }q_{1},...,\mbox{Re }q_{n})\in {\mathbb{R}}^n
{\mbox{ mod}}2\pi$.
 That is,  $F(q)$ and $G(q)$
are functions on ${\mathbb{C}}^{n}$, $2\pi$ periodic
with respect to the real part of each $q_{1},...,q_{n}$.
Although we are concerned with real
Hamiltonian systems, we will use techniques
from several complex variables theory in the analysis.
This will require  complex extensions of functions
originally defined on ${\mathbb{R}}^{n}$ to  ${\mathbb{C}}^{n}$.
Fix $\rho<1$, we define the complex extension of ${\mathbb{T}}^{n}$
\begin{eqnarray*}
{{D}}_{\rho}=\{q\in {\mathbb{C}}^{n} \vert \quad
{\mbox{Re }} q\in {\mathbb{R}}^n {\mbox{ mod }}2\pi, \quad
 \| {\mbox{Im }} q\| \leq \rho \},
\end{eqnarray*}
where 
${\mbox{Re }} q \equiv ({\mbox{Re }} q_{1} ,..., {\mbox{Re }} q_{n})$,
${\mbox{Im }} q \equiv ({\mbox{Im }} q_{1} ,..., {\mbox{Im }} q_{n})$.
We define ${\cal{A}}_{\rho}$, the set  of all complex, continuous functions
defined on ${{D}}_{\rho}$ that are analytic in the interior of
 ${{D}}_{\rho}$ and real for real values of the variables.

The following lemma gives a useful bound for the Fourier coefficients
of a function $G(q)$ with a $2\pi$ periodic  dependence on $q$.
\begin{lem}
\label{exponentialbound2}
\quad \\

Assume $G\in {\cal{A}}_{\rho}$ and  $G(q) = \sum _{k\in {\mathbb{Z}}^{n}}
g_{k}e^{ik\cdot q}$, then for every $k\in {\mathbb{Z}}^{n}$ \quad
$|g_{k}| \leq \|G \|_{\rho} e^{-|k|\rho}$,
where $|k| = \sum _{i} |k_{i}|$.
Proof see appendix $\ref{appPDETo}$.
\end{lem}

The following lemma gives the existence and uniqueness of  solutions
of a p.d.e involving quasi-periodic functions. The lemma also
gives analyticity results for the solution, $F(q)$, of the 
p.d.e given analyticity
restrictions on the known function $G(q)$.
\begin{lem}
\label{lemmaqb}
Consider the following linear partial differential
equation
\begin{equation}
\label{diffeqnone}
\sum _{i=1}^{n} \lambda_{i} \frac{\partial F}{\partial q_{i}}(q)
=G(q),
\end{equation}
where $F$ and $G$ are functions defined on the torus ${\mathbb{T}}^{n}$, and assume
$\lambda = (\lambda _{1},..., \lambda _{n}) \in \Omega _{\Gamma}$, for some
$\Gamma >0$, and $G\in {\cal{A}}_{\rho}$ for some positive $\rho<1$
with $\overline {G} =0$. Then, for some positive $\delta <\rho$,
$(\ref{diffeqnone})$ admits a unique solution $F \in {\cal{A}}_{\rho -\delta}$
with $\overline {F} =0$, and one has the estimates
\begin{eqnarray*}
&\quad &\| F \|_{\rho -\delta}
\leq \frac{\varpi }{\Gamma \delta ^{2n}} \|G \|_{\rho},\quad \quad
 \left \| \frac{\partial F}{\partial q} \right \|_{\rho -\delta}
\leq \frac{\varpi}{\Gamma \delta ^{2n+1}} \|G \|_{\rho},
\end{eqnarray*}
where
$
\varpi = 2^{4n+1} \left( \frac{n+1}{e} \right)^{n+1}$.
Proof see appendix $\ref{appPDETo}$.
\end{lem}
\section{P.D.E on a Torus with Time Dependent Coefficients}

Consider the functions $F(q,t)$ and $G(q,t)$ where
$q=(q_{1},...,q_{n})\in {\mathbb{T}}^{n}$, $t\in {\mathbb{R}}$.
We will require as before complex extensions of functions
originally defined on ${\mathbb{R}}^{n+1}$ to ${\mathbb{C}}^{n+1}$.
We define the complex extension of ${\mathbb{T}}^{n} \times
{\mathbb{R}}$
\begin{eqnarray*}
{{D}}_{\rho, \sigma}=\{ (q,t)\in {\mathbb{C}}^{n+1} \vert \quad
{\mbox{Re }}q\in {\mathbb{R}}^n {\mbox{ mod }}2 \pi, 
\| {\mbox{Im }} q \| \leq \rho, |{\mbox{Im t}}|  \leq \sigma \},
\end{eqnarray*}
where 
${\mbox{Re }} q \equiv ({\mbox{Re }} q_{1} ,..., {\mbox{Re }} q_{n})$,
${\mbox{Im }} q \equiv ({\mbox{Im }} q_{1} ,..., {\mbox{Im }} q_{n})$.
We define ${\cal{A}}_{\rho ,\sigma}$, the set  of all complex, continuous functions
defined on ${{D}}_{\rho ,\sigma
}$ that are analytic in the interior of
 ${{D}}_{\rho ,\sigma}$ and real for real values of the variables.
\\

We want show that for a given function $G(q,t)\in{\cal{A}}_{\rho, \sigma}$
satisfying certain conditions
there
exists an $F(q,t)$ which satisfies the following
\begin{equation}
\label{eqntwo}
\sum _{i=1}^{n+1} \lambda _{i} \frac{\partial F}{\partial q_{i}}(q,t) = G(q,t),
\end{equation}
where we identify $q_{n+1}$ with $t$ and set $\lambda _{n+1}=1$.

\begin{theom}
\label{maintheorem3}
\quad \\
\\
Given
\begin{eqnarray*}
\sum ^{n+1}_{i=1}\lambda _{i} \frac{\partial F}
{\partial q_{i}}(q,t)
=G(q,t),
\end{eqnarray*}
\begin{eqnarray*}
 G(q,t)=\sum _{k\in {\mathbb{Z}}^{n}} g_{k}(t)e^{ik\cdot q},
\quad \quad
 F(q,t)=\sum _{k\in {\mathbb{Z}}^{n}} f_{k}(t)e^{ik\cdot q},
\end{eqnarray*}
where $G\in{\cal{A}}_{\rho, \sigma}$ for $1>\rho>0,\quad1>\sigma>0$
and $q\in{\mathbb{T}}^{n},t=t_{R}+it_{I}\in{\mathbb{C}},t_{R},t_{I}\in{\mathbb{R}}$.
Since $G\in{\cal{A}}_{\rho, \sigma}$ it follows
$g_{k}(t)$ is analytic in the strip $|t_{I}| <\sigma$ and further assume
for any given $k$
\\
\begin{eqnarray*}
g_{k}(t) = \left \{
\begin{array}{cc}
{\cal{O}}(e^{-(\nu - \varepsilon)t_{R}}) &  (t_{R} \rightarrow \infty)\\
\quad & \quad \\
{\cal{O}}(e^{(\mu - \varepsilon)t_{R}}) &  (t_{R} \rightarrow -\infty)
\end{array}
\right.
\end{eqnarray*}
\\
where $\nu ,\mu >0$.
Then there exist $\delta >0$ and $\gamma >0$, $ \gamma <\nu $, $ \gamma <\mu $ such that
$f_{k}(t)$ is analytic in the strip $|t_{I}| <\sigma -\delta $ and satisfies
\\
\begin{eqnarray*}
f_{k}(t)= \left \{
\begin{array}{cc}
{\cal{O}}(e^{-(\nu -\varepsilon )t_{R}}){\cal{O}}(e^{-|k|\rho}) & (t_{R} \rightarrow \infty)\\
\quad & \quad \\
{\cal{O}}(e^{(\mu -\varepsilon )t_{R}}){\cal{O}}(e^{-|k|\rho}) & (t_{R} \rightarrow -\infty)
\end{array}
\right.
\end{eqnarray*}
\end{theom}
{Proof}
\quad \\
Since $F(q,t)$ and $G(q,t)$ are $2\pi$ periodic
in each  $q_{1},...,q_{n}$, one can write the
Fourier series
\begin{eqnarray*}
F(q,t)=\sum _{k\in {\mathbb{Z}}^{n}} f_{k}(t) e^{ik \cdot q},
\quad \quad
G(q,t)=\sum _{k\in {\mathbb{Z}}^{n}} g_{k}(t) e^{ik \cdot q}.
\end{eqnarray*}

Substituting these in $(\ref{eqntwo})$ results in the following
differential equation which the Fourier coefficients $f_{k}(t)$ and
$g_{k}(t)$ must satisfy
\begin{eqnarray*}
i(\tilde{\lambda} \cdot k) f_{k}(t)+ \frac{df_{k}(t)}{dt} = g_{k}(t).
\end{eqnarray*}
The solution of the differential equation is given by
\begin{equation}
\label{eqnthree}
f_{k}(t)=f_{k}(0)e^{-i(\tilde{\lambda} \cdot k)t}+ \int ^{t}_{0}
g_{k}(s) e^{i(\tilde{\lambda} \cdot k)(s-t)}ds.
\end{equation}
We can argue  $f_{k}(t)$ has a Fourier transform provided
$f_{k}(0)$ satisfies certain condition. Ultimately,
 with this condition on $f_{k}(0)$,  we will
obtain a new expression for $f_{k}(t)$ which will be better
suited to apply Fourier theory.
Assuming that $g_{k}(t)$ has a complex  Fourier transform ${\cal{G}}(\omega)$,
with $\omega=u+iv,\quad u,v\in{\mathbb{R}}$, analytic
in some strip $-\mu < -\beta < v < \beta < \nu  $, rewrite
$(\ref{eqnthree})$ in terms of the Fourier inversion formula. That is, given
the Fourier transform
\begin{equation}
{\cal{G}}_{k}(\omega)=  \int ^{\infty }_{- \infty} g_{k}(t) e^{-i \omega t}dt,
\end{equation}
and the Fourier inversion formula
\begin{eqnarray*}
g_{k}(t)= \int ^{\infty + i\beta }_{- \infty +i\beta} {\cal{G}}_{k}(\omega)e^{i \omega t}d\omega,
\end{eqnarray*}
$(\ref{eqnthree})$ becomes
\begin{eqnarray*}
f_{k}(t)=f_{k}(0)e^{-i(\tilde{\lambda} \cdot k)t}+ \int ^{t}_{0}
\left( \int ^{\infty+ i\beta }_{-\infty+ i\beta } {\cal{G}}_{k}(\omega)e^{i\omega s} d\omega \right)
 e^{i(\tilde{\lambda} \cdot k)(s-t)}ds.
\end{eqnarray*}
Interchanging integrals and integrating with respect to $s$ gives the expression
\begin{equation}
\label{eqnfour}
f_{k}(t)= f_{k}(0)e^{-i(\tilde{\lambda} \cdot k)t}
+ e^{-i(\tilde{\lambda} \cdot k)t}\int ^{\infty+ i\beta }_{-\infty+ i\beta } {\cal{G}}_{k}(\omega)
\left[ \frac{e^{i(\omega+ \tilde{\lambda} \cdot k)t}-1}{i(\omega + \tilde{\lambda}
\cdot k)} \right] d\omega.
\end{equation}
We consider an arbitrary function $s(\omega)$ with the property
$
\int^{\infty+ i\beta }_{-\infty+ i\beta } s (\omega) d\omega =
1
$
and rewrite $(\ref{eqnfour})$ as follows
\begin{eqnarray*}
f_{k}(t)= e^{-i(\tilde{\lambda} \cdot k)t}
\int ^{\infty+ i\beta }_{-\infty+ i\beta }\left( f_{k}(0)s(\omega)+
{\cal{G}}_{k}(\omega)
\left[ \frac{e^{i(\omega+ \tilde{\lambda} \cdot k)t}-1}{i(\omega + \tilde{\lambda}
\cdot k)} \right] \right) d\omega
\end{eqnarray*}
\begin{equation}
\label{eqnfive}
\quad \quad =\int ^{\infty+ i\beta }_{-\infty+ i\beta }\left(f_{k}(0)s(\omega)e^{-i(\tilde{\lambda} \cdot k
+\omega)t}
+{\cal{G}}_{k}(\omega)
\left[ \frac{1-e^{-i(\omega + \tilde{\lambda} \cdot k)t}}{i(\omega + \tilde{\lambda}
\cdot k)}\right]\right)e^{i\omega t}d\omega.
\end{equation}
Note  $(\ref{eqnfive})$ is written in the form of the Fourier inversion formula. That is
\\
\begin{equation}
\label{eqnsix}
f_{k}(t)= \int^{\infty+ i\beta }_{-\infty+ i\beta }{\cal{F}}_{k}(\omega) e^{i\omega t}d \omega,
\end{equation}
\\
where ${\cal{F}}_{k}$ the Fourier transform of $f_{k}(t)$ given by
\\
\begin{eqnarray*}
{\cal{F}}_{k}(\omega) = \int ^{\infty}_{-\infty} f_{k}(t) e^{-i\omega t} dt.
\end{eqnarray*}
\\
Consequently, from $(\ref{eqnfive})$ and $(\ref{eqnsix})$ we write
\\
\begin{equation}
\label{eqnsixone}
{\cal{F}}_{k}(\omega) = f_{k}(0)s(\omega)e^{-i(\tilde{\lambda} \cdot k+\omega)t}
+{\cal{G}}_{k}(\omega)
\left[ \frac{1-e^{-i(\omega +\tilde{\lambda} \cdot k)t}}{i(\omega + \tilde{\lambda}
\cdot k)}\right].
\end{equation}
\\
Furthermore, note  ${\cal{F}}_{k}$ is only a function of $\omega$
which implies the following
\\
\begin{equation}
\label{eqnsixtwo}
f_{k}(0)s(\omega)e^{-i(\tilde{\lambda} \cdot k+\omega)t} =
{\cal{G}}_{k}(\omega)\frac{e^{-i(\omega +\tilde{\lambda} \cdot k)t}}{i(\omega + \tilde{\lambda}
\cdot k)},
\end{equation}
\\
or
\\
\begin{equation}
\label{eqnseven}
f_{k}(0)s(\omega) =
\frac{{\cal{G}}_{k}(\omega)}{i(\omega + \tilde{\lambda}
\cdot k)},
\end{equation}
\\
and $(\ref{eqnfive})$ reduces to the following
\\
\begin{equation}
\label{eqneight}
f_{k}(t)= \int^{\infty+ i\beta }_{-\infty+ i\beta } \frac{{\cal{G}}_{k}(\omega) e^{i\omega t}}{i(\omega + \tilde{\
\lambda} \cdot k)}d\omega.
\end{equation}
\\
Note the condition on $f_{k}(0)$ given by
 $(\ref{eqnseven})$ reduces to  $(\ref{eqneight})$ for $t=0$
by integrating $(\ref{eqnseven})$ on both sides
\begin{eqnarray*}
&\quad& \int^{\infty+ i\beta }_{-\infty+ i\beta }f_{k}(0)s(\omega)d\omega=
\int^{\infty+ i\beta }_{-\infty+ i\beta }
\frac{{\cal{G}}_{k}(\omega)}{i(\omega + \tilde{\lambda}
\cdot k)}d\omega, \\
\quad \\
&\quad&
f_{k}(0)\int^{\infty+ i\beta }_{-\infty+ i\beta }s(\omega)d\omega =
\int^{\infty+ i\beta }_{-\infty+ i\beta }
\frac{{\cal{G}}_{k}(\omega)}{i(\omega + \tilde{\lambda}
\cdot k)}d\omega, \\
\quad \\
&\quad& f_{k}(0)=\int^{\infty+ i\beta }_{-\infty+ i\beta }
\frac{{\cal{G}}_{k}(\omega)}{i(\omega + \tilde{\lambda}
\cdot k)}d\omega.
\end{eqnarray*}
Now, using Lemma $\ref{lemblue}$ it follows
\begin{eqnarray*}
g_{k}(t) = \left \{
\begin{array}{cc}
{\cal{O}}(e^{-(\nu - \varepsilon)t_{R}}){\cal{O}}(e^{-|k|\rho}) &  (t_{R} \rightarrow \infty)\\
\quad & \quad \\
{\cal{O}}(e^{(\mu - \varepsilon)t_{R}}){\cal{O}}(e^{-|k|\rho}) &  (t_{R} \rightarrow -\infty)
\end{array}
\right.
\end{eqnarray*}
or
\begin{eqnarray*}
&\quad& |g_{k}(t)|\leq C_{1}e^{-(\nu -\varepsilon)t_{R}}e^{-|k|\rho}, \quad
0< c_{1} < t_{R} < \infty,\\
&\quad&\\
&\quad& |g_{k}(t)|\leq C_{2}e^{(\mu -\varepsilon)t_{R}}e^{-|k|\rho}, \quad
-\infty < t_{R} < c_{2} <0 .
\end{eqnarray*}
where the $k$ order will pass through the proofs of Theorem \ref{theomone}
and Theorem \ref{theomtwo} like a constant.
By Theorem \ref{theomone}, there exists a $\delta >0$ such that
 the complex Fourier transform of
$g_{k}(t)$ defined by
\begin{eqnarray*}
{\cal{G}}_{k}(\omega) = \int ^{\infty + i\beta}_{-\infty +i\beta}
g_{k}(t) e^{-i\omega t} dt,
\end{eqnarray*}
is analytic in the strip $-\mu +\delta  \leq  v \leq  \nu -\delta $ and satisfies
\begin{eqnarray*}
&\quad &\left|{\cal{G}}_{k}(\omega)\right| \leq \frac{2}{\sqrt{2\pi}}
e^{-(\sigma-\varepsilon)u}e^{-|k|\rho} \left[
C{c_{3}}+ \frac{C_{3}e^{-(\delta -\varepsilon)c_{3}}}{(\delta -\varepsilon)}
\right], \quad 0\leq u < \infty,\\
&\quad &\\
&\quad &\left|{\cal{G}}_{k}(\omega)\right| \leq \frac{2}{\sqrt{2\pi}}
e^{(\sigma-\varepsilon)u}e^{-|k|\rho} \left[
C{c_{3}}+ \frac{C_{3}e^{-(\delta -\varepsilon)c_{3}}}{(\delta -\varepsilon)}
\right], -\infty < u \leq 0
\end{eqnarray*}
where we have used the fact
\begin{eqnarray*}
 \int ^{c_{1}}_{0} |g_{k}(\zeta -i\eta)|e^{(\nu -\delta)\zeta} d\zeta
\leq C_{1}c_{1}
,\\
&\quad&\\
 \int _{c_{2}}^{0} |g_{k}(\zeta -i\eta)|e^{-(\mu -\delta)\zeta} d\zeta \leq C_{2}|c_{2}| ,
\end{eqnarray*}
and $c_{3}=\min (c_{1},|c_{2}|)$, $C_{3}=\max (C_{1},C_{2})$.
From $(\ref{eqnsixone})$ and $(\ref{eqnsixtwo})$ we see that  the complex Fourier
transform of $f_{k}(t)$ is given by
\begin{eqnarray*}
{\cal{F}}_{k}(\omega)= \frac{{\cal{G}}_{k}(\omega)}{i(\tilde{\lambda}\cdot k + \omega)}.
\end{eqnarray*}
\\
We will show that ${\cal{F}}_{k}(\omega)$ is exponentially small with respect to
$u$ and $k$  and analytic in two parallel strips. Clearly, since $\tilde{\lambda }\cdot
k \in {\mathbb{R}}$, we can pick a $\gamma >0$
such that
$
|\tilde{\lambda} \cdot k + \omega|> \gamma
$
and since ${\cal{G}}_{k}(\omega)$ is analytic in the strip
$-\mu <v< \nu $, ${\cal{F}}_{k}(\omega)$ will be analytic in the strips
$A(\gamma < v< \nu ) $ and $B(-\mu <v<-\gamma )$.
We have
\begin{eqnarray*}
f_{k}(t)= \int ^{\infty +i\beta}_{-\infty +i\beta}
\frac{{\cal{G}}_{k}(\omega)}{i(\tilde{\lambda} \cdot k + \omega)}e^{i\omega t}
d\omega,
\end{eqnarray*}
and the integral converges uniformly for $|t_{I}| < \sigma$. Hence $f_{k}(t)$
is analytic in this strip. Furthermore
\\
\begin{eqnarray*}
f_{k}(t)&=& \int ^{\infty +i\beta}_{-\infty +i\beta}
\frac{{\cal{G}}_{k}(\omega)}{i(\tilde{\lambda} \cdot k + \omega)}e^{i\omega t}
d\omega=
\int^{\infty}_{-\infty} \frac{{\cal{G}}_{k}(u+i\beta)}{i(\tilde{\lambda }\cdot k
+ u + i\beta)}e^{i(u+i\beta )t}du,
\end{eqnarray*}
and
\begin{eqnarray*}
|f_{k}(t)| &\leq&  \int ^{\infty}_{-\infty}
\frac{|{\cal{G}}_{k}(u+i\beta)|}{|i(\tilde{\lambda }\cdot k
+ u + i\beta)|}\left|
e^{i(u+i\beta)(t_{R}+it_{I})} \right| du\\
\quad \\
&\leq&
\frac{1}{\gamma} \int ^{\infty}_{-\infty}|{\cal{G}}_{k}(u+i\beta)|
|e^{iut_{R}}||e^{-ut_{I}}||e^{-\beta t_{R}}||e^{-i\beta t_{I}}|du\\
\quad \\
&\leq&
\frac{2}{\sqrt{2\pi}\gamma} e^{-|k|\rho}e^{-\beta t_{R}}\left[
C{c_{3}}+\frac{C_{3} e^{-(\delta -\varepsilon)c_{3}}}{(\delta -\varepsilon)} \right]
\left[ \int ^{\infty}_{0} e^{-(\sigma-\varepsilon)u}e^{-ut_{I}}du +
\int^{0}_{-\infty} e^{(\sigma-\varepsilon)u}e^{-ut_{I}}du \right]\\
\quad \\
&=&
\frac{2}{\sqrt{2\pi}\gamma} e^{-|k|\rho}e^{-\beta t_{R}}\left[
C{c_{3}}+\frac{C_{3} e^{-(\delta -\varepsilon)c_{3}}}{(\delta -\varepsilon)} \right]
\left[
\frac{e^{-(\sigma-\varepsilon +t_{I})u}}{-(\sigma-\varepsilon +t_{I})} \Big|
^{\infty}_{0} + \frac{e^{(\sigma-\varepsilon -t_{I})u}}{(\sigma-\varepsilon -t_{I})}
\Big | ^{0}_{-\infty} \right],
\end{eqnarray*}
where we have used the fact that we can pick a $\gamma >0$
such that
$
|i(\tilde{\lambda}\cdot k + u +i\beta )|=
\left[ (\tilde{\lambda}\cdot k + u)^{2}+\beta ^{2}\right]^{1/2}
\geq \beta \geq \gamma$.
For the $u$ intervals $[0,\infty )$ and $(-\infty ,0]$,  we pick
$t_{I}=-\sigma+ \delta$ and $t_{I}=\sigma-\delta$,\quad  $\delta > \varepsilon$ , respectively, so that
\quad \\
\begin{eqnarray*}
|f_{k}(t)| &\leq&
\frac{4}{\sqrt{2\pi}\gamma}\frac{ e^{-|k|\rho}e^{-\beta t_{R}}}{(\delta-\varepsilon)}\left[
C_{3}{c_{3}}+\frac{C_{3} e^{-(\delta -\varepsilon)c_{3}}}{(\delta -\varepsilon)} \right].
\end{eqnarray*}
\quad \\
Finally, we obtain the order results by taking $\beta$, which was arbitrary from
the definition of ${\cal{G}}(\omega)$, arbitrarily close to $\nu$ and
$-\mu$. Note that since $f_{k}(t)$ is only defined in the strips
$A(\gamma < v < \nu )$ and $B(-\mu < v < -\gamma)$, $\beta$ must be within
these strips; $\gamma < \beta < \nu $ and $-\mu < -\beta < -\gamma $.
By  Lemma $\ref{coeffbound}$ for some $\tilde{\delta} < \rho$
\quad \\
\begin{eqnarray*}
\|F \|_{\rho -\tilde{{\delta}},\sigma -\delta} \leq
\frac{4}{\gamma \sqrt{2\pi}}
\frac{1}{(\delta -\varepsilon)}
\left[
C{c_{3}}+
\frac{C_{3}e^{-(\delta -\varepsilon)c_{3}}}{(\delta -\varepsilon)}\right]
\left(\frac{4}{\tilde{{\delta}}} \right)^{n}.
\end{eqnarray*}
\quad \\
We can easily find estimates for the partial derivatives
of $F(q,t)$. Given
\[ F(q,t)=\sum _{k\in {\mathbb{Z}}^{n}}f_{k}(t)e^{ik\cdot q},
\quad F\in {\cal{A}}_{\rho ,\sigma}, \]
the partial derivatives of $F(q,t)$ with respect to $q_{j}$ $j\neq n+1$ are
as follow
\[
\frac{\partial F}{\partial q_{j}} =\sum _{k\in {\mathbb{Z}}^{n}}ik_{j}f_{k}(t)
e^{ik\cdot q}.\]
For some positive  $\tilde{\tilde{\delta}} <\rho$
\begin{eqnarray*}
|ik_{j}f_{k}(t)| &= & |k_{j}||f_{k}(t)|
\leq \left( \frac{1}{e\tilde{\tilde{\delta}}} \right) e^{|k|\tilde{\tilde{\delta}}}
|f_{k}(t)|\\
\quad \\
&\leq& \left( \frac{1}{e\tilde{\tilde{\delta}}} \right) e^{|k|\tilde{\tilde{\delta}}} \frac{4}{\gamma
\sqrt{2\pi}}\frac{e^{-|k|\rho}}{(\delta -\varepsilon)}
\left[
C{c_{3}}+
\frac{C_{3}e^{-(\delta -\varepsilon)c_{3}}}{(\delta -\varepsilon)}\right].
\end{eqnarray*}
By Lemma $\ref{coeffbound}$ with $\rho -\tilde{\tilde{\delta}}$ in placed of
$\rho$ and $\tilde{{\delta}}$ in place of $\delta $
we obtain the following
\[
\left \|
\frac{\partial F}{\partial q_{i}} \right \|
_{\rho -\tilde{\delta} -\tilde{\tilde{\delta}}, \sigma -\delta} \leq
\frac{4}{\gamma \sqrt{2\pi}}\frac{1}{e\tilde{\tilde{\delta}}}
\frac{1}{(\delta -\varepsilon)}
\left[
C{c_{3}}+
\frac{C_{3}e^{-(\delta -\varepsilon)c_{3}}}{(\delta -\varepsilon)}\right]
\left(\frac{4}{\tilde{{\delta}}} \right)^{n},
\]
and
$
\frac{\partial F}{\partial q_{i}} \in
{\cal{A}}_{\rho -\tilde{\delta} -\tilde{\tilde{\delta}}, \sigma}$.
Similarly, we obtain an estimate  for  the partial derivative of
$F(q,t)$ with respect to $q_{n+1}=t$ which is given by the following expression
\[
\frac{\partial F}{\partial t} =
\sum _{k\in {\mathbb{Z}}^{n}} \frac{ df_{k}(t)}{dt} e^{ik\cdot q}.
\]
Using Cauchy's Inequalities
\begin{eqnarray*}
\left| \frac{df_{k}(t)}{dt} \right| &\leq& \frac{1}{\delta }
\| f_{k} \|_{\sigma}
\leq
\frac{1}{\delta }
\frac{4}{\gamma \sqrt{2\pi}}
\frac{e^{-|k|\rho}}{(\delta -\varepsilon)}
\left[
C{c_{3}}+
\frac{C_{3}e^{-(\delta -\varepsilon)c_{3}}}{(\delta -\varepsilon)}\right].
\end{eqnarray*}
By Lemma $\ref{coeffbound}$ for some $\tilde{{\delta}} < \rho$
\[
\left \| \frac{\partial F}{\partial t} \right\|
_{\rho -\tilde{{\delta}}, \sigma-2\delta}
\leq \frac{1}{\delta}
\frac{4}{\gamma \sqrt{2\pi}}
\frac{1}{(\delta -\varepsilon)}
\left[
C{c_{3}}+
\frac{C_{3}e^{-(\delta -\varepsilon)c_{3}}}{(\delta -\varepsilon)}\right]
\left(\frac{4}{\tilde{{\delta}}} \right)^{n},
\]
and
\[
\frac{\partial F}{\partial q_{i}} \in
{\cal{A}}_{\rho  -\tilde{{\delta}}, \sigma-\delta}
\quad {\mbox{for}}\quad  i=1,...,n+1 .
\]
Let $d = \max (1/e\tilde{\tilde{\delta}}, 1/\delta)$, then
\[
\left \|
\frac{\partial F}{\partial q'} \right \|
_{\rho -\tilde{\delta} -\tilde{\tilde{\delta}}, \sigma-2\delta}
\leq
\frac{4d}{\gamma \sqrt{2\pi}}
\frac{1}{(\delta -\varepsilon)}
\left[
C{c_{3}}+
\frac{C_{3}e^{-(\delta -\varepsilon)c_{3}}}{(\delta -\varepsilon)}\right]
\left(\frac{4}{\tilde{{\delta}}} \right)^{n}.
\]
\squareforqed

\section{Iterative Lemma}
\begin{lem}(Iterative Lemma)
\label{iterativelemma1}
\quad\\

Given positive constants $\Gamma ,\rho , \sigma ,\kappa <1$,
consider the Hamiltonian $H(p',q')=U(p',q')+P(p',q')$
defined by
\begin{eqnarray*}
&U(p',q')&=a+\lambda \cdot p' + \frac {1}{2}\sum _{i,j} C_{i,j}(q')p'_{i}p'_{j}+
R(p',q'),\\
&P(p',q')&= \kappa \left(A(q')+\sum _{i}B_{i}(q')p'_{i}\right),
\end{eqnarray*}
with $\| H\|_{\rho ,\sigma } <1$, $H\in{\cal{A}}_{\rho, \sigma}$,
with some constant vector $ \lambda =(\tilde{\lambda},1)
\in {\mathbb{C}}^{n+1}$
such that $| \tilde{\lambda} | < L_{0}$ for some constant $L_{0}$,
$\tilde{\lambda } \in \Omega _{\Gamma}$,
$ A, B_{i}, C_{ij}, R \in {\cal {A}}_{\rho , \sigma}$
and $R$ is of order $\| p \| ^{3}$.
Assume $H$ is of $(C_{1},C_{2},c_{1},c_{2},\nu,\mu)p',q'$-exponential form.
Assume there are positive
constants $m,f,E_{1}<1$ such that
\begin{eqnarray*}
&\quad & 2m\|\tilde {v} \| \leq \|\tilde{C}^{*} \tilde {v} \|
 , \quad \forall \tilde {v} \in {\mathbb{C}}^{n},\\
\quad\\
&\quad &
f\| v \|\leq \left\|
\left(
\begin{array}{cc}
\overline{\overline{C}} & \lambda ^{T} \\
\lambda & 0
\end{array}
\right)v
\right\| , \quad \forall v \in {\mathbb {C}}^{n+1}, \\
\quad\\
&\quad & \| Cv \| _{\rho ,\sigma} < m^{-1} \| v \|, \quad \forall v \in {\mathbb {C}}^{n+1},\\
\quad\\
&\quad & \mbox {max} (\| A \| _{\rho ,\sigma}, \|B \| _{\rho ,\sigma})< E_{1}.
\end{eqnarray*}
One can construct positive constants $\rho_{*},\sigma_{*},m_{*},L,c_{3},\gamma$
with
$\rho _{*} < \rho , \sigma _{*} <\sigma ,
m_{*}<m,L= \frac{3}{2}L_{0}, c_{3}=\min (c_{1},|c_{2}|)$ and
\begin{eqnarray*}
&\quad&
\eta =
\frac{\Lambda }{f^{2}m^{2}\delta ^{\aleph}}\kappa E_{1},
\quad
\aleph =10n+9,
\quad
\Lambda =
\frac{3^{n}2^{2n+12}\varpi ^{3}(6n+6)^{4}(c_{3}+1)^{2}}{\rho_{*}\Gamma ^{3}
\gamma ^{2}}e^{\nu +\mu},\quad
\varpi = 2^{4n+1}\left(\frac{n+1}{e}\right)^{n+1},
\end{eqnarray*}
such that for any $\delta>0$ small enough so $\rho-4\delta > \rho _{*}$,
$\sigma -4\delta >\sigma _{*} $
and with
 $\kappa E_{1}$  small enough that
\begin{equation}
\label{etasize}
m-\frac{ 4(n+1)\eta}{ \sigma ^{2}_{*}} > m_{*},
\quad
f-\frac{(n+3)\eta}{\sigma_{*}^{2}} >f_{*},
\end{equation}
there exists a $\zeta_{A}\in {\mathbb{R}}$ and
an analytic canonical change of variables, $\phi : D_{\rho
-4\delta,\sigma-4\delta} \rightarrow D_{\rho,\sigma},
 \quad \phi \in {\cal {A}}_{\rho -4\delta,\sigma -4\delta}$,
which transforms the Hamiltonian $H$ to $H'({\cal{P}}',{\cal{Q}}')
\equiv {\cal{U}}H=
(H\circ \phi)({\cal{P}}',{\cal{Q}}')$. $H'({\cal{P}}',{\cal{Q}}')$
is of ${\cal{P}}',{\cal{Q}}'$ exponential form and
can be decomposed as the original $H$ with primed
quantities $ A', B', C',$ and $R'$
\[
H'=U'+P',
\]
\[
U'=a'+ {\lambda}' \cdot {\cal{P}}' +\frac {1}{2}
\sum _{i,j}C'_{ij}{\cal{P}}'_{i}{\cal{P}}'_{j}+
R'({\cal{P}}',{\cal{Q}}'),
\]
\[
P'= \kappa '\left(A'({\cal{Q}}')+ B'({\cal{Q}}')\cdot {\cal{P}}'\right),
\]
where
\begin{eqnarray*}
&\quad&
\lambda'=(1+\kappa E_{1}\zeta_{A})\lambda\in \Omega_{\Gamma},\quad
a'=\overline{\overline{H'}}(0),
\quad
A'({\cal{Q}}')=H'(0,{\cal{Q}}')-a',\\
&\quad&
B_{i}'({\cal{Q}}')=\frac {\partial H'}{\partial
{\cal{P}}'_{i}}(0,{\cal{Q}}')-\lambda '_{i},
\quad
C_{ij}'({\cal{Q}}')
=\frac {\partial^{2}H'}{\partial {\cal{P}}'_{i} \partial {\cal{P}}'_{j}}
(0,{\cal{Q}}'),
\quad
R'={\cal{O}}(\| {\cal{P}}' \|^{3}),
\end{eqnarray*}
and satisfies similar conditions with positive numbers
  $\rho ',\sigma ',m',f',\kappa ',E_{1}'$ all less
thans one and  $L'$ also positive with $\| H' \|_{\rho '} \leq 1$, where $A',B'_{i},C'_{ij}, R' \in {\cal{A}}_{\rho ',\sigma'}$,
\begin{eqnarray*}
&\quad& 2m'\|\tilde{v} \| \leq \| \tilde{C'}^{*} \tilde{v} \|, \quad \forall
 \tilde{v}\in {\mathbb{C}}^{n},\\
\quad\\
&\quad &
f'\| v \|\leq \left\|
\left(
\begin{array}{cc}
\overline{\overline{C}}' & \lambda ^{T} \\
\quad\\
\lambda & 0
\end{array}
\right)v
\right\|,\quad \forall v\in {\mathbb{C}}^{n+1},  \\
\quad\\
&\quad& \|C' v \|_{\rho ', \sigma '}< m '^{-1}\| v\|, \quad \forall v\in {\mathbb{C}}^{n+1},\\
\quad\\
&\quad& \max (\|A' \| _{\rho ', \sigma ' }, \|B' \|_{\rho ', \sigma ' })< E_{1}' ,
\end{eqnarray*}
where
\begin{eqnarray*}
&\quad&\rho '=\rho -4\delta >\rho _{*},
\quad
\sigma '=\sigma -4 \delta >\sigma_{*},
\quad
m'=m-\frac{4(n+1)\eta }{\sigma ^{2}_{*}}>m_{*},
\quad
f'= f- \frac{(n+3)\eta }{\sigma ^{2}_{*}},
\quad
L'= \frac{3}{2}L_{0},\\
\quad\\
&\quad&\kappa ' = \frac{\Lambda}{f'^{2}m'^{2} \delta'^{\aleph}}\kappa ^{2},
\quad
E_{1}'= \frac{\Lambda}{f'^{2}m'^{2} \delta'^{\aleph}}\frac{E_{1}}{\sigma_{*}}.
\end{eqnarray*}
Furthermore, one has for any function $F\in {\cal {A}}_{\rho ,\sigma}$,
$
\| {\cal {U}}F-F \|_{\rho ' ,\sigma '}\leq  \eta \| F \|_{\rho ,\sigma}$.
\end{lem}
{Proof}\\

We will construct a canonical transformation of the Hamiltonian,
with a generating function for this transformation denoted by
$\chi$.
The generating function is constructed in such a way  it
eliminates the part of the transformed Hamiltonian that
prevents the preservation of the invariant
structure $p'=0$. Recall the Hamiltonian
\quad

\begin{eqnarray*}
&\quad& H = U+P,\\
\quad \\
&\quad& U(p',q')=a+\lambda \cdot p' + \frac {1}{2}\sum _{i,j} C_{i,j}(q')p'_{i}p'_{j}+
R(p',q'),\\
\quad \\
&\quad&
P(p',q')=\kappa \left( A(q')+\sum _{i}B_{i}(q')p'_{i}\right).
\end{eqnarray*}
We write the transformed Hamiltonian as follows
\begin{eqnarray*}
H'={\cal {U}}H=U+P+\{ \chi , U \} +[ \{ \chi ,P \} +
{\cal {U}}H - H- \{ \chi , H \} ].
\end{eqnarray*}
Alternatively, assuming analyticity in $t$ and taking the Taylor expansion,
one has the Lie series
 \begin{eqnarray*}
{\cal{U}}H = \sum ^{\infty}_{m=0} \frac{t^{m}}{m!} L^{m}_{\chi}H,
\end{eqnarray*}
where we denote $L^{0}_{\chi} H = H$ and $L^{m}_{\chi}H =\{L^{m-1}_{\chi}H,\chi \}$
for $m\geq 1$. For the $m$th remainder of the Lie series, we use the notation
\begin{eqnarray*}
r_{m}(H,\chi ,t) = {\cal{U}}H -
\sum ^{m-1}_{l=0} \frac{t^{l}}{l!}L^{l}_{\chi}H =
\sum ^{\infty}_{l=m} \frac{t^{l}}{l!}L^{l}_{\chi}H.
\end{eqnarray*}
With this notation, the transformed Hamiltonian can be expressed as follows
\quad
\begin{eqnarray*}
H'={\cal {U}}H=U+P+\{ \chi , U \} + r_{2}(U,\chi ,1)+ r_{1}(P,\chi ,1),
\end{eqnarray*}
so that
$\quad
 \{ \chi ,P \} +
{\cal {U}}H - H- \{ \chi , H \}=
r_{2}(U,\chi ,1)+ r_{1}(P,\chi ,1).
$
We choose $\chi$ so that
$
 \{ \chi ,P \} +
{\cal {U}}H - H- \{ \chi , H \}
$
is second order in the size of the perturbation and
\begin{equation}
\label{form}
P+ \{ \chi ,U \} =c\lambda  \cdot p' +    O(\| p' \| ^{2}),
\end{equation}
which implies $P+ \{ \chi ,U \}$ does not contribute to $P'$ but
it shifts the frequency by a multiple of the original frequency.
Furthermore, one can show all conditions are met by a generating
function, similar to the one introduced by Kolmogorov, of the form
\\
\begin{equation}
\label{generatingfn}
\chi  =  X(q') +\xi \cdot q' + Y(q')\cdot p',
\end{equation}
\\
where the functions $X(q')$ and $Y(q')$ are of
$q'$-exponential form
and $\xi \in {\mathbb{R}}^{n+1}$.
A simple calculation with the appropriate definitions of
$U$ and $P$ gives
\\
\begin{eqnarray*}
 \{ \chi  , U \}&=&
 \left( -\lambda \cdot \xi
-\sum _{i} \lambda _{i} \frac {\partial X}{\partial q'_{i}}\right)
+ \sum_{j} \left [-\sum _{i} C_{ij}(q')\xi_{i}  - \sum_{i}C_{ij}(q') \left (
\frac {\partial X}{\partial q'_{i}} \right )  \right ]p'_{j}\\
&-& \sum_{i,j} \lambda _{j}\frac{\partial Y_{i}}{\partial q'_{j}}(q')p'_{i}
+ {\cal {O}}(\| p' \| ^{2}).
\end{eqnarray*}
Equivalently
\begin{eqnarray*}
 \{ \chi  , U \}  &=&
-  \Big( \lambda \cdot \xi + \lambda \cdot \partial_{q'}X(q') \Big)
+  \Big(-\xi \cdot C(q') - \partial_{q'}X(q') \cdot C(q') -\lambda \cdot \partial
_{q'}Y(q')  \Big)
\cdot p'
\end{eqnarray*}
\begin{equation}
\label{pbracket}
+\quad {\cal {O}}(\| p' \| ^{2}).\quad \quad \quad \quad \quad \quad \quad \quad
\quad \quad \quad \quad
\end{equation}
\quad \\
To obtain the form of $(\ref{form})$ we first impose the following condition
\quad \\
\begin{equation}
\label{Condition1}
-\lambda \cdot \xi
- \lambda \cdot \partial_{q'}X(q')+\kappa A(q')=0.
\end{equation}
We set
\begin{equation}
\label{doublebarHxi}
\lambda \cdot \xi =\kappa \overline{ \overline{A}},
\end{equation}
so that $(\ref{Condition1})$ becomes
\begin{equation}
 \label{xsolution}
\lambda \cdot \partial_{q'}X(q')=\kappa \left(A(q')-\overline{ \overline{A}}\right).
\end{equation}
\\
By Lemma $\ref{exponentialpart}$ and the definition of $A(q')$ we know
the right hand side of $(\ref{xsolution})$ consists of a quasiperiodic part
and a part of exponential order with respect to time.
Consequently $(\ref{xsolution})$ can be split in two
problems corresponding to the quasiperiodic part
and to the part of exponential order with respect to time  
respectively and solved as explained in
section $\ref{sectionwithsplit}$.
By lemma $\ref{lemmaqb}$ and theorem $\ref{maintheorem3}$
there exists an analytic function $X(q')$ satisfying $(\ref{xsolution})$.
Note, by assuming the perturbation of the Hamiltonian, mainly $H^{1}(p',q')$,
consists  of a quasiperiodic part
and an  exponential-order-with-respect-to-time part, we are assured $X$
 and $\partial_{q'}X$ are of the same form.
Next, for some $(\zeta_{A} ,\xi ) \in {\mathbb {R}}\times {\mathbb {R}}^{n+1}$ to be determined,
 set
$
 \overline { \overline {(
\kappa B-C\cdot \xi -C \cdot \partial _{q'}X)}} =
\kappa E_{1} \zeta _{A} \lambda$.
Consequently we have the following conditions
$
 \overline{ \overline{C}}\cdot \xi
 +\kappa E_{1} \zeta _{A}\lambda =\kappa \overline{ \overline {B}}
-\overline{\overline {C}}\cdot \overline{\overline {\partial _{q'}
X}}$,
$
 \lambda \cdot \xi  =\kappa \overline{\overline {A}}$,
or equivalently
\begin{equation}
\label{zetasolution}
\left( \begin{array}{cc}
\overline{ \overline{C}} & \lambda ^{T} \\
\lambda & 0
\end{array}
\right) \left(
\begin{array}{c}
\xi \\\kappa E_{1} \zeta_{A} \end{array}
\right)
=
\left(
\begin{array}{c}
\kappa \overline{\overline {B}}-\overline{\overline {C}}
\cdot \overline{\overline {\partial _{q'}X}} \\
\kappa \overline{\overline {A}}
\end{array}
\right).
\end{equation}
It follows from lemmas $\ref{lemmaiso}$ and $\ref{thmisob}$  there exists a solution,
$(  \xi  , \zeta_{A} )\in{\mathbb {R}}^{n+1}\times {\mathbb {R}}$,
for the matrix equation above.\\

Up to this point we have determined conditions on $X(q')$ and
$( \xi ,\zeta_{A} ) \in {\mathbb {R}}^{n+1}\times {\mathbb {R}}$ to obtain the Hamiltonian
\begin{eqnarray*}
H'(p',q')&=&
 U+\kappa  \zeta_{A}\lambda \cdot p'+ {\cal {O}}(\|p'\|^{2})
+ \Big[ (\kappa B(q')-C(q')\cdot \xi -C(q')\cdot \partial _{q'}X(q'))\\
&\quad&-(\kappa \overline{\overline {B}}-\overline{\overline {C}}\cdot \xi -\overline{\overline {C}}\cdot
\overline{\overline {\partial _{q'}X}}) \Big]\cdot p'
+ \Big(- \lambda \cdot \partial _{q'}Y(q') \Big)\cdot p'\\
&\quad&+ [ \{ \chi,P\}+{\cal{U}}H -H-\{\chi ,H \}]\\
&=& \tilde{U}+  \beta\cdot p'+\Big(- \lambda \cdot \partial _{q'}Y(q') \Big)\cdot p' +{\cal {R}}_{A},
\end{eqnarray*}
where we have set
\\
\begin{eqnarray*}
\tilde{U} &=& U+\kappa E_{1}\zeta_{A}\lambda \cdot p' + {\cal {O}}(\|p'\|^{2})\\
&=&a+ (1+\kappa E_{1}\zeta_{A} )\lambda \cdot p'+\frac{1}{2} \sum _{i,j}C_{i,j}(q')
p'_{i}p'_{j}+R(p',q')+ {\cal {O}}(\|p'\|^{2})\\
&=& a+{\lambda }' \cdot p'+\frac{1}{2} \sum _{i,j}C_{i,j}(q')
p'_{i}p'_{j}+R(p',q')+ {\cal {O}}(\|p'\|^{2}),
\end{eqnarray*}
where $R(p',q')$ is $ {\cal {O}}(\|p'\|^{3})$,
$\beta =\Big[ (\kappa B(q')-C(q')\cdot \xi -C(q')\cdot \partial _{q'}X(q'))
-(\kappa \overline{\overline {B}}-\overline{\overline {C}}\cdot \xi -\overline{\overline {C}}\cdot
\overline{\overline {\partial _{q'}X}}) \Big]$,
${\lambda}'=(1+\kappa E_{1}\zeta_{A})\lambda$
and
\begin{equation}
\label{remainder}
{\cal {R}}_{A}=
 \{ \chi,P\}+{\cal{U}}H -H-\{\chi ,H \}.
\end{equation}
\\
Next we set
$
\Big(-{\lambda}\cdot \partial _{q'}Y(q')+\beta \Big)\cdot p'=0
$
or equivalently
\\
\begin{equation}
\label{yyequationsol}
{\lambda}\cdot \partial _{q'}Y(q')=\beta.
\end{equation}
\quad \\
Note, by the definition of $B$ and $C$ and by the form of
$\partial _{q'}X$,   $\beta$ consists of a quasiperiodic part and
a part of exponential order with respect to time. Consequently,
$(\ref{yyequationsol})$ can be split into two problems and solved as
explained in section  $\ref{sectionwithsplit}$.
The final Hamiltonian after one application of the generating function is therefore
\\
\begin{eqnarray}
H' ={\cal{U}}H
= \Big[a+\lambda ' \cdot p' +\frac{1}{2} \sum_{i,j}C_{i,j}(q')p'_{i}p'_{j}+{\cal {O}}(\|p'\|^{2})
\Big]+ {\cal {R}}_{A}
=a+\lambda ' \cdot p' + {\cal {O}}(\|p'\|^{2})+ {\cal {R}}_{A}.\quad
\quad \quad  \quad
\label{finalhamil}
\end{eqnarray}
\\
where ${\cal {R}}_{A}$ is ${\cal{O}}(\kappa ^{2})$.
By lemma $\ref{lemmaRA}$ the new Hamiltonian
$H'(p',q')$ is of $p',q'$exponential form.
The formula for $X(q')$ can be obtained in terms of Fourier coefficients. Consider the following
\\
\begin{eqnarray}
\label{Hamiltonianperturbation1}
&\quad &H^{1}(p',q,t)=G(p',q)
+
T(p',q')\in {\cal{A}}_{\rho ,\sigma},
\end{eqnarray}
\begin{eqnarray*}
&\quad & G(p',q)=\sum _{k\in {\mathbb {Z}}^{n}} s^{1}_{k}(p')e^{ik \cdot q}
\in {\cal{A}}_{\rho},\quad
\quad
 T(p',q')=\sum _{k\in {\mathbb {Z}}^{n}} h^{1}_{k}(p')e^{1}_{k}(t)e^{ik \cdot q}
\in {\cal{A}}_{\rho ,\sigma}, \\
\quad \\
&\quad& X(q,t)= {\cal{Y}} (q)+ {\cal{T}}(q'),\quad \quad
 {\cal{Y}}(q)= \sum _{k\in {\mathbb {Z}}^{n}} y_{k}e^{ik \cdot q},
\quad
\quad
{\cal{T}}(q')=
\sum _{k\in {\mathbb {Z}}^{n}} x_{k}(t)e^{ik \cdot q}.
\end{eqnarray*}
We have assumed $e^{1}_{k}(t)$ is analytic in the strip $-\sigma < {\mbox{Im }}(t) <\sigma$
and satisfies
\begin{eqnarray*}
e^{1}_{k}(t) = \left \{
\begin{array}{cc}
{\cal{O}}(e^{-(\nu - \varepsilon)t_{R}}) &  (t_{R} \rightarrow \infty)\\
\quad & \quad \\
{\cal{O}}(e^{(\mu - \varepsilon)t_{R}}) &  (t_{R} \rightarrow -\infty)
\end{array}
\right.
\end{eqnarray*}
or
\begin{eqnarray*}
&\quad& |e^{1}_{k}(t)|\leq C_{0}e^{-(\nu -\varepsilon)t_{R}}, \quad
0< c_{1} < t_{R} < \infty,\\
&\quad&\\
&\quad& |e^{1}_{k}(t)|\leq C_{0}'e^{(\mu -\varepsilon)t_{R}}, \quad
-\infty < t_{R} < c_{2} <0,
\end{eqnarray*}
for some positive $C_{0},C_{0}'$ and 
\begin{eqnarray*}
&\quad&|h_{k}^{1}(0)e^{1}_{k}(t)| \leq
C_{1}e^{-(\nu -\varepsilon)t_{R}}e^{-|k|\rho}, \quad
0< c_{1} < t_{R} < \infty,\\
&\quad&\\
&\quad& |h_{k}^{1}(0)e^{1}_{k}(t)|\leq C_{2}e^{(\mu -\varepsilon)t_{R}}e^{-|k|\rho}, \quad
-\infty < t_{R} < c_{2} <0 .
\end{eqnarray*}
Recall the equation to be solved
\begin{eqnarray}
\label{eqntobesolved}
\lambda \cdot \partial _{q'} X(q') = \kappa \left(
A(q') -\overline{\overline{A}}\right),
\end{eqnarray}
and by the defined normal form
\begin{eqnarray*}
\kappa A(q')&=& \kappa (H^{1}(0,q')-\overline{\overline{H^{1}}}(0))
= \kappa \left(
\sum _{k\in {\mathbb {Z}}^{n}\backslash 0} s^{1}_{k}(0)e^{ik \cdot q}
+
\sum _{k\in {\mathbb {Z}}^{n}} h^{1}_{k}(0)e^{1}_{k}(t)e^{ik \cdot q}\right).
\end{eqnarray*}
Note
$
\overline{\overline{A}} =0$.
Solving $(\ref{eqntobesolved})$ can be done by separating the
quasiperiodic part and
the exponential-time-dependent part and solving the two parts  separately as was
presented in section four.
The quasiperiodic part is as follows
\begin{eqnarray}
\label{eqnqbs}
\tilde{\lambda} \cdot \partial _{q'} {\cal{Y}}(q)
=
\kappa \left( G(0,q)-\overline{\overline{G}}(0)\right)
=
\sum _{k\in {\mathbb {Z}}^{n}\backslash 0}
\kappa
 s^{1}_{k}(0)e^{ik \cdot q}.
\end{eqnarray}
Since we assume
$\tilde{\lambda} \in \Omega _{\Gamma}$ and since
the right hand side of $(\ref{eqnqbs})$  has zero average,
by lemma $\ref{lemmaqb}$, for some positive $\delta <\rho$
, $(\ref{eqnqbs})$ admits a unique solution
\begin{eqnarray*}
{\cal{Y}}(q)=\sum _{k\in {\mathbb {Z}}^{n}\backslash 0} y_{k}e^{ik \cdot q}
\in {\cal{A}}_{\rho -\delta},
\end{eqnarray*}
with
$
\overline {\cal{Y}}=0
$
and one has the estimates
\begin{eqnarray}
\label{ykestimate}
|y_{k} |
\leq
\frac{\kappa}{\Gamma}\left( \frac{n}{e\delta} \right)^{n}
\|(G-\overline{\overline{G}})(0) \|_{\rho}e^{-|k|(\rho -\delta)},
\end{eqnarray}
\begin{eqnarray*}
\| {\cal{Y}} \|_{\rho -\delta}\leq
\frac{\varpi \kappa }{\Gamma \delta ^{2n}}
 \|(G-\overline{\overline{G}})(0) \|_{\rho},
\quad
\quad
\left \| \frac{\partial {\cal{Y}}}{\partial q} \right \|_{\rho -\delta}
\leq
\frac{\varpi \kappa }{\Gamma \delta ^{2n+1}}
 \|(G-\overline{\overline{G}})(0) \|_{\rho},
\end{eqnarray*}
where
$\varpi =2^{4n+1}\left(
\frac{n+1}{e} \right)^{n+1}$.
The exponential-order-with-respect-to-time  part of
$(\ref{eqntobesolved})$
 is as follows
\begin{eqnarray}
\label{timeequation}
\lambda \cdot \partial _{q'}{\cal{T}}(q')=\kappa  T(0,q'),
\end{eqnarray}
which reduces to the differential equation
\begin{eqnarray*}
i(\tilde{\lambda} \cdot k)x_{k}(t) + \frac{dx_{k}(t)}{dt} =
\kappa h_{k}^{1}(0) e^{1}_{k}(t).
\end{eqnarray*}
By Theorem $\ref{theomone}$
and lemma $\ref{existence}$
, there exists a $\delta >0$ such that
 the complex Fourier transform of
$e^{1}_{k}(t)$ defined by
\begin{eqnarray*}
{\cal{E}} ^{1}_{k}(\omega)= \int ^{\infty +i\beta} _{-\infty +i\beta} e^{1}_{k}(t)
e^{-i\omega t}dt,
\end{eqnarray*}
exists and is analytic in the strip $-\mu +\delta  \leq  v \leq  \nu -\delta $
and the Fourier inversion formula is
\begin{eqnarray*}
e ^{1}_{k}(t)= \int ^{\infty } _{-\infty} {\cal{E}}^{1}_{k}(\omega)
e^{i\omega t}d\omega.
\end{eqnarray*}
We found, $(\ref{eqneight})$, $x_{k}(t)$ has the following form
\begin{eqnarray*}
x_{k}(t)=
\int ^{\infty +i\beta } _{-\infty +i\beta}
\frac{\kappa h^{1}_{k}(0) {\cal{E}}^{1}_{k}(\omega)e^{i\omega t}}{i(\omega + \tilde{\lambda}
\cdot k)} d\omega.
\end{eqnarray*}
By  Theorem  $\ref{maintheorem3}$ there exist $\delta>0$, $\tilde{\delta}>0$,
$\tilde{\tilde{\delta}}>0$, without loss of generality we set $\delta
=\tilde{\delta}=\tilde{\tilde{\delta}}$ and $\varepsilon = \delta /2$,
and
$\gamma >0$, $\gamma <\nu$, $\gamma <\mu$ such that
 $(\ref{timeequation})$ has a unique
solution given by
\begin{eqnarray*}
 {\cal{T}}(q,t)
= \sum _{k\in {\mathbb{Z}}^{n}}x_{k}(t)e^{ik\cdot q}
 \in {\cal{A}}_{\rho -\delta, \rho -\delta},
\end{eqnarray*}
and the following estimates hold
\begin{eqnarray}
\label{eqnabba}
|x_{k}(t)|\leq \frac{4\kappa }{\gamma \sqrt{2\pi}}
\left[ \frac{2C_{3}{c_{3}}}
{\delta }
+\frac{4C_{3}e^{-\frac{\delta}{2}c_{3}}}{\delta^{2}}
\right]e^{-(\nu -\frac{\delta}{2})t_{R}}e^{-|k|\rho}, \quad
0\leq t_{R} < \infty,
\end{eqnarray}
\begin{eqnarray}
\label{eqnabba1}
|x_{k}(t)|\leq \frac{4\kappa }{\gamma \sqrt{2\pi}}
\left[ \frac{2C_{3}{c_{3}}}
{\delta}
+\frac{4C_{3}e^{-\frac{\delta c_{3}}{2}}}{\delta^{2}}
\right]e^{(\mu -\frac{\delta}{2})t_{R}}e^{-|k|\rho}, \quad
-\infty < t_{R} < 0,
\end{eqnarray}
\begin{eqnarray*}
&\quad& \|{\cal{T}}\|_{\rho -{\delta},\sigma -\delta }
\leq \frac{8\kappa }{\delta \gamma \sqrt{2\pi}}
\left[
C_{3}{c_{3}}+
\frac{2} {\delta} C_{3}e^{-\frac{\delta c_{3}}{2}}\right]
\left(\frac{4}{{{\delta}}} \right)^{n},\\
\quad\\
&\quad&
\left\| \frac {\partial {\cal{T}}}{\partial  q'} \right\|
_{ \rho -\delta,\sigma -\delta }
\leq
\frac{8\kappa }{\delta ^{2}\gamma \sqrt{2\pi}}
\left[
C_{3}{c_{3}}+
\frac{2}{\delta } C_{3}e^{-\frac{\delta c_{3}}{2}}\right]
\left(\frac{4}{\tilde{{\delta}}} \right)^{n}.
\end{eqnarray*}
\quad\\
Finally we have
\quad\\
\begin{eqnarray*}
\| X \|_{\rho -\delta, \rho -\delta}
=
\frac{\varpi \kappa}{\Gamma \delta ^{2n}}
\|(G-\overline{\overline{G}})(0)\|_{\rho}
+
\frac{8\kappa }{\delta \gamma \sqrt{2\pi}}
\left[
C_{3}{c_{3}}+
\frac{2}{\delta } C_{3}e^{-\frac{\delta c_{3}}{2}}\right]
\left(\frac{4}{{{\delta}}} \right)^{n},
\end{eqnarray*}
\quad\\
\begin{eqnarray*}
\left\| \frac{\partial X}{\partial q'}\right\|
_{\rho -\delta, \rho -\delta}
=
\frac{\varpi \kappa}{\Gamma \delta ^{2n+1}}
\|(G-\overline{\overline{G}})(0)\|_{\rho}
+
\frac{8\kappa }{\delta ^{2}\gamma \sqrt{2\pi}}
\left[
C_{3}{c_{3}}+
\frac{2}{\delta } C_{3}e^{-\frac{\delta c_{3}}{2}}\right]
\left(\frac{4}{{{\delta}}} \right)^{n}.
\end{eqnarray*}
\\
\\
Next we look at the solution of
$
\lambda \cdot \partial_{q'} Y=\beta$.
Recall
\quad\\
\begin{eqnarray*}
\beta_{j}=\kappa \left( B_{j}(q') - \overline{\overline{B}}_{j}\right)
-\left[
C(q')\cdot \partial _{q'} X(q')
-
\overline{\overline{C}} \cdot \overline{\overline{\partial _{q'} X(q')}} \right]_{j}
-
\left[
C(q') \cdot \xi -\overline{\overline{C}} \cdot \xi\right]_{j}.
\end{eqnarray*}
\quad \\
First we have
\begin{eqnarray*}
B_{j} = \frac{\partial H^{1}}{\partial p'_{j}}(0,q')
= \left[
\sum _{k\in {\mathbb{Z}}^{n}} \frac{\partial s^{1}_{k}}{\partial p'_{j}}(0)
e^{ik\cdot q}
+
\sum _{k\in {\mathbb{Z}}^{n}}
\frac{\partial h^{1}_{k}}{\partial p'_{j}} (0) e^{1}_{k}(t) e^{ik\cdot q} \right]
\in {\cal{A}}_{\rho -\delta ,\sigma}
,
\end{eqnarray*}
and
\begin{eqnarray*}
\overline{\overline{B}}_{j} =\frac{\partial s^{1}_{0}}{\partial p'_{j}}(0).
\end{eqnarray*}
Substituting for the definition of $B_{j}$ we obtain
\begin{eqnarray}
\label{eqn1001}
\kappa \left(B_{j}(q') - \overline{\overline{B}}_{j}\right)=
\kappa \left[
\sum _{k\in {\mathbb{Z}}^{n}\backslash 0} \frac{\partial s^{1}_{k}}{\partial p'_{j}}(0)
e^{ik\cdot q}
+
\sum _{k\in {\mathbb{Z}}^{n}}
\frac{\partial h^{1}_{k}}{\partial p'_{j}} (0) e^{1}_{k}(t) e^{ik\cdot q} \right]\in {\cal{A}}_{\rho -\delta, \sigma}
.
\end{eqnarray}
\quad \\
Next, we write out an expression for
$C(q')\cdot \partial _{q'} X(q')
-\overline{\overline{C}} \cdot
 \overline{\overline{\partial _{q'} X(q')}}$
given the following expressions
\begin{eqnarray*}
&\quad& X(q,t)= {\cal{Y}}(q)+ {\cal{T}}(q') \in {\cal{A}}_{\rho -\delta ,\sigma -\delta},\\
\quad\\
&\quad&
{\cal{Y}}(q)= \sum _{k\in {\mathbb{Z}}^{n}} y_{k}e^{ik\cdot q}
\in {\cal{A}}_{\rho -\delta},
\quad \quad
{\cal{T}}(q')=\sum _{k\in {\mathbb{Z}}^{n}}
x_{k}(t)e^{ik\cdot q}
\in {\cal{A}}_{\rho -\delta, \sigma -\delta}
,\\
\quad \\
&\quad&
C_{ij}(q')= \frac{\partial ^{2}\tilde{H}^{0}}{\partial p'_{i} \partial p'_{j}}(0)
+
\kappa  \frac{\partial ^{2} H^{1}}{\partial p'_{i} \partial p'_{j}}(0,q')
\in {\cal{A}}_{\rho -\delta, \sigma}
,\\
\quad \\
&\quad &H^{1}(p',q,t)=G(p',q)
+
T(p',q') \in {\cal{A}}_{\rho ,\sigma}, \\
\quad \\
&\quad & G(p',q)=\sum _{k\in {\mathbb {Z}}^{n}} s^{1}_{k}(p')e^{ik \cdot q}
\in {\cal{A}}_{\rho},
\quad \quad
 T(p',q')=\sum _{k\in {\mathbb {Z}}^{n}} h^{1}_{k}(p')e^{1}_{k}(t)e^{ik \cdot q}
\in {\cal{A}}_{\rho ,\sigma}.
\end{eqnarray*}
We have completely solved for $X(q,t)$.
Recall from Theorem $\ref{maintheorem3}$, $x_{k}(t)$ is of
exponential order
with respect to $t_{R}$.
We differentiate and obtain
\begin{eqnarray*}
&\quad& \partial _{q'} X = \left(
\partial_{q'_{1}}X, ..., \partial _{q'_{n+1}}X\right)
\in {\cal{A}}_{\rho -2\delta, \sigma -2\delta}
,\\
\quad \\
&\quad&
C(q') \cdot  \partial _{q'} X =
\left( \sum_{l=1}^{n+1} C_{1,l}(q') \partial _{q'_{l}}X,...,
\sum_{l=1}^{n+1} C_{n+1,l}(q') \partial _{q'_{l}}X \right)
\in {\cal{A}}_{\rho -2\delta, \sigma -2\delta},
\end{eqnarray*}
and
\begin{eqnarray*}
 \left(C(q') \cdot  \partial _{q'} X\right)_{j} &=&
\sum _{l=1}^{n+1} C_{j,l}(q') \partial _{q'_{l}}X
\end{eqnarray*}
\begin{eqnarray*}
&=&\sum _{l=1}^{n+1} \left[
\frac{\partial ^{2}\tilde{H}^{0}}{\partial p'_{l} \partial p'_{j}}(0)
+
\kappa  \frac{\partial ^{2} H^{1}}{\partial p'_{l} \partial p'_{j}}(0,q')
\right]
\partial _{q'_{l}}X\\
\quad \\
&=&
\sum _{l=1}^{n+1} \left[
\frac{\partial ^{2}\tilde{H}^{0}}{\partial p'_{l} \partial p'_{j}}(0)
+
\kappa  \frac{\partial ^{2} H^{1}}{\partial p'_{l} \partial p'_{j}}(0,q')
\right]\partial _{q'_{l}}
\left[{\cal{Y}}(q)+ {\cal{T}}(q')\right]\\
\quad \\
&=&
\sum _{l=1}^{n+1} \left[
\frac{\partial ^{2}\tilde{H}^{0}}{\partial p'_{l} \partial p'_{j}}(0)
+
\kappa \sum _{k\in {\mathbb {Z}}^{n}} \frac{ \partial ^{2} s^{1}
_{k}}
{\partial p'_{l} \partial p'_{j}}
(0)e^{ik \cdot q}
+
\kappa \sum _{k\in {\mathbb {Z}}^{n}} 
\frac{ \partial ^{2} h^{1}
_{k}}
{\partial p'_{l} \partial p'_{j}}
(0)
e^{1}_{k}(t)e^{ik \cdot q}
\right]\\
\quad \\
&\cdot&
\partial _{q'_{l}}
\left[{\cal{Y}}(q)+ {\cal{T}}(q')\right]\\
\quad \\
&=&
\sum _{l=1}^{n} \left[
\frac{\partial ^{2}\tilde{H}^{0}}{\partial p'_{l} \partial p'_{j}}(0)
+
\kappa \sum _{k\in {\mathbb {Z}}^{n}} \frac{ \partial ^{2} s^{1}
_{k}}
{\partial p'_{l} \partial p'_{j}}
(0)e^{ik \cdot q}
+
\kappa \sum _{k\in {\mathbb {Z}}^{n}} 
\frac{ \partial ^{2} h^{1}
_{k}}
{\partial p'_{l} \partial p'_{j}}
(0)
e^{1}_{k}(t)e^{ik \cdot q}
\right]\\
\quad \\
&\cdot&
\partial _{q'_{l}}
\left[{\cal{Y}}(q)+ {\cal{T}}(q')\right]\\
\quad\\
&+&
\left[
\frac{\partial ^{2}\tilde{H}^{0}}{\partial p'_{n+1} \partial p'_{j}}(0)
+
\kappa \sum _{k\in {\mathbb {Z}}^{n}} \frac{ \partial ^{2} s^{1}
_{k}}
{\partial p'_{n+1} \partial p'_{j}}
(0)e^{ik \cdot q}
+
\kappa \sum _{k\in {\mathbb {Z}}^{n}} 
\frac{ \partial ^{2} h^{1}
_{k}}
{\partial p'_{n+1} \partial p'_{j}}
(0)
e^{1}_{k}(t)e^{ik \cdot q}
\right]\\
\quad \\
&\cdot&
\partial _{q'_{n+1}}
\left[{\cal{Y}}(q)+ {\cal{T}}(q')\right]\\
\quad \\
&=&
\sum _{l=1}^{n} \left[
\frac{\partial ^{2}\tilde{H}^{0}}{\partial p'_{l} \partial p'_{j}}(0)
\partial _{q'_{l}} {\cal{Y}}(q)
+
\kappa \sum _{k\in {\mathbb {Z}}^{n}} \frac{ \partial ^{2} s^{1}
_{k}}
{\partial p'_{l} \partial p'_{j}}
(0)e^{ik \cdot q}
\partial _{q'_{l}} {\cal{Y}}(q) \right.\\
\quad \\
&+&
\kappa \sum _{k\in {\mathbb {Z}}^{n}} 
\frac{ \partial ^{2} h^{1}
_{k}}
{\partial p'_{l} \partial p'_{j}}
(0)
e^{1}_{k}(t)e^{ik \cdot q}
\partial _{q'_{l}} {\cal{Y}}(q)
+
\frac{\partial ^{2}\tilde{H}^{0}}{\partial p'_{l} \partial p'_{j}}(0)
\partial _{q'_{l}} {\cal{T}}(q')\\
\quad \\
&+& \left.
\kappa \sum _{k\in {\mathbb {Z}}^{n}} \frac{ \partial ^{2} s^{1}
_{k}}
{\partial p'_{l} \partial p'_{j}}
(0)e^{ik \cdot q}
\partial _{q'_{l}} {\cal{T}}(q')
+
\kappa \sum _{k\in {\mathbb {Z}}^{n}} 
\frac{ \partial ^{2} h^{1}
_{k}}
{\partial p'_{l} \partial p'_{j}}
(0)
e^{1}_{k}(t)e^{ik \cdot q}
\partial _{q'_{l}} {\cal{T}}(q')
\right]
\\
\quad \\
&+&
\left[
\frac{\partial ^{2}\tilde{H}^{0}}{\partial p'_{n+1} \partial p'_{j}}(0)\partial _{q'_{n+1}}
 {\cal{T}}(q')
+
\kappa \sum _{k\in {\mathbb {Z}}^{n}} \frac{ \partial ^{2} s^{1}
_{k}}
{\partial p'_{n+1} \partial p'_{j}}
(0)e^{ik \cdot q}\partial _{q'_{n+1}}
 {\cal{T}}(q')  \right.\\
\quad  \\
&+&
\left.
\kappa \sum _{k\in {\mathbb {Z}}^{n}} 
\frac{ \partial ^{2} h^{1}
_{k}}
{\partial p'_{n+1} \partial p'_{j}}
(0)
e^{1}_{k}(t)e^{ik \cdot q}\partial _{q'_{n+1}}
 {\cal{T}}(q')
\right].
\end{eqnarray*}
At this point we will pause to examine the terms obtained in the
expression above.
We begin with the first term
\begin{eqnarray*}
\sum _{l=1}^{n} \left[
\frac{\partial ^{2}\tilde{H}^{0}}{\partial p'_{l} \partial p'_{j}}(0)
\partial _{q'_{l}} {\cal{Y}}(q) \right] &=& 
\sum _{l=1}^{n} \left[
\frac{\partial ^{2}\tilde{H}^{0}}{\partial p'_{l} \partial p'_{j}}(0)
\sum _{k\in {\mathbb{Z}}^{n}} ik_{l}y_{k}e^{ik\cdot q}\right]
=
\sum _{k\in {\mathbb{Z}}^{n}}
\left[
\sum _{l=1}^{n}
ik_{l}y_{k}
\frac{\partial ^{2}\tilde{H}^{0}}{\partial p'_{l} \partial p'_{j}}(0)
\right]e^{ik\cdot q}
\end{eqnarray*}
The second term
\begin{eqnarray*}
\sum _{l=1}^{n} \left[
\kappa \sum _{k\in {\mathbb {Z}}^{n}} \frac{ \partial ^{2} s^{1}
_{k}}
{\partial p'_{l} \partial p'_{j}}
(0)e^{ik \cdot q}
\partial _{q'_{l}} {\cal{Y}}(q) \right]
 &=&
\sum _{l=1}^{n} \left[
\kappa \sum _{k\in {\mathbb {Z}}^{n}} \frac{ \partial ^{2} s^{1}
_{k}}
{\partial p'_{l} \partial p'_{j}}
(0)e^{ik \cdot q}
\sum _{k\in {\mathbb{Z}}^{n}} ik_{l}y_{k}e^{ik\cdot q}\right]
\end{eqnarray*}
\begin{eqnarray*}
&=&
\sum _{l=1}^{n} \left[
\kappa \sum _{k\in {\mathbb {Z}}^{n}}
\left(
\sum _{m\in {\mathbb {Z}}^{n}}
i
\frac{ \partial ^{2} s^{1}
_{m}}
{\partial p'_{l} \partial p'_{j}}
(0)
k_{l}y_{k}e^{im\cdot q}
\right)
e^{ik\cdot q}
\right]\\
\quad \\
&=&
\sum _{w\in {\mathbb {Z}}^{n}}
 \left[
\kappa \sum _{l=1}^{n}
\left(
\sum _{m\in {\mathbb {Z}}^{n}}
i
\frac{ \partial ^{2} s^{1}
_{m}}
{\partial p'_{l} \partial p'_{j}}
(0)
(w-m)_{l}y_{w-m}
\right)
\right]
e^{iw\cdot q}
.
\end{eqnarray*}
The third term
\begin{eqnarray*}
\sum _{l=1}^{n} \left[
\kappa \sum _{k\in {\mathbb {Z}}^{n}} 
\frac{ \partial ^{2} h^{1}
_{k}}
{\partial p'_{l} \partial p'_{j}}
(0)
e^{1}_{k}(t)e^{ik \cdot q}
\partial _{q'_{l}} {\cal{Y}}(q) \right]
 &=&
\sum _{l=1}^{n} \left[
\kappa \sum _{k\in {\mathbb {Z}}^{n}} 
\frac{ \partial ^{2} h^{1}
_{k}}
{\partial p'_{l} \partial p'_{j}}
(0)
e^{1}_{k}(t)e^{ik \cdot q}
\sum _{k\in {\mathbb{Z}}^{n}} ik_{l}y_{k}e^{ik\cdot q}
 \right]
\end{eqnarray*}
\begin{eqnarray*}
&=&
\sum _{l=1}^{n} \left[
\kappa \sum _{k\in {\mathbb {Z}}^{n}} 
\left(
\sum _{m\in {\mathbb {Z}}^{n}}i
\frac{ \partial ^{2} h^{1}
_{m}}
{\partial p'_{l} \partial p'_{j}}
(0)
e^{1}_{m}(t)
k_{l}y_{k} e^{im\cdot q}
\right)
e^{ik\cdot q}
\right]
\\
\quad \\
&=&
\sum _{w\in {\mathbb {Z}}^{n}} 
 \left[\sum _{m\in {\mathbb {Z}}^{n}}
\left( \sum _{l=1}^{n}
i\kappa
\frac{ \partial ^{2} h^{1}
_{m}}
{\partial p'_{l} \partial p'_{j}}
(0)
e^{1}_{m}(t)
(w-m)_{l}y_{w-m} 
\right)
\right]
e^{iw\cdot q}
.
\end{eqnarray*}
The fourth term
\begin{eqnarray*}
\sum _{l=1}^{n} \left[
\frac{\partial ^{2}\tilde{H}^{0}}{\partial p'_{l} \partial p'_{j}}(0)
\partial _{q'_{l}} {\cal{T}}(q') \right]
&=&
\sum _{l=1}^{n} \left[
\frac{\partial ^{2}\tilde{H}^{0}}{\partial p'_{l} \partial p'_{j}}(0)
\sum _{k\in {\mathbb {Z}}^{n}} 
ik_{l}x_{k}(t)
e^{ik\cdot q}
 \right]
=
\sum _{k\in {\mathbb {Z}}^{n}} 
\left[
\sum _{l=1}^{n}
ik_{l}x_{k}(t)
\frac{\partial ^{2}\tilde{H}^{0}}{\partial p'_{l} \partial p'_{j}}(0)
 \right]
e^{ik\cdot q}.
\end{eqnarray*}
The fifth term
\begin{eqnarray*}
\sum _{l=1}^{n} \left[
\kappa \sum _{k\in {\mathbb {Z}}^{n}} \frac{ \partial ^{2} s^{1}
_{k}}
{\partial p'_{l} \partial p'_{j}}
(0)e^{ik \cdot q}
\partial _{q'_{l}} {\cal{T}}(q')\right]
&=&
\sum _{l=1}^{n} \left[
\kappa \sum _{k\in {\mathbb {Z}}^{n}} \frac{ \partial ^{2} s^{1}
_{k}}
{\partial p'_{l} \partial p'_{j}}
(0)e^{ik \cdot q}
\sum _{k\in {\mathbb {Z}}^{n}} 
ik_{l}x_{k}(t)
e^{ik\cdot q}
\right]
\end{eqnarray*}
\begin{eqnarray*}
&=&
\sum _{l=1}^{n} \left[
\kappa \sum _{k\in {\mathbb {Z}}^{n}}
\left(
 \sum _{m\in {\mathbb {Z}}^{n}}
i
\frac{ \partial ^{2} s^{1}
_{m}}
{\partial p'_{l} \partial p'_{j}}
(0)
k_{l}x_{k}(t)e^{im\cdot q}
\right)
e^{ik\cdot q}
\right]
\\
\quad \\
&=&
\sum _{w\in {\mathbb {Z}}^{n}}
 \left[ \sum _{m\in {\mathbb {Z}}^{n}}
\left( \sum _{l=1}^{n}
i\kappa
\frac{ \partial ^{2} s^{1}
_{m}}
{\partial p'_{l} \partial p'_{j}}
(0)
(w-m)_{l}x_{w-m}(t)
\right)
\right]
e^{iw\cdot q}
.
\end{eqnarray*}
The sixth term
\begin{eqnarray*}
\sum _{l=1}^{n} \left[
\kappa \sum _{k\in {\mathbb {Z}}^{n}} 
\frac{ \partial ^{2} h^{1}
_{k}}
{\partial p'_{l} \partial p'_{j}}
(0)
e^{1}_{k}(t)e^{ik \cdot q}
\partial _{q'_{l}} {\cal{T}}(q')
\right]
&=&
\sum _{l=1}^{n} \left[
\kappa \sum _{k\in {\mathbb {Z}}^{n}} 
\frac{ \partial ^{2} h^{1}
_{k}}
{\partial p'_{l} \partial p'_{j}}
(0)
e^{1}_{k}(t)e^{ik \cdot q}
\sum _{k\in {\mathbb {Z}}^{n}} 
ik_{l}x_{k}(t)
e^{ik\cdot q}
\right]
\end{eqnarray*}
\begin{eqnarray*}
&=&
\sum _{l=1}^{n} \left[
\kappa \sum _{k\in {\mathbb {Z}}^{n}} 
\left(
 \sum _{m\in {\mathbb {Z}}^{n}} i
\frac{ \partial ^{2} h^{1}
_{m}}
{\partial p'_{l} \partial p'_{j}}
(0)
e^{1}_{m}(t)
k_{l}x_{k}(t)e^{im\cdot q}
\right)
e^{ik\cdot q}
\right]
\\
\quad \\
&=&
\sum _{w\in {\mathbb {Z}}^{n}} 
 \left[\sum _{m\in {\mathbb {Z}}^{n}}
\left(  \sum _{l=1}^{n}
 i\kappa
\frac{ \partial ^{2} h^{1}
_{m}}
{\partial p'_{l} \partial p'_{j}}
(0)
e^{1}_{m}(t)
(w-m)_{l}x_{w-m}(t)
\right)
\right]
e^{iw\cdot q}
.
\end{eqnarray*}
The seventh term
\begin{eqnarray*}
\frac{\partial ^{2}\tilde{H}^{0}}{\partial p'_{n+1} \partial p'_{j}}(0)\partial _{q'_{n+1}}
 {\cal{T}}(q') &=&
\frac{\partial ^{2}\tilde{H}^{0}}{\partial p'_{n+1} \partial p'_{j}}(0)
\sum _{k\in {\mathbb {Z}}^{n}} 
\frac{dx_{k}}{dt}(t)
e^{ik\cdot q}
=
\sum _{k\in {\mathbb {Z}}^{n}} 
\left(
\frac{\partial ^{2}\tilde{H}^{0}}{\partial p'_{n+1} \partial p'_{j}}(0)
\frac{dx_{k}}{dt}(t)
\right)
e^{ik\cdot q}.
\end{eqnarray*}
The eight term
\begin{eqnarray*}
\kappa &\sum _{k\in {\mathbb {Z}}^{n}} &\frac{ \partial ^{2} s^{1}
_{k}}
{\partial p'_{n+1} \partial p'_{j}}
(0)e^{ik \cdot q}\partial _{q'_{n+1}}
 {\cal{T}}(q') 
=
\kappa \sum _{k\in {\mathbb {Z}}^{n}} \frac{ \partial ^{2} s^{1}
_{k}}
{\partial p'_{n+1} \partial p'_{j}}
(0)
e^{ik\cdot q}
\sum _{k\in {\mathbb {Z}}^{n}} 
\frac{dx_{k}}{dt}(t)
e^{ik\cdot q}\\
\quad\\
&=&
\sum _{k\in {\mathbb {Z}}^{n}} 
\left(
\sum _{m\in {\mathbb {Z}}^{n}} 
\kappa
\frac{ \partial ^{2} s^{1}
_{m}}
{\partial p'_{n+1} \partial p'_{j}}
(0)
\frac{dx_{k}}{dt}(t)e^{im\cdot q}
\right)
e^{ik\cdot q}
=
\sum _{w\in {\mathbb {Z}}^{n}} 
\left(
\sum _{m\in {\mathbb {Z}}^{n}} 
\kappa
\frac{ \partial ^{2} s^{1}
_{m}}
{\partial p'_{n+1} \partial p'_{j}}
(0)
\frac{dx_{w-m}}{dt}(t)
\right)
e^{iw\cdot q}.
\end{eqnarray*}
The nineth term
\begin{eqnarray*}
\kappa &\sum _{k\in {\mathbb {Z}}^{n}} &
\frac{ \partial ^{2} h^{1}
_{k}}
{\partial p'_{n+1} \partial p'_{j}}
(0)
e^{1}_{k}(t)e^{ik \cdot q}\partial _{q'_{n+1}}
 {\cal{T}}(q')
=
\kappa \sum _{k\in {\mathbb {Z}}^{n}} 
\frac{ \partial ^{2} h^{1}
_{k}}
{\partial p'_{n+1} \partial p'_{j}}
(0)
e^{1}_{k}(t)e^{ik \cdot q}
\sum _{k\in {\mathbb {Z}}^{n}} 
\frac{dx_{k}}{dt}(t)
e^{ik\cdot q}\\
\quad\\
&=&
\sum _{k\in {\mathbb {Z}}^{n}} \left(
\sum _{m\in {\mathbb {Z}}^{n}} 
\kappa
\frac{ \partial ^{2} h^{1}
_{m}}
{\partial p'_{n+1} \partial p'_{j}}
(0)
e^{1}_{m}(t)
\frac{dx_{k}}{dt}(t)e^{im\cdot q}
\right)
e^{ik\cdot q}
=
\sum _{w\in {\mathbb {Z}}^{n}} \left(
\sum _{m\in {\mathbb {Z}}^{n}} 
\kappa
\frac{ \partial ^{2} h^{1}
_{m}}
{\partial p'_{n+1} \partial p'_{j}}
(0)
e^{1}_{m}(t)
\frac{dx_{w-m}}{dt}(t)
\right)
e^{iw\cdot q}
.
\end{eqnarray*}
We finally write
\begin{eqnarray*}
(C(q') \cdot \partial_{q'}X)_{j}&=&
\sum _{k\in {\mathbb {Z}}^{n}} 
\left[
\sum _{l=1}^{n}
ik_{l}y_{k}
\frac{\partial ^{2}\tilde{H}^{0}}{\partial p'_{l} \partial p'_{j}}(0)
+
\kappa \sum _{l=1}^{n}
\left(
\sum _{m\in {\mathbb {Z}}^{n}}
i
\frac{ \partial ^{2} s^{1}
_{m}}
{\partial p'_{l} \partial p'_{j}}
(0)
(k-m)_{l}y_{k-m}
\right)
\right]
e^{ik\cdot q}
\end{eqnarray*}
\begin{eqnarray*}
&+&
\sum _{k\in {\mathbb {Z}}^{n}} 
\left[
\kappa \sum _{l=1}^{n}
\left(
\sum _{m\in {\mathbb {Z}}^{n}}i
\frac{ \partial ^{2} h^{1}
_{m}}
{\partial p'_{l} \partial p'_{j}}
(0)
e^{1}_{m}(t)
(k-m)_{l}y_{k-m} 
\right)
+
\sum _{l=1}^{n}
ik_{l}x_{k}(t)
\frac{\partial ^{2}\tilde{H}^{0}}{\partial p'_{l} \partial p'_{j}}(0)
\right.
\\
\quad \\
&+&
\kappa \sum _{l=1}^{n}
\left(
 \sum _{m\in {\mathbb {Z}}^{n}}
i
\frac{ \partial ^{2} s^{1}
_{m}}
{\partial p'_{l} \partial p'_{j}}
(0)
(k-m)_{l}x_{k-m}(t)
\right)\\
\quad \\
&+&
\kappa  \sum _{l=1}^{n}
\left(
 \sum _{m\in {\mathbb {Z}}^{n}} i
\frac{ \partial ^{2} h^{1}
_{m}}
{\partial p'_{l} \partial p'_{j}}
(0)
e^{1}_{m}(t)
(k-m)_{l}x_{k-m}(t)
\right)\\
\quad \\
&+&
\frac{\partial ^{2}\tilde{H}^{0}}{\partial p'_{n+1} \partial p'_{j}}(0)
\frac{dx_{k}}{dt}(t)
+
\sum _{m\in {\mathbb {Z}}^{n}} 
\kappa
\frac{ \partial ^{2} s^{1}
_{m}}
{\partial p'_{n+1} \partial p'_{j}}
(0)
\frac{dx_{k-m}}{dt}(t)\\
\quad \\
&+&
\left.
\sum _{m\in {\mathbb {Z}}^{n}} 
\kappa
\frac{ \partial ^{2} h^{1}
_{m}}
{\partial p'_{n+1} \partial p'_{j}}
(0)
e^{1}_{m}(t)
\frac{dx_{k-m}}{dt}(t)
\right]
e^{ik\cdot q}
\end{eqnarray*}
and
\begin{eqnarray*}
&\quad& (C(q') \cdot \partial_{q'}X)_{j}
-
(\overline{\overline{C}} \cdot \overline{\overline{\partial_{q'}X}})_{j}
=
\sum _{k\in {\mathbb {Z}}^{n} \backslash 0} 
\left[
\sum _{l=1}^{n}
ik_{l}y_{k}
\frac{\partial ^{2}\tilde{H}^{0}}{\partial p'_{l} \partial p'_{j}}(0)
\right]e^{ik\cdot q}
\end{eqnarray*}
\begin{eqnarray*}
&+&
\kappa 
\sum _{k\in {\mathbb {Z}}^{n}}
\sum _{l=1}^{n}
\left(
\sum _{m\in {\mathbb {Z}}^{n}}
i
\frac{ \partial ^{2} s^{1}
_{m}}
{\partial p'_{l} \partial p'_{j}}
(0)
(k-m)_{l}y_{k-m}
-i\frac{\partial ^{2}s_{k}^{1}}{\partial p'_{l}
\partial p'_{j}}(0)k_{l}y_{k}
\right)
e^{ik\cdot q}\\
\quad \\
&+&
\sum _{k\in {\mathbb {Z}}^{n}} 
\left[
\kappa \sum _{l=1}^{n}
\left(
\sum _{m\in {\mathbb {Z}}^{n}}i
\frac{ \partial ^{2} h^{1}
_{m}}
{\partial p'_{l} \partial p'_{j}}
(0)
e^{1}_{m}(t)
(k-m)_{l}y_{k-m} 
\right)
+
\sum _{l=1}^{n}
ik_{l}x_{k}(t)
\frac{\partial ^{2}\tilde{H}^{0}}{\partial p'_{l} \partial p'_{j}}(0)
\right.
\\
\quad \\
&+&
\kappa \sum _{l=1}^{n}
\left(
 \sum _{m\in {\mathbb {Z}}^{n}}
i
\frac{ \partial ^{2} s^{1}
_{m}}
{\partial p'_{l} \partial p'_{j}}
(0)
(k-m)_{l}x_{k-m}(t)
\right)\\
\quad \\
&+&
\kappa  \sum _{l=1}^{n}
\left(
 \sum _{m\in {\mathbb {Z}}^{n}} i
\frac{ \partial ^{2} h^{1}
_{m}}
{\partial p'_{l} \partial p'_{j}}
(0)
e^{1}_{m}(t)
(k-m)_{l}x_{k-m}(t)
\right)\\
\quad \\
&+&
\frac{\partial ^{2}\tilde{H}^{0}}{\partial p'_{n+1} \partial p'_{j}}(0)
\frac{dx_{k}}{dt}(t)
+
\sum _{m\in {\mathbb {Z}}^{n}} 
\kappa
\frac{ \partial ^{2} s^{1}
_{m}}
{\partial p'_{n+1} \partial p'_{j}}
(0)
\frac{dx_{k-m}}{dt}(t)
\end{eqnarray*}
\begin{eqnarray}
\label{eqn1002}
+
\left.
\sum _{m\in {\mathbb {Z}}^{n}} 
\kappa
\frac{ \partial ^{2} h^{1}
_{m}}
{\partial p'_{n+1} \partial p'_{j}}
(0)
e^{1}_{m}(t)
\frac{dx_{k-m}}{dt}(t)
\right]
e^{ik\cdot q}
. \quad \quad \quad \quad \quad \quad \quad \quad
\end{eqnarray}
Next we write out  the expression
$
\left(
C(q') \cdot \xi -\overline{\overline{C}} \cdot \xi \right)_{j}.
$
\begin{eqnarray*}
 \left(C(q') \cdot \xi \right)_{j} &=&
\sum _{l=1}^{n+1} C_{j,l}(q') \xi _{l}
=\sum _{l=1}^{n+1} \left[
\frac{\partial ^{2}\tilde{H}^{0}}{\partial p'_{l} \partial p'_{j}}(0)
+
\kappa  \frac{\partial ^{2} H^{1}}{\partial p'_{l} \partial p'_{j}}(0,q')
\right]
\xi_{l}\\
\quad\\
&=&
\sum _{l=1}^{n+1} \left[
\frac{\partial ^{2}\tilde{H}^{0}}{\partial p'_{l} \partial p'_{j}}(0)
+
\kappa \sum _{k\in {\mathbb {Z}}^{n}} \frac{ \partial ^{2} s^{1}
_{k}}
{\partial p'_{l} \partial p'_{j}}
(0)e^{ik \cdot q}
+
\kappa \sum _{k\in {\mathbb {Z}}^{n}} 
\frac{ \partial ^{2} h^{1}
_{k}}
{\partial p'_{l} \partial p'_{j}}
(0)
e^{1}_{k}(t)e^{ik \cdot q}
\right]\xi_{l}
\end{eqnarray*}
and
\begin{eqnarray*}
\left(
C(q') \cdot \xi -\overline{\overline{C}} \cdot \xi \right)_{j} &=&
\sum _{l=1}^{n+1} \left[
\kappa \xi_{l} \sum _{k\in {\mathbb {Z}}^{n} \backslash 0} \frac{ \partial ^{2} s^{1}
_{k}}
{\partial p'_{l} \partial p'_{j}}
(0)e^{ik \cdot q}
+
\kappa \xi_{l}\sum _{k\in {\mathbb {Z}}^{n}} 
\frac{ \partial ^{2} h^{1}
_{k}}
{\partial p'_{l} \partial p'_{j}}
(0)
e^{1}_{k}(t)e^{ik \cdot q}
\right]
\end{eqnarray*}
\begin{eqnarray}
\label{eqn1003}
=
\sum _{k\in {\mathbb {Z}}^{n} \backslash 0}
\left(
\sum _{l=1}^{n+1}
\kappa \xi_{l}
\frac{ \partial ^{2} s^{1}
_{k}}
{\partial p'_{l} \partial p'_{j}}
(0)
\right)
e^{ik \cdot q}
+
\sum _{k\in {\mathbb {Z}}^{n}} 
\left(
\sum _{l=1}^{n+1}
\kappa \xi_{l}
\frac{ \partial ^{2} h^{1}
_{k}}
{\partial p'_{l} \partial p'_{j}}
(0)
\right)
e^{1}_{k}(t)e^{ik \cdot q}.
\end{eqnarray}
We have thus separated the right hand side of
\begin{eqnarray}
\label{eqn1000}
\lambda \cdot \partial _{q'}Y = \beta,
\end{eqnarray}
into a quasiperiodic and a exponential-order-with-respect-to-time  part. We first solve the
quasiperiodic part.
Assume
\begin{eqnarray*}
Y_{j}(q')&=& {\cal{S}}_{j}(q) + {\cal{F}}_{j}(q'),
\quad
\quad
{\cal{S}}_{j}(q)=
\sum _{k\in {\mathbb{Z}}^{n}}{\cal{S}}_{k,j}e^{ik\cdot q},
\quad
\quad
{\cal{F}}_{j}(q')=
\sum _{k\in {\mathbb{Z}}^{n}}
{\cal{F}}_{k,j}(t)e^{ik\cdot q}.
\end{eqnarray*}
The j th component of $(\ref{eqn1000})$ is given by
\begin{eqnarray}
\label{eqn1000j}
\lambda \cdot \partial _{q'}Y_{j}(q') =
\sum _{i=1}^{n+1}
\lambda _{i} \partial _{q'_{i}} Y_{j}(q') = \beta _{j}.
\end{eqnarray}
From $(\ref{eqn1001})$, $(\ref{eqn1002})$
and $(\ref{eqn1003})$ we obtain the quasiperiodic part of $\beta_{j}$
\begin{eqnarray*}
\beta_{j}^{Q}(q)&=& \kappa
\sum _{k\in {\mathbb{Z}}^{n}\backslash 0} \frac{\partial s^{1}_{k}}{\partial p'_{j}}(0)
e^{ik\cdot q}+
\sum _{k\in {\mathbb {Z}}^{n} \backslash 0} 
\left[
\sum _{l=1}^{n}
ik_{l}y_{k}
\frac{\partial ^{2}\tilde{H}^{0}}{\partial p'_{l} \partial p'_{j}}(0)\right]
e^{ik\cdot q}\\
\quad\\
&\quad&+
\kappa 
\sum _{k\in {\mathbb {Z}}^{n}}
\sum _{l=1}^{n}
\left(
\sum _{m\in {\mathbb {Z}}^{n}}
i
\frac{ \partial ^{2} s^{1}
_{m}}
{\partial p'_{l} \partial p'_{j}}
(0)
k_{l}y_{k}e^{im\cdot q}e^{ik\cdot q}
-
i\frac{\partial ^{2}s_{-k}^{1}}{\partial p'_{l} \partial p'_{j}}(0)k_{l}y_{k}
\right)
\\
\quad \\
&\quad&+
\sum _{k\in {\mathbb {Z}}^{n} \backslash 0}
\left(
\sum _{l=1}^{n+1}
\kappa \xi_{l}
\frac{ \partial ^{2} s^{1}
_{k}}
{\partial p'_{l} \partial p'_{j}}
(0)
\right)
e^{ik \cdot q}\\
\quad\\
&=&
\sum _{k\in {\mathbb {Z}}^{n} \backslash 0}
\left[
\kappa
\frac{\partial s^{1}_{k}}{\partial p'_{j}}(0)
+
\sum _{l=1}^{n}
ik_{l}y_{k}
\frac{\partial ^{2}\tilde{H}^{0}}{\partial p'_{l} \partial p'_{j}}(0)
\right.\\
\quad\\
&\quad&+
\kappa
\sum _{l=1}^{n}
\sum _{
\begin{array}{c}
m\in {\mathbb {Z}}^{n}\\
m\neq -k
\end{array}
}
i
\frac{ \partial ^{2} s^{1}
_{m}}
{\partial p'_{l} \partial p'_{j}}
(0)
k_{l}y_{k}e^{im\cdot q}
+
\left.
\sum _{l=1}^{n+1}
\kappa \xi_{l}
\frac{ \partial ^{2} s^{1}
_{k}}
{\partial p'_{l} \partial p'_{j}}
(0)
\right]
e^{ik \cdot q}\\
\quad\\
&=&
\sum _{h\in {\mathbb{Z}}^{n} \backslash 0}
\left[\kappa
\frac{\partial s^{1}_{h}}{\partial p'_{j}}(0)
+
\sum _{l=1}^{n}
ih_{l}y_{h}
\frac{\partial ^{2}\tilde{H}^{0}}{\partial p'_{l} \partial p'_{j}}(0)
\right.\\
\quad\\
&\quad&+
\kappa
\sum _{l=1}^{n}
\sum _{m\in {\mathbb{Z}}^{n}}
i
\frac{ \partial ^{2} s^{1}
_{m}}
{\partial p'_{l} \partial p'_{j}}
(0)
(h-m)_{l}y_{h-m}
+
\left.
\sum _{l=1}^{n+1}
\kappa \xi_{l}
\frac{ \partial ^{2} s^{1}
_{h}}
{\partial p'_{l} \partial p'_{j}}
(0)
\right]
e^{ih \cdot q}\\
\quad\\
&=&
\sum _{k\in {\mathbb {Z}}^{n} \backslash 0}
\beta^{Q}_{j,k}
e^{ik \cdot q},
\end{eqnarray*}
where we have set
\begin{eqnarray*}
\beta^{Q}_{j,k} &=&
\kappa
\frac{\partial s^{1}_{h}}{\partial p'_{j}}(0)
+
\sum _{l=1}^{n}
ih_{l}y_{h}
\frac{\partial ^{2}\tilde{H}^{0}}{\partial p'_{l} \partial p'_{j}}(0)
+
\kappa
\sum _{l=1}^{n}
\sum _{m\in {\mathbb{Z}}^{n}}
i
\frac{ \partial ^{2} s^{1}
_{m}}
{\partial p'_{l} \partial p'_{j}}
(0)
(h-m)_{l}y_{h-m}
+
\sum _{l=1}^{n+1}
\kappa \xi_{l}
\frac{ \partial ^{2} s^{1}
_{h}}
{\partial p'_{l} \partial p'_{j}}
(0).
\end{eqnarray*}
The quasiperiodic part of $(\ref{eqn1000})$ is
\begin{eqnarray}
\label{eqn1004}
\tilde{\lambda} \cdot \partial _{q}
\left(
{\cal{S}}_{j}(q) \right)
=
\beta^{Q}_{j}(q)
=
\sum _{k\in {\mathbb {Z}}^{n} \backslash 0}
\beta^{Q}_{j,k}
e^{ik \cdot q}
\in {\cal{A}}_{\rho -2\delta}
.
\end{eqnarray}
Since we assume $\tilde{\lambda} \in \Omega _{\Gamma}$ and since the right hand side
of $(\ref{eqn1004})$ has zero average, by lemma $\ref{lemmaqb}$,
 for some positive $\delta <\rho$
, $(\ref{eqn1004})$ admits a unique solution
\begin{eqnarray*}
{\cal{S}}_{j}(q)=
\sum _{k\in {\mathbb{Z}}^{n}}{\cal{S}}_{k,j}e^{ik\cdot q} \in {\cal{A}}_{\rho -3\delta},
\end{eqnarray*}
with $\overline{{\cal{S}}_{j}} =0$ and one has the estimates
\begin{eqnarray*}
\| {\cal{S}}_{j}\|_{\rho -3\delta} \leq
\frac{\varpi}{\Gamma \delta ^{2n}}
\| {\beta}^{Q}_{j}\|_{\rho-2\delta},
\quad \quad
\left \|\frac{\partial {\cal{S}}_{j}}{\partial q}\right\|_{\rho -3\delta} \leq
\frac{\varpi}{\Gamma \delta ^{2n+1}}
\| \beta ^{Q}_{j}\|_{\rho-2\delta},
\end{eqnarray*}
where
$
\varpi = 2^{4n+1} \left(
\frac{n+1}{e} \right)^{n+1}$.
Next we want to solve the
exponential-order-with-respect-to-time  part of $(\ref{eqn1000})$.
From $(\ref{eqn1001})$, $(\ref{eqn1002})$
and $(\ref{eqn1003})$ we obtain the
exponential-time dependent part of $\beta_{j}$
\begin{eqnarray*}
\beta ^{E}_{j}(q')
&=&
\sum _{k\in {\mathbb{Z}}^{n}}
\left[
\kappa
\frac{\partial h^{1}_{k}}{\partial p'_{j}} (0)
 e^{1}_{k}(t)
\right.
\\
\quad\\
&+&
\kappa \sum _{l=1}^{n}
\left(
\sum _{m\in {\mathbb {Z}}^{n}}i
\frac{ \partial ^{2} h^{1}
_{m}}
{\partial p'_{l} \partial p'_{j}}
(0)
e^{1}_{m}(t)
(k-m)_{l}y_{k-m}
\right)
+
\sum _{l=1}^{n}
ik_{l}x_{k}(t)
\frac{\partial ^{2}\tilde{H}^{0}}{\partial p'_{l} \partial p'_{j}}(0)
\\
\quad \\
&+&
\kappa \sum _{l=1}^{n}
\left(
 \sum _{m\in {\mathbb {Z}}^{n}}
i
\frac{ \partial ^{2} s^{1}
_{m}}
{\partial p'_{l} \partial p'_{j}}
(0)
(k-m)_{l}x_{k-m}(t)
\right)\\
\quad \\
&+&
\kappa  \sum _{l=1}^{n}
\left(
 \sum _{m\in {\mathbb {Z}}^{n}} i
\frac{ \partial ^{2} h^{1}
_{m}}
{\partial p'_{l} \partial p'_{j}}
(0)
e^{1}_{m}(t)
(k-m)_{l}x_{k-m}(t)
\right)\\
\quad \\
&+&
\frac{\partial ^{2}\tilde{H}^{0}}{\partial p'_{n+1} \partial p'_{j}}(0)
\frac{dx_{k}}{dt}(t)
+
\sum _{m\in {\mathbb {Z}}^{n}}
\kappa
\frac{ \partial ^{2} s^{1}
_{m}}
{\partial p'_{n+1} \partial p'_{j}}
(0)
\frac{dx_{k-m}}{dt}(t)\\
\quad\\
&+&
\sum _{m\in {\mathbb {Z}}^{n}}
\kappa
\frac{ \partial ^{2} h^{1}
_{m}}
{\partial p'_{n+1} \partial p'_{j}}
(0)
e^{1}_{m}(t)
\frac{dx_{k-m}}{dt}(t)
\\
\quad\\
&+&
\left.
\left(
\sum _{l=1}^{n+1}
\kappa \xi_{l}
\frac{ \partial ^{2} h^{1}
_{k}}
{\partial p'_{l} \partial p'_{j}}
(0)
\right)
e^{1}_{k}(t)
\right]
e^{ik \cdot q}
=
\sum _{k\in {\mathbb{Z}}^{n}}
\beta ^{E}_{jk}(t)
e^{ik \cdot q}
.
\end{eqnarray*}
Therefore, the exponential-order-with-respect-to-time dependent part of
$(\ref{eqn1000})$ is
\begin{eqnarray}
\label{eqnexptimesolve}
\lambda \cdot
\partial _{q'}{\cal{F}}_{j}(q')
= \beta ^{E}_{j}(q')=\sum _{k\in {\mathbb{Z}}^{n}}
\beta ^{E}_{jk}(t)
e^{ik \cdot q}
\in {\cal{A}}_{\rho -2\delta, \sigma -2\delta} .
\end{eqnarray}
To solve $(\ref{eqnexptimesolve})$ we must
find an exponential bound for the time dependent Fourier coefficients
 $\beta ^{E}_{jk}(t)$. This exponential bound can be found
by bounding each of the nine time dependent terms that make up
$\beta ^{E}_{jk}(t)$.
Note, to bound the terms
involving sums over $m$ we use the bounds found
in appendix $\ref{lastappendix}$. For the first term
\begin{eqnarray*}
\left|
 \frac{\partial h^{1}_{k}}{\partial p'_{j}} (0)
 e^{1}_{k}(t) 
\right|
&\leq&
\frac{\kappa}{\delta} C_{1}e^{-(\nu -\varepsilon)t_{R}}e^{-|k|\rho}, \quad 
0< c_{1} < t_{R} < \infty,
\end{eqnarray*}
\begin{eqnarray*}
\left|
 \frac{\partial h^{1}_{k}}{\partial p'_{j}} (0)
 e^{1}_{k}(t) 
\right|
&\leq&
\frac{\kappa}{\delta}
 C_{2}e^{(\mu -\varepsilon)t_{R}}e^{-|k|\rho}, \quad 
-\infty < t_{R} < c_{2} <0 .
\end{eqnarray*}
For the second term we refer to $(\ref{ykestimate})$
\begin{eqnarray*}
\left|
\kappa 
\sum _{l=1}^{n}
\left(
\sum _{m\in {\mathbb {Z}}^{n}}i
\frac{ \partial ^{2} h^{1}
_{m}}
{\partial p'_{l} \partial p'_{j}}
(0)
e^{1}_{m}(t)
(k-m)_{l}y_{k-m}
\right)
\right|
\end{eqnarray*}
\begin{eqnarray*}
&\leq&
\sum _{l=1}^{n}
\sum _{m\in {\mathbb {Z}}^{n}}
\kappa 
\frac{\|h^{1}_{m}\|_{\rho}}{\delta ^{2}}
\left|
e^{1}_{m}(t)
(k-m)_{l}y_{k-m}
\right|\\
\quad\\
&\leq&
\sum _{m\in {\mathbb {Z}}^{n}}
\kappa 
\frac{\|h^{1}_{m}\|_{\rho}}{\delta ^{2}}
\frac{\kappa}{\Gamma}\left( \frac{n}{e\delta} \right)^{n}
\|G(0,q) -\overline{\overline{G}}(0) \|_{\rho}e^{-|k-m|(\rho -\delta)}
\left|
e^{1}_{m}(t)
\right|
\left|
k-m
\right|\\
\quad\\
&\leq&
\sum _{m\in {\mathbb {Z}}^{n}}
\kappa 
\frac{\|h^{1}_{m}\|_{\rho}}{\delta ^{2}}
\frac{\kappa}{\Gamma}\left( \frac{n}{e\delta} \right)^{n}
\|G(0,q) -\overline{\overline{G}}(0) \|_{\rho}e^{-|k-m|(\rho -\delta)}
\frac{e^{|k-m|\delta}}{e\delta}
\left|
e^{1}_{m}(t)
\right|\\
\quad\\
&\leq&
\sum _{m\in {\mathbb {Z}}^{n}}
\kappa 
\frac{1}{e\delta ^{3}}
\frac{\kappa}{\Gamma}\left( \frac{n}{e\delta} \right)^{n}
\|G(0,q) -\overline{\overline{G}}(0) \|_{\rho}e^{-|k-m|(\rho -2\delta)}
\|h^{1}_{m}\|_{\rho}\left|
e^{1}_{m}(t)
\right|
\end{eqnarray*}
so that
\begin{eqnarray*}
\left|
\kappa 
\sum _{l=1}^{n}
\left(
\sum _{m\in {\mathbb {Z}}^{n}}i
\frac{ \partial ^{2} h^{1}
_{m}}
{\partial p'_{l} \partial p'_{j}}
(0)
e^{1}_{m}(t)
(k-m)_{l}y_{k-m} 
\right)
\right|
\end{eqnarray*}
\begin{eqnarray*}
&\leq&
\sum _{m\in {\mathbb {Z}}^{n}}
\left[
\kappa 
\frac{1}{e\delta ^{3}}
\frac{\kappa}{\Gamma}\left( \frac{n}{e\delta} \right)^{n}
\|G(0,q) -\overline{\overline{G}}(0) \|_{\rho}
C_{1}
\right]
e^{-|m|\rho}
e^{-|k-m|(\rho -2\delta)}
e^{-(\nu -\frac{\delta}{2})t_{R}}
,\\
\quad\\
&\leq&
\left[
\kappa 
\frac{3^{n}}{e\delta ^{3}}
\frac{\kappa}{\Gamma}\left( \frac{n}{e\delta} \right)^{n}
\left(
\frac{1}{1-e^{-2\delta}}\right)^{n}
\|G(0,q) -\overline{\overline{G}}(0) \|_{\rho}
C_{1}
\right]
e^{-|k|(\rho -2\delta)}
e^{-(\nu -\frac{\delta}{2})t_{R}}\\
\quad\\
&\quad&  
0< c_{1} < t_{R} < \infty,
\end{eqnarray*}
\begin{eqnarray*}
\left|
\kappa 
\sum _{l=1}^{n}
\left(
\sum _{m\in {\mathbb {Z}}^{n}}i
\frac{ \partial ^{2} h^{1}
_{m}}
{\partial p'_{l} \partial p'_{j}}
(0)
e^{1}_{m}(t)
(k-m)_{l}y_{k-m} 
\right)
\right|
\end{eqnarray*}
\begin{eqnarray*}
&\leq&
\sum _{m\in {\mathbb {Z}}^{n}}
\left[
\kappa 
\frac{1}{e\delta ^{3}}
\frac{\kappa}{\Gamma}\left( \frac{n}{e\delta} \right)^{n}
\|G(0,q) -\overline{\overline{G}}(0) \|_{\rho}
C_{2}
\right]
e^{-|m|\rho}
e^{-|k-m|(\rho -2\delta)}
e^{-(\mu -\frac{\delta}{2})t_{R}}\\
\quad\\
&\leq&
\left[
\kappa 
\frac{3^{n}}{e\delta ^{3}}
\frac{\kappa}{\Gamma}\left( \frac{n}{e\delta} \right)^{n}
\left(
\frac{1}{1-e^{-2\delta}}\right)^{n}
\|G(0,q) -\overline{\overline{G}}(0) \|_{\rho}
C_{2}
\right]
e^{-|k|(\rho -2\delta)}
e^{-(\mu -\frac{\delta}{2})t_{R}}
,\\
\quad\\
&\quad&  
-\infty  < t_{R} < c_{2} <0.
\end{eqnarray*}
For the third term
\begin{eqnarray*}
\left|
\sum _{l=1}^{n}
ik_{l}x_{k}(t)
\frac{\partial ^{2}\tilde{H}^{0}}{\partial p'_{l} \partial p'_{j}}(0)
\right|
&\leq&
\sum _{l=1}^{n}
\left(
\frac{1}{e\delta}
\right)e^{|k|\delta}
\left|
\frac{\partial ^{2}\tilde{H}^{0}}{\partial p'_{l} \partial p'_{j}}(0)
\right|
|x_{k}(t)|,
\end{eqnarray*}
so that
\begin{eqnarray*}
\left|
\sum _{l=1}^{n}
ik_{l}x_{k}(t)
\frac{\partial ^{2}\tilde{H}^{0}}{\partial p'_{l} \partial p'_{j}}(0)
\right|
\end{eqnarray*}
\begin{eqnarray*}
&\leq&
\left[
\sum _{l=1}^{n}
\left(
\frac{1}{e\delta}
\right)e^{|k|\delta}
\left|
\frac{\partial ^{2}\tilde{H}^{0}}{\partial p'_{l} \partial p'_{j}}(0)
\right|
 \frac{8\kappa }{\delta \gamma \sqrt{2\pi}}
\left[ C_{3}{c_{3}}
+\frac{2}{\delta}   C_{3}e^{-\frac{\delta}{2}c_{3}}
\right]
\right]
e^{-(\nu -\frac{\delta}{2})t_{R}}e^{-|k|\rho}, \quad
0\leq t_{R} < \infty,
\end{eqnarray*}
\begin{eqnarray*}
\left|
\sum _{l=1}^{n}
ik_{l}x_{k}(t)
\frac{\partial ^{2}\tilde{H}^{0}}{\partial p'_{l} \partial p'_{j}}(0)
\right|
\end{eqnarray*}
\begin{eqnarray*}
&\leq&
\left[
\sum _{l=1}^{n}
\left(
\frac{1}{e\delta}
\right)e^{|k|\delta}
\left|
\frac{\partial ^{2}\tilde{H}^{0}}{\partial p'_{l} \partial p'_{j}}(0)
\right|
 \frac{8\kappa}{\delta \gamma \sqrt{2\pi}}
\left[ C_{3}{c_{3}}
+\frac{2}{\delta}   C_{3}e^{-\frac{\delta}{2}c_{3}}
\right]
\right]
e^{(\mu -\frac{\delta}{2})t_{R}}e^{-|k|\rho}, \quad
- \infty  \leq t_{R} <0.
\end{eqnarray*}
\quad\\
The fourth term
\quad\\
\begin{eqnarray*}
\left|
\kappa \sum _{l=1}^{n}
\left(
 \sum _{m\in {\mathbb {Z}}^{n}}
i
\frac{ \partial ^{2} s^{1}
_{m}}
{\partial p'_{l} \partial p'_{j}}
(0)
(k-m)_{l}x_{k-m}(t)
\right)
\right|
&\leq&
 \sum _{m\in {\mathbb {Z}}^{n}}
\kappa
\frac{\|s_{m}^{1} \|_{\rho}}{\delta ^{2}}
|k-m| |x_{k-m}(t)|\\
\quad\\
&\leq&
 \sum _{m\in {\mathbb {Z}}^{n}}
\kappa
\frac{\|G \|_{\rho}e^{-|m|\rho}}{\delta ^{2}}
\left(
\frac{1}{e\delta} \right)
e^{|k-m|\delta}
|x_{k-m}(t)|,
\end{eqnarray*}
so that
\begin{eqnarray*}
\left|
\kappa \sum _{l=1}^{n}
\left(
 \sum _{m\in {\mathbb {Z}}^{n}}
i
\frac{ \partial ^{2} s^{1}
_{m}}
{\partial p'_{l} \partial p'_{j}}
(0)
(k-m)_{l}x_{k-m}(t)
\right)
\right|
\end{eqnarray*}
\begin{eqnarray*}
&\leq&
 \sum _{m\in {\mathbb {Z}}^{n}}
\left[
\kappa
\left(
\frac{\|G \|_{\rho}}{e\delta ^{3}} \right)
\frac{8\kappa}{\delta \gamma \sqrt{2\pi}}
\left[ C_{3}{c_{3}}
+\frac{2}{\delta} C_{3}e^{-\frac{\delta}{2} c_{3}}
\right]
\right]
e^{-|m|\rho}
e^{-|k-m|(\rho-\delta)}
e^{-(\nu -\frac{\delta}{2})t_{R}}
\\
\quad\\
&\leq&
\left[
3^{n}\kappa
\left(
\frac{\|G \|_{\rho}}{e\delta ^{3}} \right)
\left(\frac{1}{
1-e^{-2\delta}}\right)^{n}
\frac{8\kappa}{\delta \gamma \sqrt{2\pi}}
\left[ C_{3}{c_{3}}
+\frac{2}{\delta} C_{3}e^{-\frac{\delta}{2} c_{3}}
\right]
\right]
e^{-|k|(\rho -\delta)}
e^{-(\nu -\frac{\delta}{2})t_{R}}
\\
\quad\\
,
\end{eqnarray*}
\quad\\
for 
$0\leq t_{R} < \infty,$
\begin{eqnarray*}
\left|
\kappa \sum _{l=1}^{n}
\left(
 \sum _{m\in {\mathbb {Z}}^{n}}
i
\frac{ \partial ^{2} s^{1}
_{m}}
{\partial p'_{l} \partial p'_{j}}
(0)
(k-m)_{l}x_{k-m}(t)
\right)
\right|
\end{eqnarray*}
\begin{eqnarray*}
&\leq&
 \sum _{m\in {\mathbb {Z}}^{n}}
\left[
\kappa
\left(
\frac{\|G\|}{e\delta ^{3}} \right)
\frac{8\kappa}{\delta \gamma \sqrt{2\pi}}
\left[ C_{3}{c_{3}}
+\frac{2}{\delta} C_{3}e^{-\frac{\delta}{2} c_{3}}
\right]
\right]
e^{-|m|\rho}
e^{-|k-m|(\rho -\delta)}
e^{(\mu -\frac{\delta}{2})t_{R}}
\\
\quad\\
&\leq&
\left[
3^{n}\kappa
\left(
\frac{\|G \|_{\rho}}{e\delta ^{3}} \right)
\left(\frac{1}{
1-e^{-2\delta}}\right)^{n}
\frac{8\kappa}{\delta \gamma \sqrt{2\pi}}
\left[ C_{3}{c_{3}}
+\frac{2}{\delta} C_{3}e^{-\frac{\delta}{2} c_{3}}
\right]
\right]
e^{-|k|(\rho-\delta)}
e^{-(\mu -\frac{\delta}{2})t_{R}}
,
\end{eqnarray*}
\quad\\
for
$-  \infty \leq t_{R} < 0$.\\
\quad\\
The fifth term
\begin{eqnarray*}
\left|
\kappa  \sum _{l=1}^{n}
\left(
 \sum _{m\in {\mathbb {Z}}^{n}} i
\frac{ \partial ^{2} h^{1}
_{m}}
{\partial p'_{l} \partial p'_{j}}
(0)
e^{1}_{m}(t)
(k-m)_{l}x_{k-m}(t)
\right)
\right|
\end{eqnarray*}
\begin{eqnarray*}
&\leq&
 \sum _{m\in {\mathbb {Z}}^{n}}
\kappa 
\frac{\|h^{1}_{m}\|_{\rho}}{\delta ^{2}}
|k-m|
|e^{1}_{m}(t)x_{k-m}(t)|
\leq
 \sum _{m\in {\mathbb {Z}}^{n}}
\kappa 
\frac{1}{\delta ^{2}}
\left(
\frac{1}{e\delta}
\right)
e^{|k-m|\delta}
\|h^{1}_{m}\|_{\rho}
|e^{1}_{m}(t)x_{k-m}(t)|,
\end{eqnarray*}
so that
\begin{eqnarray*}
\left|
\kappa  \sum _{l=1}^{n}
\left(
 \sum _{m\in {\mathbb {Z}}^{n}} i
\frac{ \partial ^{2} h^{1}
_{m}}
{\partial p'_{l} \partial p'_{j}}
(0)
e^{1}_{m}(t)
(k-m)_{l}x_{k-m}(t)
\right)
\right|
\end{eqnarray*}
\begin{eqnarray*}
&\leq&
 \sum _{m\in {\mathbb {Z}}^{n}}
\left[
\kappa 
\left(
\frac{1}{e\delta^{3}}
\right)
C_{1}
\frac{8\kappa}{\delta \gamma \sqrt{2\pi}}
\left[ C_{3}{c_{3}}
+\frac{2}{\delta} C_{3}e^{-\frac{\delta}{2} c_{3}}
\right]
\right]
e^{-|m|\rho}
e^{-|k-m|(\rho -\delta)}
e^{-2(\nu -\frac{\delta}{2})t_{R}}
\\
\quad\\
&\leq&
\left[
\kappa 
\left(
\frac{3^{n}}{e\delta^{3}}
\right)
\left(\frac{
1}{1-e^{-2\delta}}\right)^{n}
C_{1}
\frac{8\kappa}{\delta \gamma \sqrt{2\pi}}
\left[ C_{3}{c_{3}}
+\frac{2}{\delta} C_{3}e^{-\frac{\delta}{2} c_{3}}
\right]
\right]
e^{-|k|(\rho -\delta)}
e^{-2(\nu -\frac{\delta}{2})t_{R}}
\end{eqnarray*}
\quad\\
for $0< c_{1} < t_{R} < \infty,$
\quad\\
\begin{eqnarray*}
\left|
\kappa  \sum _{l=1}^{n}
\left(
 \sum _{m\in {\mathbb {Z}}^{n}} i
\frac{ \partial ^{2} h^{1}
_{m}}
{\partial p'_{l} \partial p'_{j}}
(0)
e^{1}_{m}(t)
(k-m)_{l}x_{k-m}(t)
\right)
\right|
\end{eqnarray*}
\begin{eqnarray*}
&\leq&
 \sum _{m\in {\mathbb {Z}}^{n}}
\left[
\kappa 
\left(
\frac{1}{e\delta ^{3}}
\right)
C_{2}
\frac{8\kappa}{\delta \gamma \sqrt{2\pi}}
\left[ C_{3}{c_{3}}
+\frac{2}{\delta} C_{3}e^{-\frac{\delta}{2} c_{3}}
\right]
\right]
e^{-|m|\rho}
e^{-|k-m|(\rho -\delta)}
e^{2(\mu -\frac{\delta}{2})t_{R}}
\\
\quad\\
&\leq&
\left[
\kappa 
\left(
\frac{3^{n}}{e\delta^{3}}
\right)
\left(\frac{
1}{1-e^{-2\delta}}\right)^{n}
C_{2}
\frac{8\kappa}{\delta \gamma \sqrt{2\pi}}
\left[ C_{3}{c_{3}}
+\frac{2}{\delta} C_{3}e^{-\frac{\delta}{2} c_{3}}
\right]
\right]
e^{-|k|(\rho -\delta)}
e^{-2(\mu -\frac{\delta}{2})t_{R}}
\end{eqnarray*}
\quad\\
for $- \infty < t_{R} < c_{2} <0$.\\
\quad\\
For the sixth, seventh and eighth term we refer to lemma $\ref{lemmatimederv}$ when
dealing with the bound of the derivative of an exponential-time
dependent function. The sixth term 
\quad\\
\begin{eqnarray*}
\left|
\frac{\partial ^{2}\tilde{H}^{0}}{\partial p'_{n+1} \partial p'_{j}}(0)
\frac{dx_{k}}{dt}(t)
\right|
&\leq&
\left|
\frac{\partial ^{2}\tilde{H}^{0}}{\partial p'_{n+1} \partial p'_{j}}(0)
\right|
\frac{e^{(\nu -\frac{\delta}{2})\delta}}{\delta}
\frac{8\kappa}{\delta \gamma \sqrt{2\pi}}
\left[ C_{3}{c_{3}}
+\frac{2}{\delta} C_{3}e^{-\frac{\delta}{2} c_{3}}
\right]
e^{-(\nu -\frac{\delta}{2})t_{R}}e^{-|k|\rho},
\end{eqnarray*}
\quad\\
for
$0\leq t_{R} < \infty$,
\quad\\
\begin{eqnarray*}
\left|
\frac{\partial ^{2}\tilde{H}^{0}}{\partial p'_{n+1} \partial p'_{j}}(0)
\frac{dx_{k}}{dt}(t)
\right|
&\leq&
\left|
\frac{\partial ^{2}\tilde{H}^{0}}{\partial p'_{n+1} \partial p'_{j}}(0)
\right|
\frac{e^{(\mu -\frac{\delta}{2})\delta}}{\delta}
\frac{8\kappa}{\delta \gamma \sqrt{2\pi}}
\left[ C_{3}{c_{3}}
+\frac{2}{\delta} C_{3}e^{-\frac{\delta}{2} c_{3}}
\right]
e^{(\mu -\frac{\delta}{2})t_{R}}e^{-|k|\rho},
\end{eqnarray*}
\quad\\
for
$-\infty \leq t_{R} < 0$.
\\
\quad\\
The seventh term
\begin{eqnarray*}
\left|
\sum _{m\in {\mathbb {Z}}^{n}} 
\kappa
\frac{ \partial ^{2} s^{1}
_{m}}
{\partial p'_{n+1} \partial p'_{j}}
(0)
\frac{dx_{k-m}}{dt}(t)
\right|
&\leq&
\sum _{m\in {\mathbb {Z}}^{n}} 
\kappa
\frac{\|s^{1}
_{m}\|_{\rho}}{\delta ^{2}}
\left|
\frac{dx_{k-m}}{dt}(t)
\right|
\leq
\sum _{m\in {\mathbb {Z}}^{n}} 
\kappa
\frac{\|G\|_{\rho}e^{-|m|\rho}}{\delta ^{2}}
\left|
\frac{dx_{k-m}}{dt}(t)
\right|,
\end{eqnarray*}
so that
\begin{eqnarray*}
\left|
\sum _{m\in {\mathbb {Z}}^{n}} 
\kappa
\frac{ \partial ^{2} s^{1}
_{m}}
{\partial p'_{n+1} \partial p'_{j}}
(0)
\frac{dx_{k-m}}{dt}(t)
\right|
\end{eqnarray*}
\begin{eqnarray*}
&\leq&
\sum _{m\in {\mathbb {Z}}^{n}} 
\left[
\kappa \|G\|_{\rho}
\frac{e^{(\nu -\frac{\delta}{2})\delta}}{\delta^{3}}
\frac{8\kappa}{\delta \gamma \sqrt{2\pi}}
\left[ C_{3}{c_{3}}
+\frac{2}{\delta} C_{3}e^{-\frac{\delta}{2} c_{3}}
\right]
 \right]
e^{-|m|\rho}
e^{-(\nu -\frac{\delta}{2})t_{R}}e^{-|k-m|\rho}
\\
\quad\\
&\leq&
\left[
3^{n}\kappa \|G\|_{\rho}
\left(\frac{1}{e\delta}\right)^{n}
\frac{e^{(\nu -\frac{\delta}{2})\delta}}{\delta^{3}}
\frac{8\kappa}{\delta \gamma \sqrt{2\pi}}
\left[ C_{3}{c_{3}}
+\frac{2}{\delta} C_{3}e^{-\frac{\delta}{2} c_{3}}
\right]
 \right]
e^{-|k|(\rho -\delta)}
e^{-(\nu -\frac{\delta}{2})t_{R}}
,
\end{eqnarray*}
\quad\\
for
$0\leq t_{R} < \infty$,
\quad\\
\begin{eqnarray*}
\left|
\sum _{m\in {\mathbb {Z}}^{n}} 
\kappa
\frac{ \partial ^{2} s^{1}
_{m}}
{\partial p'_{n+1} \partial p'_{j}}
(0)
\frac{dx_{k-m}}{dt}(t)
\right|
\end{eqnarray*}
\begin{eqnarray*}
&\leq&
\sum _{m\in {\mathbb {Z}}^{n}}
 \left[
\kappa \|G\|_{\rho}
\frac{e^{(\mu -\frac{\delta}{2})\delta}}{\delta^{3}}
\frac{8\kappa}{\delta \gamma \sqrt{2\pi}}
\left[ C_{3}{c_{3}}
+\frac{2}{\delta} C_{3}e^{-\frac{\delta}{2} c_{3}}
\right]
 \right]
e^{-|m|\rho}
e^{(\mu -\frac{\delta}{2})t_{R}}e^{-|k-m|\rho}
\\
\quad\\
&\leq&
\left[
3^{n}\kappa \|G\|_{\rho}
\left(\frac{1}{e\delta}\right)^{n}
\frac{e^{(\mu -\frac{\delta}{2})\delta}}{\delta^{3}}
\frac{8\kappa}{\delta \gamma \sqrt{2\pi}}
\left[ C_{3}{c_{3}}
+\frac{2}{\delta} C_{3}e^{-\frac{\delta}{2} c_{3}}
\right]
 \right]
e^{-|k|(\rho -\delta)}
e^{-(\mu -\frac{\delta}{2})t_{R}}
,
\end{eqnarray*}
\quad\\
for
$-\infty \leq t_{R} < 0$.
\\
\quad\\
The eighth term 
\begin{eqnarray*}
\left|
\sum _{m\in {\mathbb {Z}}^{n}} 
\kappa
\frac{ \partial ^{2} h^{1}
_{m}}
{\partial p'_{n+1} \partial p'_{j}}
(0)
e^{1}_{m}(t)
\frac{dx_{k-m}}{dt}(t)
\right|
&\leq&
\sum _{m\in {\mathbb {Z}}^{n}} 
\kappa
\frac{\|h^{1}
_{m}\|_{\rho}}{\delta ^{2}}
|e^{1}_{m}(t)|
\left|
\frac{dx_{k}}{dt}(t)
\right|
\leq
\sum _{m\in {\mathbb {Z}}^{n}} 
\kappa
\frac{1}{\delta ^{2}}
\|h^{1}_{m}\|_{\rho}
|e^{1}_{m}(t)|
\left|
\frac{dx_{k}}{dt}(t)
\right|,
\end{eqnarray*}
so that
\begin{eqnarray*}
\left|
\sum _{m\in {\mathbb {Z}}^{n}} 
\kappa
\frac{ \partial ^{2} h^{1}
_{m}}
{\partial p'_{n+1} \partial p'_{j}}
(0)
e^{1}_{m}(t)
\frac{dx_{k-m}}{dt}(t)
\right|
\end{eqnarray*}
\begin{eqnarray*}
&\leq&
\sum _{m\in {\mathbb {Z}}^{n}} 
\kappa
\frac{1}{\delta ^{2}}
 C_{1}e^{-(\nu -\frac{\delta}{2})t_{R}}e^{-|m|\rho}
\frac{e^{(\nu-\frac{\delta}{2})\delta}}{\delta}
\frac{8\kappa}{\delta \gamma \sqrt{2\pi}}
\left[ C_{3}{c_{3}}
+\frac{2}{\delta} C_{3}e^{-\frac{\delta}{2} c_{3}}
\right]
e^{-(\nu -\frac{\delta}{2})t_{R}}e^{-|k-m|\rho}\\
\quad\\
&=&
\sum _{m\in {\mathbb {Z}}^{n}} 
\left[
\kappa
 C_{1}
\frac{e^{(\nu-\frac{\delta}{2})\delta}}{\delta^{3}}
\frac{8\kappa}{\delta \gamma \sqrt{2\pi}}
\left[ C_{3}{c_{3}}
+\frac{2}{\delta} C_{3}e^{-\frac{\delta}{2} c_{3}}
\right]
\right]
e^{-|m|\rho}
e^{-|k-m|\rho}
e^{-2(\nu -\frac{\delta}{2})t_{R}}
\\
\quad\\
&\leq&
\left[
3^{n}\kappa
 C_{1}
\frac{e^{(\nu-\frac{\delta}{2})\delta}}{\delta^{3}}
\left(
\frac{
1}{e\delta}\right)^{n}
\frac{8\kappa}{\delta \gamma \sqrt{2\pi}}
\left[ C_{3}{c_{3}}
+\frac{2}{\delta} C_{3}e^{-\frac{\delta}{2} c_{3}}
\right]
\right]
e^{-|k|(\rho -\delta)}
e^{-2(\nu -\frac{\delta}{2})t_{R}}
\end{eqnarray*}
\quad\\
for $0< c_{1} < t_{R} < \infty$,
\begin{eqnarray*}
\left|
\sum _{m\in {\mathbb {Z}}^{n}} 
\kappa
\frac{ \partial ^{2} h^{1}
_{m}}
{\partial p'_{n+1} \partial p'_{j}}
(0)
e^{1}_{m}(t)
\frac{dx_{k-m}}{dt}(t)
\right|
\end{eqnarray*}
\begin{eqnarray*}
&\leq&
\sum _{m\in {\mathbb {Z}}^{n}} 
\kappa
\frac{1}{\delta ^{2}}
 C_{1}e^{(\mu -\frac{\delta}{2})t_{R}}e^{-|m|\rho}
\frac{e^{(\mu-\frac{\delta}{2})\delta}}{\delta}
\frac{8\kappa}{\delta \gamma \sqrt{2\pi}}
\left[ C_{3}{c_{3}}
+\frac{2}{\delta} C_{3}e^{-\frac{\delta}{2} c_{3}}
\right]
e^{(\mu -\frac{\delta}{2})t_{R}}e^{-|k-m|\rho}\\
\quad\\
&=&
\sum _{m\in {\mathbb {Z}}^{n}} 
\left[
\kappa
 C_{1}
\frac{e^{(\mu-\frac{\delta}{2})\delta}}{\delta^{3}}
\frac{8\kappa}{\delta \gamma \sqrt{2\pi}}
\left[ C_{3}{c_{3}}
+\frac{2}{\delta} C_{3}e^{-\frac{\delta}{2} c_{3}}
\right]
\right]
e^{-|m|\rho}
e^{-|k-m|\rho}
e^{2(\mu -\frac{\delta}{2})t_{R}}
\\
\quad\\
&\leq&
\left[
3^{n}\kappa
 C_{m}
\frac{e^{(\nu-\frac{\delta}{2})\delta}}{\delta^{3}}
\left(
\frac{
1}{e\delta}\right)^{n}
\frac{8\kappa}{\delta \gamma \sqrt{2\pi}}
\left[ C_{3}{c_{3}}
+\frac{2}{\delta} C_{3}e^{-\frac{\delta}{2} c_{3}}
\right]
\right]
e^{-|k|(\rho -\delta)}
e^{-2(\mu -\frac{\delta}{2})t_{R}}
\end{eqnarray*}
\quad\\
for $-\infty < t_{R} < c_{2} <0 .$
\\
\quad\\
The nine th term
\begin{eqnarray*}
\left|
\sum _{l=1}^{n+1}
\kappa \xi_{l}
\frac{ \partial ^{2} h^{1}
_{k}}
{\partial p'_{l} \partial p'_{j}}
(0)
e^{1}_{k}(t)
\right|
\leq
\kappa|\xi|
\frac{\|h^{1}
_{k}\|_{\rho}}{\delta^{2}}
\left|e^{1}_{k}(t)
\right|,
\end{eqnarray*}
so that
\begin{eqnarray*}
\left|
\sum _{l=1}^{n+1}
\kappa \xi_{l}
\frac{ \partial ^{2} h^{1}
_{k}}
{\partial p'_{l} \partial p'_{j}}
(0)
e^{1}_{k}(t)
\right|\leq
\kappa|\xi|
\frac{\|h^{1}
_{k}\|_{\rho}}{\delta^{2}}
 C_{1}e^{-(\nu -\frac{\delta}{2})t_{R}}e^{-|k|\rho}, \quad 
0< c_{1} < t_{R} < \infty,
\end{eqnarray*}
\begin{eqnarray*}
\left|
\sum _{l=1}^{n+1}
\kappa \xi_{l}
\frac{ \partial ^{2} h^{1}
_{k}}
{\partial p'_{l} \partial p'_{j}}
(0)
e^{1}_{k}(t)
\right|\leq
\kappa|\xi|
\frac{\|h^{1}
_{k}\|_{\rho}}{\delta^{2}}
 C_{2}e^{(\mu -\frac{\delta}{2})t_{R}}e^{-|k|\rho}, \quad 
-\infty < t_{R} < c_{2} <0 .
\end{eqnarray*}
Note we have bounded the sums on $m$ as indicated in Appendix $3$.
Putting the estimates for the nine terms we obtain the following
bound for $\beta^{E}_{j}$.
\begin{eqnarray*}
\left| \beta^{E}_{jk}(t) \right| &\leq&
\left[
\frac{\kappa}{\delta} C_{1}
e^{-|k|\rho} \right. \\
\quad\\
&+&
\left(
\kappa 
\frac{3^{n}}{e\delta ^{3}}
\frac{\kappa}{\Gamma}\left( \frac{n}{e\delta} \right)^{n}
\left(
\frac{1}{1-e^{-2\delta}}\right)^{n}
\|G(0,q) -\overline{\overline{G}}(0) \|_{\rho}
C_{1}
\right)
e^{-|k|(\rho -2\delta)}
e^{-(\nu -\frac{\delta}{2})t_{R}}
\\
\quad\\
&+&
\left(
\sum _{l=1}^{n}
\left(
\frac{1}{e\delta}
\right)
\left|
\frac{\partial ^{2}\tilde{H}^{0}}{\partial p'_{l} \partial p'_{j}}(0)
\right|
\frac{8\kappa}{\delta \gamma \sqrt{2\pi}}
\left[ C_{3}{c_{3}}
+\frac{2}{\delta} C_{3}e^{-\frac{\delta}{2} c_{3}}
\right]
e^{-|k|(\rho -\delta)}
\right)\\
\quad\\
&+&
\left(
3^{n}\kappa
\left(
\frac{\|G \|_{\rho}}{e\delta ^{3}} \right)
\left(\frac{1}{
1-e^{-2\delta}}\right)^{n}
\frac{8\kappa}{\delta \gamma \sqrt{2\pi}}
\left[ C_{3}{c_{3}}
+\frac{2}{\delta} C_{3}e^{-\frac{\delta}{2} c_{3}}
\right]
\right)
e^{-|k|(\rho -\delta)}
e^{-(\nu -\frac{\delta}{2})t_{R}}
\\
\quad\\
&+&
\left(
\kappa 
\left(
\frac{3^{n}}{e\delta^{3}}
\right)
\left(\frac{
1}{1-e^{-2\delta}}\right)^{n}
C_{1}
\frac{8\kappa}{\delta \gamma \sqrt{2\pi}}
\left[ C_{3}{c_{3}}
+\frac{2}{\delta} C_{3}e^{-\frac{\delta}{2} c_{3}}
\right]
\right)
e^{-|k|(\rho -\delta)}
e^{-2(\nu -\frac{\delta}{2})t_{R}}
\\
\quad\\
&+&
\left(
\left|
\frac{\partial ^{2}\tilde{H}^{0}}{\partial p'_{n+1} \partial p'_{j}}(0)
\right|
\frac{e^{(\nu -\frac{\delta}{2})\delta}}{\delta}
 \frac{8\kappa}{\delta \gamma \sqrt{2\pi}}
\left[ C_{3}{c_{3}}
+\frac{2}{\delta} C_{3}e^{-\frac{\delta}{2} c_{3}}
\right]
e^{-|k|\rho}\right)\\
\quad\\
&+&
\left(
3^{n}\kappa \|G\|_{\rho}
\left(\frac{1}{e\delta}\right)^{n}
\frac{e^{(\nu -\frac{\delta}{2})\delta}}{\delta^{3}}
\frac{8\kappa}{\delta \gamma \sqrt{2\pi}}
\left[ C_{3}{c_{3}}
+\frac{2}{\delta} C_{3}e^{-\frac{\delta}{2} c_{3}}
\right]
\right)
e^{-|k|(\rho -\delta)}
e^{-(\nu -\frac{\delta}{2})t_{R}}
\\
\quad\\
&+&
\left(
3^{n}\kappa
 C_{1}
\frac{e^{(\nu-\frac{\delta}{2})\delta}}{\delta^{3}}
\left(
\frac{
1}{e\delta}\right)^{n}
\frac{8\kappa}{\delta \gamma \sqrt{2\pi}}
\left[ C_{3}{c_{3}}
+\frac{2}{\delta} C_{3}e^{-\frac{\delta}{2} c_{3}}
\right]
\right)
e^{-|k|(\rho -\delta)}
e^{-2(\nu -\frac{\delta}{2})t_{R}}
\\
\quad\\
&+&
\left.
\kappa|\xi|
\frac{1}{\delta^{2}}
 C_{1}e^{-|k|\rho}\right]e^{-(\nu -\frac{\delta}{2})t_{R}},
\quad 
0< c_{1} < t_{R} < \infty,
\end{eqnarray*}
where we have used the fact $e^{-2(\nu -\frac{\delta}{2})t_{R}} \leq
e^{-(\nu -\frac{\delta}{2})t_{R}}$.
We can further simplify by using the fact
$e^{-|k|\rho} \leq e^{-|k|(\rho -\delta)} \leq e^{-|k|(\rho-2\delta)}$
so that
\begin{eqnarray*}
\left| \beta^{E}_{jk}(t) \right| &\leq&
\left[
\frac{\kappa}{\delta} C_{1}
 \right.
+
\left(
\kappa 
\frac{3^{n}}{e\delta ^{3}}
\frac{\kappa}{\Gamma}\left( \frac{n}{e\delta} \right)^{n}
\left(
\frac{1}{1-e^{-2\delta}}\right)^{n}
\|G(0,q) -\overline{\overline{G}}(0) \|_{\rho}
C_{1}
\right)\\
\quad\\
&+&
\left(
\sum _{l=1}^{n}
\left(
\frac{1}{e\delta}
\right)
\left|
\frac{\partial ^{2}\tilde{H}^{0}}{\partial p'_{l} \partial p'_{j}}(0)
\right|
\frac{8\kappa}{\delta \gamma \sqrt{2\pi}}
\left[ C_{3}{c_{3}}
+\frac{2}{\delta} C_{3}e^{-\frac{\delta}{2} c_{3}}
\right]
\right)\\
\quad\\
&+&
\left(
3^{n}\kappa
\left(
\frac{\|G \|_{\rho}}{e\delta ^{3}} \right)
\left(\frac{1}{
1-e^{-2\delta}}\right)^{n}
\frac{8\kappa}{\delta \gamma \sqrt{2\pi}}
\left[ C_{3}{c_{3}}
+\frac{2}{\delta} C_{3}e^{-\frac{\delta}{2} c_{3}}
\right]
\right)\\
\quad\\
&+&
\left(
\kappa 
\left(
\frac{3^{n}}{e\delta^{3}}
\right)
\left(\frac{
1}{1-e^{-2\delta}}\right)^{n}
C_{1}
\frac{8\kappa}{\delta \gamma \sqrt{2\pi}}
\left[ C_{3}{c_{3}}
+\frac{2}{\delta} C_{3}e^{-\frac{\delta}{2} c_{3}}
\right]
\right)\\
\quad\\
&+&
\left(
\left|
\frac{\partial ^{2}\tilde{H}^{0}}{\partial p'_{n+1} \partial p'_{j}}(0)
\right|
\frac{e^{(\nu -\frac{\delta}{2})\delta}}{\delta}
\frac{8\kappa}{\delta \gamma \sqrt{2\pi}}
\left[ C_{3}{c_{3}}
+\frac{2}{\delta} C_{3}e^{-\frac{\delta}{2} c_{3}}
\right]
\right)
\\
\quad\\
&+&
\left(
3^{n}\kappa \|G\|_{\rho}
\left(\frac{1}{e\delta}\right)^{n}
\frac{e^{(\nu -\frac{\delta}{2})\delta}}{\delta^{3}}
\frac{8\kappa}{\delta \gamma \sqrt{2\pi}}
\left[ C_{3}{c_{3}}
+\frac{2}{\delta} C_{3}e^{-\frac{\delta}{2} c_{3}}
\right]
\right)\\
\quad\\
&+&
\left(
3^{n}\kappa
 C_{1}
\frac{e^{(\nu-\frac{\delta}{2})\delta}}{\delta^{3}}
\left(
\frac{
1}{e\delta}\right)^{n}
\frac{8\kappa}{\delta \gamma \sqrt{2\pi}}
\left[ C_{3}{c_{3}}
+\frac{2}{\delta} C_{3}e^{-\frac{\delta}{2} c_{3}}
\right]
\right)\\
\quad\\
&+&
\left.
\kappa|\xi|
\frac{1}{\delta^{2}}
 C_{1}\right]e^{-|k|(\rho -2\delta)}e^{-(\nu -\frac{\delta}{2})t_{R}},
\quad 
0< c_{1} < t_{R} < \infty,
\end{eqnarray*}
similarly
\begin{eqnarray*}
\left| \beta^{E}_{jk}(t) \right| &\leq&
\left[
\frac{\kappa}{\delta} C_{2}
 \right.
+
\left(
\kappa 
\frac{3^{n}}{e\delta ^{3}}
\frac{\kappa}{\Gamma}\left( \frac{n}{e\delta} \right)^{n}
\left(
\frac{1}{1-e^{-2\delta}}\right)^{n}
\|G(0,q) -\overline{\overline{G}}(0) \|_{\rho}
C_{2}
\right)\\
\quad\\
&+&
\left(
\sum _{l=1}^{n}
\left(
\frac{1}{e\delta}
\right)
\left|
\frac{\partial ^{2}\tilde{H}^{0}}{\partial p'_{l} \partial p'_{j}}(0)
\right|
\frac{8\kappa}{\delta \gamma \sqrt{2\pi}}
\left[ C_{3}{c_{3}}
+\frac{2}{\delta} C_{3}e^{-\frac{\delta}{2} c_{3}}
\right]
\right)\\
\quad\\
&+&
\left(
3^{n}\kappa
\left(
\frac{\|G \|_{\rho}}{e\delta ^{3}} \right)
\left(\frac{1}{
1-e^{-2\delta}}\right)^{n}
\frac{8\kappa}{\delta \gamma \sqrt{2\pi}}
\left[ C_{3}{c_{3}}
+\frac{2}{\delta} C_{3}e^{-\frac{\delta}{2} c_{3}}
\right]
\right)\\
\quad\\
&+&
\left(
\kappa 
\left(
\frac{3^{n}}{e\delta^{3}}
\right)
\left(\frac{
1}{1-e^{-2\delta}}\right)^{n}
C_{2}
\frac{8\kappa}{\delta \gamma \sqrt{2\pi}}
\left[ C_{3}{c_{3}}
+\frac{2}{\delta} C_{3}e^{-\frac{\delta}{2} c_{3}}
\right]
\right)\\
\quad\\
&+&
\left(
\left|
\frac{\partial ^{2}\tilde{H}^{0}}{\partial p'_{n+1} \partial p'_{j}}(0)
\right|
\frac{e^{(\mu -\frac{\delta}{2})\delta}}{\delta}
\frac{8\kappa}{\delta \gamma \sqrt{2\pi}}
\left[ C_{3}{c_{3}}
+\frac{2}{\delta} C_{3}e^{-\frac{\delta}{2} c_{3}}
\right]
\right)
\\
\quad\\
&+&
\left(
3^{n}\kappa \|G\|_{\rho}
\left(\frac{1}{e\delta}\right)^{n}
\frac{e^{(\mu -\frac{\delta}{2})\delta}}{\delta^{3}}
\frac{8\kappa}{\delta \gamma \sqrt{2\pi}}
\left[ C_{3}{c_{3}}
+\frac{2}{\delta} C_{3}e^{-\frac{\delta}{2} c_{3}}
\right]
\right)\\
\quad\\
&+&
\left(
3^{n}\kappa
 C_{2}
\frac{e^{(\mu-\frac{\delta}{2})\delta}}{\delta^{3}}
\left(
\frac{
1}{e\delta}\right)^{n}
\frac{8\kappa}{\delta \gamma \sqrt{2\pi}}
\left[ C_{3}{c_{3}}
+\frac{2}{\delta} C_{3}e^{-\frac{\delta}{2} c_{3}}
\right]
\right)\\
\quad\\
&+&
\left.
\kappa|\xi|
\frac{1}{\delta^{2}}
 C_{2}\right]e^{-|k|(\rho -2\delta)}e^{(\mu -\frac{\delta}{2})t_{R}},
\quad 
-\infty  < t_{R} < c_{2} <0.
\end{eqnarray*}
At this point we must relate the constants used to 
bound the time dependence, $C_{1}$ and $ C_{2}$, to a constant 
which bounds the entire perturbation.
Recall that by assumption on the time dependence and 
by the quasiperiodic $q$ dependence we obtain the following 
estimates 
\begin{eqnarray*}
&\quad&|h_{k}^{1}(0)e_{k}^{1}(t)| \leq
C_{1} e^{-|k|\rho}e^{-(\nu -\frac{\delta}{2})t_{R}}, \quad 0< c_{1} \geq t_{R} < \infty,\\
\quad\\
&\quad&|h_{k}^{1}(0)e_{k}^{1}(t)| \leq
 C_{2} e^{-|k|\rho}e^{(\mu -\frac{\delta}{2})t_{R}}, \quad -\infty < t_{R} \leq c_{2} <0,
\end{eqnarray*}
and clearly
\begin{eqnarray*}
&\quad&|h_{k}^{1}(0)e_{k}^{1}(t)| \leq
C_{1} e^{-(\nu -\frac{\delta}{2})t_{R}}, \quad 0< c_{1} \geq t_{R} < \infty,\\
\quad\\
&\quad&|h_{k}^{1}(0)e_{k}^{1}(t)| \leq
 C_{2} e^{(\mu -\frac{\delta}{2})t_{R}}, \quad -\infty < t_{R} \leq c_{2} <0.
\end{eqnarray*}
We want this estimates also to include the interval
$c_{2} < t_{R} < c_{1}$. The assumption on the function
$e_{k}^{1}(t)$ in this interval was simply to be a bounded
function. Consequently, there exists a constant $E_{0}$ such that
$C_{1} \leq E_{0}$, $C_{2} \leq E_{0}$ and the following estimates hold
\begin{eqnarray*}
&\quad&|h_{k}^{1}(0)e_{k}^{1}(t)| \leq
E_{0} e^{-|k|\rho}, \quad 0< t_{R} < \infty,\\
\quad\\
&\quad&|h_{k}^{1}(0)e_{k}^{1}(t)| \leq
E_{0} e^{-|k|\rho}, \quad -\infty < t_{R} \leq 0.
\end{eqnarray*}
Now recall the total perturbation
$H^{1}(0,q') = G(q) + T(q')$ and assume $|G(q)| \leq E_{G}$. By the argument above
we have
\begin{eqnarray*}
|T(q')|&\leq& 
\sum _{k\in {\mathbb{Z}}^{n}} |h_{k}^{1}(0)|e_{k}^{1}(t)||e^{ik\cdot q}|
\leq
\sum _{k\in {\mathbb{Z}}^{n}}
E_{0}e^{-|k|\rho}e^{|k|(\rho -\delta)}\\
\quad\\
&=&
\sum _{k\in {\mathbb{Z}}^{n}}
E_{0}e^{-|k|\delta}
=2E_{0}\frac{1}{1-e^{-\delta}}=E_{T}.
\end{eqnarray*}
Finally we have 
\begin{eqnarray*}
&\quad&C_{1}\leq E_{0} \leq E_{T} \leq E_{G} + E_{T}=E\leq E_{1},\\
\quad\\
&\quad&C_{2}\leq E_{0} \leq E_{T} \leq E_{G} + E_{T}=E\leq E_{1},\\
\quad\\
&\quad&C_{3} = \max(C_{1} ,C_{2}) \leq E_{1},
\end{eqnarray*}
where $E_{1}$ is the bound on the total perturbation $H^{1}(0,q')$.
The estimates on $\beta ^{E}_{jk}(t)$ become
\begin{eqnarray*}
\left| \beta^{E}_{jk}(t) \right| &\leq&
\left[
\frac{1}{\delta}
 \right.
+
\left(
\frac{3^{n}}{e\delta ^{3}}
\frac{\kappa}{\Gamma}\left( \frac{n}{e\delta} \right)^{n}
\left(
\frac{1}{1-e^{-2\delta}}\right)^{n}
\|G(0,q) -\overline{\overline{G}}(0) \|_{\rho}
\right)\\
\quad\\
&+&
\left(
\sum _{l=1}^{n}
\left(
\frac{1}{e\delta}
\right)
\left|
\frac{\partial ^{2}\tilde{H}^{0}}{\partial p'_{l} \partial p'_{j}}(0)
\right|
\frac{8}{\delta \gamma \sqrt{2\pi}}
\left[ {c_{3}}
+\frac{2}{\delta} e^{-\frac{\delta}{2} c_{3}}
\right]
\right)\\
\quad\\
&+&
\left(
3^{n}\kappa
\left(
\frac{\|G \|_{\rho}}{e\delta ^{3}} \right)
\left(\frac{1}{
1-e^{-2\delta}}\right)^{n}
\frac{8}{\delta \gamma \sqrt{2\pi}}
\left[ {c_{3}}
+\frac{2}{\delta} e^{-\frac{\delta}{2} c_{3}}
\right]
\right)\\
\quad\\
&+&
\left(
\kappa 
\left(
\frac{3^{n}}{e\delta^{3}}
\right)
\left(\frac{
1}{1-e^{-2\delta}}\right)^{n}
C_{1}
\frac{8}{\delta \gamma \sqrt{2\pi}}
\left[ {c_{3}}
+\frac{2}{\delta} e^{-\frac{\delta}{2} c_{3}}
\right]
\right)\\
\quad\\
&+&
\left(
\left|
\frac{\partial ^{2}\tilde{H}^{0}}{\partial p'_{n+1} \partial p'_{j}}(0)
\right|
\frac{e^{(\nu -\frac{\delta}{2})\delta}}{\delta}
\frac{8}{\delta \gamma \sqrt{2\pi}}
\left[ {c_{3}}
+\frac{2}{\delta} e^{-\frac{\delta}{2} c_{3}}
\right]
\right)
\\
\quad\\
&+&
\left(
3^{n}\kappa \|G\|_{\rho}
\left(\frac{1}{e\delta}\right)^{n}
\frac{e^{(\nu -\frac{\delta}{2})\delta}}{\delta^{3}}
\frac{8}{\delta \gamma \sqrt{2\pi}}
\left[ {c_{3}}
+\frac{2}{\delta} e^{-\frac{\delta}{2} c_{3}}
\right]
\right)\\
\quad\\
&+&
\left(
3^{n}\kappa
 C_{1}
\frac{e^{(\nu-\frac{\delta}{2})\delta}}{\delta^{3}}
\left(
\frac{
1}{e\delta}\right)^{n}
\frac{8}{\delta \gamma \sqrt{2\pi}}
\left[ {c_{3}}
+\frac{2}{\delta} e^{-\frac{\delta}{2} c_{3}}
\right]
\right)\\
\quad\\
&+&
\left.
|\xi|
\frac{1}{\delta^{2}}
\right]\kappa E_{1}e^{-|k|(\rho -2\delta)}e^{-(\nu -\frac{\delta}{2})t_{R}}
 ,
\quad
0< c_{1} < t_{R} < \infty,
\end{eqnarray*}
similarly
\begin{eqnarray*}
\left| \beta^{E}_{jk}(t) \right| &\leq&
\left[
\frac{1}{\delta} 
 \right.
+
\left(
\frac{3^{n}}{e\delta ^{3}}
\frac{\kappa}{\Gamma}\left( \frac{n}{e\delta} \right)^{n}
\left(
\frac{1}{1-e^{-2\delta}}\right)^{n}
\|G(0,q) -\overline{\overline{G}}(0) \|_{\rho}
\right)\\
\quad\\
&+&
\left(
\sum _{l=1}^{n}
\left(
\frac{1}{e\delta}
\right)
\left|
\frac{\partial ^{2}\tilde{H}^{0}}{\partial p'_{l} \partial p'_{j}}(0)
\right|
\frac{8}{\delta \gamma \sqrt{2\pi}}
\left[ {c_{3}}
+\frac{2}{\delta} e^{-\frac{\delta}{2} c_{3}}
\right]
\right)\\
\quad\\
&+&
\left(
3^{n}\kappa
\left(
\frac{\|G \|_{\rho}}{e\delta ^{3}} \right)
\left(\frac{1}{
1-e^{-2\delta}}\right)^{n}
\frac{8}{\delta \gamma \sqrt{2\pi}}
\left[ {c_{3}}
+\frac{2}{\delta} e^{-\frac{\delta}{2} c_{3}}
\right]
\right)\\
\quad\\
&+&
\left(
\kappa 
\left(
\frac{3^{n}}{e\delta^{3}}
\right)
\left(\frac{
1}{1-e^{-2\delta}}\right)^{n}
C_{2}
\frac{8}{\delta \gamma \sqrt{2\pi}}
\left[ {c_{3}}
+\frac{2}{\delta} e^{-\frac{\delta}{2} c_{3}}
\right]
\right)\\
\quad\\
&+&
\left(
\left|
\frac{\partial ^{2}\tilde{H}^{0}}{\partial p'_{n+1} \partial p'_{j}}(0)
\right|
\frac{e^{(\mu -\frac{\delta}{2})\delta}}{\delta}
\frac{8}{\delta \gamma \sqrt{2\pi}}
\left[ {c_{3}}
+\frac{2}{\delta} e^{-\frac{\delta}{2} c_{3}}
\right]
\right)
\\
\quad\\
&+&
\left(
3^{n}\kappa \|G\|_{\rho}
\left(\frac{1}{e\delta}\right)^{n}
\frac{e^{(\mu -\frac{\delta}{2})\delta}}{\delta^{3}}
\frac{8}{\delta \gamma \sqrt{2\pi}}
\left[ {c_{3}}
+\frac{2}{\delta} e^{-\frac{\delta}{2} c_{3}}
\right]
\right)\\
\quad\\
&+&
\left(
3^{n}\kappa
 C_{2}
\frac{e^{(\mu-\frac{\delta}{2})\delta}}{\delta^{3}}
\left(
\frac{
1}{e\delta}\right)^{n}
\frac{8}{\delta \gamma \sqrt{2\pi}}
\left[ {c_{3}}
+\frac{2}{\delta} e^{-\frac{\delta}{2} c_{3}}
\right]
\right)\\
\quad\\
&+&
\left.
|\xi|
\frac{1}{\delta^{2}}
 \right]\kappa E_{1}e^{-|k|(\rho -2\delta)}e^{(\mu -\frac{\delta}{2})t_{R}}
 ,
\quad
-\infty  < t_{R} < c_{2} <0.
\end{eqnarray*}
We take a pause to address the analyticity issue.
We know from the functions that make up
$\beta ^{E}_{j}(q')$ that $\beta ^{E}_{j}(q')\in {\cal{A}}_{\rho-2\delta, \sigma -2\delta}$.
Our estimates above demonstrate the Fourier coefficients of 
$\beta ^{E}_{j}(q')$ depend on $k$ as $e^{-|k|(\rho -2\delta)}$ which agrees 
by Lemma $\ref{lemmaanalib}$ with our expectations.
Next we want to further simplify the estimates above by removing
dependence on $C_{1}$, $C_{2}$, $G$, $\kappa$, $\tilde{H}^{0}$, 
and $\xi$.
We first find an estimate for $\| \xi \|$. Recall from Lemma $\ref{thmisob}$ there
exists a positive constant $f$ such that
\begin{eqnarray*}
\left \| \left(
\begin{array}{cc}
\overline{\overline{C}} & \lambda ^{T} \\
\lambda & 0
\end{array}
\right)
\left(
\begin{array}{c}
\xi \\
\kappa E_{1}\zeta_{A} 
\end{array}
\right)
\right\| \geq
f \left \|\left(
\begin{array}{c}
\xi \\
\kappa E_{1}\zeta_{A} 
\end{array}
\right)
\right\| \geq f \| \xi \|.
\end{eqnarray*}
Also from $(\ref{zetasolution})$ we have
\begin{eqnarray*}
\left\|
\left(
\begin{array}{cc}
\overline{\overline{C}} & \lambda ^{T} \\
\lambda & 0
\end{array}
\right)
\left(
\begin{array}{c}
\xi \\
\kappa E_{1}\zeta_{A} 
\end{array}
\right)
\right\| &=&
\left\|
\left(
\begin{array}{c}
\kappa \overline{\overline{B}} -
\overline{\overline{C}}
\cdot
\overline{\overline{\partial _{q'}X}} \\
\kappa
\overline{\overline{A}}
\end{array}
\right)
\right\|
\end{eqnarray*}
\begin{eqnarray*}
&\leq&
\kappa \|B \|_{\rho ,\sigma}
+
\| C \cdot \partial _{q'}X \|_{\rho ,\sigma}
+ \kappa \|A \|_{\rho ,\sigma}
\leq
2\kappa E_{1} + m^{-1}\|\partial _{q'}X \|_{\rho ,\sigma}\\
\quad\\
&\leq&
2\kappa E_{1}
+
\frac{\varpi \kappa}{m\Gamma \delta ^{2n+1}}
\|G(0,q) -\overline{\overline{G}}(0)\|_{\rho}
+
\frac{8\kappa }{m\delta ^{2}\gamma \sqrt{2\pi}}
\left[
C_{3}{c_{3}}+
\frac{2}{\delta}C_{3}e^{-\frac{\delta}{2} c_{3}}\right]
\left(\frac{4}{{{\delta}}} \right)^{n}.
\end{eqnarray*}
We finally obtain
\begin{eqnarray*}
\| \xi \| \leq
\frac{1}{f}\left[
2\kappa E_{1}
+
\frac{2\varpi \kappa}{m\Gamma \delta ^{2n+1}}E_{1}
+
\frac{8\kappa }{m\delta ^{2}\gamma \sqrt{2\pi}}
\left[
{c_{3}}+
\frac{2}{\delta}e^{-\frac{\delta}{2} c_{3}}\right]
\left(\frac{4}{{{\delta}}} \right)^{n}E_{1}\right].
\end{eqnarray*}
\quad\\
With this bound we can bound $|\xi| < (n+1)\|\xi \|$.
Also note 
\begin{eqnarray*}
\| G(0,q) -\overline{\overline{G}}(0) \|_{\rho}
&\leq&
2\|G(0,q) \|_{\rho}
\leq
2\| H^{1} \|_{\rho} \leq 2E_{1},
\quad
\|G(0,q)\|_{\rho} \leq E_{1}.
\end{eqnarray*}
The bounds on $\xi$, $G$  together with 
$
\kappa C_{1} \leq \kappa E_{1} <1,\quad
\kappa C_{2} \leq \kappa E_{1} <1,
$
and
\begin{eqnarray*}
\sum _{l=1}^{n} \frac{\partial^{2} \tilde{H}^{0}}{\partial p'_{l}\partial p'_{j}}
\leq \frac{n}{\delta}\|H \|_{\rho}\leq \frac{n}{\delta},
\end{eqnarray*}
are used to reduce the bounds 
on $| \beta ^{E}_{jk}(t) |$ to the following
\begin{eqnarray*}
\left| \beta^{E}_{jk}(t) \right| &\leq&
\left[
\frac{1}{\delta} 
 \right. 
+
\left(
\frac{3^{n}}{e\delta ^{3}}
\frac{1}{\Gamma}\left( \frac{n}{e\delta} \right)^{n}
\left(
\frac{1}{1-e^{-2\delta}}\right)^{n}
\right)
+
\left(
\left(
\frac{1}{e\delta}
\right)
\frac{n}{\delta}
\frac{8}{\delta \gamma \sqrt{2\pi}}
\left[ {c_{3}}
+\frac{2}{\delta} e^{-\frac{\delta}{2} c_{3}}
\right]
\right)\\
\quad\\
&+&
\left(
3^{n}
\left(
\frac{1}{e\delta ^{3}} \right)
\left(\frac{1}{
1-e^{-2\delta}}\right)^{n}
\frac{8}{\delta \gamma \sqrt{2\pi}}
\left[ {c_{3}}
+\frac{2}{\delta} e^{-\frac{\delta}{2} c_{3}}
\right]
\right)\\
\quad\\
&+&
\left(
\left(
\frac{3^{n}}{e\delta^{3}}
\right)
\left(\frac{
1}{1-e^{-2\delta}}\right)^{n}
\frac{8}{\delta \gamma \sqrt{2\pi}}
\left[ {c_{3}}
+\frac{2}{\delta} e^{-\frac{\delta}{2} c_{3}}
\right]
\right)\\
\quad\\
&+&
\left(
\frac{1}{\delta}
\frac{e^{(\nu -\frac{\delta}{2})\delta}}{\delta}
\frac{8}{\delta \gamma \sqrt{2\pi}}
\left[ {c_{3}}
+\frac{2}{\delta} e^{-\frac{\delta}{2} c_{3}}
\right]
\right)
\\
\quad\\
&+&
\left(
3^{n}
\left(\frac{1}{e\delta}\right)^{n}
\frac{e^{(\nu -\frac{\delta}{2})\delta}}{\delta^{3}}
\frac{8}{\delta \gamma \sqrt{2\pi}}
\left[ {c_{3}}
+\frac{2}{\delta} e^{-\frac{\delta}{2} c_{3}}
\right]
\right)\\
\quad\\
&+&
\left(
3^{n}
\frac{e^{(\nu-\frac{\delta}{2})\delta}}{\delta^{3}}
\left(
\frac{
1}{e\delta}\right)^{n}
\frac{8}{\delta \gamma \sqrt{2\pi}}
\left[ {c_{3}}
+\frac{2}{\delta} e^{-\frac{\delta}{2} c_{3}}
\right]
\right)\\
\quad\\
&+&
\left.
\frac{1}{\delta^{2}}
\frac{(n+1)}{f}\left[
2
+
\frac{2\varpi }{m\Gamma \delta ^{2n+1}}
+
\frac{8}{\delta \gamma \sqrt{2\pi}}
\left[ {c_{3}}
+\frac{2}{\delta} e^{-\frac{\delta}{2} c_{3}}
\right]
\left(\frac{4}{{{\delta}}} \right)^{n}\right]
\right]\\
\quad\\
&\cdot& \kappa E_{1}e^{-|k|(\rho -2\delta)}e^{-(\nu -\frac{\delta}{2})t_{R}}\\
\quad\\
&\leq&
\frac{3^{n} 16(n+1)\varpi }{\delta ^{4n+3}\Gamma \gamma f m }
\left[
c_{3}+
\frac{2}{\delta}e^{ -\frac{\delta}{2}c_{3}}\right]
\frac{e^{(\nu -\frac{\delta}{2})\delta}}{(1-e^{-\delta})^{n}}
\kappa E_{1}e^{-|k|(\rho -2\delta)}e^{-(\nu -\frac{\delta}{2})t_{R}}
\end{eqnarray*}
for $0 < c_{1} <t_{R} <\infty$.
\quad\\
Now we use the fact  for $x \geq 0$, $e^{-x}< x+1$. Setting $x= \frac{\delta}{2}
c_{3}$ we have
\begin{eqnarray*}
{c_{3}}
+\frac{2}{\delta}e^{-\frac{\delta}{2}c_{3}}
&<& 
{c_{3}}
+
\frac{2}{\delta}\left(\frac{\delta}{2}s_{3} +1 \right)
=
2c_{3}
+\frac{2}{\delta}
<
\frac{2}
{\delta}
(c_{3}+1)
.
\end{eqnarray*}
So that
\begin{eqnarray*}
|\beta ^{E}_{jk}(t)| &\leq&
\frac{3^{n}32(n+1)\varpi }{\delta ^{4n+4}\Gamma \gamma f m }
\left(
{c_{3}+1}
\right)
\frac{e^{(\nu -\frac{\delta}{2})\delta}}{(1-e^{-\delta})^{n}}
\kappa E_{1}e^{-|k|(\rho -2\delta)}e^{-(\nu -\frac{\delta}{2})t_{R}},
\end{eqnarray*}
for $0 < c_{1} <t_{R} <\infty$.
\quad\\
We further simplify by using the fact  $\delta < 1$ and the following
\begin{eqnarray*}
&\quad& 
\frac{1}{1-e^{-2\delta}} <
\frac{1}{1-e^{-\delta}} < \frac{1}{\delta} +1<\frac{2}{\delta}, 
\quad
e^{(\nu -\varepsilon)\delta} < e^{\nu -\varepsilon}.
\end{eqnarray*}
The exponential bound of $\beta ^{E}_{jk}(t)$ is
\begin{eqnarray*}
|\beta ^{E}_{jk}(t)| &\leq&
\frac{6^{n}32(n+1)\varpi }{\delta ^{5n+4}\Gamma \gamma f m }
(c_{3}+1)
e^{\nu}
\kappa E_{1}e^{-|k|(\rho -2\delta)}e^{-(\nu -\frac{\delta}{2})t_{R}}
=C^{\beta_{1}}\kappa E_{1}e^{-|k|(\rho -2\delta)}e^{-(\nu -\frac{\delta}{2})t_{R}}
,
\end{eqnarray*}
for
$
 0 < c_{1} \leq t_{R} < \infty
$
 and
\begin{eqnarray*}
|\beta ^{E}_{jk}(t)| &\leq&
\frac{6^{n}32(n+1)\varpi }{\delta ^{5n+4}\Gamma \gamma f m }
(c_{3}+1)
e^{\mu}
\kappa E_{1}e^{-|k|(\rho -2\delta)}e^{(\mu -\frac{\delta}{2})t_{R}}
=
C^{\beta_{2}}\kappa E_{1}e^{-|k|(\rho -2\delta)}e^{(\mu -\frac{\delta}{2})t_{R}}.
\end{eqnarray*}
for
$
 -\infty < t_{R} \leq c_{2} <0.
$
Next, by Theorem $\ref{maintheorem3}$ there exists $\delta>0, \tilde{\delta}>0,
\tilde{\tilde{\delta}}>0$, without loss of generality we set
$\delta =\tilde{\delta} =\tilde{\tilde{\delta}}$ and
$\varepsilon =\delta /2 $
, and $\gamma <\nu$,
$\gamma < \mu$ such that $(\ref{eqnexptimesolve})$ has a unique
solution given by
\begin{eqnarray*}
{\cal{F}}_{j}(q') =\sum _{k\in {\mathbb{Z}}^{n}} {\cal{F}}_{jk}(t)
e^{ik\cdot q} \in {\cal{A}}_{\rho -3\delta, \sigma -3\delta}.
\end{eqnarray*}
Note, since
\begin{eqnarray*}
\sum _{k \in {\mathbb{Z}}^{n}} \beta ^{E}_{jk} (t)e^{ik\cdot q}
\in {\cal{A}}_{\rho -2\delta , \sigma -2\delta},
\end{eqnarray*}
when using the bounds obtained in theorem $\ref{maintheorem3}$
we  replace $\delta$ with $3\delta$
 and the following estimates hold
\begin{eqnarray*}
&\quad&|{\cal{F}}_{jk}(t)| \leq
\frac{8\kappa}{5 \delta \gamma \sqrt{2\pi}}
e^{-|k|(\rho -2\delta)}e^{-(\nu -\frac{5}{2}\delta)t_{R}}
\left[
C^{\beta_{1}}{c_{3}} + \frac{2}{5\delta}C^{\beta_{1}}e^{-\frac{5}{2}\delta c_{3}}
 \right]E_{1},
\quad 0\leq t_{R} < \infty,\\
\quad\\
&\quad&|{\cal{F}}_{jk}(t)| \leq
\frac{8\kappa}{5 \delta \gamma \sqrt{2\pi}}
e^{-|k|(\rho -2\delta)}e^{(\mu -\frac{5}{2}\delta)t_{R}}
\left[
C^{\beta_{2}}{c_{3}} + \frac{2}{5\delta}C^{\beta_{2}}e^{-\frac{5}{2}\delta c_{3}}
 \right]E_{1},
\quad -\infty < t_{R} \leq 0,
\\
\quad\\
&\quad&\| {\cal{F}}_{j}\|_{\rho-3\delta, \sigma -3\delta}
\leq
\frac{8\kappa}{5 \delta \gamma \sqrt{2\pi}}
\left[
C^{\beta_{3}}{c_{3}} + \frac{2}{5\delta}C^{\beta_{3}}e^{-\frac{5}{2}\delta c_{3}}
 \right]
E_{1}
\left(\frac{4}{\delta} \right)^{n},\\
\quad\\
&\quad&\left \| \frac{\partial {\cal{F}}_{j}}{\partial q'}\right\|
_{\rho-3\delta, \sigma -3\delta}
\leq
\frac{8\kappa}{5 \delta ^{2}\gamma \sqrt{2\pi}}
\left[
C^{\beta_{3}}{c_{3}} + \frac{2}{5\delta}C^{\beta_{3}}e^{-\frac{5}{2}\delta c_{3}}
 \right]
E_{1}
\left(\frac{4}{\delta} \right)^{n},\\
\end{eqnarray*}
\quad\\
where $C^{\beta_{3}}= \max (C^{\beta_{1}}, C^{\beta_{2}})$.
Now that we have obtained estimates for ${\cal{S}}_{j}(q)$, ${\cal{F}}_{j}(q')$ and
their derivatives, we can obtain estimates for $Y_{j}(q')$, $Y(q')$ and
their derivatives. Recall
$
Y_{j}(q')= {\cal{S}}_{j}(q) + {\cal{F}}_{j}(q').
$
It follows
\begin{eqnarray*}
\| Y_{j}\|_{\rho -3\delta, \sigma -3\delta} &\leq&
\frac{\varpi}{\Gamma \delta^{2n}}\|\beta ^{Q}_{j}\|_{\rho-2\delta}
+
\frac{8\kappa}{5 \delta \gamma \sqrt{2\pi}}
\left[
C^{\beta_{3}}{c_{3}} + \frac{2}{5\delta}C^{\beta_{3}}e^{-\frac{5}{2}\delta c_{3}}
 \right]
E_{1}
\left(\frac{4}{\delta} \right)^{n},
\end{eqnarray*}
\begin{eqnarray*}
\| Y\|_{\rho -3\delta, \sigma -3\delta} &\leq&
\frac{\varpi}{\Gamma \delta^{2n}}\|\beta ^{Q}_{j}\|_{\rho-2\delta}
+
\frac{8\kappa}{5 \delta \gamma \sqrt{2\pi}}
\left[
C^{\beta_{3}}{c_{3}} + \frac{2}{5\delta}C^{\beta_{3}}e^{-\frac{5}{2}\delta c_{3}}
 \right]
E_{1}
\left(\frac{4}{\delta} \right)^{n},
\end{eqnarray*}
and
\begin{eqnarray*}
\left\| \frac{\partial Y_{j}}{\partial q'} \right\|_{\rho -3\delta, \sigma -3\delta} &\leq&
\frac{\varpi}{\Gamma \delta^{2n+1}}\|\beta ^{Q}_{j}\|_{\rho-2\delta}
+
\frac{8\kappa}{5 \delta ^{2}\gamma \sqrt{2\pi}}
\left[
C^{\beta_{3}}{c_{3}} + \frac{2}{5\delta}C^{\beta_{3}}e^{-\frac{5}{2}\delta c_{3}}
 \right]
E_{1}
\left(\frac{4}{\delta} \right)^{n},
\end{eqnarray*}
\begin{eqnarray*}
\left\|\frac{\partial Y}{\partial q'}\right\|_{\rho -3\delta, \sigma -3\delta} &\leq&
\frac{\varpi}{\Gamma \delta^{2n+1}}\|\beta ^{Q}_{j}\|_{\rho-2\delta}
+
\frac{8\kappa}{5 \delta ^{2}\gamma \sqrt{2\pi}}
\left[
C^{\beta_{3}}{c_{3}} + \frac{2}{5\delta}C^{\beta_{3}}e^{-\frac{5}{2}\delta c_{3}}
 \right]
E_{1}
\left(\frac{4}{\delta} \right)^{n}.
\end{eqnarray*}
Now we apply the estimates above to the generating function
$\chi= X(q')+\xi \cdot q' +\sum_{i} Y_{i}(q')p'_{i}$ .
We have
\begin{eqnarray}
\label{1121}
 \left\| \frac {\partial \chi}{\partial p'} \right\|
 _{\rho -3\delta,\sigma -3\delta}
=\|Y \|_{\rho -3\delta,\sigma- 3\delta}
\leq
\frac{\varpi}{\Gamma \delta ^{2n}} \|\beta ^{Q}_{j}\|_{\rho-2\delta}
+
\frac{8\kappa}{5 \delta \gamma \sqrt{2\pi}}
\left[
C^{\beta_{3}}{c_{3}} + \frac{2}{5\delta}C^{\beta_{3}}e^{-\frac{5}{2}\delta c_{3}}
 \right]
E_{1}
\left(\frac{4}{\delta} \right)^{n}.
\end{eqnarray}
\quad\\
Next we must relate the constant $\| \beta ^{Q}_{j}(q) \|_{\rho -2\delta}$
and $E_{1}$.
To do this we find the following estimates for the terms 
that make up  $\| \beta ^{Q}_{j}(q) \|_{\rho -2\delta}$. The first term
\begin{eqnarray*}
\kappa \sum _{k\in {\mathbb{Z}}^{n} \backslash 0} \frac{\partial s_{k}^{1}}{
\partial p'_{j}}(0)e^{ik\cdot q}
&=&
\kappa \frac{\partial}{\partial p'_{j}}
\left(
\sum _{k\in {\mathbb{Z}}^{n}\backslash 0}
s_{k}^{1}(p')e^{ik\cdot q}
\right)
\Big | _{p'=0}
=
\kappa 
\frac{\partial}{\partial p'_{j}}
\left( G(p',q) - \overline{\overline{G}}(p')\right)
\Big | _{p'=0}
\end{eqnarray*}
and
\begin{eqnarray*}
\left| \kappa
\frac{\partial (G-\overline{\overline{G}})}{\partial p'_{j}}(0,q)\right|
&\leq&
\frac{\kappa }{\delta}
\| G(0,q) - \overline{\overline{G}}(0)\|_{\rho -\delta}
\leq
\frac{2\kappa}{\delta}E_{1}, \quad \forall q\in {\cal{D}}_{\rho -2\delta}.
\end{eqnarray*}
\quad\\
The second term
\begin{eqnarray*}
\sum _{k\in {\mathbb{Z}}^{n}\backslash 0} \sum ^{n}_{l=1}
ik_{l}y_{k}\frac{\partial ^{2} \tilde{H}^{0}}{\partial p'_{l} \partial p'_{j}}
(0)e^{ik\cdot q} &=&
\sum ^{n}_{l=1}
\frac{\partial ^{2} \tilde{H}^{0}}{\partial p'_{l} \partial p'_{j}}
\sum _{k\in {\mathbb{Z}}^{n}\backslash 0}
ik_{l}y_{k}
e^{ik\cdot q}
=
\sum ^{n}_{l=1}
\frac{\partial ^{2} \tilde{H}^{0}}{\partial p'_{l} \partial p'_{j}}
\frac{\partial {\cal{Y}}}{\partial q'_{l}}(q), 
\end{eqnarray*}
and
\begin{eqnarray*}
\left|
\sum _{k\in {\mathbb{Z}}^{n}\backslash 0} \sum ^{n}_{l=1}
ik_{l}y_{k}\frac{\partial ^{2} \tilde{H}^{0}}{\partial p'_{l} \partial p'_{j}}
(0)e^{ik\cdot q}
\right|
&\leq&
\left|
\sum ^{n}_{l=1}
\frac{\partial ^{2} \tilde{H}^{0}}{\partial p'_{l} \partial p'_{j}}
\right|
\frac{1}{\delta}\|{\cal{Y}}(q) \|_{\rho -\delta}
\leq
\left|
\sum ^{n}_{l=1}
\frac{\partial ^{2} \tilde{H}^{0}}{\partial p'_{l} \partial p'_{j}}
\right|
\frac{\varpi \kappa }{\Gamma \delta ^{2n+1}} \|G(0,q) -\overline{\overline{G}}(0) \|_{\rho}\\
\quad\\
&\leq&
\frac{2\varpi \kappa E_{1} }{\Gamma \delta ^{2n+1}}
\left|
\sum ^{n}_{l=1}
\frac{\partial ^{2} \tilde{H}^{0}}{\partial p'_{l} \partial p'_{j}}
\right|
\leq
\frac{2\varpi \kappa E_{1} }{\Gamma \delta ^{2n+1}}
\frac{n}{\delta},
\end{eqnarray*}
\quad\\
where we have used in the last inequality Cauchy's inequality and 
the fact that $\| H \|_{\rho ,\sigma }\leq 1$.
The third term
\begin{eqnarray*}
&\kappa&
\sum _{k\in {\mathbb {Z}}^{n}}
\sum _{l=1}^{n}
\left(
\sum _{m\in {\mathbb {Z}}^{n}}
i
\frac{ \partial ^{2} s^{1}
_{m}}
{\partial p'_{l} \partial p'_{j}}
(0)
k_{l}y_{k}e^{im\cdot q}e^{ik\cdot q}
-
i\frac{\partial ^{2}s_{-k}^{1}}{\partial p'_{l} \partial p'_{j}}(0)k_{l}y_{k}
\right)
\\
\quad\\
&\quad&= \kappa 
\sum _{l=1}^{n}
\left(
\sum _{k\in {\mathbb {Z}}^{n}}
ik_{l}y_{k}
e^{ik\cdot q}
\right)
\left(
\sum _{m\in {\mathbb {Z}}^{n}}
\frac{ \partial ^{2} s^{1}
_{m}}
{\partial p'_{l} \partial p'_{j}}
(0)
e^{im\cdot q}\right)
-i
\kappa 
\sum _{k\in {\mathbb {Z}}^{n}}
\sum _{l=1}^{n}
\frac{\partial ^{2}s_{-k}^{1}}{\partial p'_{l} \partial p'_{j}}(0)k_{l}y_{k}
\\
\quad\\
&\quad&=
\kappa
\sum _{l=1}^{n}
\frac{\partial  {\cal{Y}}}{\partial q'_{l}}(q)
\frac{\partial ^{2} G(p',q)}{\partial p'_{l} \partial p'_{j}}
\Big | _{p'=0}
-i
\kappa 
\sum _{k\in {\mathbb {Z}}^{n}}
\sum _{l=1}^{n}
\frac{\partial ^{2}s_{-k}^{1}}{\partial p'_{l} \partial p'_{j}}(0)k_{l}y_{k},
\end{eqnarray*}
so 
\begin{eqnarray*}
&\quad&
\left| 
\kappa
\sum _{k\in {\mathbb {Z}}^{n}}
\sum _{l=1}^{n}
\left(
\sum _{m\in {\mathbb {Z}}^{n}}
i
\frac{ \partial ^{2} s^{1}
_{m}}
{\partial p'_{l} \partial p'_{j}}
(0)
k_{l}y_{k}e^{im\cdot q}e^{ik\cdot q}
-
i\frac{\partial ^{2}s_{-k}^{1}}{\partial p'_{l} \partial p'_{j}}(0)k_{l}y_{k}
\right)\right|\\
\quad\\
&\quad& \quad \leq
\frac{n\kappa}{\delta ^{2}}
\|{\cal{Y}}(q) \|_{\rho -\delta}
\|G(0,q) \|_{\rho -\delta}
+
\kappa 
\sum _{k\in {\mathbb {Z}}^{n}}
\frac{1}{\delta}\| s_{-k}^{1} (p') \|_{\rho -\delta}
|k||y_{k}|\\
\quad\\
&\quad&\quad  \leq
\frac{n\kappa}{\delta ^{2}}
\frac{\varpi \kappa }{\Gamma \delta ^{2n}} \|G(0,q) -\overline{\overline{G}}(0) \|_{\rho}
\|G(0,q) \|_{\rho -\delta}\\
\quad\\
&\quad & \quad \quad +
\kappa 
\sum _{k\in {\mathbb {Z}}^{n}}
\frac{1}{\delta}
\|G \|_{\rho-\delta}e^{-|k|(\rho -\delta)}
\left( \frac{1}{e\delta} \right)e^{|k|\delta}
\frac{\kappa}{\Gamma}\left( \frac{n}{e\delta} \right)^{n}
\|G(0,q) -\overline{\overline{G}}(0) \|_{\rho}e^{-|k|(\rho -\delta)}
\\
\quad\\
&\quad &\quad \leq
\frac{2n\varpi \kappa ^{2}}{\Gamma \delta ^{2n+2}}
E_{1}^{2}
+
\left( \frac{\kappa^{2}}{e\delta^{2}\Gamma} \right)
\left( \frac{n}{e\delta} \right)^{n}
E_{1}^{2}
\sum _{k\in {\mathbb {Z}}^{n}}
e^{-|k|(2\rho -\delta)}\\
\quad\\
&\quad&\quad =
\frac{2n\varpi  \kappa ^{2}}{\Gamma \delta ^{2n+2}}
E_{1}^{2}
+
\left( \frac{2\kappa^{2}}{e\delta^{2}\Gamma} \right)
\left( \frac{n}{e\delta} \right)^{n}
E_{1}^{2}
\left(\frac{1}{1-e^{-(2\rho -\delta)}}\right).
\end{eqnarray*}
The fourth term
\begin{eqnarray*}
&\quad&
\left|
\sum _{k\in {\mathbb{Z}}^{n} \backslash 0}
\left(
\sum _{l=1}^{n+1}
\kappa
\xi_{l}
\frac{\partial ^{2} s_{k}^{1}}{\partial p'_{l}
\partial p'_{j}}(0)\right)
e^{ik\cdot q}
\right|
\leq
\kappa 
\sum _{l=1}^{n+1}
|\xi _{l}|
\left|
\sum _{k\in {\mathbb{Z}}^{n} \backslash 0}
\frac{\partial ^{2} s_{k}^{1}}{\partial p'_{l}
\partial p'_{j}}(0)
e^{ik\cdot q}
\right|\\
\quad\\
&\quad& \quad \leq
\kappa 
\sum _{l=1}^{n+1}
|\xi _{l}|
\left|
\frac{\partial ^{2}}{\partial p'_{l} \partial p'_{j}}
(G(p',q)-\overline{\overline{G}}(p'))\Big|  _{p'=0}
\right|
\leq
\kappa 
\sum _{l=1}^{n+1}
|\xi _{l}|\frac{1}{\delta}
\|G(0,q)-\overline{\overline{G}}(0) \|_{\rho}
\leq \frac{2\kappa E_{1}}{\delta} |\xi|,
\end{eqnarray*}
and with the four estimates above we obtain
\begin{eqnarray*}
\|\beta ^{Q}_{j}(q) \|_{\rho}
&\leq&
\frac{2\kappa}{\delta}E_{1}
+
\frac{2\varpi \kappa E_{1} }{\Gamma \delta ^{2n+1}}
\frac{n}{\delta}
+
\frac{2n\varpi  \kappa ^{2}}{\Gamma \delta ^{2n+2}}
E_{1}^{2}
 +
\left( \frac{2\kappa^{2}}{e\delta^{2}\Gamma} \right)
\left( \frac{n}{e\delta} \right)^{n}
E_{1}^{2}
\left(\frac{1}{1-e^{-(2\rho -\delta)}}\right)
+
\frac{2\kappa E_{1}}{\delta} |\xi|.
\end{eqnarray*}
Finally we have
\begin{eqnarray*}
 \left\| \frac {\partial \chi}{\partial p'} \right\|
 _{\rho -3\delta,\sigma -3\delta}
&\leq&
\frac{\varpi}{\Gamma \delta ^{2n}}
\left[
\frac{2\kappa}{\delta}E_{1}
+
\frac{2\varpi \kappa E_{1} }{\Gamma \delta ^{2n+1}}
\frac{n}{\delta}
+
\frac{2n\varpi  \kappa ^{2}}{\Gamma \delta ^{2n+2}}
E_{1}^{2}
+
\left( \frac{2\kappa^{2}}{e\delta^{2}\Gamma} \right)
\left( \frac{n}{e\delta} \right)^{n}
E_{1}^{2}
\left(\frac{1}{1-e^{-(2\rho -\delta)}}\right)
+
\frac{2\kappa E_{1}}{\delta} |\xi|
\right]\\
\quad\\
&\quad&+
\frac{8\kappa}{5 \delta \gamma \sqrt{2\pi}}
\left[
C^{\beta_{3}}{c_{3}} + \frac{2}{5\delta}C^{\beta_{3}}e^{-\frac{5}{2}\delta c_{3}}
 \right]
\left(\frac{4}{\delta} \right)^{n}
E_{1}
.
\end{eqnarray*}
Next
\begin{eqnarray*}
\left\| \frac {\partial \chi}{\partial q'} \right\|
 _{\rho -3{\delta},\sigma -3\delta}
&\leq&
\left\| \frac {\partial X}{\partial q'} \right\|
_{\rho -3\delta , \sigma -3\delta} 
+
(\rho -3\delta)(n+1)\left\|
\frac {\partial Y}{\partial q'} \right\|
_{\rho -3{\delta},\sigma -3\delta}
+\|\xi \|
_{\rho -3{\delta},\sigma -3\delta} .
\end{eqnarray*}

\quad \\
We put together the bounds for $\| \xi \|$, $\left\| \frac {\partial X}{\partial q'}
\right\|_{{\rho} -3\delta , {\sigma}-3\delta }$ and
$\left\| \frac {\partial Y}{\partial q'}
\right\|_{{\rho}-3\delta,{\sigma}-3\delta}$
and obtain the following
\quad\\
\begin{eqnarray*}
\left\| \frac {\partial \chi}{\partial q'} \right\|
_{\rho -3{\delta},\sigma -3\delta}
 &< &
\frac{2\varpi \kappa}{\Gamma \delta ^{2n+1}}
E_{1}
+
\frac{8\kappa }{\delta ^{2}\gamma \sqrt{2\pi}}
\left[
{c_{3}}+
\frac{2}{\delta } e^{-\frac{\delta c_{3}}{2}}\right]
\left(\frac{4}{{{\delta}}} \right)^{n}E_{1}
\\
\quad\\
&+&
(n+1)\frac{\varpi}{\Gamma \delta^{2n+1}}\|\beta ^{Q}_{j}\|_{\rho-2\delta}
+
\frac{8\kappa (n+1)}{5 \delta ^{2}\gamma \sqrt{2\pi}}
\left[
C^{\beta_{3}}{c_{3}} + \frac{2}{5\delta}C^{\beta_{3}}e^{-\frac{5}{2}\delta c_{3}}
 \right]
E_{1}
\left(\frac{4}{\delta} \right)^{n}
\\
\quad \\
&+&
\frac{1}{f}\left[
2\kappa E_{1}
+
\frac{2\varpi \kappa}{m\Gamma \delta ^{2n+1}}E_{1}
+
\frac{8\kappa }{m\delta ^{2}\gamma \sqrt{2\pi}}
\left[
{c_{3}}+
\frac{2}{\delta}e^{-\frac{\delta}{2} c_{3}}\right]
\left(\frac{4}{{{\delta}}} \right)^{n}E_{1}\right].
\end{eqnarray*}
\quad \\
We set $\tilde{\rho} \equiv \rho -3\delta$, $\tilde{\sigma} \equiv \sigma - 3\delta$ and
examine the estimates for
$\left\| \frac {\partial \chi}{\partial p'}
\right\|_{\tilde{\rho},\tilde{\sigma}}$ and
$\left\| \frac {\partial \chi}{\partial q'}
\right\|_{\tilde{\rho},\tilde{\sigma}}$.
Since $\delta << 1$, the first term of 
$\left\| \frac {\partial \chi}{\partial p'}
\right\|_{\tilde{\rho},\tilde{\sigma}}$,from $(\ref{1121})$, is less than the third term of 
$\left\| \frac {\partial \chi}{\partial q'}
\right\|_{\tilde{\rho},\tilde{\sigma}}$
\quad\\
\begin{eqnarray*}
\frac{\varpi}{\Gamma \delta ^{2n}}\| \beta ^{Q}_{j}(q) \|_{\rho -2\delta}
<
(n+1)\frac{\varpi}{\Gamma \delta ^{2n+1}}\| \beta ^{Q}_{j}(q) \|_{\rho -2\delta}.
\end{eqnarray*}
\quad\\
Also, the second term of 
$\left\| \frac {\partial \chi}{\partial p'}
\right\|_{\tilde{\rho},\tilde{\sigma}}$,from $(\ref{1121})$, is less than
the fourth term of
$\left\| \frac {\partial \chi}{\partial q'}
\right\|_{\tilde{\rho},\tilde{\sigma}}$
\quad\\
\begin{eqnarray*}
\frac{8\kappa }{5 \delta \gamma \sqrt{2\pi}}
\left[
C^{\beta_{3}}{c_{3}} + \frac{2}{5\delta}C^{\beta_{3}}e^{-\frac{5}{2}\delta c_{3}}
 \right]
E_{1}
<
\frac{8\kappa (n+1)}{5 \delta ^{2}\gamma \sqrt{2\pi}}
\left[
C^{\beta_{3}}{c_{3}} + \frac{2}{5\delta}C^{\beta_{3}}e^{-\frac{5}{2}\delta c_{3}}
 \right]
E_{1}
.
\end{eqnarray*}
\quad\\
Consequently we obtain the following estimate
\quad\\
\begin{eqnarray*}
\chi^{*}_{  \tilde{\rho}, \tilde {\sigma}}
&=&
\max \left( \left\| \frac {\partial \chi}{\partial q'}
\right\|_{\tilde{\rho},\tilde{\sigma}}, \left\| \frac {\partial \chi}{\partial p'} \right\|_{\tilde
{\rho},\tilde {\sigma}} \right)
=\left\| \frac {\partial \chi}{\partial q'}
\right\|_{\tilde{\rho},\tilde{\sigma}}
=
\frac{2\varpi \kappa}{\Gamma \delta ^{2n+1}}
E_{1}
+
\frac{8\kappa }{\delta ^{2}\gamma \sqrt{2\pi}}
\left[
{c_{3}}+
\frac{2}{\delta } e^{-\frac{\delta c_{3}}{2}}\right]
\left(\frac{4}{{{\delta}}} \right)^{n}E_{1}
\\
\quad\\
&\quad&
+
(n+1)\frac{\varpi}{\Gamma \delta^{2n+1}}\|\beta ^{Q}_{j}(q) \|_{\rho-2\delta}
+
\frac{8\kappa (n+1)}{5 \delta ^{2}\gamma \sqrt{2\pi}}
\left[
C^{\beta_{3}}{c_{3}} + \frac{2}{5\delta}C^{\beta_{3}}e^{-\frac{5}{2}\delta c_{3}}
 \right]
E_{1}
\left(\frac{4}{\delta} \right)^{n}
\\
\quad \\
&\quad&
+
\frac{1}{f}\left[
2\kappa E_{1}
+
\frac{2\varpi \kappa}{m\Gamma \delta ^{2n+1}}E_{1}
+
\frac{8\kappa }{m\delta ^{2}\gamma \sqrt{2\pi}}
\left[
{c_{3}}+
\frac{2}{\delta}e^{-\frac{\delta}{2} c_{3}}\right]
\left(\frac{4}{{{\delta}}} \right)^{n}E_{1}\right]
\\
\quad\\
&\leq&
\frac{2\varpi \kappa}{\Gamma \delta ^{2n+1}}
E_{1}
+
\frac{8\kappa }{\delta ^{2}\gamma \sqrt{2\pi}}
\left[
{c_{3}}+
\frac{2}{\delta } e^{-\frac{\delta c_{3}}{2}}\right]
\left(\frac{4}{{{\delta}}} \right)^{n}E_{1}
+
(n+1)\frac{\varpi}{\Gamma \delta^{2n+1}}
\left[
\frac{2\kappa}{\delta}E_{1}
+
\frac{2\varpi \kappa E_{1} }{\Gamma \delta ^{2n+1}}
\frac{n}{\delta}
+
\frac{2n\varpi  \kappa ^{2}}{\Gamma \delta ^{2n+2}}
E_{1}^{2}\right.
\\
\quad\\
&\quad&
\left.
 +
\left( \frac{2\kappa^{2}}{e\delta^{2}\Gamma} \right)
\left( \frac{n}{e\delta} \right)^{n}
E_{1}^{2}
\left(\frac{1}{1-e^{-(2\rho -\delta)}}\right)
+
\frac{2\kappa E_{1}}{\delta} |\xi|
\right]
+
\frac{8\kappa (n+1)}{5 \delta ^{2}\gamma \sqrt{2\pi}}
\left[
C^{\beta_{3}}{c_{3}} + \frac{2}{5\delta}C^{\beta_{3}}e^{-\frac{5}{2}\delta c_{3}}
 \right]
E_{1}
\left(\frac{4}{\delta} \right)^{n}
\\
\quad \\
&\quad&
+
\frac{1}{f}\left[
2\kappa E_{1}
+
\frac{2\varpi \kappa}{m\Gamma \delta ^{2n+1}}E_{1}
+
\frac{8\kappa }{m\delta ^{2}\gamma \sqrt{2\pi}}
\left[
{c_{3}}+
\frac{2}{\delta}e^{-\frac{\delta}{2} c_{3}}\right]
\left(\frac{4}{{{\delta}}} \right)^{n}E_{1}\right]
\\
\quad\\
&\leq&
\frac{2\varpi \kappa}{\Gamma \delta ^{2n+1}}
E_{1}
+
\frac{8\kappa }{\delta ^{2}\gamma \sqrt{2\pi}}
\left[
{c_{3}}+
\frac{2}{\delta } e^{-\frac{\delta c_{3}}{2}}\right]
\left(\frac{4}{{{\delta}}} \right)^{n}E_{1}
+
(n+1)\frac{\varpi}{\Gamma \delta^{2n+1}}
\left[
\frac{2\kappa}{\delta}E_{1}
+
\frac{2\varpi \kappa E_{1} }{\Gamma \delta ^{2n+1}}
\frac{n}{\delta}
+
\frac{2n\varpi  \kappa ^{2}}{\Gamma \delta ^{2n+2}}
E_{1}^{2}\right.
\\
\quad\\
&\quad&
\left.
 +
\left( \frac{2\kappa^{2}}{e\delta^{2}\Gamma} \right)
\left( \frac{n}{e\delta} \right)^{n}
E_{1}^{2}
\left(\frac{1}{1-e^{-(2\rho -\delta)}}\right) \right]\\
\quad\\
&\quad&
+
\frac{2\varpi \kappa (n+1)^{2}}{\Gamma \delta ^{2n+2}f}E_{1} 
\left[
2\kappa E_{1}
+
\frac{2\varpi \kappa}{m\Gamma \delta ^{2n+1}}E_{1}
+
\frac{8\kappa }{m\delta ^{2}\gamma \sqrt{2\pi}}
\left[
{c_{3}}+
\frac{2}{\delta}e^{-\frac{\delta}{2} c_{3}}\right]
\left(\frac{4}{{{\delta}}} \right)^{n}E_{1}\right]
\\
\quad\\
&\quad&
+
\frac{8\kappa (n+1)}{5 \delta ^{2}\gamma \sqrt{2\pi}}
\left[
C^{\beta_{3}}{c_{3}} + \frac{2}{5\delta}C^{\beta_{3}}e^{-\frac{5}{2}\delta c_{3}}
 \right]
E_{1}
\left(\frac{4}{\delta} \right)^{n}
+
\frac{1}{f}\left[
2\kappa E_{1}
+
\frac{2\varpi \kappa E_{1}}{m\Gamma \delta ^{2n+1}}
+
\frac{8\kappa E_{1}}{m\delta ^{2}\gamma \sqrt{2\pi}}
\left[
{c_{3}}+
\frac{2}{\delta}e^{-\frac{\delta}{2} c_{3}}\right]
\left(\frac{4}{{{\delta}}} \right)^{n}\right].
\end{eqnarray*}
\quad\\
There are thirteen terms in the estimate above 
which we would like to reduce to one term. We first list all the terms
\begin{eqnarray*}
&1)& \frac{2\varpi \kappa}{\Gamma
\delta ^{2n+1}}E_{1},\\
\quad\\
&2)& 
\frac{8\kappa }{\delta ^{2}\gamma \sqrt{2\pi}}
\left[
{c_{3}}+
\frac{2}{\delta } e^{-\frac{\delta c_{3}}{2}}\right]
\left(\frac{4}{{{\delta}}} \right)^{n}E_{1}
,\\
\quad\\
&3)&
\frac{2\varpi \kappa}{\Gamma
\delta ^{2n+2}}(n+1)E_{1},\\
\quad\\
&4)&
\frac{2\varpi ^{2} \kappa}{\Gamma ^{2}
\delta ^{4n+3}}n(n+1)E_{1},\\
\quad\\
&5)&
\frac{2\varpi ^{2} \kappa ^{2}}{\Gamma ^{2}
\delta ^{4n+3}}n(n+1)E_{1} ^{2},\\
\quad\\
&6)&
\frac{2\varpi \kappa ^{2}}{\Gamma ^{2}e \delta ^{3n+3}}
\left(\frac{n}{e}\right)^{n}(n+1)\left(
\frac{1}{1-e^{-(2\rho -\delta)}} \right)E_{1}^{2},\\
\quad\\
&7)&
\frac{4\varpi \kappa ^{2}}{\delta ^{2n+2}\Gamma f}(n+1)^{2}E_{1}^{2},\\
\quad\\
&8)&
\frac{4\varpi ^{2}\kappa ^{2}}{m\Gamma ^{2}\delta ^{4n+3}f}
(n+1)^{2}E_{1}^{2},\\
\quad\\
&9)&
\frac{16 \varpi \kappa ^{2}}{m\delta ^{2n+4}\Gamma f\gamma \sqrt{2\pi}}(n+1)^{2}
\left[
c_{3}+\frac{2}{\delta}e^{-\frac{\delta}{2}c_{3}} 
\right]
\left(\frac{4}{\delta}\right)^{n}E_{1}^{2},\\
\quad\\
&10)&
\frac{8\kappa(n+1)}{5\delta ^{2}\gamma \sqrt{2\pi}}
\left[
C^{\beta _{3}}c_{3}+
\frac{2}{5\delta}C^{\beta _{3}}e^{-\frac{5}{2}\delta c_{3}}\right]
\left(\frac{4}{\delta}\right)^{n}E_{1},\\
\quad\\
&11)&
\frac{2\kappa }{f}E_{1},\\
\quad\\
&12)&
\frac{2\varpi \kappa}{m\Gamma f \delta ^{2n+1}}E_{1},\\
\quad\\
&13)&
\frac{8\kappa}{fm\delta ^{2}\gamma \sqrt{2\pi}}
\left[
c_{3}+
\frac{2}{\delta}e^{-\frac{\delta}{2}c_{3}}\right]
\left(\frac{4}{\delta}\right)^{n}E_{1}
.
\end{eqnarray*}
We begin to combine terms keeping in mind
$\Gamma, \gamma, \delta, m, f, \kappa E_{1} <1$ and $\varpi >>1.$
 The sum of terms $1$ and $3$
is bounded by
\begin{eqnarray*}
\frac{2\varpi \kappa}{\Gamma \delta ^{2n+2}}(n+2)E_{1},
\end{eqnarray*}
the sum of terms $2$ and $13$
is bounded by
\begin{eqnarray*}
\frac{16 \kappa}{fm\gamma \delta^{2} \sqrt{2\pi}}
\left[
c_{3}+
\frac{2}{\delta}e^{-\frac{\delta}{2}c_{3}}\right]
\left(\frac{4}{\delta}\right)^{n}E_{1}
\end{eqnarray*}
the sum terms $1$, $3$, and $4$, knowing that $\varpi >1$ and
assuming without loss of generality that $\Gamma <1$, is bounded by
\begin{eqnarray*}
\frac{2\varpi ^{2} \kappa}{\Gamma ^{2}\delta ^{4n+3}}
(n^{2}+2n+2)E_{1},
\end{eqnarray*}
the sum of terms $11$ and $12$ is bounded by
\begin{eqnarray*}
\frac{4\varpi \kappa}{m\Gamma f \delta ^{2n+1}}E_{1},
\end{eqnarray*}
the sum of terms $1$, $3$, $4$, $11$, and $12$ is bounded by
\begin{eqnarray*}
\frac{2\varpi ^{2} \kappa}{mf\Gamma ^{2} \delta ^{4n+3}}(n^{2}+2n+4)E_{1},
\end{eqnarray*}
the sum of terms $7$ and $8$ is bounded by 
\begin{eqnarray*}
\frac{8\varpi \kappa ^{2}}{m\Gamma ^{2}f \delta ^{4n+3}}(n+1)^{2}E_{1}^{2},
\end{eqnarray*}
the sum of terms $7$, $8$, and $5$ is bounded by
\begin{eqnarray*}
\frac{\varpi^{2} \kappa^{2}}{mf\Gamma ^{2} \delta ^{4n+3}}(10n^{2}+18n +8)E_{1}^{2},
\end{eqnarray*}
the sum of terms $7$, $8$, $5$, and $6$ is bounded by
\begin{eqnarray*}
\left(1 +
\frac{1}{1-e^{-(2\rho -\delta)}}
\right)
\frac{\varpi^{2} \kappa^{2}}{mf\Gamma ^{2} \delta ^{4n+3}}(10n^{2}+18n +8)E_{1}^{2}
,
\end{eqnarray*}
where, to bound $6$, we have used the fact $\frac{2}{e}\left(
\frac{n}{e} \right)^{n} < \varpi$.
The sum of terms $1$, $3$, $4$, $11$, $12$, $7$, $8$, $5$, and $6$
is bounded by
\begin{eqnarray}
\label{eqnfh1}
\left(2 +
\frac{1}{1-e^{-(2\rho -\delta)}}
\right)
\frac{\varpi^{2} }{mf\Gamma ^{2} \delta ^{4n+3}}(10n^{2}+18n +8)\kappa E_{1}
,
\end{eqnarray}
the sum of terms $9$, $2$, and $13$ is bounded by
\begin{eqnarray*}
\frac{16 \varpi (n^{2}+2n+2) }{fm \Gamma \gamma \delta^{2n+4} \sqrt{2\pi}}
\left[
c_{3}+
\frac{2}{\delta}e^{-\frac{\delta}{2}c_{3}}\right]
\left(\frac{4}{\delta}\right)^{n}\kappa E_{1}
,
\end{eqnarray*}
and the sum of terms $9$, $2$, $13$, and $10$
is bounded by
\begin{eqnarray}
\label{eqnfh2}
\left(
1 + C^{\beta_{3}}\right)
\frac{ \varpi 16(n^{2}+2n+2) }{fm \Gamma \gamma \delta^{2n+4} \sqrt{2\pi}}
\left[
c_{3}+
\frac{2}{\delta}e^{-\frac{\delta}{2}c_{3}}\right]
\left(\frac{4}{\delta}\right)^{n}\kappa E_{1}
.
\end{eqnarray}
\quad\\
We use the fact 
\begin{eqnarray*}
c_{3}+
\frac{2}{\delta}e^{-\frac{\delta}{2}c_{3}}
< 
\frac{2}{\delta}(c_{3}+1),
\end{eqnarray*}
together with 
\begin{eqnarray*}
\frac{1}{1-e^{-(2\rho -\delta)}} &<& \frac{1}{2\rho -\delta}+1
\leq
\frac{1}{\rho -\delta} +1
\leq \frac{1}{\rho_{*}}+1< \frac{2}{\rho_{*}}
\end{eqnarray*}
and combining $(\ref{eqnfh1})$ and $(\ref{eqnfh2})$ we obtain
\begin{eqnarray*}
\chi ^{*}_{\tilde {\rho}, \tilde {\sigma}} &<&
\left(
\frac{2}{\rho_{*}} + C^{\beta_{3}}\right)
\frac{16 \varpi ^{2} (10n^{2}+18n+8) }{fm \Gamma ^{2}\gamma \delta^{4n+3} }
\frac{2}{\delta}
\left(
c_{3}+1
\right)
\left(\frac{4}{\delta}\right)^{n}\kappa E_{1}.
\end{eqnarray*}
\quad\\
Recall
\begin{eqnarray*}
&\quad& C^{\beta_{1}}=
\frac{6^{n}32(n+1)\varpi}{\delta^{5n+4}\Gamma \gamma fm}(c_{3}+1)
e^{\nu},
\quad  C^{\beta_{2}}=
\frac{6^{n}32(n+1)\varpi}{\delta^{5n+4}\Gamma \gamma fm}(c_{3}+1)
e^{\mu}.
\end{eqnarray*}
Then
\begin{eqnarray*}
C^{\beta_{3}}=\max (C^{\beta_{1}},C^{\beta_{2}})<
\frac{6^{n}32(n+1)\varpi}{\delta^{5n+4}\Gamma \gamma fm}(c_{3}+1)
e^{\nu +\mu}.
\end{eqnarray*}
\quad\\
Finally
\begin{eqnarray*}
\chi ^{*}_{\tilde {\rho}, \tilde {\sigma}} &<&
\frac{3^{n}2^{12}\varpi ^{3}(6n+6)^{3}(c_{3}+1)^{2}}
{\rho_{*} f^{2}m^{2}\Gamma ^{3} \gamma ^{2}
\delta^{9n+8}}e^{\nu +\mu}
\left(\frac{4}{\delta}\right)^{n}\kappa E_{1}
=
\frac{\delta}{6n+6}\eta,
\end{eqnarray*}
where
\begin{eqnarray}
&\quad&
\eta =
\frac{\Lambda }{f^{2}m^{2}\delta ^{\aleph}}\kappa E_{1},\\
&\quad& \aleph =10n+9,\\
&\quad&
\label{Lambdaexpression}
\Lambda =
\frac{3^{n}2^{2n+12}\varpi ^{3}(6n+6)^{4}(c_{3}+1)^{2}}{\rho_{*}\Gamma ^{3}
\gamma ^{2}}e^{\nu +\mu}.
\end{eqnarray}
With the estimates on the derivatives of the generating function
one can now estimate the analyticity domain of the flow
associated with the corresponding Hamiltonian vector field.
We have obtain the following
\begin{equation}
\label{chibeta}
\chi ^{*}_{\tilde {\rho}, \tilde {\sigma}} 
<
\frac{{\delta}}{6n+6}\eta
\end{equation}
where 
\[
\tilde{\rho }= \rho - 3{\delta},\quad
\tilde {\sigma }=
\sigma -3\delta.\]
In particular, note  by $(\ref{etasize})$ it follows $\eta <1$ since it can be
written in the form
\\
\[\eta < \sigma_{*}^{2}\frac {m-m_{*}}{4(n+1)},\]
\\
and therefore
\[\chi ^{*}_{\tilde {\rho}, \tilde {\sigma}}
\leq
\frac{{\delta}}{6n+6}\eta
\leq
\frac {{\delta}}{2}.\]
\quad \\
Consequently, Lemma $\ref{lemcanon}$ applies and  $\chi$ generates
a canonical transformation
\[\phi : D_{\rho ',\sigma '} \rightarrow D_{\rho ,\sigma}, \]
where
$\rho '= \tilde {\rho} -\delta = \rho -4\delta$ ,
and
$\sigma '= \tilde {\sigma }-\delta = \sigma- 4\delta$.
We have seen
$X \in {\cal {A}}_{\rho -3\delta ,\sigma -3\delta}$,
$Y_{i} \in  {\cal{A}} _{\rho -3\delta ,\sigma -3\delta}, \quad
\mbox{for}\quad i=1,...,n+1 $
and since
$\chi = X(q')+\xi \cdot q'+ \sum _{i} Y_{i}(q')p'_{i}$
it follows 
$\chi \in {\cal{A}}_{\rho -3\delta ,\sigma -3\delta}$.
Next, we consider the following estimates.\\
\\
\fbox {$\|H' \|_{\rho ',\sigma'} <1$}\\
\\
From the second inequality of Lemma $\ref{lemcanon}$ it follows 
$\|H' \|_{\rho ',\sigma '}
\leq  \| H \| _{\rho ,\sigma }
< 1$.
\\
\\
\fbox {$F \in A_{\rho ,\sigma }, \quad \| {\cal {U}}F-F \| _{\rho ', \sigma '}
\leq
\eta\| F \|_{\rho ,\sigma }$}\\
\\
From Lemma $\ref{lemcanon}$ we have the following estimates
\[\| {\cal {U}}F-F \|_{\tilde {\rho}-\delta  , \tilde{\sigma} -\delta}
\leq
4(n+1)\frac {\chi ^{*}_{\tilde {\rho },\tilde {\sigma}}}{\delta}\| F \|_{\tilde {\rho},
\tilde {\sigma}},
\quad \quad
\| F\|_{\tilde {\rho},\tilde {\sigma}}
\leq
\| F \| _{\rho ,\sigma}, \]
Using $(\ref{chibeta})$ we have
\[
4(n+1)\frac {\chi ^{*}_{\tilde {\rho}, \tilde {\sigma}}}{\delta}
\leq
\left( \frac {4n+4}{6n+6} \right) \eta \leq \eta ,\]
and combining these estimates we obtain
\begin{equation} \label{Udiff}
 \| {\cal {U}}F-F \|_{\rho ', \sigma '} \leq
\eta \|F \| _{\tilde {\rho} ,\tilde {\sigma }} \leq
\eta \| F \|_{\rho ,\sigma }.
\end{equation}
\\
\fbox {$m' \equiv m - \frac {4(n+1)\eta }{\sigma _{*}^{2}} $}\\
\\
The approach is to construct $m'$ so  the following
inequality hold
$ \| C'v \|_{\rho ,\sigma } \leq  m'^{-1} \| v \| ,\quad
\forall v \in {\mathbb {C}}^{n}$.
By definition it follows
\[C'_{i,j}(q') - C_{i,j}(q') =
 \frac {\partial ^{2}}{\partial p'_{i}p'_{j}}[ {\cal {U}}H - H ] (0,q').\]
Applying Cauchy's inequality, $(\ref{Udiff})$ and the fact that
$\sigma ' > \sigma _{*}$
we obtain
\[\| C'_{i,j}-C_{i,j} \| \leq
\frac {4 \| {\cal {U}} H -H \|_{\rho ' ,\sigma '}}{{\sigma'} ^{2}} \leq
\frac {4\eta }{\sigma _{*}^{2}}, \]
from which it follows
\begin{equation} \label{etaineq}
\|( C'-C)v \|_{\rho ', \sigma '} \leq
\frac {4(n+1)\eta }{\sigma _{*}^{2}}\| v \|.
\end{equation}
Using
$ C'=C+(C'-C)$
we have
\begin{equation} \label{plusminusc}
\| C'v \|_{\rho ', \sigma '} \leq
\| Cv \|_{\rho ', \sigma '} +
\|(C'-C)v \| _{\rho ', \sigma '}.
\end{equation}
Using the estimate
$\|Cv \|_{\rho ',\sigma '} \leq
\| Cv \|_{\rho ,\sigma } \leq
m^{-1}\|v \|$,
along with $(\ref{etaineq})$ and $(\ref{plusminusc})$ gives
\[ \| C'v \| _{\rho ', \sigma '} \leq
(m^{-1}+ \frac {4(n+1)\eta }{{\sigma _{*}}^{2}} )\|v \|.\]
Applying the inequality
$a^{-1}+b<(a-b)^{-1} \quad for \quad 0<b<a<1$,
the estimate becomes
\[\|C'v \|_{\rho ', \sigma '} \leq
\left( m- \frac {4(n+1)\eta}{{\sigma _{*}}^{2}} \right )^{-1}\| v\|. \]
This estimate provides the value for $m'$
\[m'= m-\frac {4(n+1)\eta}{{\sigma _{*}}^{2}}.\]
\fbox{$f'$}
\\
\\
We begin with the known condition
\\
\begin{equation}
\label{isocond}
\left\|
\left(
\begin{array}{cc}
\overline{\overline{C}} & \lambda ^{T} \\
\lambda & 0
\end{array}\right)
v \right\|  \geq f\|v \|.
\end{equation}
\\
Furthermore we have by definition
\\
\[
C'_{i,j}(q')-C_{i,j}(q')=\frac {\partial ^{2}}{\partial p'_{i}\partial p'_{j}}
[{\cal{U}}H-H](0,q').
\]
Applying Cauchy's inequality, using (\ref{Udiff}) and the fact  $\sigma '>\sigma_{*}$, gives
\[
\left\|\overline{\overline{C'}}_{i,j}-\overline{\overline{C}}_{i,j} \right\| \leq
\frac{\|{\cal{U}}H-H \|_{\rho ',\sigma '}}{\sigma '^{2}}\leq \frac{\eta}{\sigma _{*}^{2}}
\|H\|_{\rho ,\sigma}\leq \frac{\eta}{\sigma _{*}^{2}},
\]
from which it follows 
\begin{equation}
\label{C'Cdiff}
\left\| \left(\overline{\overline{C'}}-\overline{\overline{C}}\right) v\right\|\leq
\frac{(n+1)\eta}{\sigma_{*}^{2}}\|v\|.
\end{equation}
Also one has
$
\lambda '=(1+\kappa E_{1} \zeta_{A})\lambda$.
From this follows
$
\lambda '-\lambda =\kappa E_{1}\zeta _{A} \lambda$,
and we obtain the estimate
$
\|(\lambda '-\lambda )v \| \leq \kappa E_{1}
\|\zeta _{A} \| \|v \|$.
We now find an estimate for $\zeta_{A}$.
We first find a bound for
\[\left( \begin{array}{c}
\xi \\ \kappa E_{1}\zeta_{A}
\end{array}
\right).\]
We have
\[\left( \begin{array}{c}
\xi \\ \kappa E_{1}\zeta_{A}
\end{array}
\right)=
\left( \begin{array}{cc}
\overline{\overline{C}} & \lambda ^{T} \\
\lambda & 0
\end{array}\right) ^{-1}
\left(
\begin{array}{c}
\kappa \overline{\overline{B}}-\overline{\overline{C}}\cdot \overline{\overline{\partial _{q'}X}} \\
\kappa \overline{\overline{A}} \end{array}
\right),
\]
and
\\
\begin{eqnarray*}
\left\|
\left(
\begin{array}{c}
\kappa \overline{\overline{B}}-\overline{\overline{C}}\cdot \overline{\overline{\partial _{q'}X}} \\
\kappa \overline{\overline{A}} \end{array}
\right)
\right\|
&\leq &
\kappa \|B \|_{\rho ,\sigma}
+ \|C \cdot \partial _{q'} X \|_{\rho -2\delta ,\sigma-2\delta}
+\kappa \|A \|_{\rho ,\sigma}\\
\quad\\
&\leq&
2\kappa E_{1} + m^{-1} \|\partial _{q'} X \|_{\rho -2\delta ,\sigma -2\delta}
\end{eqnarray*}
\begin{eqnarray*}
&\leq&
2\kappa E_{1}
+
\frac{\varpi \kappa}{m\Gamma \delta ^{2n+1}}
\|(G-\overline{\overline{G}})(0)\|_{\rho}
+
\frac{8\kappa }{m\delta ^{2}\gamma \sqrt{2\pi}}
\left[
C_{3}{c_{3}}+
\frac{2}{\delta } C_{3}e^{-\frac{\delta c_{3}}{2}}\right]
\left(\frac{4}{{{\delta}}} \right)^{n}
\\
\quad\\
&\leq&
2\kappa E_{1}
+
\frac{\varpi \kappa E_{1}}{m\Gamma \delta ^{2n+1}}
+
\frac{8\kappa }{m\delta ^{2}\gamma \sqrt{2\pi}}
\left[
{c_{3}}+
\frac{2}{\delta } e^{-\frac{\delta c_{3}}{2}}\right]
\left(\frac{4}{{{\delta}}} \right)^{n}E_{1}.
\end{eqnarray*}
\\
We know
\[
\left\| \left(
\begin{array}{cc}
\overline{\overline{C}} & \lambda ^{T} \\
\lambda & 0
\end{array} \right) v \right\|
\geq
f\|v \|.\]
Its not hard to show
\[
\left\| \left(
\begin{array}{cc}
\overline{\overline{C}} & \lambda ^{T} \\
\lambda & 0
\end{array} \right)^{-1} v \right\|
\leq
\frac{1}{f}\|v \|
.\]
Therefore
\begin{eqnarray*}
\kappa E_{1}\| \zeta_{A} \|\leq
\left\| \left( \begin{array}{c}
\xi \\ \kappa E_{1}\zeta_{A}
\end{array}
\right)\right\|
\leq
\frac{1}{f}\left[
2
+
\frac{\varpi }{m\Gamma \delta ^{2n+1}}
+
\frac{8 }{m\delta ^{2}\gamma \sqrt{2\pi}}
\left[
{c_{3}}+
\frac{2}{\delta } e^{-\frac{\delta c_{3}}{2}}\right]
\left(\frac{4}{{{\delta}}} \right)^{n}
\right]\kappa E_{1}
\end{eqnarray*}
\begin{eqnarray}
\label{lambdadiff}
&\leq&
\frac{4^{n}11\varpi}{mf\delta^{9n+10}\Gamma \gamma}(c_{3}+1)\kappa E_{1}
= {\Upsilon}\kappa E_{1}.
\end{eqnarray}
Using $(\ref{C'Cdiff}$) and $ (\ref{lambdadiff})$ we obtain the following
\begin{eqnarray*}
\left\| \left(
\begin{array}{cc}
\overline{\overline{C}}-\overline{\overline{C'}} &
 \lambda ^{T}-\lambda '^{T} \\
\quad\\
\lambda -\lambda ' & 0
\end{array} \right)
\left(
\begin{array}{c}
v_{1} \\ v_{2}
\end{array}
\right)
\right\|
&=&
\left \|
\left(
\begin{array}{c}
(\overline{\overline{C}}-\overline{\overline{C'}})v_{1}+(\lambda ^{T}-\lambda '^{T})v_{2}\\
\quad\\
(\lambda -\lambda ')v_{2}
\end{array}
\right)
\right \|
\end{eqnarray*}
\begin{eqnarray*}
&\leq&
\left\| \left (\overline{\overline{C}}-\overline{\overline{C'}}\right )v_{1} \right\|
+
\left\| \left(\lambda ^{T}-\lambda '^{T}\right)v_{2} \right\|
+
\left \| \left (\lambda -\lambda '\right )v_{2} \right\|\\
\quad\\
&\leq&
\frac{(n+1)\eta}{\sigma ^{2}_{*}}\|v_{1}\|+
\frac{2 }{f}
\left[
2
+
\frac{\varpi }{m\Gamma \delta ^{2n+1}}
+
\frac{8 }{m\delta ^{2}\gamma \sqrt{2\pi}}
\left[
{c_{3}}+
\frac{2}{\delta } e^{-\frac{\delta c_{3}}{2}}\right]
\left(\frac{4}{{{\delta}}} \right)^{n}
\right]\kappa E_{1} \|v_{2}\|
\end{eqnarray*}
\begin{eqnarray}
\label{C'Cdiffestimate}
&\leq&
\frac{(n+1)\eta}{\sigma ^{2}_{*}}\|v\|+
\frac{4 }{f}\kappa E_{1} \|v\|
+
\frac{2 \varpi \kappa E_{1}}{fm\Gamma \delta ^{2n+1}} \|v\|
+
\frac{16 }{m\delta ^{2}\gamma \sqrt{2\pi}}
\left[
{c_{3}}+
\frac{2}{\delta } e^{-\frac{\delta c_{3}}{2}}\right]
\left(\frac{4}{{{\delta}}} \right)^{n}
\kappa E_{1} \|v\|,
\end{eqnarray}
where $v=(v_{1},v_{2})$ and clearly $\|v_{1}\| \leq \|v\|$ and
$\|v_{2}\| \leq \|v\|$.
Using $(\ref{C'Cdiffestimate})$ and  $(\ref{isocond})$
 we obtain
\quad\\
\begin{eqnarray*}
\left\|
\left(
\begin{array}{cc}
\overline{\overline{C}} & \lambda ^{T} \\
\lambda & 0
\end{array}\right)v\right\|  -
\left\| \left(
\begin{array}{cc}
\overline{\overline{C}}-\overline{\overline{C'}} & \lambda ^{T}-\lambda '^{T} \\
\lambda -\lambda ' & 0
\end{array} \right)v\right\| \quad
\end{eqnarray*}
\begin{eqnarray}
\label{CCC'diff}
&\geq& \left[
f-
\frac{(n+1)\eta}{\sigma ^{2}_{*}}
-
\left(
\frac{4 }{f}
+
\frac{2 \varpi }{fm\Gamma \delta ^{2n+1}}
\right)\kappa E_{1}
-
\frac{16\kappa E_{1} }{m\delta ^{2}\gamma \sqrt{2\pi}}
\left[
{c_{3}}+
\frac{2}{\delta } e^{-\frac{\delta c_{3}}{2}}\right]
\left(\frac{4}{{{\delta}}} \right)^{n}
\right] \|v \|.
\end{eqnarray}
For any matrix $M$ one can write
$
M'=M+(M'-M)
$ and
$
\|M' \| \geq
 \|M \| - \|M'-M \|.
$
Together with $(\ref{CCC'diff})$ we can bound from below as follows
\begin{eqnarray*}
\left\|
\left(
\begin{array}{cc}
\overline{\overline{C'}} & \lambda '^{T} \\
\lambda ' & 0
\end{array}\right)v \right\|
&\geq& \left [
f-
\frac{(n+1)\eta}{\sigma ^{2}_{*}}
-
\left(
\frac{4 }{f}
+
\frac{2 \varpi }{fm\Gamma \delta ^{2n+1}}
\right)\kappa E_{1} 
-
\frac{16 }{m\delta ^{2}\gamma \sqrt{2\pi}}
\left[
{c_{3}}+
\frac{2}{\delta } e^{-\frac{\delta c_{3}}{2}}\right]
\left(\frac{4}{{{\delta}}} \right)^{n}\kappa E_{1}
\right ] \|v \|\\
\quad\\
&\geq &
\left(
f-
\frac{(n+1)\eta}{\sigma ^{2}_{*}}
- \eta
- \eta \right)\|v \|
\geq
\left(
f-
\frac{(n+3)\eta}{\sigma ^{2}_{*}}
\right)\|v \|=f'\| v \|.
\end{eqnarray*}
With the assumption $\max ( \|A \|_{\rho ,\sigma }, \| B \|_{\rho ,\sigma})<E_{1}$,
the following estimate holds
\begin{equation}
\|P \|_{\rho ,\sigma } = \kappa \| A + B \cdot p' \|
_{\rho ,\sigma }<  (n+2)\kappa E_{1}.
\end{equation}
We want for the new perturbation $P'$ a similar expression namely
\begin{equation}
\|P' \|_{\rho ' ,\sigma ' } =\kappa ' \| A' + B' \cdot p' \|
_{\rho ' ,\sigma ' }< (n+2)\kappa ' E_{1}'.
\end{equation}
\\
\fbox{$\max ( \|A' \|_{\rho ' ,\sigma ' },
\| B'\|_{\rho ' ,\sigma ' })\leq E_{1}'$}
\\

To arrive at this estimate we must first find estimates for
$\kappa '\| A' \| _{\rho ', \sigma '}, \kappa '\| B' \|_{\rho ', \sigma '}$
and identify the values for $\kappa '$ and $E_{1}'$.
We know 
\[ \kappa 'A'(q')=H'(0,q')-a'=H'(0,q')-\overline{\overline{H'}}(0),\]
\[H'(p',q')=
U+P+\{ \chi ,U \} +
 [\{ \chi ,P \}+{\cal {U}}H-H-\{ \chi ,H \} ]. \]
By our choice of $\chi $ we obtained\\
\begin{eqnarray*}
H'(p',q')=
a+(1+\varepsilon \zeta_{A})\lambda \cdot p'+{\cal {O}}(\| p' \|^{2})+[ \{ \chi ,P \}+
 {\cal {U}}H-H- \{\chi ,H \} ],
\end{eqnarray*}
\begin{eqnarray*}
\kappa 'A'(q')&=&H'(0,q')-a'
\end{eqnarray*}
\begin{eqnarray*}
&=&
a+
 {\cal {U}}H-H- \{\chi ,H \} ](0,q') -
a
-
\overline{\overline{
[ \{ \chi ,P \}+
 {\cal {U}}H-H- \{\chi ,H \} ]}}(0)
\\
\quad\\
&=& \Big(\{\chi ,P \}+ {\cal {U}}H-H- \{\chi ,H \}\Big)(0,q')
-\left(
\overline{\overline{
 \{ \chi ,P \}+
 {\cal {U}}H-H- \{\chi ,H \} }}\right)(0)
 .
\end{eqnarray*}
Therefore we obtain the estimate
$ \kappa '\| A' \| _{\rho ', \sigma '} \leq
2\| \{ \chi ,P \}+ {\cal {U}}H-H-\{ \chi ,H \} \|_{\rho ', \sigma '}$.
Using the previously obtained estimates
\begin{eqnarray*}
&\quad & \| \{\chi ,f \} \| _{\rho -\delta ,\sigma -\delta } \leq
2(n+1) \left( \frac {\chi ^{*}_{\rho ,\sigma }}{\delta} \right) \|f \|_{\rho ,\sigma},
\\
\quad \\
&\quad &\| {\cal {U}}f-f -\{\chi ,f \} \|_{\rho -\delta ,\sigma -\delta} \leq
32(n+1)^{2}\left( \frac {\chi ^{*}_{\rho ,\sigma }}{\delta } \right)^{2}
\|f \|_{\rho ,\sigma},
\\
\quad \\
&\quad &\| P \|_{\tilde {\rho}, \tilde {\sigma}} \leq
 \kappa \|A \|_{\tilde {\rho}, \tilde {\sigma}} +
(n+1)\kappa \| B \|_{\tilde {\rho}, \tilde {\sigma}} \leq
(n+2)\kappa  E_{1} ,
\\
\quad \\
&\quad & \|H \|_{\tilde {\rho}, \tilde {\sigma}}<1 ,
\\
\quad \\
&\quad &\rho ' = \tilde {\rho} -\delta =
\rho -4\delta ,
\\
\quad \\
&\quad & \sigma '= \tilde {\sigma} -\delta =
\sigma -4\delta ,
\\
\quad \\
&\quad &\frac {\chi ^{*}_{\tilde {\rho}, \tilde {\sigma}}}{\delta} \leq
\frac {\eta }{6n+6} ,
\end{eqnarray*}
we obtain the following
\begin{eqnarray*}
&\quad &\| \{\chi ,P \}+ {\cal {U}}H-H-\{ \chi,H \} \|_{\rho ',\sigma '}
 \leq \| \{ \chi ,P \} \|_{\rho ',\sigma '}+
\| {\cal{U}}H-H-\{\chi ,H \} \|_{\rho ',\sigma '} \\
\\
&\quad & \leq \| \{ \chi ,P \} \|_{\tilde {\rho }-\delta,\tilde {\sigma }-\delta}+
\| {\cal{U}}H-H-\{\chi ,H \} \|_{\tilde {\rho }-\delta,\tilde {\sigma }-\delta} \\
\\
&\quad & \leq 2(n+1) \left( \frac {\chi ^{*}_{\tilde {\rho },\tilde {\sigma}}}{\delta} \right)
\|P\|_{\tilde {\rho },\tilde {\sigma}} +
32(n+1)^{2}\left( \frac {\chi ^{*}_{\tilde {\rho },\tilde {\sigma}}}{\delta} \right)^{2}
\|H \|_{\tilde {\rho },\tilde {\sigma}}\\
\\
&\quad & \leq 2(n+1) \left( \frac {\chi ^{*}_{\tilde {\rho },\tilde {\sigma}}}{\delta} \right)
(n+2) \kappa E_{1} +
32(n+1)^{2}\left( \frac {\chi ^{*}_{\tilde {\rho },\tilde {\sigma}}}{\delta} \right)^{2}\\
\\
&\quad & < 2(n+2)^{2}\kappa  E_{1}
\left( \frac {\chi ^{*}_{\tilde {\rho },\tilde {\sigma}}}{\delta} \right)
 +
32(n+1)^{2}\left( \frac {\chi ^{*}_{\tilde {\rho },\tilde {\sigma}}}{\delta} \right)^{2}
\end{eqnarray*}
\begin{equation}
\label{estremainder}
 \leq [1+32(n+1)^{2}]\left( \frac {\eta }{6n+6} \right)^{2}
 \leq \left[ \frac {1+32(n+1)^{2}}{(6n+6)^{2}} \right]
 \eta ^{2}
\leq
\eta ^{2} .
\end{equation}
Where we have used the 
\[2(n+2)^{2} \kappa E_{1} \leq
\frac {\eta }{6n+6}.\]
\\
We then have
\begin{eqnarray}
\label{kappaA}
\kappa '\|A'\| _{\rho ' ,\sigma '} <
 \eta^{2}=\left(\frac{ \Lambda}{f^{2}m^{2}\delta^{\aleph}}\right) ^{2}\kappa ^{2} E_{1}^{2}.
\end{eqnarray}
\\
Next recall
\begin{eqnarray*}
\kappa ' B'_{i}(q') &=&
\frac {\partial H'}{\partial p'_{i}}(0,q')-\lambda _{i}
=
\frac {\partial H'}{\partial p'_{i}}(0,q')-
 \frac {\partial \tilde{H}^{0}}{\partial p'_{i}}(0).
\end{eqnarray*}
Using Cauchy's inequality we obtain
\begin{eqnarray*}
\kappa ' \|B' \|_{\rho ', \sigma '} &\leq&
\frac {1}{\sigma '} \|(H'-\tilde{H}^{0})(0) \|_{\rho ', \sigma '}
=\frac {1}{\sigma '} \|(H'-a)(0) \|_{\rho ', \sigma '}
=\frac {1}{\sigma '} \|\{ \chi ,P \}+{\cal{U}}H-H-\{\chi ,H \} \|_{\rho ',\sigma '}
\end{eqnarray*}
\begin{eqnarray}
\label{kappaB}
&< & \frac {\eta^{2} }{\sigma '}<\frac {\eta^{2} }{\sigma *}
 =\left(\frac{\Lambda}{f^{2}m^{2}\delta^{\aleph}}\right) ^{2} \frac{ \kappa ^{2}
 E_{1}^{2}}{\sigma _{*}} .
\end{eqnarray}
Finally with $(\ref{kappaA})$ and $(\ref{kappaB})$ we set
\begin{eqnarray*}
&\quad& \kappa ' = \frac{\Lambda}{f^{2}m^{2}\delta^{\aleph}}  \kappa^{2},
\quad
\quad
E_{1}' = \frac{\Lambda}{f^{2}m^{2}\delta^{\aleph}} \frac{E_{1}^{2}}{\sigma _{*}},
\end{eqnarray*}
and clearly
$
\max (\|A' \|_{\rho ', \sigma '}, \|B'\|_{\rho ',\sigma '})\leq E_{1}'.
$
\\

Every time the iterative lemma is applied an equation of the
form of  $(\ref{zetasolution})$ must be solved.
The $\zeta_{A}$ that is chosen at the first step must
satisfy the equation of the form of  $ (\ref{zetasolution})$ for
all other steps. Therefore at the step $k+1$ we have
\quad\\
\begin{equation}
\label{iterlambda}
\lambda ^{k+1}=(1+ \kappa ^{k} E_{1}^{k}\zeta_{A})\lambda ^{k}
=(1+ \kappa ^{k} E_{1}^{k}\zeta_{A})(1+\zeta_{k})\lambda ^{0}
=(1+\zeta_{k+1})\lambda ^{0}.
\end{equation}
\quad\\
Therefore we need an estimate for $1+\zeta_{k+1}$.
\\
\\
\fbox{$1+\zeta '$}
\\
\\
Assuming
$
\frac{1}{2}<1+\zeta < \frac{3}{2}$,
with $(\ref{lambdadiff})$ and
$
1+\zeta ' = (1+ \kappa  E_{1}\zeta_{A})(1+\zeta )$
it follows
\[
1 + \zeta -\kappa E_{1}\zeta_{A}(1+\zeta)
<1+\zeta ' <
1+\zeta +\kappa E_{1}\zeta_{A}(1+\zeta).
\]
As will be shown later,we can  assume $\kappa $ to be  small enough so that
\[
\frac{1}{2} <1 + \zeta -\frac{3}{2}\kappa E_{1}{\Upsilon}
<1+\zeta ' <
1+\zeta +\frac{3}{2}\kappa E_{1}{\Upsilon}
< \frac{3}{2},
\]
where $\Upsilon$ is as in $(\ref{lambdadiff})$.
\\
\\
\fbox{$L'$}
\\
\\
We know
$
\tilde{ \lambda }' = (1+   \zeta ')\tilde {\lambda }^{0}.
$
Then it follows 
\[
| \tilde{\lambda}' |=|(1+  \zeta ') \tilde{\lambda}^{0} |< \frac{3}{2}L _{0}=L'.
\]
\section{Conclusion of the Proof}

We now describe the set up for several steps in the application of
the iterative lemma.\\
\quad\\
\\
\fbox{Zeroth Step}\\
\\
We begin with the given positive constants\\
\\
\[\Gamma, \rho _{0},\sigma _{0},m_{0},\kappa _{0} <1,\quad
\rho _{*}<\rho _{0}, \quad \sigma_{*}<\sigma_{0} \quad
m_{*}<m_{0},\quad L_{0} , \quad E_{1}^{0}\]
\\
and
\[
- \frac{1}{2}< \kappa _{0} E_{1}^{0} \zeta _{A_{0}} < \frac{1}{2},
\quad
f_{0}=\left |\frac{1}{2}|2m_{0}-L_{0}|-\frac{\kappa_{0}E_{1}^0}{\sigma^2_0} \right |.
\]
\\
We assume a Hamiltonian of the form described above defined on $D_{\rho _{0}, \sigma _{0}}$
\\
\begin{eqnarray*}
&\quad &H_{0}(p',q')=U_{0}(p',q')+P_{0}(p',q')\\
\\
&\quad &U_{0}(p',q')=
a^{0}+\lambda ^{0} \cdot p' +
\frac {1}{2}\sum C^{0}_{i,j}(q')p'_{i}p'_{j}+R^{0}(p',q') \\
\\
&\quad & P_{0}(p',q')=
\kappa _{0} A^{0}(q')+
\kappa _{0} \sum B^{0}_{i}(q')p'_{i}
\end{eqnarray*}
where all the functions are in ${\cal{A}}_{\rho _{0},\sigma _{0}}$, $\tilde{\lambda}
 _{0} \in 
\Omega _{\Gamma}$, with $| \tilde {\lambda}_{0} |<L_{0}$
and the following bounds hold\\
\begin{eqnarray*}
&\quad &f_{0}\| v\| \leq \left\|  \left(
\begin{array}{cc}
\overline{\overline{C^{0}}} & \lambda ^{T}\\
\lambda & 0 
\end{array}
  \right)v \right \|, \quad \forall v\in {\mathbb{C}}^{n+1},\\ 
\\
&\quad &\|C^{0}v \|_{\rho _{0},\sigma _{0}} <
m_{0}^{-1} \|v \|, \quad \forall v\in {\mathbb {C}}^{n+1},\\
\\
&\quad & \mbox {max} (\|A^{0} \|_{\rho _{0},\sigma _{0}},\|B^{0} \|_{\rho _{0},\sigma _{0}})
<
E_{1}^{0}.
\end{eqnarray*}
We apply the iterative lemma.
\\
\\
\fbox{First Step}\\
\\
Applying the iterative lemma for any $\delta_{0} >0$ such that
\begin{eqnarray*}
\rho_{0}  -4\delta_{0}> \rho_{*},\quad
\sigma _{0}-4\delta_{0} >\sigma _{*},
\end{eqnarray*}
and
$\kappa  _{0} E_{1}^{0}$
small enough that
\begin{eqnarray*}
m_{0}-4(n+1)\eta_{0} /\sigma ^{2}_{*} > m_{*},\quad
f_{0}-\frac{(n+3)\eta_{0}}{\sigma^2_{*}} >0,
\end{eqnarray*}
 there exists a canonical
transformation
\\
\[\phi _{1}:D_{\rho_{0} -4\delta_{0},\sigma _{0}-4\delta_{0}}
\rightarrow D_{\rho_{0},\sigma _{0}}, \quad 
\phi _{1} \in {\cal{A}}_{\rho_{0} -4\delta_{0},\sigma _{0}-4\delta_{0}}\]
\\
that transforms the Hamiltonian into
\\
\[H_{1}=H_{0} \circ \phi _{1} 
=U_{1}(p',q')+P_{1}(p',q'),\]
\\
which can be decomposed as done previously\\
\begin{eqnarray*}
&\quad &U_{1}(p',q')=
a^{1}+\lambda ^{1} \cdot p' +
\frac {1}{2}\sum C^{1}_{i,j}(q')p'_{i}p'_{j}+R^{1}(p',q')\\
\\
&\quad & P_{1}(p',q')=
\kappa _{1}A^{1}(q')+
\kappa _{1}\sum B^{1}_{i}(q')p'_{i}
\end{eqnarray*}
with the same $\Gamma$ and  with the new constants
\begin{eqnarray*}
&\quad& \rho _{1}=\rho_{0} -4\delta_{0}>\rho_{*},\quad
\sigma _{1}=\sigma_{0}-4\delta_{0}>\sigma_{*},\quad
m_{1}=m_{0}-4(n+1)\frac {\eta_{0}}{\sigma^{2}_{*}}>m_{*},\quad
f_{1}=f_{0}-\frac{(n+3)\eta _{0}}{\sigma _{*}^{2}},\\
\quad\\
&\quad& L_{1}=\frac{3}{2}L_{0},\quad
\eta_{1} =\frac{\Lambda}{f^{2}_{0}m^{2}_{0}\delta_{0}^{\aleph}}
\kappa_{0}E_{1}^{0},\quad
\kappa _{1}= \frac{\Lambda}{f^{2}_{0}m^{2}_{0}\delta _{0}^{\aleph}}{ \kappa_{0}^{2}},\quad 
E_{1}^{1} = \frac{\Lambda}{ f^{2}_{0}m^{2}_{0}\delta _{0}^{\aleph} }
\frac{ (E_{1}^{0})^{2}}{\sigma ^{*}} 
\end{eqnarray*}
with all functions analytic on $D_{\rho _{1},\sigma _{1}}$, and
bounds
\\
\begin{eqnarray*}
&\quad &f_{1}\| v\|\leq \left \| 
\left(
\begin{array}{cc}
\overline{\overline{C^{1}}} & \lambda ^{T}\\
\lambda & 0
\end{array}
\right)
 v \right\|, \quad \forall v \in {\mathbb{C}}^{n+1}\\
\\
&\quad &\|C^{1}v \|_{\rho _{1},\sigma _{1}} <
m_{1}^{-1} \|v \|, \quad \forall v\in {\mathbb {C}}^{n+1}\\
\\
&\quad & \mbox {max} (\|A^{1} \|_{\rho _{1},\sigma _{1}},\|B^{1} \|_{\rho _{1},\sigma _{1}})
<
E _{1}^{1}.
\end{eqnarray*}
\\
\\
From the iterative lemma we know the following estimate holds
\\
\[
1 -\Upsilon _{0} \kappa _{0} E_{1}^{0}<
1+ \zeta _{1}=1+\kappa _{0}E_{1}^{0}\zeta _{A_{0}} < 1  + \Upsilon _{0} \kappa _{0}E_{1}^{0} .
\]
\\
Later we will show that in fact
\\
\begin{equation}
\label{zetaestone}
\frac{1}{2}<1 -\Upsilon _{0}\kappa _{0} E_{1}^{0}<
1+ \zeta _{1} < 1  + \Upsilon _{0}\kappa _{0} E_{1}^{0}< \frac{3}{2}.
\end{equation}
\\
Using $(\ref{iterlambda})$ and $(\ref{zetaestone})$ we have
\begin{eqnarray*}
|\tilde{\lambda} _{1} |= |(1+ \zeta_{1})\tilde{\lambda}_{0}|
\leq (1+ |\zeta_{1} |)|\tilde{\lambda} _{0} |
<\frac{3}{2}L_{0}=L_{*}.
\end{eqnarray*}
\\
Moreover, for $f\in {\cal{A}}_{\rho _{1},\sigma _{1}}$, we have
$
\|{\cal {U}}_{1}f-f \|_{\rho_{1},\sigma _{1}} \leq
\eta_{0} \|f \|_{\rho_{0},\sigma _{0}}$,
where
${\cal {U}}_{1}f \equiv f \circ \phi _{1}$.
\\
\quad\\
\fbox {Second Step}\\
\\
Applying the iterative lemma for any $\delta_{1} >0$ such that
\begin{eqnarray*}
 \rho_{1}  -4\delta_{1}> \rho_{*},\quad
 \sigma _{1}-4\delta_{1} >\sigma _{*},
\end{eqnarray*}
 and $\kappa _{1}E_{1}^1$
so small that
\begin{eqnarray*}
&\quad&
m_{1}-4(n+1)\eta_{1} /\sigma ^{2}_{*} > m_{*},\quad
f_{1}-\frac{(n+3)\eta_{1}}{\sigma ^2_*} >0,
\end{eqnarray*}
 there exists a canonical
transformation
\\
\[\phi _{2}:D_{\rho_{1} -4\delta_{1},\sigma _{1}-4\delta_{1}}
\rightarrow D_{\rho_{1},\sigma _{1}}, \quad 
\phi _{2} \in {\cal{A}}_{\rho_{1} -4\delta_{1},\sigma _{1}-4\delta_{1}}\]
\\
that transforms the Hamiltonian into
\\
\[H_{2}=H_{1} \circ \phi _{2} 
=U_{2}(p',q')+P_{2}(p',q'),\]
\\
which can be decomposed as  done previously\\
\begin{eqnarray*}
&\quad &U_{2}(p',q')=
a^{2}+\lambda ^{2} \cdot p' +
\frac {1}{2}\sum C^{2}_{i,j}(q')p'_{i}p'_{j}+R^{2}(p',q')\\
\\
&\quad & P_{2}(p',q')=
\kappa _{2}A^{2}(q')+
\kappa _{2}\sum B^{2}_{i}(q')p'_{i}
\end{eqnarray*}
with the same $\Gamma$ and  with the new constants
\begin{eqnarray*}
&\quad&  \rho _{2}=\rho_{1} -4\delta_{1}>\rho_{*},\quad
\sigma _{2}=\sigma_{1}-4\delta_{1}>\sigma_{*},\quad
m_{2}=m_{1}-4(n+1)\frac {\eta_{1}}{\sigma^{2}_{*}}>m_{*},\quad
f_{2}=f_{1}-\frac{(n+3)\eta _{1}}{\sigma _{*}^{2}},\\
\quad\\
&\quad& L_{2}=\frac{3}{2}L_{0},\quad
\eta_{2} =\frac{\Lambda}{f^{2}_{1}m^{2}_{1}\delta_{1}^{\aleph}}
\kappa_{1}E_{1}^{1},\quad
\kappa _{2}= \frac{\Lambda}{f^{2}_{1}m^{2}_{1}\delta _{1}^{\aleph}}{ \kappa_{1}^{2}},\quad
E_{1}^{2} = \frac{\Lambda}{ f^{2}_{1}m^{2}_{1}\delta _{1}^{\aleph} }
\frac{ (E_{1}^{1})^{2}}{\sigma ^{*}}, 
\end{eqnarray*}
with all functions analytic on $D_{\rho _{1},\sigma _{1}}$, and
bounds
\\
\begin{eqnarray*}
&\quad &f_{2}\| v\|\leq \left \| 
\left(
\begin{array}{cc}
\overline{\overline{C^{2}}} & \lambda ^{T}\\
\lambda & 0
\end{array}
\right)
 v \right\|, \quad \forall v \in {\mathbb{C}}^{n+1},\\
\\
&\quad &\|C^{2}v \|_{\rho _{2},\sigma _{2}} <
m_{2}^{-1} \|v \|, \quad \forall v\in {\mathbb {C}}^{n+1},\\
\\
&\quad & \mbox {max} (\|A^{2} \|_{\rho _{2},\sigma _{2}},\|B^{2} \|_{\rho _{2},\sigma _{2}})
<
E _{1}^{2}.
\end{eqnarray*}
\\
\\
From the iterative lemma we know the following estimate holds
\\
\[
1-\kappa _{0}E_{1}^{0}\zeta _{A_{0}} -\frac{3}{2}\zeta _{1} \kappa _{1} E_{1}^{1}<
1+ \zeta _{2} < 1+ \kappa_{0}E_{1}^{0}\zeta _{A_{0}} + \frac{3}{2}\zeta _{1}\kappa _{1} E_{1}^{1}.
\]
\\
Later we will show that in fact
\\
\begin{equation}
\label{zetaestone}
\frac{1}{2}<1-\kappa_{0}E_{1}^{0}\zeta _{A} -\frac{3}{2}\Upsilon _{1}\kappa _{1} E_{1}^{1}<
1+ \zeta _{2} < 1+ \kappa_{0}E_{1}^{0}\zeta _{A} + \frac{3}{2}\Upsilon _{1}\kappa _{1} E_{1}^{1}< \frac{3}{2}.
\end{equation}
\\
Using $(\ref{iterlambda})$ and $(\ref{zetaestone})$ we have
\begin{eqnarray*}
|\tilde{\lambda} _{2} |&=& |(1+ \zeta_{2})\tilde{\lambda}_{0}|
\leq (1+ |\zeta_{2} |)|\tilde{\lambda} _{0} |
< \frac{3}{2}L_{0}=L_{*}.
\end{eqnarray*}
\\
Moreover, for $f\in {\cal{A}}_{\rho _{2},\sigma _{2}}$, we have
$
\|{\cal {U}}_{2}f-f \|_{\rho_{2},\sigma _{2}} \leq
\eta_{1} \|f \|_{\rho_{1},\sigma _{1}}$,
where
${\cal {U}}_{2}f \equiv f \circ \phi _{2}$.\\
\quad\\
\fbox{$(k+1)^{th}$ Step.}\\
\\
We repeatedly apply the iterative lemma $k+1$ times.
We choose $\delta_{k}$ at each step small enough so that\\
\begin{eqnarray*}
&\quad&\rho_{k}  -4\delta_{k}> \rho_{*},
\quad
\quad
\sigma _{k}-4\delta_{k} >\sigma _{*},
\end{eqnarray*}
and
$\kappa  _{k} E_{1}^{k}$
 small enough  that
\begin{eqnarray*}
&\quad&
m_{k}-4(n+1)\eta_{k} /\sigma ^{2}_{*} > m_{*},
\quad
\quad
f_{k}-\frac{(n+3)\eta _{k}}{\sigma _{*}^{2}}  >0.
\end{eqnarray*}
There exists a canonical
transformation
\\
\[\phi _{k+1}:D_{\rho_{k} -4\delta_{k},\sigma _{k}-4\delta_{k}}
\rightarrow D_{\rho_{k},\sigma _{k}}, \quad
\phi _{k+1} \in {\cal{A}}_{\rho_{k} -4\delta_{k},\sigma _{k}-4\delta_{k}},\]
\\
that transforms the Hamiltonian into
\\
\[H_{k+1}=H_{k} \circ \phi _{k+1}
=U_{k+1}(p',q')+P_{k+1}(p',q'),\]
\\
which can be decomposed as  done previously\\
\begin{eqnarray*}
&\quad &U_{k+1}(p',q')=
a^{k+1}+\lambda ^{k+1} \cdot p' +
\frac {1}{2}\sum C^{k+1}_{i,j}(q')p'_{i}p'_{j}+R^{k+1}(p',q'),\\
\\
&\quad & P_{k+1}(p',q')=
\kappa _{k+1}A^{k+1}(q')+
\kappa _{k+1}\sum B^{k+1}_{i}(q')p'_{i},
\end{eqnarray*}
with the same $\Gamma$ and  with the new constants
\begin{eqnarray*}
&\quad&  \rho _{k+1}=\rho_{k} -4\delta_{k}>\rho_{*},
\quad
\sigma _{k+1}=\sigma_{k}-4\delta_{k}>\sigma_{*},
\quad
m_{k+1}=m_{k}-4(n+1)\frac {\eta_{k}}{\sigma^{2}_{*}}>m_{*},\\
\quad\\
&\quad& f_{k+1}=f_{k}-\frac{(n+3)\eta _{k}}{\sigma _{*}^{2}},
\quad
L_{k+1}=\frac{3}{2}L_{0},
\quad
\eta_{k+1} =\frac{\Lambda}{f^{2}_{k}m^{2}_{k}\delta_{k}^{\aleph}}
\kappa_{k}E_{1}^{k},\\
\quad\\
&\quad& \kappa _{k+1}= \frac{\Lambda}{f^{2}_{k}m^{2}_{0}\delta _{k}^{\aleph}}{ \kappa_{k}^{2}},
\quad
 E_{1}^{k+1} = \frac{\Lambda}{ f^{2}_{k}m^{2}_{k}\delta _{k}^{\aleph} }
\frac{ (E_{1}^{k})^{2}}{\sigma ^{*}},
\end{eqnarray*}
with all functions analytic on $D_{\rho _{k+1},\sigma _{k+1}}$, with
bounds
\\
\begin{eqnarray*}
&\quad &f_{k+1}\| v\|\leq \left \|
\left(
\begin{array}{cc}
\overline{\overline{C^{k+1}}} & \lambda ^{T}\\
\lambda & 0
\end{array}
\right)
 v \right\|, \quad \forall v \in {\mathbb{C}}^{n+1},\\
\\
&\quad &\|C^{k+1}v \|_{\rho _{k+1},\sigma _{k+1}} <
m_{k+1}^{-1} \|v \|, \quad \forall v\in {\mathbb {C}}^{n+1},\\
\\
&\quad & \mbox {max} (\|A^{k+1} \|_{\rho _{k+1},\sigma _{k+1}},\|B^{k+1} \|_{\rho _{k+1},
\sigma _{k+1}})
<
E _{1}^{k+1}.
\end{eqnarray*}
\\
\\
From the iterative lemma we know the following estimate holds
\\
\[
1-\kappa_{0}E_{1}^{0}\zeta _{A_{0}}- \cdot \cdot \cdot
-\frac{3}{2}\zeta _{k-1}\kappa _{k-1} E_{1}^{k-1}
  -\frac{3}{2}\zeta _{k}\kappa _{k} E_{1}^{k}<
1+ \zeta _{k+1} < 1+ \kappa_{0}E_{1}^{0}\zeta _{A_{0}}
+\cdot \cdot \cdot +
\frac{3}{2}\zeta_{k-1}\kappa _{k-1} E_{1}^{k-1}
 + \frac{3}{2}\zeta _{k}\kappa _{k} E_{1}^{k}.
\]
\\
Later we will show that in fact
\\
\begin{equation}
\label{zetaestone}
\frac{1}{2}<
1-\kappa_{0}E_{1}^{0}\zeta _{A_{0}}- \cdot \cdot \cdot
  -\frac{3}{2}\Upsilon _{k}\kappa _{k} E_{1}^{k}
<
1+ \zeta _{k+1} < 1+
 \kappa_{0}E_{1}^{0}\zeta _{A_{0}}
+\cdot \cdot \cdot  + \frac{3}{2}\Upsilon _{k}\kappa _{k} E_{1}^{k}
< \frac{3}{2}.
\end{equation}
\\
Using $(\ref{iterlambda})$ and $(\ref{zetaestone})$ we have
\begin{eqnarray*}
|\tilde{\lambda} _{k+1} |&=& |(1+ \zeta_{k+1})\tilde{\lambda}_{0}|
\leq (1+ |\zeta_{k+1} |)|\tilde{\lambda} _{0} |
< \frac{3}{2}L_{0}=L_{*}.
\end{eqnarray*}
\\
Moreover, for $f\in {\cal{A}}_{\rho _{k+1},\sigma _{k+1}}$, we have
$
\|{\cal {U}}_{k+1}f-f \|_{\rho_{k+1},\sigma _{k+1}} \leq
\eta_{k} \|f \|_{\rho_{k},\sigma _{k}}$,
where
${\cal {U}}_{k+1}f \equiv f \circ \phi _{k+1}$.
\\
\\
\fbox{Limit as $ k \rightarrow \infty$}\\
\\
In the limit case we formally obtain the Hamiltonian
\[H_{\infty}(p',q')=U_{\infty}(p',q')+P_{\infty}(p',q'),\]
which can be decomposed as before
\begin{eqnarray*}
&\quad &U_{\infty}(p',q')=
a^{\infty}+\lambda ^{\infty} \cdot p' +
\frac {1}{2}\sum C^{\infty}_{i,j}(q')p'_{i}p'_{j}+R^{\infty}(p',q'),\\
\\
&\quad & P_{\infty}(p',q')=
\varepsilon _{\infty}A^{\infty}(q')+
\varepsilon _{\infty} \sum B^{\infty}_{i}(q')p'_{i},
\end{eqnarray*}
with constants
$\lambda ^{\infty}, \quad \rho_{\infty}, \quad \sigma _{\infty},\quad m_{\infty},
\quad f_{\infty},\quad  L_{\infty}, \quad \kappa _{\infty},\quad E_{1}^{\infty}$
where all functions are analytic on $D_{\rho _{\infty},\sigma _{\infty}}$ with bounds
\begin{eqnarray*}
&\quad &f_{\infty}\| v \| \leq \left\| \left(
\begin{array}{cc}
\overline{\overline{C^{\infty}}} & \lambda ^{T}\\
\lambda & 0
\end{array}
\right) v \right\|, \quad \forall v \in {\mathbb{C}}^{n+1},\\
\\
&\quad &\|C^{\infty}v \|_{\rho _{\infty},\sigma _{\infty}} <
m_{\infty}^{-1} \|v \|, \quad \forall v\in {\mathbb {C}}^{n+1},\\
\\
&\quad & \mbox {max} (\|A^{\infty} \|_{\rho _{\infty},\sigma _{\infty}},\|B^{\infty} \|_{\rho _{\infty},\sigma _{\infty}})
<
E_{1}^{\infty}.
\end{eqnarray*}
For the expressions in the limit $k\rightarrow \infty$ to have meaning it is
necessary to show
\begin{eqnarray*}
\rho _{\infty} > \rho _{*},\quad
\sigma _{\infty}>\sigma _{*},\quad
m_{\infty} > m_{*},\quad
L_{\infty}=\frac{3}{2}L_{0}=L_{*},\quad
f_{\infty}> f_{*}
\end{eqnarray*}
which will allow to specify $\rho _{*}$, $\sigma _{*}$ and $m_{*}$, and
will lead to
$\kappa _{\infty} = 0$,
i.e. the perturbation vanishes in the limit.\\
\\
It is also necessary to show that the sequence of canonical transformations
defined by\\
\[\widehat {\phi} _{k}: D_{\rho _{k},\sigma _{k}} \rightarrow
D_{\rho _{0},\sigma _{0}}, \quad
\widehat {\phi} _{k} \equiv \phi _{1} \circ  \cdot \cdot \cdot \circ \phi _{k},\]
where
\[\phi _{k}:D_{\rho _{k},\sigma _{k}} \rightarrow D_{\rho _{k-1},\sigma _{k-1}},\]
converges to an analytic, canonical transformation.
\\
We have seen the following recursion formulas amongst the different constants
\begin{eqnarray*}
&\quad& \rho _{k+1} =
\rho _{k}-4\delta_{k},
\quad
 \sigma _{k+1} =
\sigma _{k}-4\delta_{k},
\quad
m_{k+1} =
 m_{k}-4(n+1)\frac {\eta_{k}}{\sigma ^{2}_{*}},
\quad
f_{k+1}= f_{k}-\frac{(n+3)\eta_{k}}{\sigma_{*}^{2}}
\end{eqnarray*}
\begin{eqnarray}
&\quad& L_{k+1}= \frac{3}{2}L_{0},
\quad \kappa _{k+1} =
\frac{\Lambda }{f^{2}_{k}m^{2}_{k}\delta _{k}^{\aleph}}
  \kappa _{k}^{2},
\quad
E_{1}^{k+1} =\frac{\Lambda }{
f^{2}_{k}m^{2}_{k}\delta _{k}^{\aleph}}
\frac{ (E_{1}^{k})^{2}}{\sigma  {*}},
\quad
 \kappa _{k+1}E_{1}^{k+1} = \frac{\eta _{k}^{2}}{\sigma *} \label{vepsilonex}.
\end{eqnarray}
We also have $\kappa _{k}$, $E_{1}^{k}$ and $\eta_{k}$ related by
\begin{eqnarray}
 \label{etaexp}
\eta_{k}=\frac{ \Lambda }{f^{2}_{k}m^{2}_{k}\delta _{k}^{\aleph}}
 \kappa _{k} E_{1}^{k}.
\end{eqnarray}
We can use the expression for $\kappa_{k+1} E_{1}^{k+1}$ to eliminate
$\eta_{k}$ and thus obtain the following
\begin{eqnarray}
&\quad& \rho _{k+1} =
\rho _{k}-4\delta_{k} \label{rhoex},\\
&\quad& \sigma _{k+1} =
\sigma _{k}-4\delta_{k} \label{sigmaex},\\
&\quad& m_{k+1}=
m_{k}-\frac{4(n+1)}{\sigma_{*}^{3/2}}\left(\kappa _{k+1} E_{1}^{k+1} \right)^{1/2}
 \label{mexpression},\\
& \quad& f_{k+1}=f_{k}-\frac{(n+3)}{\sigma_{*}^{3/2}}\left(\kappa _{k+1} E_{1}^{k+1} \right)^{1/2},\\
&\quad& L_{k+1}=\frac{3}{2}L_{0} \label{Lexp},\\
&\quad& \kappa  _{k+1} E_{1}^{k+1}=
\frac{1}{\sigma *}\left(\frac{\Lambda _{k}\kappa _{k} E_{1}^{k}}
{f^{2}_{k}m^{2}_{k}\delta _{k}^{\aleph}}
\right)^{2}.
\end{eqnarray}
Relations $(\ref{rhoex})$ through $(\ref{Lexp})$ can be summed
to give
\begin{eqnarray}
&\quad& \rho _{\infty} =
\rho_{0} -4 \sum^{\infty}_{k=0}\delta _{k} >\rho _{*}, \label{rhomainfty}\\
&\quad& \sigma_{\infty}=
\sigma_{0}-4\sum ^{\infty}_{k=0}\delta _{k} > \sigma_{*}, \label{sigmainfty}\\
&\quad& L_{\infty}=\frac{3}{2}L_{0}=L_{*},\\
&\quad& f_{\infty}=f_{0}-\frac{(n+3)}{\sigma_{*}^{2}}
\sum ^{\infty}_{k=0}\left(\kappa  _{k+1} E_{1}^{k+1}\right)
^{1/2}>f_{*}, \label{finfty}\\
&\quad& m_{\infty} =
m_{0}- \frac {4(n+1)}{\sigma^{2}_{*}} \sum^{\infty}_{k=0}
\left(\kappa  _{k+1} E_{1}^{k+1}\right)
^{1/2} >m_{*}.\label{minfty}
\end{eqnarray}
Note that using $(\ref{vepsilonex})$ we can rewrite $(\ref{etaexp})$ as follows
\begin{eqnarray}
\label{deltadef}
\delta_{k}^{2\aleph}= \frac{\Lambda ^{2}}{m^{4}_{k}f_{k}^{4}\sigma _{*}}
\frac{\kappa^{2}_{k}(E_{1}^{k})^{2}}{\kappa _{k+1}E_{1}^{k+1}}.
\end{eqnarray}
Defining the sequence $\{\kappa_{k} E_{1}^{k}\}$ serves to define the
sequence $\{\delta_{k} \}$.
Now we must make an appropriate choice of the sequence $\{\varepsilon _{k}\}$.
\\
\\
First we assume  the sequence $\{\kappa_{k}E_{1}^{k} \}$ has the form
\[
\kappa_{k} E_{1}^{k}=C_{k}\kappa_{0} E_{1}^{0},
\quad
k=0,1,2...(C_{0}=1).\]
The choice of  sequence $\{\kappa_{k}E_{1}^{k}\}$ is important since
it will result in bounds for $\varepsilon_{0} E_{1}^{0}$ which is
the size of the initial perturbation.
We define the following series
\begin{eqnarray}
&\quad& \sum^{\infty}_{k=0} s_{k}=
\sum ^{\infty}_{k=0} \left( \frac{C^{2}_{k}}{C_{k+1}} \right)^{\frac{1}{2\alpha}}
\equiv
s,
\label{sumone}
\\
&\quad& \sum^{\infty}_{k=0}t_{k}=
\sum^{\infty}_{k=0}C^{\frac{1}{2}}_{k+1} \equiv
t.
\label{sumtwo}
\end{eqnarray}
With the above notation $(\ref{rhomainfty})$ can be rewritten as
\\
\begin{eqnarray}
\label{epzrho}
\rho_{0}-\rho_{*} >
4\left( \frac {\Lambda ^{2}}{m^{4}_{*}f_{*}^{4}\sigma _{*}} \right)^{\frac{1}{2\aleph}}
(\kappa_{0}E_{1}^{0})^{1/2\aleph}  s\quad
{\mbox{or}}
\quad 
\kappa_{0} E_{1}^{0} <
 \left( \frac {\rho_{0}-\rho_{*}}{4s} \right)^{2\aleph}
 \frac {m^{4}_{*}f_{*}^{4}\sigma _{*}}{\Lambda^{2}},
\end{eqnarray}
where we have used the fact  $m_{*}<m_{k}$ (which gives a more
stringent bound).\\
\\
With the above notation $(\ref{sigmainfty})$ can be rewritten as
\\
\begin{eqnarray}
\label{epzs}
\sigma_{0}-\sigma_{*} >
4\left( \frac {\Lambda ^{2}}{m^{4}_{*}f_{*}^{4}\sigma _{*}} \right)^{\frac{1}{2\aleph}}
(\kappa_{0}E_{1}^{0})^{1/2\aleph}  s
\quad
{\mbox{or}}
\quad
\kappa_{0} E_{1}^{0} <
 \left( \frac {\sigma_{0}-\sigma_{*}}{4s} \right)^{2\aleph}
 \frac {m^{4}_{*}f_{*}^{4}\sigma _{*}}{\Lambda^{2}}.
\end{eqnarray}
Similarly $(\ref{minfty})$ can be rewritten as
\begin{eqnarray*}
m_{0}-m_{*}> \frac{4(n+1)}{\sigma^{2}_{*}}(\kappa_{0}E_{1}^{0})^{1/2}t
>
\frac{4(n+1)}{\sigma^{2}_{0}}(\kappa_{0}E_{1}^{0})^{1/2}t
\end{eqnarray*}
or
\begin{eqnarray}
\label{epzm}
\kappa_{0} E_{1}^{0}<
\frac{\sigma^{4}_{0}}{16}\frac {(m_{0}-m_{*})^{2}}{(n+1)^{2}t^{2}},
\end{eqnarray}
and $(\ref{finfty})$ as
\begin{eqnarray*}
f_{0}-f_{*}&>&\frac{(n+3)}{\sigma_{*}^{2}}(\kappa_{0}E_{1}^{0})^{1/2}t
>
\frac{(n+3)}{\sigma_{0}^{2}}(\kappa_{0}E_{1}^{0})^{1/2}t
\end{eqnarray*}
or
\begin{equation}
\label{epsilonfest}
\kappa_{0} E_{1}^{0}
< \frac{\sigma_{0}^{4}}{(n+3)^{2}} \left( \frac{f_{0}-f_{*}}{t} \right)^{2}.
\end{equation}
For the sequence $\{C_{k} \}$ we make the choice
\begin{equation}
\label{epsilondef}
 C_{k}=
2^{-2\aleph k},
\end{equation}
so  $(\ref{sumone})$ and $(\ref{sumtwo})$ become
\begin{eqnarray}
&\quad& s=
\sum ^{\infty}_{k=0} \left( \frac{C^{2}_{k}}{C_{k+1}} \right)^{\frac{1}{2\aleph}}
=
2\sum^{\infty}_{k=0}2^{-2k}=
2\left( \frac {1}{1-\frac{1}{2}} \right)=4,
\label{sumonet}
\\
&\quad& t=
\sum^{\infty}_{k=0}C^{\frac{1}{2}}_{k+1} =
2^{-\aleph}\left( \frac{1}{1-2^{-\aleph}} \right)=
\frac {1}{2^{\aleph}-1}<
1.
\label{sumtwot}
\end{eqnarray}
We therefore obtain four estimates for $\varepsilon_{0}$ from $(\ref{epzrho})$
$(\ref{epzs})$, $(\ref{epzm})$, and  $(\ref{epsilonfest})$
\begin{eqnarray}
&\quad&
\kappa_{0} E_{1}^{0}<
 \left( \frac {\rho_{0}-\rho_{*}}{4s} \right)^{2\aleph}
 \frac {m^{4}_{*}f_{*}^{4}\sigma _{*}}{\Lambda^{2}},
\label{eprhot}
\\
&\quad&
\kappa_{0} E_{1}^{0}<
 \left( \frac {\sigma_{0}-\sigma_{*}}{4s} \right)^{2\aleph}
 \frac {m^{4}_{*}f_{*}^{4}\sigma _{*}}{\Lambda^{2}},
\label{epsigmat}
\\
&\quad&
\kappa_{0} E_{1}^{0}<
\frac{\sigma^{4}_{0}}{16}\frac {(m_{0}-m_{*})^{2}}{(n+1)^{2}t^{2}}
 ,
\label{epmt}
\\
&\quad&
\kappa_{0} E_{1}^{0}<
\frac{\sigma_{0}^{4}}{(n+3)^{2}} \left( \frac{f_{0}-f_{*}}{t} \right)^{2}.
\label{epff}
\end{eqnarray}
We now make the following choices
\begin{eqnarray*}
 \rho_{*}= \frac{\rho_{0}}{2},\quad
 m_{*}=\frac{m_{0}}{2},
\quad f_{*}= \frac{f_{0}}{2}
\mbox{ and }
 \sigma_{*}=\frac{\sigma_{0}}{2},
\end{eqnarray*}
so  $(\ref{epsigmat})$ and $(\ref{eprhot})$ become
\begin{eqnarray}
\label{finalpersize}
\kappa_{0} E_{1}^0<
 \left( \frac {\sigma_{0}}{32} \right)^{2\aleph }
\frac {m^{4}_{0}f^{4}_{0}\sigma_{*}}{2^{8}\Lambda^{2}},
\end{eqnarray}
\begin{eqnarray}
\kappa_{0} E_{1}^0<
\left( \frac{\rho_{0}}{32} \right)^{2 \aleph }\frac{m^{4}_{0}f^{4}_{0}\sigma_{*}}
{2^{8}\Lambda^{2}}.
\end{eqnarray}
Since $\sigma _{0} < \rho_{0}$
\[
 \left( \frac {\sigma_{0}}{32} \right)^{2\aleph }
\frac {m^{4}_{0}f^{4}_{0}\sigma_{*}}{2^{8}\Lambda^{2}}
<
\left( \frac{\rho_{0}}{32} \right)^{2 \aleph }\frac{m^{4}_{0}f^{4}_{0}\sigma_{*}}
{2^{8}\Lambda^{2}}.
\]
Also we see
\begin{eqnarray*}
&\quad&
\frac{\sigma^{4}_{0}}{16}\frac {(m_{0}-m_{*})^{2}}{(n+1)^{2}t^{2}}
=
\frac{\sigma^{4}_{0}}{2^{6}}\frac {m_{0}^{2}}{(n+1)^{2}},
\quad
\quad
\frac{\sigma_{0}^{4}}{(n+3)^{2}} \left( \frac{f_{0}-f_{*}}{t} \right)^{2}
=
\frac{\sigma_{0}^{4}}{4(n+3)^{2}} f_{0}^{2},
\end{eqnarray*}
and is not hard to see
\begin{eqnarray*}
 \left( \frac {\sigma_{0}}{32} \right)^{2\aleph }
\frac {m^{4}_{0}f^{4}_{0}\sigma_{*}}{2^{8}\Lambda^{2}}
<
\frac{\sigma^{4}_{0}}{2^{6}}\frac {m_{0}^{2}}{(n+1)^{2}},
\quad
\quad
 \left( \frac {\sigma_{0}}{32} \right)^{2\aleph }
\frac {m^{4}_{0}f^{4}_{0}\sigma_{*}}{2^{8}\Lambda^{2}}
<
\frac{\sigma_{0}^{4}}{4(n+3)^{2}} f_{0}^{2}.
\end{eqnarray*}
Therefore $(\ref{finalpersize})$
  gives the maximum upper bound for the size of the perturbation,
$\kappa_{0} E_{1}^{0}$.
\\
\\
Now we develop the estimates for $\zeta_{k+1}$ with the assumption
at the zeroth step $-1/2< \zeta_{A} < 1/2$ and then take the limit as
$k \rightarrow \infty$.
\\
\\
Assuming at the zeroth step $-1/2< \zeta_{A_{0}} = \zeta_{A} <1/2$,
at the first step we obtain
\begin{eqnarray*}
1+\zeta_{1} &<& 1 + \kappa_{0}E_{1}^{0}\zeta_{A_{0}}.
\end{eqnarray*}
At the second step we obtain
\begin{eqnarray*}
1+\zeta_{2}&<& 1+\zeta_{1} + \frac{3}{2}\zeta_{A_{1}} \kappa_{1}E_{1}^{1}
\quad <\quad 1+\kappa_{0}E_{1}^{0}\zeta_{A_{0}} +\frac{3}{2}\zeta_{A_{1}} \kappa_{1}E_{1}^{1} .
\end{eqnarray*}
Continuing in this manner we obtain in the $(k+1)$ step
\begin{eqnarray*}
1+\zeta_{k+1} &<& 1 +\kappa_{0}E_{1}^{0}\zeta_{A_{0}} 
+ \frac{3}{2}
\sum ^{k}_{i=1} \zeta_{A_{i}} \kappa _{i}E_{1}^{i} .
\end{eqnarray*}
One obtains a similar expression  for the lower bound and combining it with
the upper bound we have the following
\begin{equation}
\label{zetaone}
1-\kappa_{0}E_{1}^{0}\zeta_{A_{0}} -\frac{3}{2} \overline{\kappa} <
1+\zeta_{k+1} < 1+ \kappa_{0}E_{1}^{0}\zeta_{A_{0}} + \frac{3}{2} \overline{
\kappa },
\end{equation}
where $\overline{\kappa} =\sum_{i=1}^{k} \zeta_{A_{i}}\kappa _{i}E_{1}^{i}$.\\
\\
Now we look for a condition on $\overline{\kappa}$. Using the estimate
on $|\zeta_{A}|$ we have
\begin{eqnarray*}
1+\kappa_{0}E_{1}^{0}|\zeta_{A_{0}} | + \frac{3}{2}| \overline{\kappa}| &\leq &
1+  \kappa_{0}E_{1}^{0}\Upsilon_{0} +\frac{3}{2} | \overline{\kappa}|
\leq 1+ \frac{3}{2} \overline{\varepsilon} ,
\end{eqnarray*}
where $\overline{\varepsilon}=\sum_{i=0}^{k} \Upsilon_{i}\kappa _{i}E_{1}^{i}$.
Similarly for the lower bound
\begin{eqnarray*}
1-\kappa_{0}E_{1}^{0}|\zeta_{A_{0}} | - \frac{3}{2}| \overline{\kappa}| &>&
1- \kappa_{0}E_{1}^{0} \Upsilon_{0} - \frac{3}{2}| \overline{\kappa}|
\geq 1-  \frac{3}{2}  \overline{\varepsilon}.
\end{eqnarray*}
Therefore expression $(\ref{zetaone})$ becomes
\begin{equation}
1-\frac{3}{2} \overline{\varepsilon}
< 1+ \zeta_{k+1} < 1+
\frac{3}{2} \overline{\varepsilon},
\end{equation}
so that
\begin{equation}
|\zeta_{k+1} | < \frac{3}{2} \overline{\varepsilon}.
\end{equation}
We want the following to hold
\begin{equation}
\frac{3}{2} \overline{\varepsilon}
< \frac{1}{2}.
\end{equation}
Therefore we want $\overline{\varepsilon}$ to be a converging series with the
above upper bound.
\\
\\
With the previous definitions of $\kappa _{k} E_{1}^{k}, C_{k}, \Upsilon _{k}$
 the following estimates hold
\begin{eqnarray*}
\lim _{k\rightarrow \infty}
\frac{3}{2}\overline{\varepsilon}&=&
\frac{3}{2} \sum ^{\infty}_{k=0} \Upsilon_{k}\kappa _{k}E_{1}^{k}
\leq
\frac{3}{2}\left(
\frac{4^{n}11\varpi}{\Gamma \gamma m_{*}f_{*}}
\right)
\sum _{k=0}^{\infty}
\frac{\kappa _{k}E_{1}^{k}}{\delta_{k}^{\aleph}}.
\end{eqnarray*}
\\
Using $(\ref{deltadef})$ in the estimate above we obtain
\begin{eqnarray*}
|\zeta_{\infty}|
&\leq&
\frac{3}{2}\left(
\frac{4^{n}11\varpi}{\Gamma \gamma m_{*}f_{*}}
\right)
\sum _{k=0}^{\infty}
\frac{m_{k}^{2}f_{k}^{2}\sqrt{\sigma _{*}}}
{\Lambda} \frac{
(\kappa_{k+1}E_{1}^{k+1})^{1/2}}{\kappa_{k}E_{1}^{k}}
\kappa_{k}E_{1}^{k}\\
\quad\\
&\leq&
\frac{3}{2}\left(
\frac{4^{n}11\varpi}{\Gamma \gamma m_{*}f_{*}}
\right)
\frac{m_{0}^{2}f_{0}^{2}\sqrt{\sigma _{*}}}
{\Lambda}
\sum _{k=0}^{\infty}(\kappa_{k+1}E_{1}^{k+1})^{1/2}
\\
\quad\\
&=&
\frac{3}{2}\left(
\frac{4^{n}11\varpi}{\Gamma \gamma m_{*}f_{*}}
\right)
\frac{m_{0}^{2}f_{0}^{2}\sqrt{\sigma _{*}}}
{\Lambda}(\kappa _{0} E_{1}^{0})^{1/2}t.
\end{eqnarray*}
\\
We therefore require
\begin{eqnarray*}
\frac{3}{2}\left(
\frac{4^{n}11\varpi}{\Gamma \gamma m_{*}f_{*}}
\right)
\frac{m_{0}^{2}f_{0}^{2}\sqrt{\sigma _{*}}}
{\Lambda}(\kappa _{0} E_{1}^{0})^{1/2}<\frac{1}{2},
\end{eqnarray*}
and we obtain another restriction on the perturbation
\begin{eqnarray}
\label{perboundzeta}
\kappa_{0} E_{1}^{0}
<
\frac{1}{9}
\left(\frac{\Gamma \gamma m_{*}f_{*}}{4^{n}11\varpi}
\right)^{2}
\left(
\frac{\Lambda}{m_{0}^{2}f_{0}^{2}\sqrt{\sigma_{*}}}
\right)^{2}.
\end{eqnarray}
\\
We now compare $(\ref{perboundzeta})$ and $(\ref{finalpersize})$ and
choose the smallest of these bounds  to be the bound on $\kappa_{0}E_{1}^{0}$.
After substituting for $\Lambda$ in $(\ref{perboundzeta})$ we obtain
\begin{eqnarray}
\frac{2^{22}\varpi^{4}(6n+6)^{8}(c_{3}+1)^{4}E^{2(\nu +\mu)}}
{\rho_{*}^{2} \Gamma ^{4}\gamma ^{2}(11)^{2}9\sqrt{\sigma_{*}}},
\end{eqnarray}
and after substituting $\Lambda$, $(\ref{Lambdaexpression})$,
in $(\ref{finalpersize})$
we obtain
\begin{eqnarray}
\label{f11}
\left(\frac{\sigma_{0}}{32} \right)^{2(10n+9)}
\frac{m_{0}^{4}f_{0}^{4}\sigma_{*}}
{2^8}
\frac{\rho_{*}^{2}\Gamma^{6}\gamma ^{4}}{3^{2n}2^{4n+24}\varpi^{6}
(6n+6)^{8}(c_{3}+1)^{4}e^{2(\nu +\mu)}}.
\end{eqnarray}
A simple inspection reveals $(\ref{f11})$ is smaller
and therefore the upper bound on the perturbation
$\kappa_{0} E_{1}^{0}$.
With $(\ref{f11})$ we obtain the bound
\begin{eqnarray*}
|\zeta_{\infty}|&\leq&
\frac{3}{2}\left(
\frac{4^{n}11\varpi}{\Gamma \gamma m_{*}f_{*}}
\right)
\frac{m_{0}^{2}f_{0}^{2}\sqrt{\sigma _{*}}}
{\Lambda}(\kappa _{0} E_{1}^{0})^{1/2}
\\
\quad\\
&\leq&
\left(\frac{\sigma_{0}}
{32}\right)^{\aleph}\frac{3m_{0}^{3}f_{0}^{3}\sigma_{*}\Gamma^{2}\gamma \rho_{*}}
{3^n2^{15}\varpi^{2}(6n+6)^{4}(c_{3}+1)^{2}e^{\nu+\mu}\Lambda}.
\end{eqnarray*}
\section{Time Aperiodic Perturbation
Tending to a Quasi-periodic Time Perturbation}

In this section we consider
a nearly integrable system generated by a Hamiltonian
consisting of an integrable part and a small perturbation.
As before we consider a real valued nearly-integrable
Hamiltonian in action-angle variables of the form
\begin{eqnarray*}
H(p,q,t)= H^{0}(p)+\kappa H^{1}(p,q,t),
\end{eqnarray*}
where $p=(p_{1},...,p_{n})   \in B \subset
  {\mathbb{R}}^{n}$, $q=(q_{1},...,q_{n})\in T^{n}$ are,
respectively, the action and angle variables,
and $\kappa \in {\mathbb{R}} $ is a small perturbation parameter.
The form of the perturbation is as follows
\begin{eqnarray*}
&\quad& H^{1}(p,q,t)=Q(p,q,t)+E(p,q,t),\\
\quad\\
&\quad& Q(p,q,t)=\sum _{k\in{\mathbb{Z}}^{n+m}}
g_{k}e^{ik\cdot(q,\theta t)},\\
\quad\\
&\quad& E(p,q,t)= \sum _{k\in{\mathbb{Z}}^{n}}f_{k}(p)e_{k}(t)e^{ik\cdot q},
\end{eqnarray*}
where $Q(p,q,t)$ is time  quasi-periodic , with
$\theta =(\theta_{1},...,\theta_{m})$ as the vector of basic frequencies.
$f(p)$ is a bounded function and $e_{k}(t)$ decays
exponentially to zero
as time goes to infinity. Consequently,
 as time goes to infinity the entire perturbation tends
to a $t$-quasi-periodic function.

The result of the theorem will be  similar
to the KAM-type result obtained previously in this paper.
Mainly, we will prove the preservation of cylinders of the
form ${\mathbb{T}}^{n}\times {\mathbb{R}}$ where the tori 
${\mathbb{T}}^{n}$ can be identified  with the tori
of the integrable system generated by  $H^{0}(p)$.
We transform  the non-autonomous system into
an autonomous system  making each
of $\theta _{1}t,...,\theta _{m}t$ a dependent variable $q_{n+1},...,
q_{n+m}$ respectively. We will use either notation throughout
depending on  need to clarify. We add conjugate variables
$\tau=(\tau_{1},...,\tau_{m})$ which are needed to obtain
a Hamiltonian form. We will sometimes use the
notation $p'=(p,\tau)$ and $q'=(q,\theta t)$. The Hamiltonian takes
the form
\begin{eqnarray*}
 H(p,q,t)&=&\theta \cdot \tau +
H^{0}(p)+\kappa H^{1}(p,q,t)
= \tilde{H}^{0}(p,\tau )+\kappa H^{1}(p,q,t),
\end{eqnarray*}
and Hamilton's equations are given by
\begin{eqnarray*}
&\quad& \dot p=-\frac{\partial H}{\partial q}=-\frac{\partial H^{1}}{\partial q},\\
\quad\\
&\quad& \dot \tau =-\frac{\partial H}{\partial (\theta t)}=
-\frac{\partial H^{1}}{\partial (\theta t)},\\
\quad\\
&\quad& \dot q= \frac{\partial H}{\partial p}=\frac{\partial H^{0}}{\partial
p}+ \frac{\partial H^{1}}{\partial p},\\
\quad\\
&\quad& \dot{(\theta t)} =\frac{\partial H}{\partial \tau} =\theta.
\end{eqnarray*}
The angular frequencies are given by
$
\lambda (p') =\frac{\partial \tilde{H}^{0}}{\partial p'}(p')=
(\frac{\partial H^{0}}{\partial p_{1}},...,
\frac{\partial H^{0}}{\partial p_{n}},\theta_{1},...,\theta_{m}).
$
We will often use the notation $\lambda =
(\tilde{\lambda}, \theta)$ where $\tilde{\lambda}
=(\frac{\partial H^{0}}{\partial p_{1}},...,
\frac{\partial H^{0}}{\partial p_{n}})$.
We define the complex extension of ${\mathbb{R}}^{n}\times
{\mathbb{R}}^{m} \times T^{n} \times {\mathbb{R}}^{m}$ as follows
\begin{eqnarray*}
{{D}}_{\rho, \sigma ,p_{0}}=\{
(p,\tau ,p, \theta t)\in {\mathbb{C}}^{2n+2m}|
\quad \|p-p_{0}\|\leq \rho, \|\tau \|\leq \sigma ,
{\mbox{Re }}q\in {\mathbb{R}}^n {\mbox{ mod }}2\pi,
\|{\mbox{Im }}q \|\leq \rho , |{\mbox{Im }}t| \leq \sigma \},
\end{eqnarray*}
and define ${\cal{A}}_{\rho , \sigma ,p_{0}}$ as the set of all complex
continuous functions functions defined on ${{D}}_{\rho, \sigma ,p_{0}}$,
analytic in the interior of ${{D}}_{\rho, \sigma ,p_{0}}$ and
real for real values of the variables. The normal form
is again given as the Taylor expansion of the Hamiltonian
about $p'=0$. Namely
\begin{equation}
H(p,\tau ,q, \theta t
)=a+\lambda \cdot p'+  A(q')+ B(q')\cdot p'
 +\frac {1}{2} \sum^{n+m}
_{i,j}  C_{i,j}(q')p'_{i}p'_{j} +R(p',q'),
\end{equation}
where
\begin{eqnarray*}
&a& =\quad \overline{\overline{H(0)}}=
 \overline{\overline{\tilde{H}^{0}}}(0)+
\kappa\overline{\overline{H^{1}}}(0)
=\tilde{H}^{0}(0)+
\kappa \overline{\overline{H^{1}}}(0),\\
\quad \\
& \kappa A(q')&=\quad H(0,q')-a= \kappa(  H^{1}(0,q,\theta t)-
\overline{\overline{H^{1}}}(0)),\\
\quad \\
& \kappa B_{i}(q')&=\quad \frac {\partial H}{\partial p'_{i}}(0,q')-\lambda _{i}=
\kappa  \frac {\partial H^{1}}{\partial p'_{i}}
(0,q, \theta t), \\
\quad \\
&C_{ij}(q')&=\quad \frac {\partial ^{2}H}{\partial p'_{i}\partial p'_{j}}(0,q')
=\frac {\partial ^{2} \tilde {H}^{0}}{\partial p'_{i}\partial p'_{j}}(0)+
\kappa \frac {\partial ^{2}  {H}^{1}}{\partial p'_{i}\partial p'_{j}}(0,q,\theta t),
\end{eqnarray*}
and
\[a\in \mathbb {R},A,B_{i},C_{i,j},R\in A_{\rho , \sigma}, R={\cal{O}}(\| p'^{3}\|).\]
The matrix ${\tilde{C}}^{*}$ will be defined as before
\begin{eqnarray*}
{\tilde{C}}^{*}_{i,j}=\frac{\partial ^{2} \tilde{H}^{0}}
{\partial p_{j} \partial p_{j}}(0),\quad i,j=1,...,n,
\end{eqnarray*}
and the same non-degeneracy condition will hold
\begin{eqnarray*}
{\mbox{det}}\left(\frac{\partial ^{2} \tilde{H}^{0}}
{\partial p_{j} \partial p_{j}}(0)\right) \neq 0,
\end{eqnarray*}
or
$
d\|\tilde v \|\leq \|{\tilde{C}}^{*}\tilde v \|\leq
d^{-1}\|\tilde v \|, \quad \forall \tilde v \in {\mathbb{C}}^{n},
$
for some positive constant $d$.
As before, this non-degeneracy condition will imply an
isoenergetic non-degeneracy condition as  proven
in lemma $\ref{lemmaiso}$. In other words, although the matrix
$C_{i,j}^{*}=  \frac{\partial ^{2} \tilde{H}^{0}}
{\partial p'_{j} \partial p'_{j}}(0) \quad i,j=1,...,n,n+1,...,n+m
$ is a bigger matrix, the matrix ${\mathbb{I}}$ defined as
\begin{eqnarray*}
{\mathbb{I}}=\left(
\begin{array}{cc}
C^{*} & \lambda ^{T} \\
\lambda & 0
\end{array}
\right)=
\left(
\begin{array}{ccc}
\tilde{C}^{*} & 0 & \tilde{\lambda} ^{T}\\
0 & 0 & \theta ^{T} \\
\tilde{\lambda} & \theta & 0
\end{array}
\right),
\end{eqnarray*}
can be shown to be nonsingular and therefore a lower bound can
be found for $\|{\mathbb{I}}v\|_{\rho, \sigma },\quad
\forall v\in {\mathbb{C}}^{n+m+1}$. Using this lower bound, a similar
constant, $f$, can be found to bound the expression
\begin{eqnarray*}
\left\|
\left(
\begin{array}{cc}
\overline{\overline{C}} & \lambda ^{T}\\
\lambda & 0
\end{array}
\right)v\right\|,
\end{eqnarray*}
as was done in lemma $\ref{thmisob}$. Ultimately we will need
the implied invertibility of the matrix above to solve
completely for the generating function.

Because the form of the Hamiltonian
perturbation  in this section
is different from the one consider previously in the paper,
we will re-define  what we mean by $p',q'$-exponential form.
In this section, functions of  $p',q'$-exponential form will be understood
to have the following form
\begin{eqnarray*}
&\quad& F(p',q')=f(p')+r(p',q')+g(p',q'),\\
\quad\\
&\quad& r(p',q')=\sum _{k\in {\mathbb{Z}}^{n+m}}
s_{k}(p')e^{ik\cdot (q, \theta t)},\\
\quad\\
&\quad& g(p',q')=\sum _{k\in{\mathbb{Z}}^{n}}
h_{k}(p')e_{k}(t)e^{ik\cdot q},
\end{eqnarray*}
where $F(p',q')\in{\cal{A}}_{\rho ,\sigma}$ and as before
$e_{k}(t)$ is of exponential order with respect to time.
Clearly the only difference in the definition is the quasi-
periodic time dependence added to the term $r(p',q')$.
Before examining how this modification will affect the lemmas
proven in section
$\ref{sectp'q'exform}$
we prove a lemma concerning
the new form of $r(p',q')$.
\begin{lem}
\label{newlemmaforexform}
\quad\\
Given a function $r(p',q')\in {\cal{A}}_{\rho, \sigma}$
of the form
\begin{eqnarray*}
r(p',q')=\sum _{k\in {\mathbb{Z}}^{n+m}}
s_{k}(p')e^{ik\cdot (q, \theta t)},
\end{eqnarray*}
 it follows
$\overline{\overline{r}}=s_{0}(p')$.\\
\quad\\
{Proof}:
\quad\
We define the vector $k=(h_{1},h_{2})\in {\mathbb{Z}}^{n+m}$
where $h_{1}\in{\mathbb{Z}}^{n}$ and $h_{2}\in{\mathbb{Z}}^{m}$ and
\begin{eqnarray*}
\overline{\overline{r}} &=&
\lim _{T\rightarrow \infty}
\frac{1}{2T}\int^{T}_{-T}\int _{T^{n}}
\sum _{k\in {\mathbb{Z}}^{n+m}}
s_{k}(p')e^{ik\cdot (q, \theta t)}dqdt\\
\quad\\
&=& \sum _{h_{2}\in {\mathbb{Z}}^{m}}
s_{(0,h_{2})}(p')\lim _{T\rightarrow \infty}
\frac{1}{2T}\int^{T}_{-T}
e^{ih_{2}\cdot \theta t}dt\\
\quad\\
&=&
s_{0}(p')+ \sum_{h_{2}\in {\mathbb{Z}}^{m}\backslash 0}
s_{(0,h_{2})}(p')\lim _{T\rightarrow \infty}
\frac{1}{2T}\int^{T}_{-T}
e^{ih_{2}\cdot \theta t}dt\\
\quad\\
&=&
s_{0}(p')+ \sum_{h_{2}\in {\mathbb{Z}}^{m}\backslash 0}
s_{(0,h_{2})}(p')\lim _{T\rightarrow \infty}
\frac{sin(h_{2}\cdot \theta T)}{2T}
=s_{0}(p')
\quad \squareforqed
\end{eqnarray*}
\end{lem}
The result of lemma $\ref{lemmaforg}$ still holds.
With the result of lemma
$\ref{newlemmaforexform}$,
 lemma $\ref{exponentialpart}$  still holds. That is,
given
\begin{eqnarray*}
F(p',q')=\sum _{k\in {\mathbb{Z}}^{n+m}}
s_{k}(p')e^{ik\cdot (q, \theta t)}
+\sum _{k\in{\mathbb{Z}}^{n}}
h_{k}(p')e_{k}(t)e^{ik\cdot q},
\end{eqnarray*}
it follows
\begin{eqnarray*}
F(0,p')-\overline{\overline{F}}=
\sum _{k\in {\mathbb{Z}}^{n+m}\backslash 0}
s_{k}(0)e^{ik\cdot (q, \theta t)}
+\sum _{k\in{\mathbb{Z}}^{n}}
h_{k}(0)e_{k}(t)e^{ik\cdot q}.
\end{eqnarray*}
The results and proofs of lemmas comparable to
lemmas  $\ref{expoformderp}$ and
$\ref{expoformderq}$ do not change and follow the same arguments.
A lemma comparable to lemma $\ref{lemmatimederv}$ concerning the
time derivative of a function $F(p',q')$ of $p',q'$-exponential
form would follow the same line of arguments but one should
take in consideration the time derivative of the
function $r(p',q')=\sum _{k\in {\mathbb{Z}}^{n+m}}
s_{k}(p')e^{ik\cdot (q, \theta t)}$. Since such time derivative
results in a $q$ periodic time quasi-periodic function,
the total form of $\partial F(p',q')/\partial t$ will be
of $p',q'$-exponential
form and the domain of analyticity would be  as stated in
lemma $\ref{lemmatimederv}$.
Finally, the added time quasi-periodic dependence of
$r(p',q')$ will not change the result of lemma $\ref{lempqform}$
 concerning the $p',q'$-exponential form
for the Poisson bracket $\{ \chi,F\}$. The arguments of the proof
would follow the same and some sums would have to change from
$\sum_{j=1}^{n}$ to $\sum_{j=1}^{n+1}$ to account for
time derivatives that in lemma $\ref{lempqform}$ are zero
but now must have to be added.

The statement and proof of a  lemma comparable to the
 iterative lemma $\ref{iterativelemma1}$
 would follow
exactly the same up to the generating function used
and the definition of the perturbation
$(\ref{Hamiltonianperturbation1})$. The generating
function  will have almost the  same form
\begin{eqnarray*}
&\quad& \chi= X(q')+\xi \cdot q' + Y(q')\cdot p', \\
\quad\\
&\quad& X(q')={\cal{Y}}(q')+{\cal{T}}(q'),\\
\quad\\
&\quad& {\cal{Y}}(q')= \sum _{k\in{\mathbb{Z}}^{n+m}}
y_{k}e^{ik\cdot(q,\theta t)},
\quad
\quad {\cal{T}}(q')=\sum _{k\in {\mathbb{Z}}^{n}}x_{k}(t)e^{ik\cdot q},\\
\quad\\
&\quad& Y_{j}(q')={\cal{S}}_{j}(q')+{\cal{F}}_{j}(q'),\\
\quad\\
&\quad& {\cal{S}}_{j}(q')=\sum _{k\in {\mathbb{Z}}^{n+m}}
{\cal{S}}_{k,j}e^{ik\cdot(q,\theta t)},
\quad
\quad {\cal{F}}_{j}(q')=\sum _{k\in {\mathbb{Z}}^{n}}
{\cal{F}}_{k,j}(t)e^{ik\cdot q},
\end{eqnarray*}
where now ${\cal{Y}}(q')$ and ${\cal{S}}_{j}(q')$ depend
quasi-periodically on time.
The new perturbation
must have the following form
\begin{eqnarray*}
&\quad& H^{1}(p',q')=G(p',q')+T(p',q') \in {\cal{A}}_{\rho ,\sigma},\\
\quad\\
&\quad& G(p',q')=\sum _{k\in {\mathbb{Z}}^{n+m}}s_{k}^{1}(p')
e^{ik\cdot(p,\theta t)} \in {\cal{A}}_{\rho ,\sigma},\\
\quad\\
&\quad& T(p',q')=\sum_{k\in {\mathbb{Z}}^{n}}h_{k}^{1}(p')e_{k}^{1}(t)
e^{ik\cdot q} \in {\cal{A}}_{\rho, \sigma},
\end{eqnarray*}
where $e_{k}^{1}(t)$ will be assumed to have
the same properties as before. Solving for $X(q')$
involves solving the equation
$
\lambda \cdot \partial_{q'}X(q') 
=\kappa \left(A(q') -\overline{\overline{A}} \right),
$
which can be separated into a quasi-periodic part and
an exponential-order-with-respect-to-time part. These two problems
are given by
\begin{eqnarray}
\label{eqnquasinew}
&\quad&\lambda \cdot \partial _{q'}{\cal{Y}}(q')=\kappa
\left( G(0,q')-\overline{\overline{G}}(0)\right),
\end{eqnarray}
\begin{eqnarray}
\label{eqnexponew}
&\quad&\lambda \cdot \partial _{q'}{\cal{T}}(q')=\kappa
T(0,q'),
\end{eqnarray}
respectively. Solving $(\ref{eqnquasinew})$
will follow as before but now we must make use of
the diophantine condition on $\lambda \in \Omega_{\Gamma}$ rather
than just on $\tilde{\lambda}$.
As before we set
\quad\\
\begin{eqnarray*}
 \overline { \overline {(
\kappa B-C\cdot \xi -C \cdot \partial _{q'}X)}} =
\kappa E_{1} \zeta _{A} \lambda.
\end{eqnarray*}
\quad\\
This leads to the conditions
\begin{eqnarray*}
&\quad& \overline{ \overline{C}}\cdot \xi
 +\kappa E_{1} \zeta _{A}\lambda =\kappa \overline{ \overline {B}}
-\overline{\overline {C}}\cdot \overline{\overline {\partial _{q'}
X}},
\quad
\quad \lambda \cdot \xi  =\kappa \overline{\overline {A}},
\end{eqnarray*}
or equivalently
\begin{equation}
\label{eqnnono}
\left( \begin{array}{cc}
\overline{ \overline{C}} & \lambda ^{T} \\
\lambda & 0
\end{array}
\right) \left(
\begin{array}{c}
\xi \\\kappa E_{1} \zeta_{A} \end{array}
\right)
=
\left(
\begin{array}{c}
\kappa \overline{\overline {B}}-\overline{\overline {C}}
\cdot \overline{\overline {\partial _{q'}X}} \\
\kappa \overline{\overline {A}}
\end{array}
\right).
\end{equation}
\quad\\
Since we are interested in in solving the above
equation for the unknowns $\xi$ and $\zeta_{A}$, we would like
the matrix on the left hand side to be nonsingular.
Using the fact the matrix ${\mathbb{I}}$ is nonsingular
together with a lemma comparable to lemma
$(\ref{thmisob})$ it follows easily  the matrix
on the left hand side of $(\ref{eqnnono})$ is nonsingular.
The solution of $(\ref{eqnexponew})$ follows exactly as shown
previously in the paper.
Next we look at the solution of
\begin{eqnarray}
\label{diffeqnbeta2}
\lambda \cdot \partial _{q'}Y =\beta,
\end{eqnarray}
where as before $\beta$ is defined as
\begin{eqnarray*}
\beta &=&\Big[ (\kappa B(q')-C(q')\cdot \xi -C(q')\cdot \partial _{q'}X(q'))
-(\kappa \overline{\overline {B}}-\overline{\overline {C}}\cdot \xi -\overline{\overline {C}}\cdot
\overline{\overline {\partial _{q'}X}}) \Big].
\end{eqnarray*}
Since the definition of $H^{1}(p',q')$ has a quasi-periodic
time dependence not present before, $B_{j}(q'),C(q')$ and
$X(q')$ in terms of $H^{1}$ will change. Consequently the terms
that make up $\beta$ will change.
For instance
\begin{eqnarray*}
\kappa \left(B_{j}(q') - \overline{\overline{B}}_{j}\right)=
\kappa \left[
\sum _{k\in {\mathbb{Z}}^{n+m}\backslash 0}
 \frac{\partial s^{1}_{k}}{\partial p'_{j}}(0)
e^{ik\cdot (q,\theta t)}
+
\sum _{k\in {\mathbb{Z}}^{n}}
\frac{\partial h^{1}_{k}}{\partial p'_{j}} (0) e^{1}_{k}(t) e^{ik\cdot q} \right]
.
\end{eqnarray*}
Furthermore, when calculating
$(C(q')\cdot \partial _{q'}X(q'))_{j}$ three more terms must be added
to this expression
\begin{eqnarray*}
&\quad& \frac{\partial ^{2}\tilde{H}^{0}}{\partial p'_{n+1}
 \partial p'_{j}}(0)\partial _{q'_{n+1}}
 {\cal{Y}}(q'),
\quad \kappa \sum _{k\in {\mathbb {Z}}^{n}} \frac{ \partial ^{2} s^{1}
_{k}}
{\partial p'_{n+1} \partial p'_{j}}
(0)e^{ik \cdot q}\partial _{q'_{n+1}}
 {\cal{Y}}(q'),
\quad \kappa \sum _{k\in {\mathbb {Z}}^{n}}
\frac{ \partial ^{2} h^{1}
_{k}}
{\partial p'_{n+1} \partial p'_{j}}
(0)
e^{1}_{k}(t)e^{ik \cdot q}\partial _{q'_{n+1}}
 {\cal{Y}}(q').
\end{eqnarray*}
Previously these terms would dropout of the expression
since $\partial _{q'_{n+1}}
 {\cal{Y}}(q')=0$. The solution of $(\ref{diffeqnbeta2})$ splits into
a quasi-periodic part and an exponential-with -respect-to-time part
given by the following
\begin{eqnarray}
\label{eqnofhell12}
&\quad& \lambda \cdot \partial _{q'}{\cal{S}}_{j}(q') =\beta ^{Q}_{j}(q')=
\sum _{k\in {\mathbb{Z}}^{n+m}\backslash 0
}\beta ^{Q}_{j,k}e^{ik\cdot(q,\theta t)}
\in {\cal{A}}_{\rho -2\delta, \sigma -2\delta},
\end{eqnarray}
\begin{eqnarray}
\label{eqnofhell13}
&\quad& \lambda \cdot \partial _{q'}{\cal{F}}_{j}(q')=
\beta^{E}_{j}(q')=\sum _{k\in {\mathbb{Z}}^{n}}
\beta^{E}_{j,k}(t)e^{ik\cdot q}\in {\cal{A}}_{\rho -2\delta,
\sigma -2\delta},
\end{eqnarray}
where $\beta ^{Q}_{j}(q')$ and $\beta^{E}_{j}(q')$ are obtained from
the expressions for
$\kappa B(q')-\kappa \overline{\overline {B}}$,
$C(q')\cdot \xi-\overline{\overline {C}}\cdot \xi$,
and
$C(q')\cdot \partial _{q'}X(q'))-\overline{\overline {C}}\cdot
\overline{\overline {\partial _{q'}X}}$.
As done previously in the paper
one can solve $(\ref{eqnofhell12})$ and $(\ref{eqnofhell13})$
to obtain $Y_{j}(q')$ with all the necessary bounds. The rest of the
bounds and the iterative process are obtained as before. The main
result is described in the following theorem.
\begin{theom}
\quad \\

Consider the Hamiltonian  $H(p',q')=H^{0}(p)  +
\kappa H^{1}(p,q,t) + \tau=\tilde{H} ^{0}(p')+  \kappa H^{1}(p',q')$
of $(C_{1},C_{2},c_{1},c_{2},\nu,\mu)p',q'$-exponential form
defined on ${{B}}
\times {\mathbb{T}^{n}}$ where
$\tilde{H}^{0}(p')=H^{0}(p)+\tau$ and $\kappa << 1$ is a
small perturbation parameter.
Assume the perturbation has the following form
\begin{eqnarray*}
&\quad& H^{1}(p',q')=Q(p',q')+E(p',q'),\\
\quad\\
&\quad& Q(p',q')=
 \sum_{k\in{\mathbb{Z}}^{n+m}}
s_{k}^{1}(p')e^{ik\cdot (q,\theta t)},
\quad
\quad E(p',q')=
 \sum_{k\in{\mathbb{Z}}^{n}}
h_{k}^{1}(p')e_{k}^{1}(t)e^{ik\cdot q}.
\end{eqnarray*}
Fix $p'_{0} \in {{B}}$ and denote
\begin{eqnarray*}
\lambda = \lambda (p'_{0})=
\left( \frac{\partial H^{0}}{\partial p_{1}},...,
\frac{\partial H^{0}}{\partial p_{n}},\theta_{1},...,\theta_{m}\right)
=
 \frac{\partial \tilde{H}^{0}}{\partial p'}(p'_{0}),
\quad
\tilde{C}^{*}_{ij}= \frac{\partial ^{2} \tilde{H}^{0}}{\partial p_{i} \partial p_{j}}(p'_{0}) \quad i,j=1,...n,
\end{eqnarray*}
\begin{eqnarray*}
C^{*}_{ij}= \frac{\partial ^{2} \tilde{H}^{0}}{\partial p'_{i} \partial p'_{j}}(p'_{0}) \quad i,j=1,...,n+m,
\quad
{\mathbb{I}}= \left(
\begin{array}{cc}
C^{*} & \lambda ^{T} \\
\lambda & 0
\end{array}
\right).
\end{eqnarray*}
Assume there exists positive numbers $\Gamma ,\gamma , \rho , \sigma
,d,\kappa $, all less than one and $L$,  such that $\rho > \sigma $ and
\begin{eqnarray*}
\begin{array}{cc}
 {\lambda} \in \Omega _{\Gamma},\quad |\tilde{\lambda} |<L,\quad
 H^{0},H^{1} \in {\cal{A}}_{\rho, \sigma , p_{0}},\quad
 d\| \tilde{v} \| \leq \| \tilde{C} ^{*} \tilde{v} \| \leq
d^{-1} \| \tilde{v} \|, \quad  \forall \tilde{v} \in {\mathbb {C}}^{n},
\end{array}
\end{eqnarray*}
 moreover, $\|H \|_{\rho, \sigma , p_{0}}<1$, (the last condition
will imply the isoenergetic condition on ${\mathbb{I}}$).
\\
\\
Then there exists  positive numbers $E,\kappa ',E',\zeta ',
\rho'$ and $\sigma '$, such that if
$
\|H^{1} \|_{\rho ,\sigma , p_{0}} \leq E
$
one can construct a canonical analytic change of variables
\begin{eqnarray*}
\begin{array}{ccc}
\phi :D_{\rho ', \sigma '} & \rightarrow & D_{\rho ,\sigma ,p_{0}},
\quad \quad  \quad \quad \quad  \\
\quad\\
({\cal{P}}',{\cal{Q}}') & \mapsto & \phi ({\cal{P}}',{\cal{Q}}') = (p',q'),
\end{array}
\end{eqnarray*}
with $\phi \in {\cal{A}}_{\rho ', \sigma '}$, which brings the Hamiltonian
$H$ into the form $H' = H \circ \phi$
given by
\begin{eqnarray*}
H'({\cal{P}}',{\cal{Q}}') =(H\circ \phi )({\cal{P}}',{\cal{Q}}')= a'
+(1+  \kappa ' E' \zeta ') \lambda
\cdot {\cal{P}}'+{\cal{O}}(\| {\cal{P}}' \|^{2}),
\end{eqnarray*}
where $a',\zeta ' \in {\mathbb{R}}$.
The change of variables is near identity in the sense that
$
\| \phi -identity \|_{\rho ', \sigma} \rightarrow 0 \mbox{ as }
\| H^{1} \|_{\rho ,\sigma , p_{0}} \rightarrow 0.
$
In the new coordinates Hamilton's Equations are given by
\begin{eqnarray*}
&\quad& \dot{{\cal{Q}}}'= \frac{\partial H'}{\partial {\cal{P}}'}({\cal{P}}',{\cal{Q}}')=
 (1+\kappa 'E' \zeta ')\lambda+
{\cal{O}}(\|{\cal{P}}' \|),\quad
\quad
\dot{{\cal{P}}}'=-\frac{\partial H'}{\partial {\cal{Q}}'}({\cal{P}}',{\cal{Q}}')=
{\cal{O}}(\|{\cal{P}}' \|^{2}).
\end{eqnarray*}
The solutions are given by
\begin{eqnarray*}
&\quad & {\cal{Q}}'(t)=(1+ \kappa 'E' \zeta ') \lambda t   +{\cal{Q}}'_{0}
\quad
\quad  {\cal{P}}'(t)={\cal{P}}'_{0}= ({\cal{P}}_{0},\tilde{\tau}_{0}) = 0 .
\end{eqnarray*}
We can write out explicitly  these solutions
${\cal{Q}}'(t)=
({\cal{Q}}_{1}(t),...,{\cal{Q}}_{n}(t),{\cal{T}}(t))=
(1+ \kappa 'E' \zeta ') \lambda t   +{\cal{Q}}'_{0}
=(1+ \kappa 'E' \zeta ')(\tilde{\lambda},1)t  +{\cal{Q}}'_{0}$
which indicates in the new coordinate system the phase space
${\mathbb{T}}^{n}\times{\mathbb{R}}\times
{\mathbb{R}}^n \times {\mathbb{R}}$ is foliated by
invariant infinite cylinders each sustaining quasi-periodic motions
identified by the frequency $\tilde{\lambda}({\cal{P}}_{0})$ and evolving in
the $t$ direction. Note  under the transformation,
the time variable ${\cal{T}}(t)=(1+ \kappa 'E' \zeta ')t$ has shifted
by a constant proportional to the size of the perturbation.
\end{theom}
\section{Conclusions}
In this paper we looked at time 
aperiodic  perturbations of nearly integrable Hamiltonian
systems.
To prove a KAM-type theorem we used the Lie series formalism in an attempt to
construct a series of generating Hamiltonians from which
we  found a series of near identity
canonical transformations. This series of canonical transformations
is used  to put the original Hamiltonian in a normal form  which
has a smaller perturbation at each step of the iterative process.
Since we considered an aperiodic time perturbation, the Fourier series
methods normally used to solve for the generating function no longer apply.
We therefore made use of Fourier transform techniques which allowed us
to understand how the analyticity domain of the Hamiltonian
changes through the iterative
process and how fast the perturbation shrinks.
We considered the general form of the Hamiltonian $H(p,\tau,q,t)=H^{0}(p,\tau)+
\kappa H^{1}(p,\tau,q,t)$ where $\tau$ is conjugate to $t$. In the
first case considered the perturbation decayed exponentially in time
to a term depending only on angles. In the
second case we added to this exponentially decaying term a time quasiperiodic
term. In both cases we saw the time aperiodicity had to decay exponentially
in order to have control over the analyticity of the Hamiltonian. Moreover,
since the $p',q'$-exponential form of the perturbation is essential
in solving the partial differential equations that give the generating
function, it was necessary to prove the $p',q'$-exponential form
of the perturbation is preserved after each step of the iterative
process.

\appendix

\section{Properties of Analytic Functions on $\mathbb{T}^{n} \times \mathbb {C}$}

As mentioned before, techniques from multi-variable complex analysis are
used in this paper.
For example,  a generating Hamiltonian with converging estimates can be obtained via
Fourier  methods assuming appropriate
domains of analyticity for the perturbation of the Hamiltonian.
 Consequently we derive some important properties of analytic
functions on  $\mathbb{T}^{n} \times \mathbb {C}$.
We recall a common Fourier expansion for a complex analytic function
depending periodically
on the  the variables $q_{1},...,q_{2}$.
Consider a multi-periodic function $f$ analytic on some open
strip of ${\mathbb{C}}^{n}$ with periods $L_{1},...,L_{n}>0$, it is
possible to represent $f$  by the formula
$
f(q_{1},...,q_{n})= \sum _{k\in {\mathbb{Z}}^{n}} c_{k} e^{i(k_{1}\omega _{1}q_{1}+
\cdot \cdot \cdot + k_{n}\omega _{n}q_{n})}
$
where the coefficients $c_{k} \in {\mathbb{C}}$ have  the relation
$c_{k}= \overline {c}_{-k}$, and
$
c_{k}= \int ^{L_{1}}_{0} \frac{dq_{1}}{L_{1}}...\int ^{L_{n}}_{0} \frac{dq_{n}}{L_{n}}
e^{-i(\omega _{1}k_{1}q_{1}+ ... +\omega _{n}k_{n}q_{n})}f(q_{1},...,q_{n}),
$
where $\omega _{i} =2\pi /L_{i}, \quad i=1,...,n$.

A function also  depending aperiodically on a
variable $t\in \mathbb {C
}$  can be  expressed similarly
in a Fourier expansion.
Let $F(q,t) \in {\cal {A}}_{\rho ,\sigma , p_{0}}$ be real value, continuous,
and $2\pi$ periodic in each  $q_{i}$. One writes
$
F(q,t)=\sum _{k\in {\mathbb {Z}}^{n}}f_{k}(t)e^{iq\cdot k}$,
where
$
f_{k}(t)=\frac {1}{(2\pi)^{n}}\int ^{2\pi } _{0} \cdot \cdot \cdot \int ^{2\pi }_{0}
F(q,t)e^{-iq\cdot k}dq_{1}\cdot \cdot \cdot dq_{n}$.
The following lemma gives a useful bound for the Fourier coefficients of
 a function $F(q,t)$
depending $2\pi$ periodically on $q$ and aperiodically on $t$.
\begin{lem}
\label{lemmaanalib}
\quad \\

For $F\in {\cal{A}}_{\rho ,\sigma }$ and given $F(q,t)= \sum _{k\in \mathbb{Z}^{n}}f_{k}(t)
e^{ik\cdot q}$, then for every $k\in \mathbb{Z}^{n}$ and for all $t\in {\mathbb{C}}$ we have
$|f_{k}(t)| \leq \| F(t) \|_{\rho}e^{-|k|\rho }$,
where $|k|=\sum _{i} |k_{i}|$.\\
\quad\\
{Proof} :
From the previous argument we can write
\\
\[f_{k}(t)=\frac {1}{(2\pi)^{n}}\int  _{{\mathbb{T}}^{n}}
F(q_{1},...,q_{n},t)e^{-i\sum_{j}k_{j}q_{j}}dq_{1}\cdot \cdot \cdot dq_{n}.\]
\\
Since $F\in {\cal{A}}_{\rho ,\sigma}$, we can shift integration paths as follows
\\
\\
\[f_{k}(t)=\frac {1}{(2\pi)^{n}}\int  _{{\mathbb{T}}^{n}}F(q_{1}-i\frac {k_{1}}{|k_{1}|}\rho
,...,q_{n}-i\frac {k_{n}}{|k_{n}|}\rho ,t)
\{ \prod _{j=1}^{n} e^{-ik_{j}(q_{j}-i\frac {k_{j}}{|k_{j}|}\rho )} \}
dq_{1}...dq_{n}. \]
\\
Note in case $k_{i}=0$ we don't shift the path in that direction.
\\
\\
Then the following estimate its clear
\begin{eqnarray*}
|f_{k}(t) | &\leq & \|F(t) \|_{\rho } e^{-(\sum |k_{j}|)\rho}\frac {1}{(2\pi )^{n}}\int _{{\mathbb{T}}^{n}}
 e^{-ik \cdot q}dq
\leq  \|F(t) \|_{\rho }e^{-|k|\rho}.
\end{eqnarray*}
(Note $\|F(t) \|_{\rho }$ is a function of $t$).
\squareforqed
\end{lem}

The following lemma is useful to  construct a function in the form
of a Fourier series with a specified
strip of analyticity given the Fourier coefficients satisfy an exponentially
decaying bound. It will become evident
 later how this will enable us to construct a generating function
in the form of a Fourier series whose coefficients have a specified
form depending directly on the perturbation of the Hamiltonian.
\begin{lem}\label{coeffbound}
\quad \\
Suppose for a  positive real constant  $\rho < 1$ and
every $k\in \mathbb {Z}^{n}$ one has
\[ |f_{k}(t)| \leq C(t)e^{-|k|\rho}, \quad C(t)\geq 0\quad,  \forall t\in {\mathbb{C}}, \]
with $f_{k}(t)\in {\cal{A}}_{\sigma}$, $\sup _{{{D}}_{\sigma}}(C(t))=C$,
and consider the function
\[F(q,t)=\sum _{k\in \mathbb {Z}^{n}} f_{k}(t)e^{ik\cdot q}. \]
For any positive $\delta$ with $\delta < \rho$ one has
$ F\in {\cal{A}}_{\rho - \delta, \sigma }$
and
$\|F \|_{\rho -\delta, \sigma } \leq C 
\left( \frac {4}{\delta } \right) ^{n}$.\\
\quad\\
{Proof} :
Let $\| \mbox {Im } q \| \leq \rho - \delta $ then
\begin{eqnarray*}
\| F \| _{\rho - \delta , \sigma } &=& \sup _{D_{\rho -\delta ,
\sigma }} \left| \sum _{k \in \mathbb {Z}^{n}} f_{k}(t)e^{ik\cdot q} \right|
\leq  C\sum _{k \in \mathbb {Z}^{n}} e^{-|k|\rho }e^{|k|(\rho - \delta)} 
= C\sum _{k \in \mathbb {Z}^{n}} e^{-|k|\delta}\\
\\
&\leq & C2^{n}\sum _{\tiny { {\begin{array} {cc} k \in \mathbb {Z}^{n} \\ 
\\
{k_{1}\geq 0 ...k_{n} \geq 0} \end{array}}}}e^{-\delta \sum k_{j}}
= C2^{n} \left( \sum ^{\infty}_{k=0} e^{-\delta k} \right) ^{n} 
=C2^{n} \left( \frac {1}{1-e^{-\delta}} \right) ^{n}.
\end{eqnarray*}
For any positive $\delta <1$, we have $ \frac {\delta}{1-e^{-\delta}}<2$ which implies
$\frac {1}{1-e^{-\delta}}< \frac {2}{\delta}, $ from which follows that
\\
\[\| F \| _{\rho -\delta ,\sigma } \leq C2^{n} \left( \frac {2}{\delta}
\right) ^{n} = C\left( \frac {4}{\delta} \right)^{n}. \]
\\
The bound holds for any finite sum of the terms $f_{k}(t)e^{ik\cdot q} \in {\cal{A}
}_{\rho -\delta, \sigma}$, and
any such  finite sum is analytic in $D_{\rho - \delta , \sigma }. $ Therefore,
$F$ is the limit of a uniformly convergent sequence of analytic functions and thus
also analytic.
\squareforqed
\end{lem}
\begin{lem}
\label{exponentialinq}
\quad \\

For any $K,s,\delta >0$ one has the inequality
$K^{s} \leq \left( \frac {s}{e\delta} \right)^{s}e^{K\delta}$.\\
\quad \\
{Proof}:
For any $x\in \mathbb {R}$ we have
$x\leq e^{x-1}$ . 
Let $x=\frac {K\delta}{s}$, then
\[\frac {K\delta}{s} \leq e^{(\frac {K\delta}{s}-1)}\]
so that 
\[ K \leq \frac {s}{\delta e} e^{\frac {K\delta}{s}}\quad
{\mbox {or}}\quad 
K^{s} \leq \left( \frac {s}{\delta e} \right) ^{s} e^{K\delta} \quad
\squareforqed \]
\end{lem}

\section{Bounds on the Kolmogorov Normal Form in
terms of the original Hamiltonian}
\label{appbounds}
\quad\\
\textbf{Proof of Lemma ${\mathbf{\ref{boundlemmaone}}}$}\\
\quad\\
Clearly
\[C^{*}=
\left(
\begin{array}{cc}
\tilde {C}^{*} & 0 \\
0     & 0
\end{array}
\right).\]
Assuming $d \| \tilde{v} \| \leq \| \tilde {C}^{*}\tilde{v} \|\leq d^{-1}\| \tilde{v}\| $
where
\\
\[\tilde {C}^{*}\tilde{v}=(\sum_{j} \tilde {C}^{*}_{1j}\tilde{v}_{j},...,\sum_{j}
 \tilde {C}^{*}_{nj}\tilde{v}_{j}), \]
it follows
\[\| \tilde {C}^{*}\tilde{v} \|= \max_{i} {\left| \sum_{j} \tilde {C}^{*}_{ij}\tilde{v}_{j}\right| }. \]
Also, given 
$v=(\tilde{v} ,w)=(\tilde {v}_{1},...,\tilde{v}_{n},w) \quad \mbox{ for } \tilde{v}\in \mathbb
{C}^{n},w\in \mathbb{C}, \mbox{ and } v\in \mathbb {C}^{n+1}$
it follows
\[ {C}^{*}{v}=(\sum_{j} \tilde {C}^{*}_{1j}\tilde{v}_{j},...,\sum_{j}
 \tilde {C}^{*}_{nj}\tilde{v}_{j},0) \]
and
\[\| \tilde {C}^{*}v \|= \max_{i} {\left| \sum_{j} \tilde {C}^{*}_{ij}v_{j}\right| }. \]
This implies 
$\| C^{*}v \| = \| \tilde {C}^{*} \tilde{v} \| $
and therefore
$\| C^{*}v \| = \| \tilde{C}^{*}\tilde {v} \|\leq  d^{-1}\| \tilde{v}\|,  \quad \forall v\in {\mathbb{C}}^{n+1}\mbox{ and } \forall \tilde{v} \in\mathbb {C}^{n}.$
\\
Finally, since
$\| \tilde {v} \|=\max (|\tilde{v}_{1}|,...,|\tilde{v}_{n}|) \leq \max
 (|\tilde{v}_{1}|,...,|\tilde{v}_{n}|,|w|)=\| v \|,$
it follows 
$\| C^{*}v \|\leq d^{-1}\| v \|$ . \quad \squareforqed \\
\quad\\
We  summarize some results and give some new definitions.
We defined previously  the $(n\times n)$ matrix $\tilde {C}^{*}= \partial ^{2}_{p} \tilde{H}^{0}(0)$  
and assumed 
$
d\|\tilde {v} \|\leq \|\tilde {C}^{*}\tilde{v} \| \leq d^{-1}\|\tilde {v} \|
\quad \tilde {v}\in {\mathbb {C}}^{n}$.
We defined the $(n+1) \times (n+1)$ matrix $C^{*}=\partial ^{2}_{p'} \tilde{H}^{0}(0)$
and obtained the upper bound 
$
\| {C}^{*}{v} \| \leq d^{-1}\| {v} \|
\quad  {v}\in {\mathbb {C}}^{n+1}$.
We defined the $(n+1) \times (n+1)$ matrix $C=\partial ^{2}_{p'} \tilde{H}^{0}(0)+
\kappa \partial^{2}_{p'}H^{1}(0,q')$.
We define the $(n+2 \times n+2)$ isoenergetic matrix
\begin{eqnarray*}
{\mathbb {I}}= \left(
\begin{array}{cc}
C^{*} & \lambda ^{T} \\
\lambda & 0
\end{array}
\right)
=\left(
\begin{array}{ccc}
\tilde{C}^{*} & 0 & (\tilde{\lambda} )^{T} \\
0 & 0 & 1\\
\tilde{\lambda}  & 1 & 0
\end{array}
\right)
\end{eqnarray*}
where $\lambda = (\tilde{\lambda } ,1),\lambda \in {\mathbb {C}}^{n+1}$ and $\tilde{\lambda }
  \in {\mathbb {C}}^{n}$.\\
\quad\\
\textbf{Proof of Lemma ${\mathbf{\ref{boundlemmatwo}}}$}\\
\quad\\
\fbox{$1.\quad \mbox {max}(\|A \|_{\rho ,\sigma }, \| B \| _{\rho ,\sigma }) \leq 
2 \frac {E}{\sigma}$.}
\\
\\
We assume for the original Hamiltonian 
$\| H^{1} \|_{\rho ,\sigma } \leq E$.
Recall 
\begin{eqnarray*}
A(q')=  (H^{1}(0,q')-\overline{\overline{H^{1}}}(0)), \quad
B(q')=  \frac {\partial H^{1}}{\partial p'} (0,q').
\end{eqnarray*}
It follows
$ \| A \|_{\rho ,\sigma } \leq 2  \| H^{1} \| _{\rho ,\sigma }$.
Applying Cauchy's inequality and the fact  $\rho > \sigma $ we obtain the bound
\[ \| B \| _{\rho ,\sigma } \leq  \frac {1 }{\sigma } \| H^{1} \| _{\rho ,\sigma}.\]
Therefore
\begin{eqnarray*}
\| A \|_{\rho ,\sigma } &\leq & 2 E < \frac {2 E}{\sigma } \quad (\sigma < 1),
\quad
\| B \| _{\rho ,\sigma} \leq  \frac { E}{\sigma } < \frac {2 E}{\sigma }
\end{eqnarray*}
from which follows
\[ \mbox {max} (\| A \|_{\rho ,\sigma },\| B \| _{\rho ,\sigma } ) < \frac {2 E}{\sigma }
=E_{1}.\]
\\
\\
\fbox{$2.\quad\| Cv \| _{\rho ,\sigma } \leq m^{-1} \| v \|$.}
\\
\\ 
By definition we have
\begin{eqnarray*}
C_{i,j}(q')= \frac {\partial ^{2} H}{\partial p'_{i}p'_{j}}(0,q'),\quad
C^{*}_{i,j}(0)= \frac {\partial ^{2} \tilde {H}^{0}}{\partial p'_{i}p'_{j}}(0).
\end{eqnarray*}
Therefore
\[C-C^{*}= \Big {\{}  \kappa \frac {\partial ^{2} {H}^{1}}{\partial p'_{i}p'_{j}}(0,q') \Big {\}} ,\]
and it follows 
\\
\[ \|(C-C^{*})v \|_{\rho ,\sigma} \leq (n+1)\|C-C^{*} \|_{\rho , \sigma } \| v \|.\]
\\
Applying Cauchy's inequality we obtain
\\
\[ \|(C-C^{*})v \|_{\rho ,\sigma} \leq\frac {2(n+1)}{\sigma ^{2}}\kappa 
 \| H^{1} \|_{\rho , \sigma }
\| v \| \leq \frac {2(n+1)}{\sigma ^{2}}\kappa  E\| v \|. \]
\\
Recall from the statement of  theorem $\ref{firststatementofthm}$
\[\kappa E=\left(\frac{\sigma}{32} \right)^{2\aleph}
\frac{d^4f^4\sigma_{*}}{2^{12}\Lambda ^2}\]
One can verify  (using $d<1$ )
\begin{eqnarray*}
2\frac{(n+1)}{\sigma^{2}} \kappa E =
2\frac{(n+1)}{\sigma^{2}}
\left(\frac{\sigma}{32} \right)^{2\aleph}
\frac{d^4f^4\sigma_{*}}{2^{12}\Lambda ^2}
<\frac{d}{2},
\end{eqnarray*}
from which  follows
\[ \| (C-C^{*})v \|_{\rho ,\sigma } \leq \frac {d}{2} \| v \|.\]
Now recall 
$\| C^{*} v \| \leq d^{-1}\| v \|, \quad \forall v \in \mathbb {C}^{n+1}$.
Writing
$C=C^{*}+(C-C^{*})$,
we obtain
\begin{eqnarray*}
\|Cv \|_{\rho ,\sigma }& \leq & \|C^{*} v \| + \|(C-C^{*})v \|_{\rho,\sigma} 
\leq  (d^{-1}+\frac {d}{2}) \| v \| 
\leq  2d^{-1} \| v \|,
\end{eqnarray*}
taking $m=\frac {d}{2}$, the result follows.
\squareforqed\\
\quad\\
The next Lemma will be used in the proof of the Lemma on the
Isoenergetic Nondegeneracy Condition.
\\
\begin{lem}
Given two positive functions $f(x)$ and $g(x)$ then
\[\max _{x} (f-g) \geq \max _{x}(f)- \max _{x}(g).\]
\end{lem}
Proof:\\
\quad\\
Let $g_{0}$ be fixed. Then clearly
\[ \max _{x} (f-g_{0}) = \max _{x}(f)- g_{0}.\]
Next set $g_{0}=\max(g)$. It follows 
\[g_{0}\geq g \quad {\mbox {and}}\quad f-g_{0}\leq f-g.\]
Consequently
\[\max _{x}(f-g) \geq \max _{x}(f-g_{0})=\max _{x} (f)-g_{0}=\max _{x} (f)-\max _{x}(g).\]
\squareforqed\\
\quad\\
\textbf{Proof of Lemma ${\mathbf{\ref{lemmaiso}}}$}\\
\quad\\
Let $v=(\tilde{v},v_{1},v_{2})$ where $\tilde{v}\in {\mathbb{C}}^{n}$,$v_{1},v_{2}\in {\mathbb{C}}$.
The following estimates follow
\begin{eqnarray*}
&\quad&
\left\|
\left(
\begin{array}{ccc}
\tilde{C}^{*} & 0 & \tilde{\lambda} ^{T} \\
0 & 0 & 1\\
\tilde{\lambda}  & 1 & 0
\end{array}\right)
\left(\begin{array}{c}
\tilde{v} \\ v_{1} \\ v_{2}
\end{array} \right)
\right\|_{\rho ,\sigma}=
\max \Big(|v_{2}|,|\tilde{\lambda} \cdot \tilde{v}+v_{1}|,
|\sum_{j}\tilde{C}^{*}_{i,j}v_{j}+v_{2}\tilde{\lambda}_{i}|\Big)\\
&\quad& \quad \geq \max_{i} \Big(|\sum_{j}\tilde{C}^{*}_{i,j}v_{j}+v_{2}\tilde{\lambda}_{i}|\Big)
\geq \max_{i} \Big |\Big(|\sum_{j}\tilde{C}^{*}_{i,j}v_{j}|-|v_{2}
\tilde{\lambda}_{i}|\Big)\Big|
\geq \Big |\max_{i} \Big(|\sum_{j}\tilde{C}^{*}_{i,j}v_{j}|\Big)
-\max_{i}\Big(|v_{2}\tilde{\lambda}_{i}|\Big) \Big|\\
&\quad & \quad \geq \Big |d\|v \|-L\|v \|\Big | 
\geq\frac{1}{2}|d-L|\|v \|
\end{eqnarray*}
and we set $l=\frac{1}{2}|d-L|$.\quad
\squareforqed\\
\quad\\
\textbf{Proof of Lemma ${\mathbf{\ref{thmisob}}}$}\\
\quad\\
Recall 
\begin{eqnarray*}
{\mathbb{I}}=
\left(
\begin{array}{cc}
C^{*} & \lambda ^{T}\\
\lambda & 0
\end{array}
\right)
=
\left(
\begin{array}{cc}
C(q') & \lambda ^{T}\\
\lambda & 0
\end{array}
\right)
+
\left(
\begin{array}{cc}
-\kappa \partial ^{2}_{p'}H^{1}(0,q') & 0\\
0  & 0
\end{array}
\right).
\end{eqnarray*}
Then clearly
\begin{eqnarray*}
&\quad&
\left\| 
\left(
\begin{array}{cc}
\overline{\overline{C}} & \lambda ^{T}\\
\lambda & 0
\end{array}
\right)v \right\| _{\rho ,\sigma}
\geq \left |
\left\| 
\left(
\begin{array}{cc}
C^{*} & \lambda ^{T}\\
\lambda & 0
\end{array}
\right)v
\right\|_{\rho ,\sigma}
-
\left\| 
\left(
\begin{array}{cc}
\kappa \overline{\overline{ \partial ^{2}_{p'}H^{1}}}(0) & 0\\
0  & 0
\end{array}
\right)v
\right\|_{\rho ,\sigma} \right| \\
&\quad& \quad \geq\left |
l\| v \|_{\rho ,\sigma}-
\left\| 
\left(
\begin{array}{cc}
\kappa \overline{\overline {\partial ^{2}_{p'}H^{1}}}(0) & 0\\
0  & 0
\end{array}
\right) \right\|_{\rho ,\sigma} \| v \|_{\rho ,\sigma}\right| 
\geq\left |
l\| v \|_{\rho ,\sigma}-
\Big\| \Big(\kappa \overline{\overline{\partial ^{2}_{p'}H^{1}}}(0)\Big) \Big\|_{{\rho}, {\sigma}}
\|v \|_{\rho ,\sigma}\right|\\
&\quad &\quad \geq\left |
l\| v \|_{\rho ,\sigma}-
\Big\| \Big(\kappa  {\partial ^{2}_{p'}H^{1}}(0,q')\Big) \Big\|_{ {\rho}, {\sigma}}
\|v \|_{\rho ,\sigma}\right| 
\geq
\left| l-\frac{\kappa  E }{\sigma ^{2}} \right|\|v \|_{\rho ,\sigma}.
\end{eqnarray*}
where we use in the last line Cauchy's inequality and the fact 
$
\| {H^{1}}(0,q') \|_{ {\rho}, {\sigma}}
\leq  E$. 
\quad \squareforqed

\section{Lemmas for Canonical Transformations }

In what follows we present several estimates
relevant to canonical changes of variables defined by the
Lie series method.
For
$ \chi(p',q') \in {\cal{A}}_{\rho ,\sigma} $
define
\[\chi ^{*}_{\rho ,\sigma} \equiv
 \max \left( \left\| \frac {\partial \chi}{\partial p'}
 \right\|_{\rho ,\sigma }, \left\| \frac {\partial \chi}{\partial q'} \right\|_{\rho ,\sigma}
\right). \]
\begin{lem}
\label{poissonb}
\quad \\

For any $\chi , f \in {\cal{A}}_{\rho ,\sigma}$ and positive $\delta < \sigma <\rho$ we have
\[1. \quad \| \{\chi ,f \} \| _{\rho -\delta ,\sigma -\delta}
\leq
2(n+1) \left(\frac {\chi ^{*}_{\rho ,\sigma}}{\delta} \right) \| f \| _{\rho ,\sigma}.
\quad \quad  \quad  \]
\[2.\quad \|  \{\chi , \{ \chi ,f \} \} \|_{\rho -\delta ,\sigma -\delta}
\leq
8(n+1)^{2}\left( \frac {\chi ^{*}_{\rho ,\sigma }}{\delta} \right) ^{2} \| f \|_{\rho ,\sigma}.\]
Proof :
\[\quad\]
\fbox{$\| \{\chi ,f \} \| _{\rho -\delta ,\sigma -\delta} 
\leq
2(n+1) \left(\frac {\chi ^{*}_{\rho ,\sigma}}{\delta} \right) \| f \| _{\rho ,\sigma}$}
\\
\\
By definition
\[ \{ \chi ,f \} =
\sum _{i} \left[ \frac {\partial \chi}{\partial p'_{i}}
\frac {\partial f}{\partial q'_{i}}-\frac  {\partial \chi}{\partial q'_{i}}
\frac {\partial f}{\partial p'_{i}}  \right]. \]
It follows 
\begin{eqnarray*}
|\{ \chi ,f \}| 
&\leq& 
\sum _{i} \left[ \left|\frac {\partial \chi}{\partial p'_{i}} \right|
\left|\frac {\partial f}{\partial q'_{i}}\right|+\left|\frac  {\partial \chi}{\partial q'_{i}}
\right| \left|\frac {\partial f}{\partial p'_{i}}\right|  \right]
\leq 
\chi ^{*}_{\rho ,\sigma }
\sum _{i} \left[ 
\left|\frac {\partial f}{\partial q'_{i}}\right|+
 \left|\frac {\partial f}{\partial p'_{i}}\right|  \right]
\leq 
\chi ^{*}_{\rho ,\sigma }
\frac {2(n+1)}{\delta} \| f \|_{\rho ,\sigma}
\end{eqnarray*}
which implies
\[
\|\{ \chi ,f \}\| _{\rho ,\sigma} \leq 2(n+1)\left( \frac {\chi ^{*}_{\rho ,\sigma }}{\delta}
\right)\| f \| _{\rho ,\sigma }.
\]
\\
\\
\fbox{$\|  \{\chi , \{ \chi ,f \} \} \|_{\rho -\delta ,\sigma -\delta} 
\leq
8(n+1)^{2}\left( \frac {\chi ^{*}_{\rho ,\sigma }}{\delta} \right) ^{2} \| f \|_{\rho ,\sigma}$}
\\
\\
 We have by definition
\[\{ \chi ,\{ \chi ,f \} \}=
\sum _{i} \left[ \frac {\partial \chi}{\partial p'_{i}} \frac {\partial}{
\partial q'_{i}}\{ \chi ,f \} - \frac {\partial \chi }{\partial q'_{i}}
\frac {\partial}{\partial p'_{i}}\{ \chi ,f \} \right] \]
\\
\begin{eqnarray*}
&=&\sum_{i}\left[ 
\frac {\partial \chi}{\partial p'_{i}}\frac {\partial}{
\partial q'_{i}} \sum _{j} \left[
 \frac {\partial \chi}{\partial p'_{j}}
\frac {\partial f}{\partial q'_{j}}-\frac  {\partial \chi}{\partial q'_{j}}
\frac {\partial f}{\partial p'_{j}}  \right]
-
\frac {\partial \chi}{\partial q'_{i}}\frac {\partial}{\partial p'_{i}}
\sum _{j} \left[ \frac {\partial \chi}{\partial p'_{j}}
\frac {\partial f}{\partial q'_{j}}-\frac  {\partial \chi}{\partial q'_{j}}
\frac {\partial f}{\partial p'_{j}}  \right] \right ] \\
\\
\\
&=& \sum _{i,j} \left [ 
\frac {\partial \chi }{\partial p'_{i}} \left [
\frac{\partial ^{2} \chi }{\partial q'_{i}\partial p'_{j}} \frac {\partial f}{\partial q'_{j}}
+
\frac {\partial \chi }{\partial p'_{j}}\frac {\partial ^{2} f}{\partial q'_{i} \partial q'_{j}}
-
\frac{\partial ^{2} \chi }{\partial q'_{i}\partial q'_{j}}\frac {\partial f}{\partial p'_{j}}
-
\frac {\partial \chi }{\partial q'_{i}}\frac{\partial ^{2} f }{\partial q'_{i}\partial p'_{j}}
\right] \right.
\\
\\
\\
&\quad &-\frac {\partial \chi }{\partial q'_{i}} \left [ \left.
\frac{\partial ^{2} \chi }{\partial p'_{i}\partial p'_{j}} \frac {\partial f}{\partial q'_{j}}
+
\frac {\partial \chi }{\partial p'_{j}}\frac {\partial ^{2} f}{\partial p'_{i} \partial q'_{j}}
-
\frac{\partial ^{2} \chi }{\partial p'_{i}\partial q'_{j}}\frac {\partial f}{\partial p'_{j}}
-
\frac {\partial \chi }{\partial q'_{i}}\frac{\partial ^{2} f }{\partial p'_{i}\partial p'_{j}}
\right ]  \right ]
\\
\end{eqnarray*}
There are eight expressions in the sum, each can be estimated using Cauchy's inequality.
The first two expressions are estimated below and the rest can be estimated the same way.
\[ \left \| \frac {\partial \chi}{\partial p'_{i}}
\frac {\partial ^{2} \chi }{\partial q'_{i} \partial p'_{j}}
\frac {\partial f}{\partial q'_{j}} \right \| _{\rho -\delta ,\sigma -\delta}
\leq
\left( \frac {\chi ^{*}_{\rho ,\sigma}}{\delta} \right)^{2} \|f \|_{\rho , \sigma},\]
\\
\[ \left \| \frac {\partial \chi}{\partial p'_{i}}
\frac {\partial  \chi }{ \partial p'_{j}}
\frac {\partial ^{2} f}{\partial q'_{i} \partial q'_{j}} \right \| _{\rho -\delta ,\sigma -\delta}
\leq
\left( \frac {\chi ^{*}_{\rho ,\sigma}}{\delta} \right)^{2} \|f \|_{\rho , \sigma}.\]
\\
Each of the eight expressions in the summation represents $(n+1)^{2}$ terms. Therefore 
the estimate gives
\\
\[ \| \{ \chi , \{ \chi ,f \} \} \| _{\rho -\delta ,\sigma -\delta }
\leq
8(n+1)^{2}\left( \frac {\chi ^{*}_{\rho ,\sigma}}{\delta} \right)^{2} \|f \|_{\rho , \sigma}.\]
\squareforqed
\end{lem}
We now review the Lie series method in which
the desired canonical transformation $\phi$ is given by the time one flow
of a Hamiltonian $\chi$ called the generating function.
If one evaluates a function $f$ on the solutions of
Hamilton's equations generated by $\chi$ one can write
$\frac{df}{dt}=\{\chi,f\}$. That same functions $f$ defined on the
phase space
is transformed $f \mapsto {\cal{U}}f=f\circ \phi$. Therefore
this transformation of $f$ can be written
as ${\cal{U}}f=f+\{\chi ,f\}+\frac{1}{2}
\{\chi,\{\chi, f\}\}+...$ which is exactly
the Taylor expansion of $f$. The transformation of a function can thus
be expressed entirely in terms of the generating Hamiltonian
$\chi$ without knowing the flow. The following lemma
gives the conditions under which such a canonical transformation exists
and several useful estimates. These estimates are key in
the iterative lemma where the size of the perturbation
after applying the canonical transformation must be known.
\begin{lem}
\label{lemcanon}
\quad \\
Suppose $\chi (p',q') \in {\cal{A}}_{\rho ,\sigma}$ and its derivatives also belong to
${\cal{A}}_{\rho ,\sigma}$ for given positive numbers $\rho , \sigma $. Consider the
corresponding Hamiltonian system
\begin{eqnarray*}
\dot {p'}= - \frac {\partial \chi }{\partial q'},\quad \quad
\dot {q'}= \frac {\partial \chi }{\partial p'}.
\end{eqnarray*}
Assume that for some positive $\delta < \sigma $, $\delta <\rho$
\[\chi ^{*}_{\rho ,\sigma }
 \equiv
\max \left( \left \| \frac {\partial \chi }{\partial p'} \right \|_{\rho ,\sigma},
\left \| \frac {\partial \chi }{\partial q'} \right \| _{\rho ,\sigma } \right)
<
\frac {\delta }{2}. \]
Then, for all initial conditions $(P',Q') \in {{D}}_{\rho -\delta ,\sigma -\delta}$,
the image of a point $(P',Q')\in {{D}}_{\rho -\delta ,\sigma -\delta}$
 is contained in ${{D}}_{\rho,\sigma}$ , thus defining a canonical
analytic transformation
\[\phi : D_{\rho -\delta ,\sigma -\delta} \rightarrow D_{\rho ,\sigma },
\quad \phi \in {\cal{A}}_{\rho -\delta ,\sigma -\delta}. \]
The operator
\begin{eqnarray*}
{\cal {U}} :& {\cal{A}}_{\rho ,\sigma } \rightarrow {\cal{A}}_{\rho -\delta ,\sigma -\delta},&\\
&f \rightarrow {\cal {U}} f \equiv f \circ \phi,&
\end{eqnarray*}
is then well defined and one has the estimates\\
\\
1.
$ \| \phi - \mbox { identity } \|_{\rho -\delta ,\sigma -\delta}
\leq
\chi ^{*}_{\rho ,\sigma }$,\\
\quad\\
2.
$ \| {\cal {U}} f \|_{\rho -\delta ,\sigma -\delta}
\leq
\| f \| _{\rho ,\sigma }$,\\
\quad\\
3.
$ \| {\cal {U}} f -f\|_{\rho -\delta ,\sigma -\delta}
\leq
4(n+1) \frac {\chi ^{*}_{\rho ,\sigma }}{\delta}\| f \| _{\rho ,\sigma }$,\\
\quad\\
4.
$ \| {\cal {U}} f -f - \{\chi ,f \} \|_{\rho -\delta ,\sigma -\delta}
\leq
32(n+1)^{2}\left( \frac {\chi ^{*}_{\rho ,\sigma }}{\delta}\right)^{2}\| f \| _{\rho ,\sigma }$.\\
\quad \\
Proof :

One first shows the existence of solutions through  an arbitrary initial condition
$(P',Q') \in D_{\rho -\delta ,\sigma -\delta}$. This follows if the vector field is
Lipschitz throughout $D_{\rho -\delta ,\sigma -\delta}$. Using Cauchy's inequality
in the domain $D_{\rho -\delta /2 ,\sigma -\delta /2}$, the Lipschitz constant
for the vector field
\begin{eqnarray*}
\dot {p'}= - \frac {\partial \chi }{\partial q'},\quad 
\dot {q'}= \frac {\partial \chi }{\partial p'},
\end{eqnarray*}
can be estimated as
\\
\[\left \| \frac {\partial ^{2} \chi }{\partial q' \partial p' } \right \| _{
\rho -\delta /2 ,\sigma -\delta /2}
\leq
4(n+1)\frac {\chi ^{*}_{\rho ,\sigma }}{\delta}.\]
\\
\\
Next we argue that this local solution exists for $t=1$, and does not leave 
$D_{\rho ,\sigma }$ in that time. By assumption, the maximum velocity of the 
vector field is bounded by
\[\max (\| \dot {p'} \| ,\| \dot {q'} \| )
\leq 
\chi ^{*}_{\rho ,\sigma }
<
\frac {\delta }{2}. \]
\\
Hence the maximum distance that a point in $D_{\rho -\delta , \sigma -\delta}$
can move in time $t=1$ is $\frac {\delta }{2}$. Therefore
\\
\[ \phi ^{t} : D_{\rho -\delta ,\sigma -\delta} \rightarrow 
D_{\rho -\delta /2 ,\sigma -\delta /2} \subset D_{\rho ,\delta} ,\]
\\
Next we prove the estimates.\\
\\
\\
\fbox{$1.\quad \|\phi - \mbox { identity } \|_{\rho -\delta ,\sigma -\delta} 
\leq 
\chi ^{*}_{\rho ,\sigma }.$}
\\
\\
This follows from the mean value theorem
\\
\begin{eqnarray*}
\mbox {For }\quad 0\leq t \leq 1,
 \quad \| \phi - \mbox { identity } \|_{\rho -\delta ,\sigma -\delta} 
&=& \| \phi^{t=1} - \phi ^{t=0} \|_{\rho -\delta ,\sigma -\delta}
\leq  \| \dot {\phi ^{t}} \|(1-0) \leq \chi ^{*}_{\rho ,\sigma}. 
\end{eqnarray*}
\fbox{$2.\quad \| {\cal {U}} f \|_{\rho -\delta ,\sigma -\delta} 
\leq 
\| f \| _{\rho ,\sigma }.$}
\\
\\
This is trivial,
$\| {\cal {U}} f \|_{\rho -\delta ,\sigma -\delta} 
\equiv
\| f \circ \phi \|_{\rho -\delta ,\sigma -\delta}  $
but
$ \phi : D_{\rho -\delta ,\sigma -\delta} \rightarrow 
D_{\rho ,\delta} $
which implies
$\| f \circ \phi \|_{\rho -\delta ,\sigma -\delta}
\leq 
\| f \| _{\rho ,\sigma }$.\\
\quad \\
\fbox{$3.\quad \| {\cal {U}} f -f\|_{\rho -\delta ,\sigma -\delta} 
\leq 
4(n+1) \frac {\chi ^{*}_{\rho ,\sigma }}{\delta}\| f \| _{\rho ,\sigma }.$}
\\
\\
By definition we have
${\cal {U}} f 
\equiv
f \circ \phi^{t=1}$. 
Taylor expanding in $t$ gives
\\
\[f \circ \phi^{t}
=
f \circ \phi ^{0} 
+
\frac {d}{dt}(f \circ \phi ^{t}) |_{t=t'},
\quad 0\leq t' \leq 1,\]
\\
from which follows
\\
\[ {\cal {U}} f =
f
+
\frac {df}{dt}|_{t=t'}\quad
{\mbox{or}}\quad 
{\cal {U}} f -f =
\frac {df}{dt}|_{t=t'} =
\{ \chi ,f \} |_{t=t'}.\]
As argued above,
\\
\[\phi (P',Q')\in D_{\rho -\delta /2 ,\sigma -\delta /2},
\quad 0 < \delta < 1,
\quad (P',Q')\in D_{\rho -\delta ,\sigma -\delta}.\]
\\
Hence, using the estimate in Lemma \ref{poissonb}, with $\delta $ in the Lemma 
replaced by $\frac {\delta }{2}$, we obtain
\\
\[\| {\cal {U}} f -f \|_{\rho -\delta ,\sigma -\delta} \leq 4(n+1)\left(
\frac {\chi ^{*}_{\rho ,\sigma }}{\delta} \right) \| f \|_{\rho ,\sigma }. \]
\\
\\
\fbox{$4.\quad \| {\cal {U}} f -f - \{\chi ,f \} \|_{\rho -\delta ,\sigma -\delta} 
\leq 
32(n+1)^{2}\left( \frac {\chi ^{*}_{\rho ,\sigma }}{\delta}\right)^{2}\| f \| _{\rho ,\sigma }.$}
\\
\\
\\
Taylor expanding in $t$
\\
\[f \circ \phi ^{t} = f \circ \phi ^{0} 
+
\frac {df}{dt}t
+
\frac {1}{2} \frac {d^{2}f}{dt^{2}}t^{2}+...,\]
\\
and applying the mean value theorem gives
\\
\[  {\cal {U}} f -f - \frac {df}{dt} |_{t=t''} =
\frac {1}{2} \frac {d^{2}f}{dt^{2}}|_{t=t''} =
\frac {1}{2} \{ \chi , \{ \chi ,f \} \}|_{t=t''}.\]
\\
As above, 
\\
\[\phi (P',Q')\in D_{\rho -\delta /2 ,\sigma -\delta /2},
\quad 0 < \delta < 1,
\quad (P',Q')\in D_{\rho -\delta ,\sigma -\delta}.\]
\\
Using this and Lemma \ref{poissonb}, with $\delta $ in the Lemma replaced by
$\frac {\delta }{2}$, gives
\\
\[\| {\cal {U}} f -f - \{\chi ,f \} \|_{\rho -\delta ,\sigma -\delta} 
\leq
 32(n+1)^{2}\left( \frac {\chi ^{*}_{\rho ,\sigma }}{\delta} \right) ^{2} \| f \|_{\rho ,\sigma}.  \]
\squareforqed
\end{lem}


\section{P.D.E on a Torus}
\label{appPDETo}
\textbf{Proof of Lemma ${\mathbf{\ref{exponentialbound2}}}$}\\
\quad\\
We write the coefficients of the Fourier series as follows
\\
\begin{eqnarray*}
g_{k}= \frac{1}{(2\pi)^{n}} \int ^{\pi}_{-\pi}\cdot \cdot \cdot
\int ^{\pi}_{-\pi}
G(q) e^{-ik\cdot q}dq_{1}\cdot \cdot \cdot dq_{n}.
\end{eqnarray*}
Since $G\in {\cal{A}}_{\rho}$, we can shift the path of integration
as follows
\begin{eqnarray*}
g_{k}&=&
\frac{1}{(2\pi)^{n}} \int ^{\pi}_{-\pi}\cdot \cdot \cdot
\int ^{\pi}_{-\pi}
G(q_{1}-i\frac{k_{1}}{|k_{1}|}\rho,...,q_{n} -i \frac{k_{n}}{|k_{n}|}\rho)
\{ \prod_{j=1}^{n}  e^{-ik_{j}(q_{j}-i\frac{k_{j}}{|k_{j}|}\rho)} \}dq_{1}\cdot \cdot 
\cdot dq_{n}\\
&=&\frac{e^{-|k|\rho} }{(2\pi)^{n} }
\int ^{\pi}_{-\pi} \cdot \cdot
\cdot \int ^{\pi}_{-\pi} 
G(q_{1}- i\frac{k_{1}}{|k_{1}|}\rho,...,q_{n}-i\frac{k_{n}}{|k_{n}|}\rho)
\cdot e^{-ik\cdot q}
dq
\end{eqnarray*}
(If $k_{i}+0$ we don't shift in that direction)
and
$
|g_{k}| \leq e^{-|k|\rho}\| G(q) \|_{\rho }$.
\squareforqed
\\
\quad\\
\textbf{Proof of Lemma ${\mathbf{\ref{lemmaqb}}}$}
\quad\\
We begin by expanding $F$ and $G$ in Fourier Series
\begin{eqnarray*}
&\quad & F(q) =\sum _{k\in {\mathbb{Z}}^{n}} f_{k} e^{ik\cdot q},
\quad
G(q)= \sum _{k\in {\mathbb{Z}}^{n}\backslash 0}
g_{k}e^{ik\cdot q},
\end{eqnarray*}
substituting in $(\ref{diffeqnone})$ gives
\begin{eqnarray*}
\sum _{i} \lambda _{i} \sum _{k\in {\mathbb{Z}}^{n}}
ik_{i} f_{k}e^{ik\cdot q} =
 \sum _{k\in {\mathbb{Z}}^{n}\backslash 0}
g_{k} e^{ik\cdot q},
\end{eqnarray*}
or 
\begin{eqnarray*}
 \sum _{k\in {\mathbb{Z}}^{n}} i \lambda \cdot k f_{k}
e^{ik\cdot q} =
 \sum _{k\in {\mathbb{Z}}^{n}\backslash 0}
g_{k} e^{ik\cdot q},
\end{eqnarray*}
which implies
\begin{eqnarray*}
f_{k}= \frac{g_{k}}{i\lambda \cdot k}, \quad k\in {\mathbb{Z}}^{n}\backslash 0.
\end{eqnarray*}
From this expression we see why we must require $\overline {G} =0$ in order to 
construct a solution.
Recall that $\lambda \in \Omega _{\Gamma}$ so that $\lambda \cdot k \neq 0$.
We next prove convergence of 
\begin{eqnarray*}
 \sum _{k\in {\mathbb{Z}}^{n}\backslash 0}
f_{k}e^{ik\cdot q} =
 \sum _{k\in {\mathbb{Z}}^{n}\backslash 0}
\frac{g_{k}}{i\lambda \cdot k}e^{ik\cdot q} .
\end{eqnarray*}
From lemma $\ref{exponentialbound2}$ since $G \in {\cal{A}}_{\rho}$
it follows
$
|g_{k}| \leq \| G \|_{\rho} e^{-|k|\rho}$,
and the following estimates, using $\| k \| \leq |k|$, hold
\begin{eqnarray*}
|\lambda \cdot  k| \geq \Gamma  \| k \| ^{-n},
\quad
|\lambda \cdot  k| \geq \Gamma  | k | ^{-n}.
\end{eqnarray*}
Using lemma $\ref{exponentialinq}$ gives
\begin{eqnarray*}
|k|^{n} \leq \left( \frac{n}{e\delta} \right)^{n}
e^{|k|\delta}.
\end{eqnarray*}
Hence, we use these estimates to estimate the Fourier coefficients as follows
\begin{eqnarray*}
|f_{k}|= \left|
\frac{g_{k}}{\lambda \cdot k} \right|
&\leq& 
\frac{ \|G \|_{\rho}}{\Gamma} |k|^{n}e^{-|k|\rho}
\leq
\frac{ \|G \|_{\rho}}{\Gamma}
\left(\frac{n}{e\delta} \right)^{n} e^{-|k|(\rho -\delta)}
=Ce^{-|k|(\rho -\delta)}
\end{eqnarray*}
where
\begin{eqnarray*}
C= \frac{ \|G \|_{\rho}}{\Gamma}
\left(\frac{n}{e\delta} \right)^{n}.
\end{eqnarray*}
Using lemma $\ref{coeffbound}$ replacing 
$\rho$ with $\rho -\delta$ we obtain
\begin{eqnarray*}
F\in {\cal{A}}_{\rho -2\delta}, \quad 
\| F \|_{\rho -2\delta} \leq C \left( 
\frac{4}{\delta} \right)^{n}.
\end{eqnarray*}
Next we get the first estimate on the norm of $F$ in terms 
of the norm of $G$. The previous estimate
can be rewritten as
\begin{eqnarray*}
\|F \|_{\rho -2\delta} 
&\leq& 
\Gamma ^{-1} \left(
\frac{16n}{e4\delta ^{2}} \right)^{n}
\|G \|_{\rho}
= \left( \frac{16n}{e} \right)^{n}
\frac{\|G \|_{\rho}}{ \Gamma (2\delta)^{2n}}.
\end{eqnarray*}
We have
\begin{eqnarray*}
\varpi = 
2^{4n+1}\left(\frac{n+1}{e} \right)^{n+1} \geq \left( 
\frac{16n}{e} \right)^{n},
\end{eqnarray*}
since 
\begin{eqnarray*}
\left(\frac{n+1}{e} \right)^{n+1} \geq
\left( 
\frac{n}{e} \right)^{n}
\end{eqnarray*}
Using this inequality, the estimate can be written as
\begin{eqnarray*}
\|F\|_{\rho -2\delta} \leq
\frac{\varpi}{\Gamma (2\delta)^{2n}} \|G \|_{\rho}.
\end{eqnarray*}
Letting $\delta \rightarrow \frac{\delta}{2}$ gives the result.

Finally, we prove the last estimate in the lemma. Recall
that 
\begin{eqnarray*}
F(q)= \sum _{k\in {\mathbb{Z}}^{n} \backslash 0}
\frac{g_{k}}{i\lambda \cdot k}e^{ik\cdot q},
\end{eqnarray*}
then, formally, we have
\begin{eqnarray*}
\frac{\partial F}{\partial q_{j}} &=&
\sum _{k\in {\mathbb{Z}}^{n} \backslash 0}
\frac{g_{k}}{i\lambda \cdot k}
(ik_{j})e^{ik\cdot q}
=
\sum _{k\in {\mathbb{Z}}^{n} \backslash 0}
\frac{g_{k}}{\lambda \cdot k}
(k_{j})e^{ik\cdot q}.
\end{eqnarray*}
We denote the Fourier coefficients of $\frac{\partial F}{\partial q_{j}}$
by
$
h_{k_{j}}=\frac{g_{k}}{\lambda \cdot k}
(k_{j}), \quad j=1,...,n$.
Therefore
$
|h_{k_{j}}| \leq \Gamma ^{-1} \|G \|_{\rho} |k|^{n+1} 
e^{-|k|\rho}$.
From lemma $\ref{exponentialinq}$ we have
\begin{eqnarray*}
|h_{k_{j}}| \leq
\Gamma ^{-1} \|G \|_{\rho}
\left(
\frac{n+1}{e\delta} \right)^{n+1}
e^{-|k|(\rho -\delta)}.
\end{eqnarray*}
Form lemma $\ref{coeffbound}$ we have
\begin{eqnarray*}
\left \| \frac{\partial F}{\partial q} \right \|_{\rho -2\delta}
\leq C \left( \frac{4}{\delta} \right)^{n},
\end{eqnarray*}
where
\begin{eqnarray*}
C= \frac{1}{\Gamma \delta ^{n+1}} \left(
\frac{n+1}{e} \right)^{n+1} \|G \|_{\rho},
\end{eqnarray*}
so
\begin{eqnarray*}
\left \| \frac{\partial F}{\partial q} \right \|
_{\rho -2\delta} \leq 
\frac{4^{n}}{\Gamma  \delta ^{n+1}}
\left(
\frac{n+1}{e} \right)^{n+1}\|G \|_{\rho}.
\end{eqnarray*}
Let $\delta \rightarrow \frac{\delta}{2}$, then the estimate
takes the form
\begin{eqnarray*}
\left \| \frac{\partial F}{\partial q} \right \|
_{\rho -\delta} \leq 
\frac{4^{n}2^{2n+1}}{\Gamma  \delta ^{n+1}}
\left(
\frac{n+1}{e} \right)^{n+1}\|G \|_{\rho}. \quad \squareforqed
\end{eqnarray*}

\section{Fourier Transform Lemmas}
We present some important results concerning a function and its
Fourier transform defined on complex domains.
\begin{lem}
\label{theomone}
\quad \\
\quad \\
Suppose $f(z)$, $z=x+iy$, is analytic for
$-a<y<b$ where $a>0,b>0$.\\
In any strip in the interior of  $-a<y<b$, let
\begin{eqnarray*}
f(z)= \left\{
\begin{array}{cc}
{\cal{O}}\left(e^{-(\nu-\varepsilon)x}\right) & (x \rightarrow \infty)\\
\quad & \quad \\
{\cal{O}}\left(e^{(\mu-\varepsilon)x}\right)\quad & (x \rightarrow -\infty)
\end{array}
\right.,
\end{eqnarray*}
for every positive $\varepsilon$, where $\nu >0, \mu >0$.
Then there exists a $\delta >\varepsilon$ such that $ F(\omega)$, $\omega =u+iv$, defined by
\begin{eqnarray*}
F(\omega)= \frac{1}{\sqrt{2\pi}}\int ^{\infty}_{-\infty}
f(\zeta)e^{-i\omega \zeta}d\zeta,
\end{eqnarray*}
is analytic  in the strip  $ -\mu +\delta  \leq v\leq \nu -\delta  $, satisfies
\begin{eqnarray*}
F(\omega)= \left\{
\begin{array}{cc}
{\cal{O}}\left(e^{-(b-\varepsilon)u}\right) & (u \rightarrow \infty)\\
\quad & \quad \\
{\cal{O}}\left(e^{(a-\varepsilon)u}\right)\quad & (u \rightarrow -\infty)
\end{array}
\right.,
\end{eqnarray*}
and
\begin{equation}
\label{eqntenn}
f(z)=\frac{1}{\sqrt{2\pi}}\int^{\infty}_{-\infty}F(\omega)e^{iz\omega}d\omega,
\end{equation}
for every $z$ in the strip $-a<y<b$.\\
\quad\\
{Proof}:
\quad
We have 
\begin{eqnarray*}
F(\omega) = \frac{1}{\sqrt{2\pi}}\int ^{\infty}_{-\infty} f(\zeta)
e^{-i\zeta \omega}d\zeta,
\end{eqnarray*}
to find the domain of analyticity of $F(\omega)$ we consider the sequence
of functions $\{ F_{n}(\omega) \}_{n=1}^{n=\infty}$ where
\begin{eqnarray*}
F_{n}(\omega) =\frac{1}{\sqrt{2\pi}}\int ^{n}_{-n} f(\zeta) e^{-i\zeta \omega }d\zeta
\end{eqnarray*}
and show that $\{ F_{n}(\omega) \}_{n=1}^{n=\infty}$ converges to $F(\omega)$ uniformly.
Consequently, since each of $F_{n}(\omega)$ is analytic in the strip
$-\mu <v< \nu $, uniform convergence will imply $F(\omega)$ is analytic
in some strip to be determined.
Let $n=\max ( c_{1}, |c_{2}|)$ where $c_{1}$ and $c_{2}$ are the constants in the
definition of a function of exponential order in the real limit. Then, taking
integrals along the real line,
\begin{eqnarray*}
\left| F_{n}(\omega) - F(\omega) \right|
&=& \left|\frac{1}{\sqrt{2\pi}}\int ^{n}_{-n} f(\zeta) e^{-i\zeta \omega }d\zeta
-\frac{1}{\sqrt{2\pi}}\int ^{\infty}_{-\infty} f(\zeta)
e^{-i\zeta \omega}d\zeta \right|\\
\quad \\
&=& \left|
\frac{1}{\sqrt{2\pi}}\int ^{\infty}_{n} f(\zeta) e^{-i\zeta \omega }d\zeta
-\frac{1}{\sqrt{2\pi}}\int ^{n}_{-\infty} f(\zeta)
e^{-i\zeta \omega}d\zeta \right|\\
\quad \\
&\leq & \frac{1}{\sqrt{2\pi}}\left[
\int ^{\infty}_{n}|f(\zeta)||e^{-i\zeta \omega }|d\zeta
+
\int ^{n}_{-\infty}| f(\zeta)|
|e^{-i\zeta \omega}|d\zeta\right]\\
\quad \\
&=& 
\frac{1}{\sqrt{2\pi}}\left[
\int ^{\infty}_{n}|f(\zeta)|
|e^{-i\zeta u}||e^{\zeta v}|d\zeta+
\int ^{n}_{-\infty}| f(\zeta)|
|e^{-i\zeta u}||e^{\zeta v}|d\zeta\right]\\
\quad \\
&=&
\frac{1}{\sqrt{2\pi}}\left[
\int ^{\infty}_{n}|f(\zeta)|e^{\zeta v}d\zeta+
\int ^{n}_{-\infty}| f(\zeta)|
e^{\zeta v}
d\zeta\right]\\
\quad \\
&\leq  &
\frac{1}{\sqrt{2\pi}}\left[
C_{1}\int ^{\infty}_{n}
e^{-(\nu -\varepsilon)\zeta}e^{\zeta v}d\zeta+
C_{2}\int ^{n}_{-\infty}
e^{(\mu -\varepsilon)\zeta}e^{\zeta v}
d\zeta\right]\\
\quad \\
&=&
\frac{1}{\sqrt{2\pi}}\left[
\frac{C_{1}e^{-(\nu -\varepsilon -v)\zeta}}{-(\nu -\varepsilon -v)}
\Big| ^{\infty}_{n}+ 
\frac{C_{2}e^{(\mu -\varepsilon +v)\zeta}}{\mu -\varepsilon +v}
\Big|  _{-\infty}^{-n}\right].
\end{eqnarray*}
The maximum finite bound for the expression above can be obtained by
picking for  the zeta  intervals $[n, \infty)$ and $(-\infty , -n]$,
$v=\nu - \delta $ and $v=-\mu + \delta$ respectively where
$\delta > \varepsilon$. Note that, if $v$ is picked anywhere
between $(\nu - \delta , \nu ]$ or $ [-\mu , -\mu +\delta )$,
the expression above becomes unbounded. On the other hand if
$v$ is chosen anywhere on $-\mu +\delta \leq v \leq \nu -\delta$
the expression above is bounded.
Therefore 
\begin{eqnarray*}
\left| F_{n}(\omega) - F(\omega) \right| \leq 
\frac{2C_{3}}{\sqrt{2\pi}}\frac{e^{-(\delta -\varepsilon)n}}{(\delta -\varepsilon)} < \epsilon
\end{eqnarray*}
for $-\mu +\delta \leq v \leq \nu -\delta$
 where $C_{3}= \max (C_{1}, C_{2})$. We can solve for $n$
\begin{eqnarray*}
-(\delta -\varepsilon)n< \ln 
\left[ \frac{\sqrt{2\pi}}{2C_{3}}(\delta -\varepsilon)
\epsilon \right]
\end{eqnarray*}
so
\begin{eqnarray*}
n > \ln \left[ \frac{\sqrt{2\pi}}{2C_{3}}(\delta -\varepsilon)
\epsilon \right]^{-\frac{1}{(\delta -\varepsilon)}}.
\end{eqnarray*}
We may take 
\begin{eqnarray*}
N(\epsilon)=\ln \left[ \frac{\sqrt{2\pi}}{2C_{3}}(\delta -\varepsilon)
\epsilon \right]^{-\frac{1}{(\delta -\varepsilon)}}.
\end{eqnarray*}
 Therefore,  for any $\epsilon >0$ we 
have defined a $N(\epsilon)$, independent of $\omega$, such that $|F(\omega)
-F_{n}(\omega)|< \epsilon$, for all $n>N$.
Consequently the series $\{F_{n}(\omega)\}$ converges uniformly to 
$F(\omega)$  for $-\mu + \delta  \leq v\leq \nu -\delta $.
Hence $F(\omega)$ is analytic in this strip.
Next we prove the order results for $F(\omega)$.
By applying Cauchy's theorem we may take the integral along any line 
of the strip parallel to the real axis. Thus
\begin{eqnarray*}
|F(\omega)|&= & \left| \frac{1}{\sqrt{2\pi}}\int ^{\infty}_{-\infty}
f(\xi -i\eta)e^{-i(\xi -i\eta)(u+iv)}d\xi \right|\\
\quad \\
&\leq&  \frac{1}{\sqrt{2\pi}}\int ^{\infty}_{-\infty}
|f(\xi -i\eta)||e^{-i\xi u}||e^{\xi v}||e^{-\eta u}||e^{-i\eta v}|d\xi\\
\quad \\
&= & \frac{e^{-\eta u}}{\sqrt{2\pi}}\int ^{\infty}_{-\infty}
|f(\xi -i\eta)|e^{\xi v}d\xi\\
\quad \\
&=& \frac{e^{-\eta u}}{\sqrt{2\pi}}\left[ \int ^{c_{1}}_{0}
|f(\xi -i\eta)|e^{\xi v}d\xi +\int ^{\infty}_{c_{1}}
|f(\xi -i\eta)|e^{\xi v}d\xi  \right.\\
\quad \\
&+& \left.
\int ^{0}_{c_{2}}
|f(\xi -i\eta)|e^{\xi v}d\xi +
\int ^{c_{2}}_{-\infty}
|f(\xi -i\eta)|e^{\xi v}d\xi
\right]\\
\quad \\
&\leq& \frac{e^{-\eta u}}{\sqrt{2\pi}}\left[
C_{c_{1}}(v) + 
C_{1}\int ^{\infty}_{c_{1}} e^{-(\nu -\varepsilon)\xi}e^{\xi v}d\xi+
C_{c_{2}}(v)+
C_{2}\int ^{c_{2}}_{-\infty}
e^{(\mu-\varepsilon)\xi}e^{\xi v}d\xi \right]\\
\quad \\
&=& \frac{e^{-\eta u}}{\sqrt{2\pi}}
\left[
C_{c_{1}}(v) + C_{c_{2}}(v)+
\frac{C_{1}e^{-(\nu -\varepsilon -v)\xi}}{-(\nu -\varepsilon -v)}
\Big| ^{\infty}_{c_{1}}+ 
\frac{C_{2}e^{(\mu -\varepsilon +v)\xi}}{\mu -\varepsilon +v}
\Big|  _{-\infty}^{c_{2}}\right],
\end{eqnarray*}
where 
\begin{eqnarray*}
&\quad& C_{c_{1}}(v)=  \int ^{c_{1}}_{0}
|f(\xi -i\eta)|e^{\xi v}d\xi,
\quad C_{c_{2}}(v)=  \int ^{c_{2}}_{0}
|f(\xi -i\eta)|e^{\xi v}d\xi.
\end{eqnarray*}
\\
For the interval $[c_{1},\infty)$ pick $v=\nu -\delta $ with
$\varepsilon < \delta$ such that
\begin{eqnarray*}
\frac{e^{-(\nu -\varepsilon -v)\xi}}{-(\nu -\varepsilon -v)}
\Big| ^{\infty}_{c_{1}}
 \leq \frac{e^{-(\delta -\varepsilon)c_{1}}}{\delta -\varepsilon}.
\end{eqnarray*}
\\
Similarly for the interval $(-\infty , c_{2}]$ pick $v= -\mu  +\delta$
and with $\varepsilon <\delta $  we have
\begin{eqnarray*}
\frac{e^{(\mu -\varepsilon +v)\xi}}{\mu -\varepsilon +v}
\Big|  _{-\infty}^{c_{2}} \leq 
\frac{e^{(\delta -\varepsilon)c_{2}}}{\delta -\varepsilon}.
\end{eqnarray*}
\\
We bound $C_{c_{1}}(v)$ and $C_{c_{2}}(v)$ by choosing 
$v=\nu - \delta$ and $v=-\mu +\delta$ respectively so that
\begin{eqnarray*}
C_{c_{1}}(v)&\leq& \int^{c_{1}}_{0} |f(\xi -i\eta)|e^{(\nu -\delta)\xi}d\xi =C_{c_{1}},\\  
C_{c_{2}}(v)&\leq& \int_{c_{2}}^{0} |f(\xi -i\eta)|e^{(-\mu +\delta)\xi}d\xi =C_{c_{2}}.
\end{eqnarray*}
Consequently, with $c_{3}=\min (c_{1},|c_{2}|)$, $C_{3}=\max (C_{1},C_{2})$ and
$C_{c_{3}}=max(C_{c_{1}},C_{c_{2}})$
\begin{eqnarray*}
|F(\omega )|\leq \frac{2e^{-\eta u}}{\sqrt{2\pi}}\left[
C_{c_{3}}+
\frac{C_{3}e^{-(\delta -\varepsilon)c_{3}}}{(\delta -\varepsilon)}\right],
\end{eqnarray*}
\\
and by taking $\eta $ arbitrarily near to $-a$ and $b$ the order results
for $F(\omega)$ follow.
$(\ref{eqntenn})$ can be proved directly by the theorem of 
residues.
Let $-a < \alpha < y < \beta < b$. 
Then
\begin{eqnarray*}
\frac{1}{\sqrt{2\pi}}\int ^{\infty}_{0}F(\omega)
e^{iz\omega}d\omega &=&
\frac{1}{2\pi}\int^{\infty}_{0}e^{iz\omega}d\omega
\int ^{-i\alpha + \infty}_{-i\alpha-\infty}f(\zeta)
e^{-i\zeta\omega}d\zeta
=\frac{1}{2\pi}
\int ^{-i\alpha + \infty}_{-i\alpha-\infty}f(\zeta)d\zeta
\int^{\infty}_{0}e^{-i(\zeta -z)\omega}d\omega\\
&=&\frac{1}{2\pi i}\int ^{-i\alpha + \infty}_{-i\alpha-\infty}
\frac{f(\zeta)}{ \zeta -z}d\zeta
\end{eqnarray*}
Similarly,
\begin{eqnarray*}
\frac{1}{\sqrt{2\pi}}\int ^{0}_{-\infty}F(\omega)
e^{iz\omega}d\omega =
\frac{1}{2\pi i}\int ^{i\beta + \infty}_{i\beta -\infty}
\frac{f(\zeta)}{z- \zeta}d\zeta.
\end{eqnarray*}
Now we consider the following counter clockwise  contour integral along a rectangle
\begin{eqnarray*}
\int_{C}\frac{f(\zeta)}{\zeta - z}d\zeta =
\int ^{i\beta - \rho}_{i\beta +\rho}\frac{f(\zeta)}{\zeta - z}d\zeta 
+\int _{L_{5}}\frac{f(\zeta)}{\zeta - z}d\zeta +
\int _{L_{6}}\frac{f(\zeta)}{\zeta - z}d\zeta
+
\int ^{i\alpha + \rho}_{i\alpha -\rho}\frac{f(\zeta)}{\zeta - z}d\zeta 
+\int _{L_{2}}\frac{f(\zeta)}{\zeta - z}d\zeta 
+\int _{L_{3}}\frac{f(\zeta)}{\zeta - z}d\zeta .
\end{eqnarray*}
\begin{center}
\begin{picture}(400,200)(-25,-25)
\put (0,45){\line(1,0){300}}
\put (150,-30){\line(0,1){170}}
\put (290,45){\vector(1,0){10}}
\put (150,140){\vector(0,1){20}}
\put (50,-20){\dashbox{3}(200,135)}
\put (303,43){$x$}
\put (153,155){$y$}
\put (35,38){$-\rho$}
\put (252,38){$\rho$}
\put (154,-32){$L_{1}$}
\put (115,45){\circle*{3}}
\put (115,52){$z$}
\put (253,75){$L_{3}$}
\put (36,75){$L_{5}$}
\put (154,122){$L_{4}$}
\put (141,120){$\beta$}
\put (253,5){$L_{2}$}
\put (36,5){$L_{6}$}
\put (134,-32){$-\alpha$}
\end{picture}
\end{center}
\quad\\
\quad\\
We show the contribution from each of $L_{2},L_{3},L_{5},L_{6}$
is zero.
For $L_{2}$ we use the following parameterization
$
\zeta (y) =\rho+iy; \quad -\alpha \leq y \leq 0
$
and since $f(\zeta)$ must be bounded for all $\zeta$ by some real number $A$
\begin{eqnarray*}
\left |\frac{f(\zeta)}{\zeta -z}\right |\leq \frac{A}
{|\rho + i y - z |}
\end{eqnarray*}
so
\begin{eqnarray*}
\lim_{\rho \rightarrow \infty}
\left| \int _{L_{2}} \frac{f(\zeta)}{\zeta -z} d\zeta \right| \leq
\lim_{\rho \rightarrow \infty} \frac{A\alpha}
{|\rho + i y - z |}
=0.
\end{eqnarray*}
\\
A similar procedure with $0\leq  y\leq \beta$ in the parameterization gives
\\
\begin{eqnarray*}
\lim_{\rho \rightarrow \infty}
\left| \int _{L_{3}}\frac{f(\zeta)}{\zeta -z} d\zeta \right| \leq
\lim_{\rho \rightarrow \infty} \frac{A\alpha} {|\rho + i y - z |}
=0.
\end{eqnarray*}
\\
For $L_{5}$ we use the following parameterization
$
\zeta(y)=-\rho + iy, \quad 0\leq y\leq  \beta$.
and
\begin{eqnarray*}
\left |\frac{f(\zeta)}{\zeta -z}\right |\leq \frac{A}
{|-\rho + i y - z |}
\end{eqnarray*}
so
\begin{eqnarray*}
\lim_{\rho \rightarrow \infty}
\left| \int _{L_{5}} \frac{f(\zeta)}{\zeta -z} d\zeta \right| \leq
\lim_{\rho \rightarrow \infty} \frac{A\alpha}
{|-\rho + i y - z |}
=0.
\end{eqnarray*}
\\
A similar procedure with $-\alpha \leq  y\leq 0$ in the parameterization gives
\\
\begin{eqnarray*}
\lim_{\rho \rightarrow \infty}
\left| \int _{L_{6}} \frac{f(\zeta)}{\zeta -z} d\zeta \right| \leq
\lim_{\rho \rightarrow \infty} \frac{A\alpha}
{|-\rho + i y - z |}
=0.
\end{eqnarray*}
Therefore
\begin{eqnarray*}
\int_{C}\frac{f(\zeta)}{\zeta - z}d\zeta =
\int ^{i\beta - \rho}_{i\beta +\rho}\frac{f(\zeta)}{\zeta - z}d\zeta +
\int ^{-i\alpha + \rho}_{-i\alpha -\rho}\frac{f(\zeta)}{\zeta - z}d\zeta 
\end{eqnarray*}
and
\begin{eqnarray*}
f(z)&=& \frac{1}{\sqrt{2\pi}}\int ^{\infty}_{-\infty}
F(\omega)e^{-iz\omega}d\omega
=
\frac{1}{2\pi i}\left[\int ^{i\beta +\infty}_{i\beta -\infty}
\frac{f(\zeta)}{z-\zeta }d\zeta +
\int ^{-i\alpha +\infty}_{-i\alpha -\infty}
\frac{f(\zeta)}{ \zeta -z}d\zeta
\right]\\
\quad \\
&=& \frac{1}{2\pi i}\left[
\int _{C}\frac{f(\zeta)}{\zeta -z}d\zeta\right] =
\frac{1}{2\pi i}\left[ 2\pi i f(z) \right]
= f(z).
\end{eqnarray*}
\squareforqed

\end{lem}
\begin{lem}
\label{theomtwo}
\quad \\
\quad \\
Suppose $F(\omega)$, $\omega=u+iv$, is analytic, regular for
$-\nu <v<\mu$ where \\ $\nu >0, \mu >0$.
In any strip in the interior to $-\nu <v<\mu$, let
\begin{eqnarray*}
F(\omega)= \left\{
\begin{array}{cc}
{\cal{O}}\left(e^{-(b-\varepsilon)u}\right) & (u \rightarrow \infty)\\
\quad & \quad \\
{\cal{O}}\left(e^{(a-\varepsilon)u}\right)\quad & (u \rightarrow -\infty)
\end{array}
\right.,
\end{eqnarray*}
for every positive $\varepsilon$, where $a>0,b>0$. Then there exists a $\delta >0$
such that $f(z)$,
$z=x+iy$, defined by
\begin{eqnarray*}
f(z)=\frac{1}{\sqrt{2\pi}}\int ^{\infty}_{-\infty}
F(\zeta)e^{iz\zeta}d\zeta,
\end{eqnarray*}
is analytic for $-b+\delta \leq y\leq a-\delta $, satisfies
\begin{eqnarray*}
f(z)= \left\{
\begin{array}{cc}
{\cal{O}}\left(e^{-(\mu -\varepsilon)x}\right) & (x \rightarrow \infty)\\
\quad & \quad \\
{\cal{O}}\left(e^{(\nu -\varepsilon)x}\right)\quad & (x \rightarrow -\infty)
\end{array}
\right.,
\end{eqnarray*}
and
\begin{eqnarray*}
F(\omega)= \frac{1}{\sqrt{2\pi}}\int ^{\infty}_{-\infty}
f(z)e^{-i\omega z}dz,
\end{eqnarray*}
for every $\omega$ in the strip $-\nu < v< \mu$.\\
\quad\\
{Proof}:
\quad
We have 
\begin{eqnarray*}
f(z)=\frac{1}{\sqrt{2\pi}}\int ^{\infty}_{-\infty}
F(\zeta)e^{iz\zeta}d\zeta,
\end{eqnarray*}
\quad \\
and define the sequence $\{f_{n}(z)\}$ where
\begin{eqnarray*}
f_{n}(z)=
\frac{1}{\sqrt{2\pi}}\int ^{n}_{-n}
F(\zeta)e^{iz\zeta}d\zeta.
\end{eqnarray*}
\quad \\
Similar to Theorem $\ref {theomone}$ we can show that this sequence 
converges uniformly to $f(z)$ for $-b+\delta \leq y\leq a-\delta $.
And since each of $f_{n}(z)$ is analytic for 
$-b+\delta \leq y\leq a-\delta $, it follows that
 $f(z)$ is analytic in this strip.
Next we prove the order results on $f(z)$. Applying Cauchy's theorem we may take the integral along
any line of the strip parallel to the real axis.
Thus
\begin{eqnarray*}
|f(z)|&=&\left| \frac{1}{\sqrt{2\pi}} \int ^{\infty}
_{-\infty} F(\xi + i\eta)e^{i(x+iy)(\xi+i\eta)}d\xi\right|\\
\quad \\
&\leq& \frac{1}{\sqrt{2\pi}}
\int ^{\infty}
_{-\infty} |F(\xi + i\eta)| |e^{ix\xi}||e^{-x\eta}||e^{-y\xi}|
|e^{-iy\eta}|d\xi\\
\quad \\
&\leq& \frac{e^{-x\eta}}{\sqrt{2\pi}} \int ^{\infty}
_{-\infty} |F(\xi + i\eta)|e^{-y\xi}d\xi\\
\quad \\
&=& 
\frac{e^{-x\eta}}{\sqrt{2\pi}}\left[
\int ^{c_{1}}_{0} |F(\xi + i\eta)|e^{-y\xi}d\xi+
\int ^{\infty}_{c_{1}} |F(\xi + i\eta)|e^{-y\xi}d\xi
\right.\\
\quad \\
&+& \left.
\int^{0}_{c_{2}} |F(\xi + i\eta)|e^{-y\xi}d\xi
+\int^{c_{2}}_{-\infty} |F(\xi + i\eta)|e^{-y\xi}d\xi\right]\\
\quad \\
&\leq& 
\frac{e^{-x\eta}}{\sqrt{2\pi}}\left[
C_{c_{1}}(y)+
C_{1}\int ^{\infty}_{c_{1}}e^{-(b-\varepsilon)\xi }e^{-y\xi}d\xi+
C_{c_{2}}(y)+ C_{2}\int^{c_{2}}_{-\infty}e^{(a-\varepsilon)\xi }e^{-y\xi}d\xi\right]\\
\quad \\
&=& \frac{e^{-x\eta}}{\sqrt{2\pi}}\left[
C_{c_{1}}(y)+
C_{c_{2}}(y)+
\frac{C_{1}e^{-(b-\varepsilon +y)\xi}}{-(b-\varepsilon +y)}
\Big| ^{\infty}_{c_{1}}+
\frac{C_{2}e^{(a-\varepsilon -y)\xi}}{(a-\varepsilon -y)}
\Big| ^{c_{2}}_{-\infty} \right].
\end{eqnarray*}
\\
To bound $C_{c_{1}}(y)$ and $C_{c_{2}}(y)$ we chose $y=-b+\delta$ 
and $y=a-\delta$ respectively so that
\begin{eqnarray*}
C_{c_{1}}(y)&\leq &
\int ^{c_{1}}_{0} |F(\xi + i\eta)|e^{(b-\delta)\xi}d\xi=C_{c_{1}} ,\\
C_{c_{2}}(y)&\leq &\int^{0}_{c_{2}} |F(\xi + i\eta)|e^{-(a-\delta)\xi}d\xi=C_{c_{2}} .
\end{eqnarray*}
For the  $\xi$ intervals $[c_{1}, \infty)$ and $(-\infty , c_{2}]$ we pick
$y=- b+\delta$ and $y=a-\delta$ respectively with $\delta > \varepsilon$
and obtain with $c_{3}= \min (c_{1} ,|c_{2}|)$, $C_{3}=\max (C_{1}, C_{2})$, and
$C_{c_{3}}=\max (C_{c_{1}}, C_{c_{2}})$
\begin{eqnarray*}
|f(z)| \leq
\frac{2e^{-x\eta}}{\sqrt{2\pi }}\left[
C_{c_{3}}+\frac{C_{3}e^{-(\delta -\varepsilon)c_{3}}}{(\delta -\varepsilon)}\right]
\end{eqnarray*}
and by choosing $\eta$ arbitrarily close to $-\nu$ and $\mu$ the order results follow.
The rest of the theorem can be proven with the theory of residues
exactly like the previous theorem except that the contour integral
will be taken clockwise.
\squareforqed

\end{lem}
\begin{lem}
\label{lemblue}
\quad \\

Given $G(q,t)\in {\cal{A}}_{\rho}$ with
\begin{eqnarray*}
G(q,t)= \sum _{k\in {\mathbb{Z}}^{n}} g_{k}(t)e^{ik\cdot q},
\end{eqnarray*}
and if
\begin{eqnarray*}
g_{k}(t)=\left\{
\begin{array}{cc}
{\cal{O}}(e^{-(b-\varepsilon)t_{R}}) & (t_{R} \rightarrow \infty)\\
\quad & \quad \\
{\cal{O}}(e^{(a-\varepsilon)t_{R}}) & (t_{R} \rightarrow -\infty)\\
\end{array},
\right.
\end{eqnarray*}
then
\begin{eqnarray*}
g_{k}(t)=\left\{
\begin{array}{cc}
{\cal{O}}(e^{-(b-\varepsilon)t_{R}}){\cal{O}}(e^{-|k|\rho}) & (t_{R} \rightarrow \infty)\\
\quad & \quad \\
{\cal{O}}(e^{(a-\varepsilon)t_{R}}){\cal{O}}(e^{-|k|\rho}) & (t_{R} \rightarrow -\infty)
\end{array}.
\right.
\end{eqnarray*}
{Proof}:\\
\quad\\
\begin{eqnarray*}
g_{k}(t)= \frac{1}{(2\pi)^{n}}\int ^{\pi}_{-\pi} \cdot \cdot
\cdot \int ^{\pi}_{-\pi} G(q,t)e^{-ik\cdot q}
dq
\end{eqnarray*}
\\
where the integral is along real axis.
Lift the integral as follows
\\
\begin{eqnarray*}
g_{k}(t)= \frac{1 }{(2\pi)^{n}}\int ^{\pi}_{-\pi} \cdot \cdot
\cdot \int ^{\pi}_{-\pi} 
G(q_{1}- i\frac{k_{1}}{|k_{1}|}\rho,...,q_{n}-i\frac{k_{n}}{|k_{n}|}\rho
)
\cdot \Big \{ \prod ^{n}_{j=1}e^{-ik_{j}(q_{j}-i\frac{k_{j}}{|k_{j}|} \rho)} \Big\}
dq 
\end{eqnarray*}
\begin{eqnarray*}
g_{k}(t)=\frac{e^{-|k|\rho} }{(2\pi)^{n} }
\int ^{\pi}_{-\pi} \cdot \cdot
\cdot \int ^{\pi}_{-\pi} 
G(q_{1}- i\frac{k_{1}}{|k_{1}|}\rho,...,q_{n}-i\frac{k_{n}}{|k_{n}|}\rho)
\cdot e^{-ik\cdot q}
dq 
\end{eqnarray*}
and
$
|g_{k}(t)| \leq e^{-|k|\rho}\| G(q,t) \|_{\rho }$.
We know
\begin{eqnarray*}
\| G(q,t) \|_{\rho }= \sup _{q\in {\cal{D}}_{\rho}} \left| G(q,t) \right|.
\end{eqnarray*}
Let $q^{*}$ be arbitrary in the interior of the strip $|q _{I} |\leq  \rho$. Then by the
above and the assumption on the order of $g_{k}(t)$
\begin{eqnarray*}
\left| G(q^{*},t) \right| &=&\left| \sum _{k\in {\mathbb{Z}}^{n}} g_{k}(t)e^{ik\cdot q^{*}}\right|
\leq  \sum _{k\in {\mathbb{Z}}^{n}} C e^{-At_{R}} e^{-|k|\rho}|e^{ik\cdot q^{*}}|
\leq Ce^{-At_{R}}\sum _{k\in {\mathbb{Z}}^{n}}e^{-|k|\rho}e^{-k\cdot q^{*}_{I}}
\leq  Ce^{-At_{R}}\sum _{k\in {\mathbb{Z}}^{n}}e^{-|k|(\rho-q^{*}_{I})}\\
\quad \\
&\leq & Ce^{-At_{R}}\sum _{k\in {\mathbb{Z}}^{n}}e^{-|k|(\rho-(\rho -\epsilon))}
\leq  Ce^{-At_{R}}\sum _{k\in {\mathbb{Z}}^{n}}e^{-|k|\epsilon}
\leq  Ce^{-At_{R}}2^{n}\sum _{\tiny{\begin{array}{c}
k\in {\mathbb{Z}}^{n}\\
k_{1},...k_{n}\geq 0
\end{array}}}e^{-\epsilon \sum k_{j}}\\
\quad\\
&=&Ce^{-At_{R}}2^{n}\left(\sum ^{\infty}_{k=0}e^{-\epsilon k} \right)^{n}
=Ce^{-At_{R}}2^{n}\left( \frac{1}{1-e^{-\epsilon}} \right)^{n},
\end{eqnarray*}
where $A$ is $b -\varepsilon$ or $-a+\varepsilon$. 
\squareforqed

\end{lem}

The following lemma shows that under certain conditions we
are assured the existence of the complex Fourier transform.
\begin{lem}
\label{existence}
\quad \\

Suppose $f(t)$, $t=t_{R}+it_{I}$, is analytic for $-a<t_{I}<b$ where $a>0,b>0$.
In any strip in the interior of $-a<t_{I}<b$, let
\begin{eqnarray*}
f(t)= \left\{
\begin{array}{cc}
{\cal{O}}\left(e^{-(\nu-\varepsilon)t_{R}}\right) & (t_{R} \rightarrow \infty)\\
\quad & \quad \\
{\cal{O}}\left(e^{(\mu-\varepsilon)t_{R}}\right)\quad & (t_{R} \rightarrow -\infty)
\end{array}
\right.,
\end{eqnarray*}
for every positive $\varepsilon$, where $\nu >0, \mu >0$.
Assume the Fourier transform exists
\begin{eqnarray*}
F(\omega)= \frac{1}{\sqrt{2\pi}}\int ^{\infty}_{-\infty}
f(\zeta)e^{-i\omega \zeta}d\zeta <\infty.
\end{eqnarray*}
Then the complex Fourier transform defined as
\begin{eqnarray*}
F^{c}(\omega)= \frac{1}{\sqrt{2\pi}}\int ^{\infty +i\beta}_{-\infty + i\beta}
f(\zeta)e^{-i\omega \zeta}d\zeta,
\end{eqnarray*}
with $-a < \beta < b$ exists and
$
F^{c}(\omega)=F(\omega)$.\\
\quad\\
{Proof}:
\quad
We consider the counter clockwise contour integral on the rectangle $R$ show below
\quad \\
\begin{center}
 \begin{picture}(400,200)(-25,-25)
\put (0,5){\line(1,0){300}}
\put (150,-30){\line(0,1){170}}
\put (290,5){\vector(1,0){10}}
\put (150,140){\vector(0,1){20}}
\put (50,5){\dashbox{3}(200,115)}
\put (303,3){$t_{R}$}
\put (153,155){$t_{I}$}
\put (37,-3){$-\rho$}
\put (250,-3){$\rho$}
\put (154,-12){$L_{1}$}
\put (253,65){$L_{3}$}
\put (36,65){$L_{4}$}
\put (158,127){$L_{2}$}
\put (140,125){$\beta$}
\end{picture}
\end{center}
Since $f(t)$ is analytic for $-a<t_{I}<b$ we know
the contour integral is zero. Consequently
\begin{eqnarray*}
-\int_{L_{2}} = \int _{L_{1}} + \int _{L_{3}} + \int _{L_{4}}.
\end{eqnarray*}
We analyze the integral along $L_{3}$ using the parameterization
$t=\rho + it_{I}$ such that
\begin{eqnarray*}
\int _{L_{3}}f_{k}(t)e^{-i\omega t}dt &=&
\int ^{\beta}_{0}
f_{k}(\rho +it_{I})e^{-i\omega(\rho +it_{I})}idt_{I}
= \int ^{\beta}_{0}
f_{k}(\rho +it_{I}) e^{-iu\rho}e^{v\rho}e^{ut_{I}}e^{ivt_{I}}idt_{I}
\end{eqnarray*}
and
\begin{eqnarray*}
\left| \int _{L_{3}}f_{k}(t)e^{-i\omega t}dt\right| &\leq&
\int  ^{\beta}_{0}
|f_{k}(\rho +it_{I})|e^{v\rho}e^{ut_{I}}dt_{I}.
\end{eqnarray*}
Note although
\begin{eqnarray*}
|f_{k}(t)|\leq C_{1} e^{-(\nu -\varepsilon)t_{R}} \quad c_{1}\leq t_{R} <\infty,
\end{eqnarray*}
we can choose a constant $C_{3}$ large enough such that 
 \begin{eqnarray*}
|f_{k}(t)|\leq C_{3} e^{-(\nu -\varepsilon)t_{R}} \quad 0 \leq t_{R} <\infty,
\end{eqnarray*}
and 
\begin{eqnarray*}
\left| \int _{L_{3}}f_{k}(t)e^{-i\omega t}dt\right| &\leq&
\int  ^{\beta}_{0}
C_{3} e^{-(\nu -\varepsilon)\rho}
 e^{v\rho}e^{ut_{I}}dt_{I}.
\end{eqnarray*}
By Theorem $\ref{theomone}$, $F(\omega)$ is analytic inside 
the strip $-\mu + \delta \leq v \leq \nu -\delta$ where $\varepsilon < \delta$
so that
\begin{eqnarray*}
\left| \int _{L_{3}}f_{k}(t)e^{-i\omega t}dt\right| &\leq&
C_{3} e^{-(\nu -\varepsilon)\rho}
e^{(\nu -\delta)\rho}e^{|u|\beta }\beta .
\end{eqnarray*}
Clearly
\begin{eqnarray*}
\lim _{\rho \rightarrow \infty} 
\left| \int _{L_{3}}f_{k}(t)e^{-i\omega t}dt\right| =0.
\end{eqnarray*}
Similarly for $L_{4}$ using the parameterization $t=-\rho +it_{I}$
\begin{eqnarray*}
\int _{L_{e}}f_{k}(t)e^{-i\omega t}dt &=&
\int _{\beta}^{0}
f_{k}(-\rho +it_{I})e^{-i\omega(-\rho +it_{I})}idt_{I}
= \int _{\beta}^{0}
f_{k}(-\rho +it_{I}) e^{iu\rho}e^{-v\rho}e^{ut_{I}}e^{ivt_{I}}idt_{I}
\end{eqnarray*}
and
\begin{eqnarray*}
\left| \int _{L_{4}}f_{k}(t)e^{-i\omega t}dt\right| &\leq&
\int  _{\beta}^{0}
|f_{k}(-\rho +it_{I})|e^{-v\rho}e^{ut_{I}}dt_{I}.
\end{eqnarray*}
With a large enough constant $C_{4}$
\begin{eqnarray*}
\left| \int _{L_{4}}f_{k}(t)e^{-i\omega t}dt\right| &\leq&
\int  _{\beta}^{0}
C_{4} e^{-(\mu -\varepsilon)\rho}
 e^{-v\rho}e^{ut_{I}}dt_{I} 
\leq
C_{4} e^{-(\mu -\varepsilon)\rho}
e^{(\mu -\delta)\rho}e^{|u|\beta }\beta 
\end{eqnarray*}
and clearly
\begin{eqnarray*}
\lim _{\rho \rightarrow \infty} 
\left| \int _{L_{4}}f_{k}(t)e^{-i\omega t}dt\right| =0.
\end{eqnarray*}
It follows
\begin{eqnarray*}
\lim _{\rho \rightarrow \infty}
-\int_{L_{2}} =
\lim _{\rho \rightarrow \infty}
\int_{L_{1}}
\end{eqnarray*}
or
\begin{eqnarray*}
F(\omega)= \frac{1}{\sqrt{2\pi}}\int ^{\infty}_{-\infty}
f(\zeta)e^{-i\omega \zeta}d\zeta =
\frac{1}{\sqrt{2\pi}}\int ^{\infty +i\beta}_{-\infty + i\beta}
f(\zeta)e^{-i\omega \zeta}d\zeta
=
F^{c}(\omega). \quad \squareforqed
\end{eqnarray*}

\end{lem}

\section{Properties of $p',q'-$Exponential Form functions}
\label{sectp'q'exform}
In this section we study some characteristics of functions of the form
\begin{eqnarray*}
&\quad& F(p',q')=f(p')+r(p',q')+ g(p',q'),\\
\quad \\
&\quad& r(p',q')= \sum _{k\in {\mathbb{Z}}^{n}}
s_{k}(p')e^{ik\cdot q},
\quad \\
&\quad& g(p',q')= \sum _{k\in {\mathbb{Z}}^{n}}
h_{k}(p')e_{k}(t)e^{ik\cdot q},
\end{eqnarray*}
where $F(p',q') \in {\cal{A}}_{\rho, \sigma}$ and
 $e_{k}(t)$ is of exponential order with respect to time.
We will refer to these functions as  functions of $p',q'-$exponential form.
The Hamiltonian under consideration will be  of $p',q'-$exponential form.
There are several things we want to know about these functions
to carry out the iterative process of the KAM proof.
First, we want  to be able to extract from
$F(p',q')$ the quasiperiodic and exponential-order-with-respect-to-time parts
as  functions of
$F(p',q')$ itself. The importance of this will become evident in the
next section.
Also, we want to show the derivative with respect to
$p'_{j}$ or $q'_{j}$, $j=1,...,n+1$,  of a $p',q'-$exponential-form
function is another $p',q'-$exponential-form function. First we prove
an important characteristic of $g(p',q')$, the function
which is of  exponential order with respect to time.
\begin{lem}
\label{lemmaforg}
\quad \\

Given a function $g(p',q') \in {\cal{A}}_{\rho , \sigma}$ of the form
\begin{eqnarray*}
g(p',q')= \sum _{k\in {\mathbb{Z}}^{n}} h_{k}(p')e_{k}(t)e^{ik\cdot q},
\end{eqnarray*}
where $e_{k}(t)$ is of exponential order with respect to time,
$
\overline{\overline{g}} (0) =0$.\\
\quad\\
{Proof}:
\quad
We simply show the time average of $g(p',q')$ is zero
\begin{eqnarray*}
\lim _{T \rightarrow \infty} \frac{1}{T}
\int ^{T}_{-T} g(p',q') dt =
\sum _{k\in {\mathbb{Z}}^{n}}
h_{k}(p')
\lim _{T \rightarrow \infty} \frac{1}{T}
\int ^{T}_{-T} e_{k}(t) dt e^{ik\cdot q} =0.\quad 
\squareforqed
\end{eqnarray*}
\end{lem}
Next we show how to express $g(0,q')$ in terms of
$F(p',q')$.
\quad \\
\begin{lem}
\label{exponentialpart}
\quad \\

Given a function $F(p',q')\in {\cal{A}}_{\rho ,\sigma}$
of $p',q'$-exponential form as defined previously,
$
T(0,q')=F(0,q') - \overline{\overline{F}}(0)
$
represents the $q$-quasiperiodic and exponential-order-with-respect-to-time  
parts of $F(0,q')$. Moreover $T(0,q')$ has zero average.\\
\quad\\
{Proof}:
\quad
First 
\[
\overline{\overline{F}}(0)=\overline{\overline{f}}(0)
+ \overline{\overline{r}}(0)
+\overline{\overline{g}}(0)=
f(0)+ s_{0}(0).\]
Therefore
\begin{eqnarray*}
F(0,q')-\overline{\overline{F}}(0)=
f(0) +r(0,q')+g(0,q')-f(0)-s_{0}(0)
=\sum _{k\in {\mathbb{Z}}^{n} \backslash 0}s_{k}(0)e^{ik\cdot q}+
g(0,q').\quad 
\squareforqed
\end{eqnarray*}
\end{lem}
Next we will show the derivative with respect to
$p'_{j}$ or $q'_{j}$, $j=1,...,n+1$,  of a $p'-q'-$exponential-form
function is another $p'-q'-$exponential-form function.
It is easy to see this is true for the derivative with respect to
$p'_{j}$ $j=1,...,n+1$.
\quad \\
\begin{lem}
\label{expoformderp}
\quad \\

Let $F(p',q')$ be defined as before. Then
$\frac{\partial F}{\partial p'_{j}}(p',q')$ $j=1,...,n+1$ are of
$p',q'$-exponential-form and $\frac{\partial F}{\partial p'_{j}}(p',q')
\in {\cal{A}}_{\rho ,\sigma}$.\\
\quad\\
{Proof}:
\quad
First we write out the derivative
\begin{eqnarray*}
\frac{\partial F(p',q')}{\partial p'_{j}}
&=&
\frac{\partial f(p')}{\partial p'_{j}}
+
\frac{\partial r(p',q')}{\partial p'_{j}}
+
\frac{\partial g(p',q')}{\partial p'_{j}}
=
\frac{\partial f(p')}{\partial p'_{j}}
+
\sum _{k\in {\mathbb{Z}}^{n}}
\frac{\partial s_{k}(p')}{\partial p'_{j}}
e^{ik\cdot q}
+
\sum _{k\in {\mathbb{Z}}^{n}}
\frac{\partial h_{k}(p')}{\partial p'_{j}}
e_{k}(t)e^{ik\cdot q}.
\end{eqnarray*}
\quad \\
Furthermore we know , as consequence of Cauchy's Integral formula,
that if a function is analytic in some domain then all
its derivatives are analytic in the same domain. \quad  \squareforqed\\
\end{lem}
\quad\\
The next lemma show  the derivative 
of a $p'-q'$-exponential-form
function with respect to $q'_{j}$, $j=1,...,n$,
is another $p'-q'$-exponential-form function.
\begin{lem}
\label{expoformderq}
\quad \\

Let $F(p',q')$ be defined as before. Then
$\frac{\partial F}{\partial q'_{j}}(p',q')$, $j=1,...,n$,
are of
$p',q'-$exponential-form
 and $\frac{\partial F}{\partial q'_{j}}(p',q')
\in {\cal{A}}_{\rho ,\sigma}$.\\
\quad\\
{Proof}:
\quad
First we write out the derivative for $j=1,...,n$
\quad \\
\begin{eqnarray*}
\frac{\partial F(p',q')}{\partial q'_{j}}
&=&
\frac{\partial r(p',q')}{\partial q'_{j}}
+
\frac{\partial g(p',q')}{\partial q'_{j}}
=
\sum _{k\in {\mathbb{Z}}^{n}}
ik_{j}s_{k}(p')e^{ik\cdot q}
+ 
\sum _{k\in {\mathbb{Z}}^{n}}
ik_{j}h_{k}(p')e_{k}(t)e^{ik\cdot q}
\end{eqnarray*}
\quad \\
Furthermore we know , as consequence of Cauchy's Integral formula,
that if a function is analytic in some domain then all
its derivatives are analytic in the same domain. \quad  \squareforqed
\end{lem}
Recall a function $e_{k}(t)$ is said to be of 
exponential order with respect to time if the following holds
\begin{eqnarray*}
e_{k}(t)= 
\left\{
\begin{array}{cc}
{\cal{O}}(e^{-(\nu -\varepsilon)t_{R}}) & ;\quad 0\leq c_{1} < t_{R} < \infty\\
\quad \\
{\cal{O}}(e^{(\mu -\varepsilon)t_{R}}) & ;\quad -\infty < t_{R} < c_{2} <0.
\end{array}
\right. 
\end{eqnarray*}
The following lemma examines 
the derivative of a $p'-q'-$exponential-form
function with respect to $q'_{n+1}=t$.

\begin{lem}
\label{lemmatimederv}
\quad \\

Let $F(p',q')$ be defined as before. Then for some positive $\delta >\sigma$,
$\frac{\partial F}{\partial t}(p',q') \in {\cal{A}}_{\rho , \sigma -\delta}$
is of $p',q'-$exponential-form. Moreover for the $n$th derivative
we have the following estimates
\begin{eqnarray*}
|e_{k}^{(n)}(t)|&\leq&
\frac{C_{1}n!}{ \delta ^{n}}e^{(\nu -\varepsilon)\delta}
e^{-(\nu -\varepsilon)t_{R}}, \quad 0< c_{1} < t_{R} < \infty,
\end{eqnarray*}
\begin{eqnarray*}
|e_{k}^{(n)}(t)|&\leq&
\frac{C_{2}n!}{ \delta ^{n}}e^{-(\mu -\varepsilon)\delta}
e^{(\mu -\varepsilon)t_{R}}, \quad -\infty < t_{R} < c_{2}<0,
\end{eqnarray*}
for all $t\in {\cal{D}}_{\rho ,\sigma -\delta}$.\\
\quad\\
{Proof}:
\quad
We will prove this using Cauchy's Integral formula.
Since $e_{k}(t)$ is exponentially small with respect to time
we have
\begin{eqnarray*}
&\quad& |e_{k}(t)|\leq C_{1}e^{-(\nu -\varepsilon)t_{R}}, \quad 
0< c_{1} < t_{R} < \infty,\\
&\quad&\\
&\quad& |e_{k}(t)|\leq C_{2}e^{(\mu -\varepsilon)t_{R}}, \quad 
-\infty < t_{R} < c_{2} <0 .
\end{eqnarray*}
\quad \\
Since $e_{k}(t)$ is analytic in the interior of the  strip
$-\sigma \leq t_{Im} \leq \sigma$ we have
\begin{eqnarray*}
e_{k}^{(n)}(t) =\frac{n!}{2\pi i}
\int _{C} \frac{e_{k}(\omega)}{(\omega -t)^{n+1}}d\omega
\end{eqnarray*}
where $C$ is the circle of radius $\delta$ centered at $t$.
We used the parameterization $\omega = t - \delta e^{i \theta}$
$0 \leq \theta \leq 2\pi$ and obtain
\begin{eqnarray*}
e_{k}^{(n)}(t) =\frac{n!}{2\pi i}
\int _{C} \frac{e_{k}(\omega)}{(\omega -t)^{n+1}}d\omega=
\int ^{2\pi}_{0}\frac{e_{k}(t_{R}-\delta cos \theta
+i(t_{I}- \delta sin \theta))}{(\delta e^{i\theta})^{n+1}}
i\delta e^{i\theta}d\theta
\end{eqnarray*}
and for $0< c_{1} < t_{R} < \infty$
\begin{eqnarray*}
|e_{k}^{(n)}(t)|&\leq&
\frac{n!}{2\pi }
\int ^{2\pi}_{0}
\frac{|e_{k}(t_{R}-\delta cos \theta
+i(t_{I}- \delta sin \theta))|}{(\delta)^{n+1}}\delta d\theta
\leq 
\frac{n!}{2\pi \delta ^{n} }
\int ^{2\pi}_{0}
 C_{1}e^{-(\nu -\varepsilon)(t_{R}-\delta cos \theta)}d\theta
\leq 
\frac{C_{1}n!}{ \delta ^{n}}e^{(\nu -\varepsilon)\delta}
e^{-(\nu -\varepsilon)t_{R}}.
\end{eqnarray*}
Therefore 
\begin{eqnarray*}
|e_{k}^{(n)}(t)|&\leq&
\frac{C_{1}n!}{ \delta ^{n}}e^{(\nu -\varepsilon)\delta}
e^{-(\nu -\varepsilon)t_{R}}, \quad 0< c_{1} < t_{R} < \infty,
\end{eqnarray*}
for all $t\in {\cal{D}}_{\rho ,\sigma -\delta}$.
Similarly
for $-\infty < t_{R} < c_{2}<0$ with the parameterization
$\omega = t + \delta e^{i \theta}$
we obtain
\begin{eqnarray*}
|e_{k}^{(n)}(t)|&\leq&
\frac{C_{2}n!}{ \delta ^{n}}e^{-(\mu -\varepsilon)\delta}
e^{(\mu -\varepsilon)t_{R}}.
\end{eqnarray*}
 \quad \squareforqed
\end{lem}
We want to show the product of two quasiperiodic functions
is quasiperiodic.
\begin{lem}
\quad\\

Let $F(q)$, $G(q)$ $\in {\cal{A}}_{\rho}$ be periodic with respect to
$q \in {\mathbb{R}}$.
It follows
$
F(q)G(q)= H(q)
$
is periodic with respect to $q$ and $H(q)$ can be expressed
as
\begin{eqnarray*}
\sum _{k\in {\mathbb{Z}}} c_{k}e^{ikq},
\end{eqnarray*}
where $c_{k}$
is defined in terms of the series expressions for $F$ and $G$.\\
\quad\\
{Proof}:
\quad
First we prove that the series expression for $F$ and $G$ converge absolutely
to the functions for  $q \in {\cal{D}}_{\rho}$.
Given
\begin{eqnarray*}
&\quad&F(q) \sim \sum _{k\in {\mathbb{Z}}} a_{k} e^{ikq} \in {\cal{A}}_{\rho},
\quad G(q) \sim \sum _{k\in {\mathbb{Z}}} b_{k} e^{ikq} \in {\cal{A}}_{\rho}
\end{eqnarray*}
we have
\begin{eqnarray*}
|a_{k}| \leq 
\|F \|_{\rho} e^{-|k|\rho}.
\end{eqnarray*}
Furthermore, given $|q_{I}| \leq \rho -\delta$ for some positive $\delta$
\begin{eqnarray*}
|e^{ik(q_{R} +iq_{I})}| \leq e^{-kq_{I}} \leq e^{|k|(\rho -\delta)}.
\end{eqnarray*}
Therefore
\begin{eqnarray*}
|a_{k}e^{ikq}| \leq \| F \|_{\rho} e^{-|k|\rho} e^{|k|(\rho -\delta)}
\leq
\| F \|_{\rho}e^{-|k|\delta}.
\end{eqnarray*}
A similar estimate holds for $|b_{k}e^{ikq}|$.
Consequently, $\sum _{k\in {\mathbb{Z}}} a_{k} e^{ikq}$  and
$\sum _{k\in {\mathbb{Z}}} b_{k} e^{ikq}$
converge absolutely to $F(q)$ and $G(q)$ respectively for $q\in {\cal{D}}_{\rho}$.
Next we look at the product
$H(q)$ given by
\begin{eqnarray*}
H(q)&=& \left( \sum _{k\in {\mathbb{Z}}}a_{k}e^{ikq} \right)
\left( \sum _{k\in {\mathbb{Z}}}b_{k}e^{ikq} \right)
=
\left( \sum _{k=-\infty}^{k=\infty}a_{k}e^{ikq} \right)
\left( \sum _{k=-\infty}^{k=\infty}b_{k}e^{ikq} \right)
=
\sum _{n=-\infty}^{n=\infty}
\left(
\sum _{k=-\infty}^{k=\infty}
a_{k}e^{ikq}b_{n}e^{iq(n)}\right)\\
\quad\\
&=&
\sum _{n\in {\mathbb{Z}}}\left(\sum _{k\in {\mathbb{Z}}}
a_{k}b_{n}e^{ikq} \right) e^{inq}
=
\sum _{n\in {\mathbb{Z}}}c_{n} e^{inq} \in {\cal{A}}_{\rho}. \quad \squareforqed
\end{eqnarray*}
\end{lem}
A similar result holds for quasiperiodic functions and can be proven similarly
as above.
Given
\begin{eqnarray*}
F(q) =\sum _{k\in {\mathbb{Z}}^{n}} a_{k}e^{ik\cdot q},\quad
G(q) =\sum _{k\in {\mathbb{Z}}^{n}} b_{k}e^{ik\cdot q}.
\end{eqnarray*}
It follows
\begin{eqnarray*}
H(q)=F(q)G(q) = \sum _{k\in {\mathbb{Z}}^{n}}
\left( 
\sum _{n\in {\mathbb{Z}}^{n}}
a_{n}b_{k}e^{in\cdot q} \right)e^{ik\cdot q}
=
\sum _{k\in {\mathbb{Z}}^{n}}c_{k}
e^{ik\cdot q}.
\end{eqnarray*}

\begin{lem}
\label{lempqform}
\quad \\

Let $F(p',q')\in {\cal{A}}_{\rho ,\sigma}$ be of the form
\begin{eqnarray*}
&\quad& F(p',q')=f(p')+ r(p',q')+ g(p',q'),\\
\quad \\
&\quad& r(p',q')= \sum _{k\in {\mathbb{Z}}^{n}}
s_{k}(p')e^{ik\cdot q},
\quad \\
&\quad& g(p',q')= \sum _{k\in {\mathbb{Z}}^{n}}
h_{k}(p')e_{k}(t)e^{ik\cdot q},
\end{eqnarray*}
where $e_{k}(t)$ is of exponential order with respect to time.
Let $\chi(p',q')\in {\cal{A}}_{\rho ' ,\sigma '}$, $\rho '<\rho $ and
$\sigma' <\sigma$, be of the form
\begin{eqnarray*}
\chi(p',q')=X(q')+ \xi \cdot q' + Y(q')\cdot p',
\end{eqnarray*}
where
\begin{eqnarray*}
&\quad& X(q,t)= {\cal{Y}} (q)+ {\cal{T}}(q'),
\quad
 {\cal{Y}}(q)= \sum _{k\in {\mathbb {Z}}^{n}} y_{k}e^{ik \cdot q},
\quad
{\cal{T}}(q')=
\sum _{k\in {\mathbb {Z}}^{n}} x_{k}(t)e^{ik \cdot q}\\
\quad \\
&\quad&Y_{j}(q')= {\cal{S}}_{j}(q) + {\cal{F}}_{j}(q'),
\quad
{\cal{S}}_{j}(q)=
\sum _{k\in {\mathbb{Z}}^{n}}{\cal{S}}_{k,j}e^{ik\cdot q},
\quad
{\cal{F}}_{j}(q')=
\sum _{k\in {\mathbb{Z}}^{n}}
{\cal{F}}_{k,j}(t)e^{ik\cdot q},
\end{eqnarray*}
and where $\xi$ is a constant vector.
It follows, given some positive $\delta < \sigma '$,
$\{\chi , F\}(p',q')\in {\cal{A}}_{\rho ' -\delta  ,\sigma '-\delta}$ is  of $p',q'$-exponential form.\\
\quad\\
{Proof}:
\quad


We write out the expression
\quad \\
\begin{eqnarray*}
\{\chi , F\}&=& \sum_{j=1}^{n+1} \left[ \frac{\partial \chi}{\partial p'_{j}}
\frac{\partial F}{\partial q'_{j}} -
\frac{\partial \chi}{\partial q'_{j}}
\frac{\partial F}{\partial p'_{j}} \right]
\end{eqnarray*}
\begin{eqnarray*}
&=& 
\sum_{j=1}^{n+1} \left[
\left(Y_{j}(q')\right)
\left(
\frac{\partial r}{\partial q'_{j}}(p',q')+
\frac{\partial g}{\partial q'_{j}}(p',q')\right)
-
\left( \frac{\partial X}{\partial q'_{j}}(q')+ \xi_{j}
+\sum_{l=1}^{n+1}\frac{\partial Y_{l}}{\partial q'_{j}}(q')p'_{l} \right)\right.\\
\quad\\
&\cdot & \left.
\left( \frac{\partial f}{\partial p'_{j}}(p') +
\sum_{k\in {\mathbb{Z}}^{n}} \frac{\partial s_{k}}{\partial p'_{j}}(p')
e^{ik\cdot q}
+
\sum_{k\in {\mathbb{Z}}^{n}} \frac{\partial h_{k}}{\partial p'_{j}}(p')
e_{k}(t)e^{ik\cdot q} \right) \right]\\
\quad\\
&=& 
\sum_{j=1}^{n+1} \left[
\left(
 {\cal{S}}_{j}(q) + {\cal{F}}_{j}(q')
\right)
\left(
\frac{\partial r}{\partial q'_{j}}(p',q')+
\frac{\partial g}{\partial q'_{j}}(p',q')\right)
-
\left( \frac{\partial }{\partial q'_{j}}\Big(
 {\cal{Y}} (q)+ {\cal{T}}(q')\Big)+ \xi_{j}
+\sum_{l=1}^{n+1}\frac{\partial }{\partial q'_{j}}
\Big(
 {\cal{S}}_{l}(q) + {\cal{F}}_{l}(q')
\Big)
p'_{l} \right)\right.\\
\quad\\
&\cdot & \left.
\left( \frac{\partial f}{\partial p'_{j}}(p') +
\sum_{k\in {\mathbb{Z}}^{n}} \frac{\partial s_{k}}{\partial p'_{j}}(p')
e^{ik\cdot q}
+
\sum_{k\in {\mathbb{Z}}^{n}} \frac{\partial h_{k}}{\partial p'_{j}}(p')
e_{k}(t)e^{ik\cdot q} \right) \right]\\
\quad\\
&=& 
\sum_{j=1}^{n+1} \left[
 {\cal{S}}_{j}(q)
\frac{\partial r}{\partial q'_{j}}(p',q')
+
 {\cal{S}}_{j}(q)
\frac{\partial g}{\partial q'_{j}}(p',q')
 +
 {\cal{F}}_{j}(q')
\frac{\partial r}{\partial q'_{j}}(p',q')
+
 {\cal{F}}_{j}(q')
\frac{\partial g}{\partial q'_{j}}(p',q')
\right. \\
\quad \\
&-&
\left.
\left( 
\frac{\partial {\cal{Y}} }{\partial q'_{j}}
 (q)
+
 \frac{\partial {\cal{T}} }{\partial q'_{j}}
 (q')
+ \xi_{j}
+
\sum_{l=1}^{n+1}
\frac{\partial  {\cal{S}}_{l}}{\partial q'_{j}}
(q)p'_{l}
 +
\sum_{l=1}^{n+1}
\frac{\partial {\cal{F}}_{l} }{\partial q'_{j}}
 (q')
p'_{l} \right)\right.\\
\quad\\
&\cdot & \left.
\left( \frac{\partial f}{\partial p'_{j}}(p') +
\sum_{k\in {\mathbb{Z}}^{n}} \frac{\partial s_{k}}{\partial p'_{j}}(p')
e^{ik\cdot q}
+
\sum_{k\in {\mathbb{Z}}^{n}} \frac{\partial h_{k}}{\partial p'_{j}}(p')
e_{k}(t)e^{ik\cdot q} \right) \right]\\
\quad\\
&=& 
\sum_{j=1}^{n+1} \left[
 {\cal{S}}_{j}(q)
\frac{\partial r}{\partial q'_{j}}(p',q')
+
 {\cal{S}}_{j}(q)
\frac{\partial g}{\partial q'_{j}}(p',q')
 +
 {\cal{F}}_{j}(q')
\frac{\partial r}{\partial q'_{j}}(p',q')
+
 {\cal{F}}_{j}(q')
\frac{\partial g}{\partial q'_{j}}(p',q')
\right. \\
\quad \\
&-&
\left( 
\frac{\partial {\cal{Y}} }{\partial q'_{j}}
 (q)
\frac{\partial f}{\partial p'_{j}}(p')
+
\frac{\partial {\cal{Y}} }{\partial q'_{j}}
 (q)
\sum_{k\in {\mathbb{Z}}^{n}} \frac{\partial s_{k}}{\partial p'_{j}}(p')
e^{ik\cdot q}
+
\frac{\partial {\cal{Y}} }{\partial q'_{j}}
 (q)
\sum_{k\in {\mathbb{Z}}^{n}} \frac{\partial h_{k}}{\partial p'_{j}}(p')
e_{k}(t)e^{ik\cdot q} 
\right.
\\
\quad \\
&\quad&
+
 \frac{\partial {\cal{T}} }{\partial q'_{j}}
 (q')
\frac{\partial f}{\partial p'_{j}}(p')
+
 \frac{\partial {\cal{T}} }{\partial q'_{j}}
 (q')
\sum_{k\in {\mathbb{Z}}^{n}} \frac{\partial s_{k}}{\partial p'_{j}}(p')
e^{ik\cdot q}
+
 \frac{\partial {\cal{T}} }{\partial q'_{j}}
 (q')
\sum_{k\in {\mathbb{Z}}^{n}} \frac{\partial h_{k}}{\partial p'_{j}}(p')
e_{k}(t)e^{ik\cdot q} \\
\quad \\
&\quad&
+
\xi_{j}
\frac{\partial f}{\partial p'_{j}}(p')
+
\xi_{j}
\sum_{k\in {\mathbb{Z}}^{n}} \frac{\partial s_{k}}{\partial p'_{j}}(p')
e^{ik\cdot q}
+
\xi_{j}
\sum_{k\in {\mathbb{Z}}^{n}} \frac{\partial h_{k}}{\partial p'_{j}}(p')
e_{k}(t)e^{ik\cdot q} \\
\quad\\
&\quad&
+
\sum_{l=1}^{n+1}
\frac{\partial  {\cal{S}}_{l}}{\partial q'_{j}}
(q)p'_{l}
\frac{\partial f}{\partial p'_{j}}(p')
+
\sum_{l=1}^{n+1}
\frac{\partial  {\cal{S}}_{l}}{\partial q'_{j}}
(q)p'_{l}
\sum_{k\in {\mathbb{Z}}^{n}} \frac{\partial s_{k}}{\partial p'_{j}}(p')
e^{ik\cdot q}
+
\sum_{l=1}^{n+1}
\frac{\partial  {\cal{S}}_{l}}{\partial q'_{j}}
(q)p'_{l}
\sum_{k\in {\mathbb{Z}}^{n}} \frac{\partial h_{k}}{\partial p'_{j}}(p')
e_{k}(t)e^{ik\cdot q}\\
\quad\\
&\quad&
+
\sum_{l=1}^{n+1}
\frac{\partial {\cal{F}}_{l} }{\partial q'_{j}}
 (q')
p'_{l}
\frac{\partial f}{\partial p'_{j}}(p')
+
\sum_{l=1}^{n+1}
\frac{\partial {\cal{F}}_{l} }{\partial q'_{j}}
 (q')
p'_{l}
\sum_{k\in {\mathbb{Z}}^{n}} \frac{\partial s_{k}}{\partial p'_{j}}(p')
e^{ik\cdot q}
\left.
\left.
+
\sum_{l=1}^{n+1}
\frac{\partial {\cal{F}}_{l} }{\partial q'_{j}}
 (q')
p'_{l}
\sum_{k\in {\mathbb{Z}}^{n}} \frac{\partial h_{k}}{\partial p'_{j}}(p')
e_{k}(t)e^{ik\cdot q}
\right)
\right].
\end{eqnarray*}
We now must write out each of the nineteen expressions above
and rewrite the entire expression in $p'-q'-$exponential form.
The first term
\quad \\
\begin{eqnarray*}
\sum_{j=1}^{n+1}
 {\cal{S}}_{j}(q)
\frac{\partial r}{\partial q'_{j}}(p',q')
&=&
\sum_{j=1}^{n+1}
\left[
\sum _{k\in {\mathbb{Z}}^{n}}{\cal{S}}_{k,j}e^{ik\cdot q}
\frac{\partial }{\partial q'_{j}}\left(
\sum _{k\in {\mathbb{Z}}^{n}}
s_{k}(p')e^{ik\cdot q}\right)\right]
=
\sum_{j=1}^{n}
\left[
\sum _{k\in {\mathbb{Z}}^{n}}
{\cal{S}}_{k,j}e^{ik\cdot q}
\sum _{k\in {\mathbb{Z}}^{n}}
ik_{j}
s_{k}(p')e^{ik\cdot q}\right]\\
\quad \\
&=&
\sum_{j=1}^{n}
\left[
\sum _{k\in {\mathbb{Z}}^{n}}
\left(
\sum _{m\in {\mathbb{Z}}^{n}}
i
{\cal{S}}_{m,j}
k_{j}
s_{k}(p')e^{im\cdot q}
\right)
e^{ik\cdot q}
\right]
=
\sum _{k\in {\mathbb{Z}}^{n}}
\left(
\sum _{m\in {\mathbb{Z}}^{n}}
\sum_{j=1}^{n}
i
{\cal{S}}_{m,j}
k_{j}
s_{k}(p')e^{im\cdot q}
\right)
e^{ik\cdot q}\\
\quad\\
&=&
\sum _{w\in {\mathbb{Z}}^{n}}
\left(
\sum _{m\in {\mathbb{Z}}^{n}}
\sum_{j=1}^{n}
i
{\cal{S}}_{m,j}
k_{j}
s_{w-m}(p')
\right)
e^{iw\cdot q}.
\end{eqnarray*}
The second term
\begin{eqnarray*}
\sum_{j=1}^{n+1}
 {\cal{S}}_{j}(q)
\frac{\partial g}{\partial q'_{j}}(p',q')
&=&
\sum_{j=1}^{n+1}
\left[
\sum _{k\in {\mathbb{Z}}^{n}}{\cal{S}}_{k,j}e^{ik\cdot q}
\frac{\partial }{\partial q'_{j}}\left(
\sum _{k\in {\mathbb{Z}}^{n}}
h_{k}(p')e_{k}(t)e^{ik\cdot q}
\right)\right]
\end{eqnarray*}
\begin{eqnarray*}
&=&
\sum_{j=1}^{n}
\left[
\sum _{k\in {\mathbb{Z}}^{n}}
{\cal{S}}_{k,j}e^{ik\cdot q}
\sum _{k\in {\mathbb{Z}}^{n}}
ik_{j}
h_{k}(p')e_{k}(t)e^{ik\cdot q}
\right]
+
\sum _{k\in {\mathbb{Z}}^{n}}
{\cal{S}}_{k,n+1}e^{ik\cdot q}
\sum _{k\in {\mathbb{Z}}^{n}}
h_{k}(p')
\frac{de_{k}}{dt}(t)e^{ik\cdot q}\\
\quad \\
&=&
\sum_{j=1}^{n}
\left[
\sum _{k\in {\mathbb{Z}}^{n}}
\left(
\sum _{m\in {\mathbb{Z}}^{n}}
i{\cal{S}}_{m,j}
k_{j}
h_{k}(p')e_{k}(t)e^{im\cdot q}
\right)
e^{ik\cdot q}
\right]
+
\sum _{k\in {\mathbb{Z}}^{n}}
\left(
\sum _{m\in {\mathbb{Z}}^{n}}
{\cal{S}}_{m,n+1}
h_{k}(p')
\frac{de_{k}}{dt}(t)e^{im\cdot q}
\right)
e^{ik\cdot q}\\
\quad\\
&=&
\sum _{k\in {\mathbb{Z}}^{n}}
\left(
\sum _{m\in {\mathbb{Z}}^{n}}
\sum_{j=1}^{n}
i{\cal{S}}_{m,j}
k_{j}
h_{k}(p')e_{k}(t)e^{im\cdot q}
\right)
e^{ik\cdot q}
+
\sum _{k\in {\mathbb{Z}}^{n}}
\left(
\sum _{m\in {\mathbb{Z}}^{n}}
{\cal{S}}_{m,n+1}
h_{k}(p')
\frac{de_{k}}{dt}(t)e^{im\cdot q}
\right)
e^{ik\cdot q}\\
\quad\\
&=&
\sum _{w\in {\mathbb{Z}}^{n}}
\left(
\sum _{m\in {\mathbb{Z}}^{n}}
\sum_{j=1}^{n}
i{\cal{S}}_{m,j}
k_{j}
h_{w-m}(p')e_{w-m}(t)
\right)
e^{iw\cdot q}
+
\sum _{w\in {\mathbb{Z}}^{n}}
\left(
\sum _{m\in {\mathbb{Z}}^{n}}
{\cal{S}}_{m,n+1}
h_{w-m}(p')
\frac{de_{w-m}}{dt}(t)
\right)
e^{iw\cdot q}.
\end{eqnarray*}
The third term
\begin{eqnarray*}
\sum_{j=1}^{n+1}
 {\cal{F}}_{j}(q')
\frac{\partial r}{\partial q'_{j}}(p',q')
&=&
\sum_{j=1}^{n+1}
\left[
\sum _{k\in {\mathbb{Z}}^{n}}
{\cal{F}}_{k,j}(t)e^{ik\cdot q}
\frac{\partial }{\partial q'_{j}}
\left(
\sum _{k\in {\mathbb{Z}}^{n}}
s_{k}(p')e^{ik\cdot q}
\right)
\right]
\end{eqnarray*}
\begin{eqnarray*}
&=&
\sum_{j=1}^{n}
\left[
\sum _{k\in {\mathbb{Z}}^{n}}
{\cal{F}}_{k,j}(t)e^{ik\cdot q}
\sum _{k\in {\mathbb{Z}}^{n}}
ik_{j}
s_{k}(p')e^{ik\cdot q}
\right]
=
\sum_{j=1}^{n}
\left[
\sum _{k\in {\mathbb{Z}}^{n}}
\left(
\sum _{m\in {\mathbb{Z}}^{n}}
i{\cal{F}}_{m,j}(t)
k_{j}
s_{k}(p')e^{im\cdot q}
\right)
e^{ik\cdot q}
\right]\\
\quad\\
&=&
\sum _{k\in {\mathbb{Z}}^{n}}
\left(
\sum _{m\in {\mathbb{Z}}^{n}}
\sum_{j=1}^{n}
i{\cal{F}}_{m,j}(t)
k_{j}
s_{k}(p')e^{im\cdot q}
\right)
e^{ik\cdot q}
=
\sum _{w\in {\mathbb{Z}}^{n}}
\left(
\sum _{m\in {\mathbb{Z}}^{n}}
\sum_{j=1}^{n}
i{\cal{F}}_{m,j}(t)
(w-m)_{j}
s_{w-m}(p')
\right)
e^{iw\cdot q}
.
\end{eqnarray*}
The fourth term
\begin{eqnarray*}
\sum_{j=1}^{n+1}
 {\cal{F}}_{j}(q')
\frac{\partial g}{\partial q'_{j}}(p',q')
&=&
\sum_{j=1}^{n+1}
\left[
\sum _{k\in {\mathbb{Z}}^{n}}
{\cal{F}}_{k,j}(t)e^{ik\cdot q}
\frac{\partial }{\partial q'_{j}}
\left(
\sum _{k\in {\mathbb{Z}}^{n}}
h_{k}(p')e_{k}(t)e^{ik\cdot q}
\right)
\right]
\end{eqnarray*}
\begin{eqnarray*}
&=&
\sum_{j=1}^{n}
\left[
\sum _{k\in {\mathbb{Z}}^{n}}
{\cal{F}}_{k,j}(t)e^{ik\cdot q}
\sum _{k\in {\mathbb{Z}}^{n}}
ik_{j}
h_{k}(p')e_{k}(t)e^{ik\cdot q}
\right]
+
\sum _{k\in {\mathbb{Z}}^{n}}
{\cal{F}}_{k,n+1}(t)e^{ik\cdot q}
\sum _{k\in {\mathbb{Z}}^{n}}
h_{k}(p')
\frac{de_{k}}{dt}(t)e^{ik\cdot q}
\\
\quad\\
&=&
\sum_{j=1}^{n}
\left[
\sum _{k\in {\mathbb{Z}}^{n}}
\left(
\sum _{m\in {\mathbb{Z}}^{n}}
i{\cal{F}}_{m,j}(t)
k_{j}
h_{k}(p')e_{k}(t)e^{im\cdot q}
\right)
e^{ik\cdot q}
\right]
+
\sum _{k\in {\mathbb{Z}}^{n}}
\left(
\sum _{m\in {\mathbb{Z}}^{n}}
{\cal{F}}_{m,n+1}(t)
h_{k}(p')
\frac{de_{k}}{dt}(t)e^{im\cdot q}
\right)
e^{ik\cdot q}\\
\quad\\
&=&
\sum _{k\in {\mathbb{Z}}^{n}}
\left(
\sum _{m\in {\mathbb{Z}}^{n}}
\sum_{j=1}^{n}
i{\cal{F}}_{m,j}(t)
k_{j}
h_{k}(p')e_{k}(t)e^{im\cdot q}
\right)
e^{ik\cdot q}
+
\sum _{k\in {\mathbb{Z}}^{n}}
\left(
\sum _{m\in {\mathbb{Z}}^{n}}
{\cal{F}}_{m,n+1}(t)
h_{k}(p')
\frac{de_{k}}{dt}(t)e^{im\cdot q}
\right)
e^{ik\cdot q}\\
\quad\\
&=&
\sum _{w\in {\mathbb{Z}}^{n}}
\left(
\sum _{m\in {\mathbb{Z}}^{n}}
\sum_{j=1}^{n}
i{\cal{F}}_{m,j}(t)
(w-m)_{j}
h_{w-m}(p')e_{w-m}(t)
\right)
e^{iw\cdot q}
+
\sum _{w\in {\mathbb{Z}}^{n}}
\left(
\sum _{m\in {\mathbb{Z}}^{n}}
{\cal{F}}_{m,n+1}(t)
h_{w-m}(p')
\frac{de_{w-m}}{dt}(t)
\right)
e^{iw\cdot q}
.
\end{eqnarray*}
The fifth term
\begin{eqnarray*}
\sum_{j=1}^{n+1}
\left[
\frac{\partial {\cal{Y}} }{\partial q'_{j}}
 (q)
\frac{\partial f}{\partial p'_{j}}(p')
\right]
=
\sum_{j=1}^{n+1}
\left[
\frac{\partial }{\partial q'_{j}}
\left(
\sum _{k\in {\mathbb {Z}}^{n}} y_{k}e^{ik \cdot q}
\right)
\frac{\partial f }{\partial p'_{j}}(p')
\right]
=
\sum_{j=1}^{n}
\left[
\sum _{k\in {\mathbb {Z}}^{n}} 
ik_{j}y_{k}e^{ik \cdot q}
\frac{\partial f }{\partial p'_{j}}(p')
\right]
=
\sum _{k\in {\mathbb {Z}}^{n}} 
\left(
\sum_{j=1}^{n}
ik_{j}y_{k}
\frac{\partial f }{\partial p'_{j}}(p')
\right)
e^{ik \cdot q}.
\end{eqnarray*}
The sixth term
\begin{eqnarray*}
\sum_{j=1}^{n+1}
\left[
\frac{\partial {\cal{Y}} }{\partial q'_{j}}
 (q)
\sum_{k\in {\mathbb{Z}}^{n}} \frac{\partial s_{k}}{\partial p'_{j}}(p')
e^{ik\cdot q}
\right]
&=&
\sum_{j=1}^{n+1}
\left[
\frac{\partial }{\partial q'_{j}}
\left(
\sum _{k\in {\mathbb {Z}}^{n}} y_{k}e^{ik \cdot q}
\right)
\sum_{k\in {\mathbb{Z}}^{n}} \frac{\partial s_{k}}{\partial p'_{j}}(p')
e^{ik\cdot q}
\right]
\end{eqnarray*}
\begin{eqnarray*}
&=&
\sum_{j=1}^{n}
\left[
\sum _{k\in {\mathbb {Z}}^{n}} 
ik_{j}y_{k}e^{ik \cdot q}
\sum_{k\in {\mathbb{Z}}^{n}}
 \frac{\partial s_{k}}{\partial p'_{j}}(p')
e^{ik\cdot q}
\right]
=
\sum_{j=1}^{n}
\left[
\sum _{k\in {\mathbb {Z}}^{n}} 
\left(
\sum _{m\in {\mathbb {Z}}^{n}} 
im_{j}y_{m}
 \frac{\partial s_{k}}{\partial p'_{j}}(p')e^{im\cdot q}
\right)
e^{ik\cdot q}
\right]\\
\quad \\
&=&
\sum _{k\in {\mathbb {Z}}^{n}} 
\left(
\sum _{m\in {\mathbb {Z}}^{n}} 
\sum_{j=1}^{n}
im_{j}y_{m}
 \frac{\partial s_{k}}{\partial p'_{j}}(p')e^{im\cdot q}
\right)
e^{ik\cdot q}
=
\sum _{w\in {\mathbb {Z}}^{n}} 
\left(
\sum _{m\in {\mathbb {Z}}^{n}} 
\sum_{j=1}^{n}
im_{j}y_{m}
 \frac{\partial s_{w-m}}{\partial p'_{j}}(p')
\right)
e^{iw\cdot q}
.
\end{eqnarray*}
The seventh term
\begin{eqnarray*}
\sum_{j=1}^{n+1}
\left[
\frac{\partial {\cal{Y}} }{\partial q'_{j}}
 (q)
\sum_{k\in {\mathbb{Z}}^{n}} \frac{\partial h_{k}}{\partial p'_{j}}(p')
e_{k}(t)e^{ik\cdot q} 
\right]
&=&
\sum_{j=1}^{n+1}
\left[
\frac{\partial }{\partial q'_{j}}
\left(
\sum _{k\in {\mathbb {Z}}^{n}} y_{k}e^{ik \cdot q}
\right)
\sum_{k\in {\mathbb{Z}}^{n}} \frac{\partial h_{k}}{\partial p'_{j}}(p')
e_{k}(t)e^{ik\cdot q} 
\right]
\end{eqnarray*}
\begin{eqnarray*}
&=&
\sum_{j=1}^{n}
\left[
\sum _{k\in {\mathbb {Z}}^{n}} 
ik_{j}y_{k}e^{ik \cdot q}
\sum_{k\in {\mathbb{Z}}^{n}}
 \frac{\partial h_{k}}{\partial p'_{j}}(p')
e_{k}(t)e^{ik\cdot q} 
\right]
=
\sum_{j=1}^{n}
\left[
\sum _{k\in {\mathbb {Z}}^{n}} 
\left(
\sum _{m\in {\mathbb {Z}}^{n}} 
im_{j}y_{m}
 \frac{\partial h_{k}}{\partial p'_{j}}(p')
e_{k}(t)e^{im\cdot q}
\right)
e^{ik\cdot q} 
\right]
\\
\quad\\
&=&
\sum _{k\in {\mathbb {Z}}^{n}} 
\left(
\sum _{m\in {\mathbb {Z}}^{n}} 
\sum_{j=1}^{n}
im_{j}y_{m}
 \frac{\partial h_{k}}{\partial p'_{j}}(p')
e_{k}(t)e^{im\cdot q}
\right)
e^{ik\cdot q}
=
\sum _{w\in {\mathbb {Z}}^{n}} 
\left(
\sum _{m\in {\mathbb {Z}}^{n}} 
\sum_{j=1}^{n}
im_{j}y_{m}
 \frac{\partial h_{w-m}}{\partial p'_{j}}(p')
e_{w-m}(t)
\right)
e^{iw\cdot q}
.
\end{eqnarray*}
The eight term
\begin{eqnarray*}
\sum_{j=1}^{n+1}
\left[
 \frac{\partial {\cal{T}} }{\partial q'_{j}}
 (q')
\frac{\partial f}{\partial p'_{j}}(p')
\right]
&=&
\sum_{j=1}^{n+1}
\left[
 \frac{\partial }{\partial q'_{j}}
\left(
\sum _{k\in {\mathbb {Z}}^{n}} x_{k}(t)e^{ik \cdot q}
\right)
\frac{\partial f}{\partial p'_{j}}(p')
\right]
\end{eqnarray*}
\begin{eqnarray*}
&=&
\sum_{j=1}^{n}
\left[
\sum _{k\in {\mathbb {Z}}^{n}}
ik_{j} x_{k}(t)e^{ik \cdot q}
\frac{\partial f}{\partial p'_{j}}(p')
\right]
+
\sum _{k\in {\mathbb {Z}}^{n}}
\frac{dx_{k}}{dt}(t)
e^{ik \cdot q}
\frac{\partial f}{\partial p'_{n+1}}(p')
\\
\quad\\
&=&
\sum _{k\in {\mathbb {Z}}^{n}}
\left(
\sum_{j=1}^{n}
ik_{j} x_{k}(t)
\frac{\partial f}{\partial p'_{j}}(p')
\right)
e^{ik \cdot q}
+
\sum _{k\in {\mathbb {Z}}^{n}}
\left(
\frac{dx_{k}}{dt}(t)
\frac{\partial f}{\partial p'_{n+1}}(p')
\right)
e^{ik \cdot q}
.
\end{eqnarray*}
The ninth term
\begin{eqnarray*}
\sum_{j=1}^{n+1}
\left[
 \frac{\partial {\cal{T}} }{\partial q'_{j}}
 (q')
\sum_{k\in {\mathbb{Z}}^{n}} \frac{\partial s_{k}}{\partial p'_{j}}(p')
e^{ik\cdot q}
\right]
&=&
\sum_{j=1}^{n+1}
\left[
 \frac{\partial }{\partial q'_{j}}
\left(
\sum _{k\in {\mathbb {Z}}^{n}}
 x_{k}(t)e^{ik \cdot q}
\right)
\sum_{k\in {\mathbb{Z}}^{n}}
 \frac{\partial s_{k}}{\partial p'_{j}}(p')
e^{ik\cdot q}
\right]
\end{eqnarray*}
\begin{eqnarray*}
&=&
\sum_{j=1}^{n}
\left[
\sum _{k\in {\mathbb {Z}}^{n}}
ik_{j} x_{k}(t)e^{ik \cdot q}
\sum_{k\in {\mathbb{Z}}^{n}}
 \frac{\partial s_{k}}{\partial p'_{j}}(p')
e^{ik\cdot q}
\right]
+
\sum _{k\in {\mathbb {Z}}^{n}}
\frac{dx_{k}}{dt}(t)
e^{ik \cdot q}
\sum_{k\in {\mathbb{Z}}^{n}}
 \frac{\partial s_{k}}{\partial p'_{n+1}}(p')
e^{ik\cdot q}
\\
\quad\\
&=&
\sum_{j=1}^{n}
\left[
\sum _{k\in {\mathbb {Z}}^{n}}
\left(
\sum _{m\in {\mathbb {Z}}^{n}}
im_{j} x_{m}(t)
 \frac{\partial s_{k}}{\partial p'_{j}}(p')e^{im\cdot q}
\right)
e^{ik \cdot q}
\right]
+
\sum _{k\in {\mathbb {Z}}^{n}}
\left(
\sum _{m\in {\mathbb {Z}}^{n}}
\frac{dx_{m}}{dt}(t)
 \frac{\partial s_{k}}{\partial p'_{n+1}}(p')e^{im\cdot q}
\right)
e^{ik\cdot q}
\\
\quad\\
&=&
\sum _{k\in {\mathbb {Z}}^{n}}
\left(
\sum _{m\in {\mathbb {Z}}^{n}}
\sum_{j=1}^{n}
im_{j} x_{m}(t)
 \frac{\partial s_{k}}{\partial p'_{j}}(p')e^{im\cdot q}
\right)
e^{ik \cdot q}
+
\sum _{k\in {\mathbb {Z}}^{n}}
\left(
\sum _{m\in {\mathbb {Z}}^{n}}
\frac{dx_{m}}{dt}(t)
 \frac{\partial s_{k}}{\partial p'_{n+1}}(p')e^{im\cdot q}
\right)
e^{ik\cdot q}
\\
\quad\\
&=&
\sum _{w\in {\mathbb {Z}}^{n}}
\left(
\sum _{m\in {\mathbb {Z}}^{n}}
\sum_{j=1}^{n}
im_{j} x_{m}(t)
 \frac{\partial s_{w-m}}{\partial p'_{j}}(p')
\right)
e^{iw \cdot q}
+
\sum _{w\in {\mathbb {Z}}^{n}}
\left(
\sum _{m\in {\mathbb {Z}}^{n}}
\frac{dx_{m}}{dt}(t)
 \frac{\partial s_{w-m}}{\partial p'_{n+1}}(p')
\right)
e^{iw\cdot q}
.
\end{eqnarray*}
The tenth term
\begin{eqnarray*}
\sum_{j=1}^{n+1}
\left[
 \frac{\partial {\cal{T}} }{\partial q'_{j}}
 (q')
\sum_{k\in {\mathbb{Z}}^{n}}
 \frac{\partial h_{k}}{\partial p'_{j}}(p')
e_{k}(t)e^{ik\cdot q}
\right]
&=&
\sum_{j=1}^{n+1}
\left[
 \frac{\partial }{\partial q'_{j}}
\left(
\sum _{k\in {\mathbb {Z}}^{n}}
 x_{k}(t)e^{ik \cdot q}
\right)
\sum_{k\in {\mathbb{Z}}^{n}}
 \frac{\partial h_{k}}{\partial p'_{j}}(p')
e_{k}(t)e^{ik\cdot q}
\right]
\end{eqnarray*}
\begin{eqnarray*}
&=&
\sum_{j=1}^{n}
\left[
\sum _{k\in {\mathbb {Z}}^{n}}
ik_{j} x_{k}(t)e^{ik \cdot q}
\sum_{k\in {\mathbb{Z}}^{n}}
 \frac{\partial h_{k}}{\partial p'_{j}}(p')
e_{k}(t)e^{ik\cdot q}
\right]
\\
\quad\\
&+&
\sum _{k\in {\mathbb {Z}}^{n}}
\frac{dx_{k}}{dt}(t)
e^{ik \cdot q}
\sum_{k\in {\mathbb{Z}}^{n}}
 \frac{\partial h_{k}}{\partial p'_{n+1}}(p')
e_{k}(t)e^{ik\cdot q}\\
\quad\\
&=&
\sum_{j=1}^{n}
\left[
\sum _{k\in {\mathbb {Z}}^{n}}
\left(
\sum _{m\in {\mathbb {Z}}^{n}}
im_{j} x_{m}(t)
 \frac{\partial h_{k}}{\partial p'_{j}}(p')
e_{k}(t)e^{im\cdot q}
\right)
e^{ik \cdot q}
\right]
\\
\quad\\
&+&
\sum _{k\in {\mathbb {Z}}^{n}}
\left(
\sum _{m\in {\mathbb {Z}}^{n}}
\frac{dx_{m}}{dt}(t)
 \frac{\partial h_{k}}{\partial p'_{n+1}}(p')
e_{k}(t)e^{im\cdot q}
\right)
e^{ik \cdot q}
\\
\quad\\
&=&
\sum _{k\in {\mathbb {Z}}^{n}}
\left(
\sum _{m\in {\mathbb {Z}}^{n}}
\sum_{j=1}^{n}
im_{j} x_{m}(t)
 \frac{\partial h_{k}}{\partial p'_{j}}(p')
e_{k}(t)e^{im\cdot q}
\right)
e^{ik \cdot q}
\\
\quad\\
&+&
\sum _{k\in {\mathbb {Z}}^{n}}
\left(
\sum _{m\in {\mathbb {Z}}^{n}}
\frac{dx_{m}}{dt}(t)
 \frac{\partial h_{k}}{\partial p'_{n+1}}(p')
e_{k}(t)e^{im\cdot q}
\right)
e^{ik \cdot q}
\\
\quad\\
&=&
\sum _{w\in {\mathbb {Z}}^{n}}
\left(
\sum _{m\in {\mathbb {Z}}^{n}}
\sum_{j=1}^{n}
im_{j} x_{m}(t)
 \frac{\partial h_{w-m}}{\partial p'_{j}}(p')
e_{w-m}(t)
\right)
e^{iw \cdot q}
\\
\quad\\
&+&
\sum _{w\in {\mathbb {Z}}^{n}}
\left(
\sum _{m\in {\mathbb {Z}}^{n}}
\frac{dx_{m}}{dt}(t)
 \frac{\partial h_{w-m}}{\partial p'_{n+1}}(p')
e_{w-m}(t)
\right)
e^{iw \cdot q}
.
\end{eqnarray*}
The eleventh term
\begin{eqnarray*}
\sum_{j=1}^{n+1}
\left[
\xi_{j}
\frac{\partial f}{\partial p'_{j}}(p')
\right].
\end{eqnarray*}
The twelve-th term
\begin{eqnarray*}
\sum_{j=1}^{n+1}
\left[
\xi_{j}
\sum_{k\in {\mathbb{Z}}^{n}}
 \frac{\partial s_{k}}{\partial p'_{j}}(p')
e^{ik\cdot q}
\right]
=
\sum_{k\in {\mathbb{Z}}^{n}}
\left(
\sum_{j=1}^{n+1}
\xi_{j}
 \frac{\partial s_{k}}{\partial p'_{j}}(p')
\right)
e^{ik\cdot q}.
\end{eqnarray*}
The thirteenth term
\quad\\
\begin{eqnarray*}
\sum_{j=1}^{n+1}
\left[
\xi_{j}
\sum_{k\in {\mathbb{Z}}^{n}} 
\frac{\partial h_{k}}{\partial p'_{j}}(p')
e_{k}(t)e^{ik\cdot q}
\right]
=
\sum_{k\in {\mathbb{Z}}^{n}} 
\left(
\sum_{j=1}^{n+1}
\xi_{j}
\frac{\partial h_{k}}{\partial p'_{j}}(p')
e_{k}(t)
\right)
e^{ik\cdot q}.
\end{eqnarray*}
The fourteenth term
\quad\\
\begin{eqnarray*}
\sum_{j=1}^{n+1}
\left[
\sum_{l=1}^{n+1}
\frac{\partial  {\cal{S}}_{l}}{\partial q'_{j}}
(q)p'_{l}
\frac{\partial f}{\partial p'_{j}}(p')
\right]
&=&
\sum_{j=1}^{n+1}
\left[
\sum_{l=1}^{n+1}
\frac{\partial }{\partial q'_{j}}
\left(
\sum _{k\in {\mathbb{Z}}^{n}}
{\cal{S}}_{k,l}e^{ik\cdot q}
\right)
p'_{l}
\frac{\partial f}{\partial p'_{j}}(p')
\right]
\\
\quad\\
&=&
\sum_{j=1}^{n}
\left[
\sum_{l=1}^{n+1}
\sum _{k\in {\mathbb{Z}}^{n}}
ik_{j}
{\cal{S}}_{k,l}e^{ik\cdot q}
p'_{l}
\frac{\partial f}{\partial p'_{j}}(p')
\right]\\
\quad\\
&=&
\sum _{k\in {\mathbb{Z}}^{n}}
\left(
\sum_{j=1}^{n}
\sum_{l=1}^{n+1}
ik_{j}
{\cal{S}}_{k,l}
p'_{l}
\frac{\partial f}{\partial p'_{j}}(p')
\right)
e^{ik\cdot q}.
\end{eqnarray*}
The fifthteenth term
\begin{eqnarray*}
\sum_{j=1}^{n+1}
\left[
\sum_{l=1}^{n+1}
\frac{\partial  {\cal{S}}_{l}}{\partial q'_{j}}
(q)p'_{l}
\sum_{k\in {\mathbb{Z}}^{n}} \frac{\partial s_{k}}{\partial p'_{j}}(p')
e^{ik\cdot q}
\right]
&=&
\sum_{j=1}^{n+1}
\left[
\sum_{l=1}^{n+1}
\frac{\partial }{\partial q'_{j}}
\left(
\sum _{k\in {\mathbb{Z}}^{n}}
{\cal{S}}_{k,l}e^{ik\cdot q}
\right)
p'_{l}
\sum_{k\in {\mathbb{Z}}^{n}} \frac{\partial s_{k}}{\partial p'_{j}}(p')
e^{ik\cdot q}
\right]
\end{eqnarray*}
\begin{eqnarray*}
&=&
\sum_{j=1}^{n}
\left[
\sum_{l=1}^{n+1}
\sum _{k\in {\mathbb{Z}}^{n}}
ik_{j}
{\cal{S}}_{k,l}e^{ik\cdot q}
p'_{l}
\sum_{k\in {\mathbb{Z}}^{n}}
 \frac{\partial s_{k}}{\partial p'_{j}}(p')
e^{ik\cdot q}
\right]\\
\quad\\
&=&
\sum_{j=1}^{n}
\left[
\sum_{l=1}^{n+1}
\left(
\sum_{k\in {\mathbb{Z}}^{n}}
\left(
\sum_{m\in {\mathbb{Z}}^{n}}
im_{j}
{\cal{S}}_{m,l}
p'_{l}
 \frac{\partial s_{k}}{\partial p'_{j}}(p')e^{im\cdot q}
\right)
e^{ik\cdot q}
\right)\right]
\\
\quad\\
&=&
\sum_{k\in {\mathbb{Z}}^{n}}
\left[
\sum_{j=1}^{n}
\sum_{l=1}^{n+1}
\left(
\sum_{m\in {\mathbb{Z}}^{n}}
im_{j}
{\cal{S}}_{m,l}
p'_{l}
 \frac{\partial s_{k}}{\partial p'_{j}}(p')e^{im\cdot q}
\right)
\right]
e^{ik\cdot q}
\\
\quad\\
&=&
\sum_{w\in {\mathbb{Z}}^{n}}
\left[
\sum_{j=1}^{n}
\sum_{l=1}^{n+1}
\left(
\sum_{m\in {\mathbb{Z}}^{n}}
im_{j}
{\cal{S}}_{m,l}
p'_{l}
 \frac{\partial s_{w-m}}{\partial p'_{j}}(p')
\right)
\right]
e^{iw\cdot q}.
\end{eqnarray*}
The sixteenth term
\quad\\
\begin{eqnarray*}
\sum_{j=1}^{n+1}
\left[
\sum_{l=1}^{n+1}
\frac{\partial  {\cal{S}}_{l}}{\partial q'_{j}}
(q)p'_{l}
\sum_{k\in {\mathbb{Z}}^{n}} \frac{\partial h_{k}}{\partial p'_{j}}(p')
e_{k}(t)e^{ik\cdot q}
\right]
&=&
\sum_{j=1}^{n+1}
\left[
\sum_{l=1}^{n+1}
\frac{\partial }{\partial q'_{j}}
\left(
\sum _{k\in {\mathbb{Z}}^{n}}
{\cal{S}}_{k,l}e^{ik\cdot q}
\right)
p'_{l}
\sum_{k\in {\mathbb{Z}}^{n}} \frac{\partial h_{k}}{\partial p'_{j}}(p')
e_{k}(t)e^{ik\cdot q}
\right]
\end{eqnarray*}
\begin{eqnarray*}
&=&
\sum_{j=1}^{n}
\left[
\sum_{l=1}^{n+1}
\sum _{k\in {\mathbb{Z}}^{n}}
ik_{j}
{\cal{S}}_{k,l}e^{ik\cdot q}
p'_{l}
\sum_{k\in {\mathbb{Z}}^{n}}
 \frac{\partial h_{k}}{\partial p'_{j}}(p')
e_{k}(t)e^{ik\cdot q}
\right]\\
\quad\\
&=&
\sum_{j=1}^{n}
\left[
\sum_{l=1}^{n+1}
\left(
\sum _{k\in {\mathbb{Z}}^{n}}
\left(
\sum _{m\in {\mathbb{Z}}^{n}}
im_{j}
{\cal{S}}_{m,l}
p'_{l}
 \frac{\partial h_{k}}{\partial p'_{j}}(p')
e_{k}(t)e^{im\cdot q}
\right)
e^{ik\cdot q}\right)
\right]\\
\quad\\
&=&
\sum _{k\in {\mathbb{Z}}^{n}}
\left[
\sum_{j=1}^{n}
\sum_{l=1}^{n+1}
\left(
\sum _{m\in {\mathbb{Z}}^{n}}
im_{j}
{\cal{S}}_{m,l}
p'_{l}
 \frac{\partial h_{k}}{\partial p'_{j}}(p')
e_{k}(t)e^{im\cdot q}
\right)
\right]
e^{ik\cdot q}
\\
\quad\\
&=&
\sum _{w\in {\mathbb{Z}}^{n}}
\left[
\sum_{j=1}^{n}
\sum_{l=1}^{n+1}
\left(
\sum _{m\in {\mathbb{Z}}^{n}}
im_{j}
{\cal{S}}_{m,l}
p'_{l}
 \frac{\partial h_{w-m}}{\partial p'_{j}}(p')
e_{w-m}(t)
\right)
\right]
e^{iw\cdot q}
.
\end{eqnarray*}
The seventeenth term
\quad\\
\begin{eqnarray*}
\sum_{j=1}^{n+1}
\left[
\sum_{l=1}^{n+1}
\frac{\partial {\cal{F}}_{l} }{\partial q'_{j}}
 (q')
p'_{l}
\frac{\partial f}{\partial p'_{j}}(p')
\right]
&=&
\sum_{j=1}^{n+1}
\left[
\sum_{l=1}^{n+1}
\frac{\partial  }{\partial q'_{j}}
\left(
\sum _{k\in {\mathbb{Z}}^{n}}
{\cal{F}}_{k,l}(t)e^{ik\cdot q}
\right)
p'_{l}
\frac{\partial f}{\partial p'_{j}}(p')
\right]
\end{eqnarray*}
\begin{eqnarray*}
&=&
\sum_{j=1}^{n}
\left[
\sum_{l=1}^{n+1}
\sum _{k\in {\mathbb{Z}}^{n}}
ik_{j}
{\cal{F}}_{k,l}(t)e^{ik\cdot q}
p'_{l}
\frac{\partial f}{\partial p'_{j}}(p')
\right]\\
\quad\\
&+&
\sum_{l=1}^{n+1}
\sum _{k\in {\mathbb{Z}}^{n}}
\frac{d{\cal{F}}_{k,l}}{dt}(t)
e^{ik\cdot q}
p'_{l}
\frac{\partial f}{\partial p'_{n+1}}(p')
\\
\quad\\
&=&
\sum _{k\in {\mathbb{Z}}^{n}}
\left[
\sum_{j=1}^{n}
\sum_{l=1}^{n+1}
ik_{j}
p'_{l}
\frac{\partial f}{\partial p'_{j}}(p')
{\cal{F}}_{k,l}(t)
\right]
e^{ik\cdot q}\\
\quad\\
&+&
\sum _{k\in {\mathbb{Z}}^{n}}
\left(
\sum_{l=1}^{n+1}
p'_{l}
\frac{\partial f}{\partial p'_{n+1}}(p')
\frac{d{\cal{F}}_{k,l}}{dt}(t)
\right)
e^{ik\cdot q}.
\end{eqnarray*}
The eighteenth term
\begin{eqnarray*}
\sum_{j=1}^{n+1}
\left[
\sum_{l=1}^{n+1}
\frac{\partial {\cal{F}}_{l} }{\partial q'_{j}}
 (q')
p'_{l}
\sum_{k\in {\mathbb{Z}}^{n}} \frac{\partial s_{k}}{\partial p'_{j}}(p')
e^{ik\cdot q}
\right]
\end{eqnarray*}
\begin{eqnarray*}
&=&
\sum_{j=1}^{n+1}
\left[
\sum_{l=1}^{n+1}
\frac{\partial  }{\partial q'_{j}}
\left(
\sum _{k\in {\mathbb{Z}}^{n}}
{\cal{F}}_{k,l}(t)e^{ik\cdot q}
\right)
p'_{l}
\sum_{k\in {\mathbb{Z}}^{n}} \frac{\partial s_{k}}{\partial p'_{j}}(p')
e^{ik\cdot q}
\right]
\\
\quad\\
&=&
\sum_{j=1}^{n}
\left[
\sum_{l=1}^{n+1}
\sum _{k\in {\mathbb{Z}}^{n}}
ik_{j}
{\cal{F}}_{k,l}(t)e^{ik\cdot q}
p'_{l}
\sum_{k\in {\mathbb{Z}}^{n}}
 \frac{\partial s_{k}}{\partial p'_{j}}(p')
e^{ik\cdot q}
\right]\\
\quad\\
&+&
\sum_{l=1}^{n+1}
\sum _{k\in {\mathbb{Z}}^{n}}
\frac{d{\cal{F}}_{k,l}}{dt}(t)
e^{ik\cdot q}
p'_{l}
\sum_{k\in {\mathbb{Z}}^{n}}
 \frac{\partial s_{k}}{\partial p'_{n+1}}(p')
e^{ik\cdot q}\\
\quad\\
&=&
\sum_{j=1}^{n}
\left[
\sum_{l=1}^{n+1}
\sum _{k\in {\mathbb{Z}}^{n}}
\left(
\sum _{m\in {\mathbb{Z}}^{n}}
im_{j}
{\cal{F}}_{m,l}(t)
p'_{l}
\frac{\partial s_{k}}{\partial p'_{j}}(p')e^{im\cdot q}
\right)
e^{ik\cdot q}
\right]
\\
\quad\\
&+&
\sum_{l=1}^{n+1}
\sum _{k\in {\mathbb{Z}}^{n}}
\left(
\sum _{m\in {\mathbb{Z}}^{n}}
\frac{d{\cal{F}}_{m,l}}{dt}(t)
p'_{l}
 \frac{\partial s_{k}}{\partial p'_{n+1}}(p')e^{im\cdot q}
\right)
e^{ik\cdot q}\\
\quad\\
&=&
\sum _{k\in {\mathbb{Z}}^{n}}
\left[
\sum_{j=1}^{n}
\sum_{l=1}^{n+1}
\left(
\sum _{m\in {\mathbb{Z}}^{n}}
im_{j}
{\cal{F}}_{m,l}(t)
p'_{l}
\frac{\partial s_{k}}{\partial p'_{j}}(p')e^{im\cdot q}
\right)
\right]
e^{ik\cdot q}\\
\quad\\
&+&
\sum _{k\in {\mathbb{Z}}^{n}}
\left(
\sum _{m\in {\mathbb{Z}}^{n}}
\sum_{l=1}^{n+1}
\frac{d{\cal{F}}_{m,l}}{dt}(t)
p'_{l}
 \frac{\partial s_{k}}{\partial p'_{n+1}}(p')e^{im\cdot q}
\right)
e^{ik\cdot q}
\\
\quad\\
&=&
\sum _{w\in {\mathbb{Z}}^{n}}
\left[
\sum_{j=1}^{n}
\sum_{l=1}^{n+1}
\left(
\sum _{m\in {\mathbb{Z}}^{n}}
im_{j}
{\cal{F}}_{m,l}(t)
p'_{l}
\frac{\partial s_{w-m}}{\partial p'_{j}}(p')
\right)
\right]
e^{iw\cdot q}\\
\quad\\
&+&
\sum _{w\in {\mathbb{Z}}^{n}}
\left(
\sum _{m\in {\mathbb{Z}}^{n}}
\sum_{l=1}^{n+1}
\frac{d{\cal{F}}_{m,l}}{dt}(t)
p'_{l}
 \frac{\partial s_{w-m}}{\partial p'_{n+1}}(p')
\right)
e^{iw\cdot q}
.
\end{eqnarray*}
The nineteenth term
\begin{eqnarray*}
\sum_{j=1}^{n+1}
\left[
\sum_{l=1}^{n+1}
\frac{\partial {\cal{F}}_{l} }{\partial q'_{j}}
 (q')
p'_{l}
\sum_{k\in {\mathbb{Z}}^{n}} \frac{\partial h_{k}}{\partial p'_{j}}(p')
e_{k}(t)e^{ik\cdot q}
\right]
\end{eqnarray*}
\begin{eqnarray*}
&=&
\sum_{j=1}^{n+1}
\left[
\sum_{l=1}^{n+1}
\frac{\partial  }{\partial q'_{j}}
\left(
\sum _{k\in {\mathbb{Z}}^{n}}
{\cal{F}}_{k,l}(t)e^{ik\cdot q}
\right)
p'_{l}
\sum_{k\in {\mathbb{Z}}^{n}} \frac{\partial h_{k}}{\partial p'_{j}}(p')e_{k}(t)
e^{ik\cdot q}
\right]\\
\quad\\
&=&
\sum_{j=1}^{n}
\left[
\sum_{l=1}^{n+1}
\sum _{k\in {\mathbb{Z}}^{n}}
ik_{j}
{\cal{F}}_{k,l}(t)e^{ik\cdot q}
p'_{l}
\sum_{k\in {\mathbb{Z}}^{n}}
 \frac{\partial h_{k}}{\partial p'_{j}}(p')e_{k}(t)
e^{ik\cdot q}
\right]\\
\quad\\
&+&
\sum_{l=1}^{n+1}
\sum _{k\in {\mathbb{Z}}^{n}}
\frac{d{\cal{F}}_{k,l}}{dt}(t)
e^{ik\cdot q}
p'_{l}
\sum_{k\in {\mathbb{Z}}^{n}}
 \frac{\partial h_{k}}{\partial p'_{n+1}}(p')e_{k}(t)
e^{ik\cdot q}\\
\quad\\
&=&
\sum_{j=1}^{n}
\left[
\sum_{l=1}^{n+1}
\sum _{k\in {\mathbb{Z}}^{n}}
\left(
\sum _{m\in {\mathbb{Z}}^{n}}
im_{j}
{\cal{F}}_{m,l}(t)
p'_{l}
\frac{\partial h_{k}}{\partial p'_{j}}(p')e_{k}(t)e^{im\cdot q}
\right)
e^{ik\cdot q}
\right]
\\
\quad\\
&+&
\sum_{l=1}^{n+1}
\sum _{k\in {\mathbb{Z}}^{n}}
\left(
\sum _{m\in {\mathbb{Z}}^{n}}
\frac{d{\cal{F}}_{m,l}}{dt}(t)
p'_{l}
 \frac{\partial h_{k}}{\partial p'_{n+1}}(p')e_{k}(t)e^{im\cdot q}
\right)
e^{ik\cdot q}\\
\quad\\
&=&
\sum _{k\in {\mathbb{Z}}^{n}}
\left[
\sum_{j=1}^{n}
\sum_{l=1}^{n+1}
\left(
\sum _{m\in {\mathbb{Z}}^{n}}
im_{j}
{\cal{F}}_{m,l}(t)
p'_{l}
\frac{\partial h_{k}}{\partial p'_{j}}(p')e_{k}(t)e^{im\cdot q}
\right)
\right]
e^{ik\cdot q}\\
\quad\\
&+&
\sum _{k\in {\mathbb{Z}}^{n}}
\left(
\sum _{m\in {\mathbb{Z}}^{n}}
\sum_{l=1}^{n+1}
\frac{d{\cal{F}}_{m,l}}{dt}(t)
p'_{l}
 \frac{\partial h_{k}}{\partial p'_{n+1}}(p')e_{k}(t)e^{im\cdot q}
\right)
e^{ik\cdot q}
\\
\quad\\
&=&
\sum _{w\in {\mathbb{Z}}^{n}}
\left[
\sum_{j=1}^{n}
\sum_{l=1}^{n+1}
\left(
\sum _{m\in {\mathbb{Z}}^{n}}
im_{j}
{\cal{F}}_{m,l}(t)
p'_{l}
\frac{\partial h_{w-m}}{\partial p'_{j}}(p')e_{w-m}(t)
\right)
\right]
e^{iw\cdot q}\\
\quad\\
&+&
\sum _{w\in {\mathbb{Z}}^{n}}
\left(
\sum _{m\in {\mathbb{Z}}^{n}}
\sum_{l=1}^{n+1}
\frac{d{\cal{F}}_{m,l}}{dt}(t)
p'_{l}
 \frac{\partial h_{w-m}}{\partial p'_{n+1}}(p')e_{w-m}(t)
\right)
e^{iw\cdot q}
.
\end{eqnarray*}
All the nineteen terms have one of three forms.
Terms one, five, six, twelve, fourteen, and fifteen
have the form
\begin{eqnarray*}
\sum_{k\in {\mathbb{Z}}^{n}}
f_{k}(p')e^{ik\cdot q}.
\end{eqnarray*}
Terms two, three , four, seven, eight, nine, ten, thirteen, sixteen, 
seventeen, eighteen and nineteen have the form
\begin{eqnarray*}
\sum_{k\in {\mathbb{Z}}^{n}}
g_{k}(p')h_{k}(t)
e^{ik\cdot q}.
\end{eqnarray*}
Term eleven is simply a function of $p'$, say $L(p')$.
Adding all terms together we see the final result,
$\{ \chi,F \}$, is of 
$p'-q'-$exponential-form.
Finally we have have the following analyticity domains
\quad \\
\begin{eqnarray*}
&\quad& Y_{j}(q')\in {\cal{A}}_{\rho ', \sigma '},
\quad \frac{\partial r}{\partial q'_{j}}(p',q')\in {\cal{A}}_{\rho -\delta , \sigma },
\quad \frac{\partial g}{\partial q'_{j}}(p',q')\in {\cal{A}}_{\rho, \sigma -\delta},
\quad \frac{\partial X}{\partial q'_{j}}(q')\in {\cal{A}}_{\rho ', \sigma '-\delta},\\
\quad \\
&\quad& \sum_{l=1}^{n+1}\frac{\partial Y_{l}}{\partial q'_{j}}(q')p'_{l}
\in {\cal{A}}_{\rho ', \sigma ' -\delta},
\quad \frac{\partial f}{\partial p'_{j}}(p')\in{\cal{A}}_{\rho,\sigma},
\quad \sum_{k\in {\mathbb{Z}}^{n}} \frac{\partial s_{k}}{\partial p'_{j}}(p')
e^{ik\cdot q}\in {\cal{A}}_{\rho, \sigma },\\
\quad \\
&\quad& \sum_{k\in {\mathbb{Z}}^{n}} \frac{\partial h_{k}}{\partial p'_{j}}(p')
e_{k}(t)e^{ik\cdot q}\in {\cal{A}}_{\rho, \sigma }
\end{eqnarray*}
for the functions that make up 
the expression for $\{ \chi , F\}(p',q')$.
Since these functions are multiplied together in a number
of ways and then summed  it follows, given $\rho '< \rho$
and $\sigma '
<\sigma$,
$\{ \chi , F\}(p',q')\in {\cal{A}}_{\rho '-\delta , \sigma ' -\delta}$ and
is of $p'-q'-$exponential form.
\quad \squareforqed


\end{lem}
\begin{lem}
\label{lemmaRA}
\quad\\
Let $H(p',q')=U(p',q')+P(p',q')$ with $H$ and $P$ being
of $p',q'$-exponential form.
Define
$
{\cal{R}}_{A}=
r_{2}(U,\chi ,1)+ r_{1}(P,\chi ,1)$,
where
\begin{eqnarray*}
r_{m}(H,\chi ,t) = {\cal{U}}H -
\sum ^{m-1}_{l=0} \frac{t^{l}}{l!}L^{l}_{\chi}H =
\sum ^{\infty}_{l=m} \frac{t^{l}}{l!}L^{l}_{\chi}H,
\end{eqnarray*}
$L^{0}_{\chi}H=H$ and $L^{m}_{\chi}H=\{L^{m-1}_{\chi}H,H\}$
for $m\geq 1$.
${\cal{R}}_{A}$ is of $p',q'$exponential form.\\
\quad\\
{Proof}:\\
\quad\\
We have
\begin{eqnarray*}
{\cal{R}}_{A}=
 \{ \chi ,P \} +
{\cal {U}}H - H- \{ \chi , H \}=
r_{2}(U,\chi ,1)+ r_{1}(P,\chi ,1),
\end{eqnarray*}
where
\begin{eqnarray*}
r_{m}(H,\chi ,t) = {\cal{U}}H -
\sum ^{m-1}_{l=0} \frac{t^{l}}{l!}L^{l}_{\chi}H =
\sum ^{\infty}_{l=m} \frac{t^{l}}{l!}L^{l}_{\chi}H.
\end{eqnarray*}
We thus have
\begin{eqnarray*}
{\cal{R}}_{A}=
 \{ \chi ,P \} +
r_{2}(H,\chi ,1).
\end{eqnarray*}
By lemma $\ref{lempqform}$ 
with $\chi \in {\cal{A}}_{\rho ', \sigma '}$ and
for some positive $\delta < \sigma '$
it follows $\{ \chi ,P \} \in {\cal{A}}_{\rho '-\delta ,\sigma ' -\delta}$
is of  $p'-q'$exponential form. 
Next we examine
\begin{eqnarray*}
r_{2}(H,\chi ,1) = 
\sum ^{\infty}_{l=2} \frac{1}{l!}L^{l}_{\chi}H.
\end{eqnarray*}
Clearly by inductively applying lemma $\ref{lempqform}$ 
$r_{2}(H,\chi ,1)$ is of   $p'-q'$exponential form in some domain,
${\cal{D}}_{\rho *, \sigma *}$, to
be determine.
We set
\begin{eqnarray*}
\rho *= \rho - \sum_{i=0}^{\infty} \delta_{i},
\\
\quad\\
\sigma *= \sigma - \sum_{i=0}^{\infty} \delta_{i}.
\end{eqnarray*}
Each time we apply lemma $\ref{lempqform}$, 
we can choose $\delta_{i}$ arbitrarily small.
In particular
we can set $\delta_{i} = \delta_{i-1}/2$ for $i=1,2,...$
with $\delta_{0} =\delta$.
Therefore
\begin{eqnarray*}
\rho *= \rho - \sum_{i=0}^{\infty} \delta \left(\frac{1}{2} \right)^{i}
=\rho -2\delta
,
\\
\quad\\
\sigma *= \sigma - \sum_{i=0}^{\infty} \delta \left(\frac{1}{2} \right)^{i}
=\sigma -2\delta
.
\end{eqnarray*}
We finally have
${\cal{R}}_{A} \in {\cal{A}}_{\rho *, \sigma *}$ is of 
  $p'-q'$exponential form. \quad \squareforqed

\end{lem}
\section{Rossby Wave Flow}
\label{Rossby}
The Rossby wave flow is generated by a Hamiltonian of the form
\begin{eqnarray*}
H(X,Y,t)&=&H^{0}(X,Y,t)+\varepsilon H^{1}(X,Y,t)\\
\quad\\
&=&A\sin k_{0}(X-c_{0}t)\sin l_{0}Y +\varepsilon H^{1}(X,Y,t),
\end{eqnarray*}
where $\varepsilon >0$, $A$ is the maximum velocity in the
$y$-direction, $(k_{0},l_{0})$ is the wave number vector, and
$c_{0}$ is the phase speed of the primary wave in the $x$-direction. In a reference frame moving with the primary wave the transformation
\begin{eqnarray*}
x=X-c_{0}t, \quad y=Y,
\end{eqnarray*}
yields  the Hamiltonian
\begin{eqnarray*}
H(x,y,t)&=&H^{0}(x,y,t)+\varepsilon H^{1}(x,y,t)\\
\quad\\
&=&-c_{0}y+A\sin k_{0}x \sin l_{0}y +\varepsilon H^{1}(x,y,t).
\end{eqnarray*}
In this frame of reference the vector field generated by the
Hamiltonian $H(x,y,t)$ has the following form:
\begin{eqnarray*}
&\quad& \dot x=c_{0}-Al_{0}\sin k_{0}x \cos l_{0}y -\varepsilon \frac{
\partial H^{1}}{\partial y}(x,y,t),\\
\quad\\
&\quad& \dot y =Ak_{0} \cos k_{0}x \sin l_{0}y +\varepsilon
\frac{\partial H^{1}}{\partial x}(x,y,t).
\end{eqnarray*}
We will consider the perturbation $H^{1}(x,y,t)=xy \mbox{ sech} ^2 at$ where
$a$ is the decay rate.

The near integrable form of this problem makes it a
candidate to apply the theorem presented in this paper provided
the conditions of the hypothesis are satisfied. The first step
in applying the theorem is to transform the Hamiltonian to action-angle
variables. Clearly, by the Liouville-Arnold Theorem $\cite{Arnold89}$ one can
construct locally for the one degree of freedom system
generated by the Hamiltonian $H^{0}(x,y)$
a symplectic coordinate transformation to action-angle variables.
The integrable system
is characterized by
two heteroclinic connections  between saddle-type equilibrium
points on the curves $y=0$, $y=\pi$ respectively and by the
symmetry trajectory
starting at $(0,\pi /2)$ shown in Figure $1$. These three structures divide
the phase space in four regions. Action-angle variables can be
found locally for each of these four regions as we will now describe.
The action variable $p$ is defined
as
\begin{eqnarray}
\label{actiondefined}
p=\frac{1}{2\pi}\int_{H} xdy,
\end{eqnarray}
which is the area of the region inside the level set
with Hamiltonian $H$ divided by $2\pi$.
Figure $2$ shows the actions for the system
in Figure $1$ starting from the right at $p=0$ for the
elliptic equilibrium point in region $I$ and rising to a maximum at
$H=-\pi /4$ for the symmetry trajectory. Similarly for regions $III$
and $IV$ the action falls from a maximum at the symmetry trajectory
to zero for the equilibrium
point in region $IV$. Note $I(H)$ is invertible in each of the intervals
corresponding to  regions $I,II,III$ and $IV$.
To define the angle $\theta (x,y)$, let $L$ denote a straight curve emanating
from the elliptic equilibrium point in region $I$ to the elliptic equilibrium
point in region $IV$. We denote solutions of the integrable vector field
starting on $L$ by $(x(t,s),y(t,s))$ where $x(0,s)=x_{0}(s)$ and
$y(0,s)=y_{0}(s)$ so for any point $(x,y)$ on the orbit $(x(t,s),y(t,s))$,
$t=t(x,y)$ is the time it takes for the solution starting at $(x_{0}(s),
y_{0}(s))$ to reach $(x,y)$. Given $T(H)$ is the period of the periodic
orbit with constant $H$, we define the angle variable, $\theta (x,y)$ as
\begin{eqnarray}
\label{thetadefined}
\theta (x,y)=2\pi \frac{t(x,y)}{T(H)},
\end{eqnarray}
where $(x,y)\in H=$ constant. Clearly by this definition action-angle
variables can not be defined on the heteroclinic connections.
\begin{figure}
\begin{center}
\epsfig{file=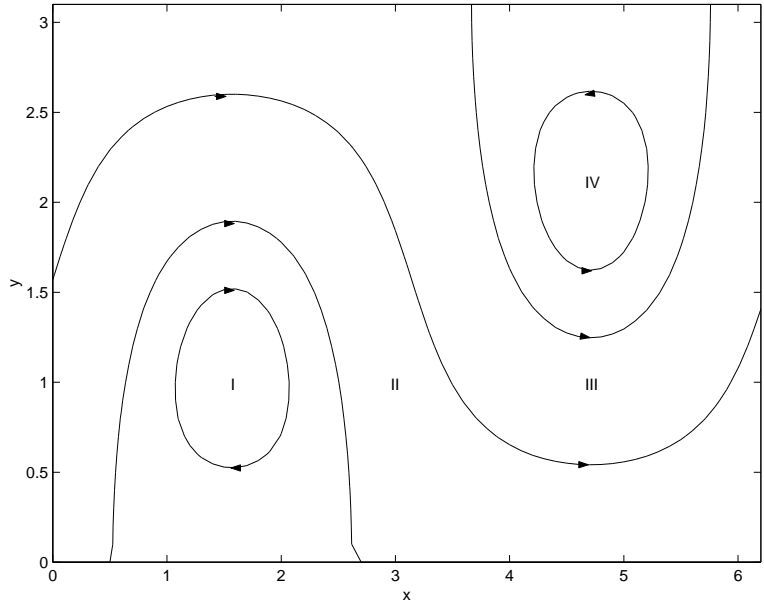,width=4in} \caption{Phase space of
the vector field generated by the Hamiltonian  $H=-c_{0}y+A\sin
k_{0}x \sin l_{0}y$ with $A=k_{0}= l_{0}=1.0$ and $c_{0}=0.5$.}
\end{center}
\end{figure}

\begin{figure}
\begin{center}
\epsfig{file=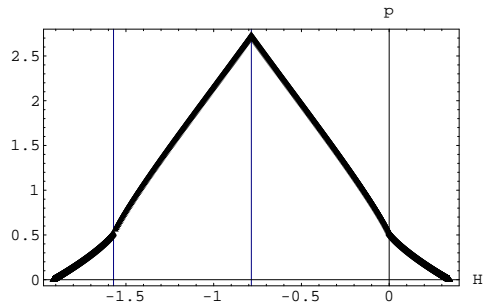,width=2.5in} \caption{Action
map for each of the four regions starting from the right
$I,II,III,IV$.}
\end{center}
\end{figure}
\begin{figure}[h]
\begin{center}
\epsfig{file=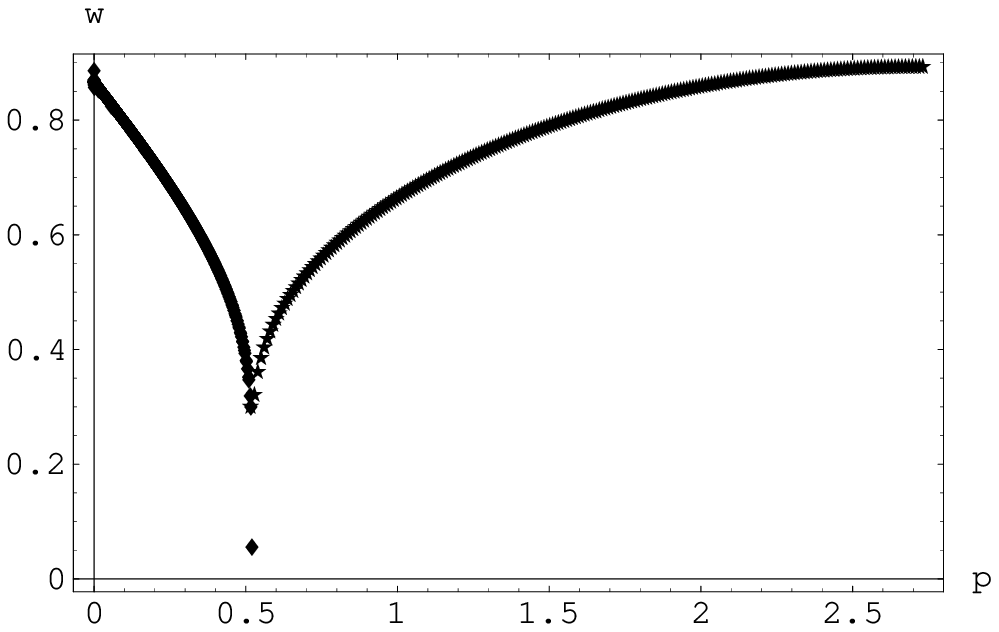,width=2.5in}
\caption{Frequency Map.}
\label{pahsespace}
\end{center}
\end{figure}

Next we  show the Hamiltonian in action-angle variables is analytic.
\begin{lem}
The transformation to action-angle variables is analytic.
\end{lem}
\begin{proflem}
We begin by transforming the integral part of the Hamiltonian $H^0(x,y)$
in region $I$.
In region $I$ $(\ref{actiondefined})$ has the following form
\begin{eqnarray}
\label{actioninI}
p(H^0)=
\frac{1}{\pi}\int ^{y_{\max}}_{y_{\min}} x(y,H^0)dy
=
\frac{1}{\pi}\int ^{y_{\max}}_{y_{\min}}\arcsin
\left[\frac{y+H^0}{\sin y} -\frac{\pi}{2}\right]dy,
\end{eqnarray}
where the $\pi/2$ term translates
the integral to the $y$-axis and $y_{\max}$, $y_{\min}$ are
the values of $y$ where the level set intersects the $y$ axis.
We consider a complex extension of the variables $H^0$, $H_{C}=H^0+iH_{Im}$,
and $y$, $y_{C}=y+iy_{Im}$, to complex strips and for sake of
simplicity do not specify their width.
The integral in $(\ref{actioninI})$
becomes a contour integral along the real axis with  end points
$y_{\max}$, $y_{\min}$.
Let ${\cal{Y}}=(Y_{min},Y_{max})$
denote the interval of y values in region $I$.
Since $T(y,H^0)$ and $T_{H^0}(y,H^0)$ are continuous for $H_{C}$ in a complex
extension, $D$, of $H^0({\cal{Y}})$ and $T(y,H^0)$ and $T_{H^0}(y,H^0)$ are
continuous for $y_{C}$ on the contour consisting of the real interval
$(y_{\max},y_{\min})$,
it follows $p(H^0)$ is analytic in $D$ $\cite{Dettman65}$.
Furthermore, by the inverse function theorem, except for the
elliptic equilibria where $p=0$, $H^0(p)$ is analytic on some strip
of $p$.
Similarly, the proof for regions
$II,III,IV$ follows.

For the angle variable in region $I$ $(\ref{thetadefined})$
becomes
\begin{eqnarray*}
\theta(x,y)=\frac{2\pi}{T(H)}\int _{y_{\min}}^{y}
\frac{dy}{\sqrt{\sin ^2 y -(H^0+\frac{y}{2})^2}},\quad
T(H)=2\int _{y_{\min}}^{y_{\max}}
\frac{dy}{\sqrt{\sin ^2 y -(H^0+\frac{y}{2})^2}}.
\end{eqnarray*}
By the same arguments as for the action transformation and defining
a complex extension of $x$, $x_{C}=x+ix_{Im}$,
it follows $\theta(x,y)$ is analytic for some strip of
$x_{C}$ and $y_{C}$.

When transforming
the perturbation $H^1(x,y,t)=xy\mbox{ sech }^2 at$
to the action-angle variables found for $H^0$,
it is clear that the time dependent term ${\mbox{ sech }}^2 at $ is the same
after the transformation.
By the Liouville-Arnold Theorem the action-angle transformation
$(x,y)\rightarrow (p(H^0),\theta(x,y))$
is invertible, $(p,\theta)\rightarrow (x(p,\theta),y(p,\theta))$,
and by the inverse function theorem $x(p,\theta)$ and $y(p,\theta)$ are
analytic. \squareforqed
\end{proflem}
Note, since the time dependence of the perturbation is $\mbox{ sech }^2at$,
the perturbation is of exponential order with respect to time as
required by the hypothesis of the theorem.

Finally the nondegeneracy condition
given by $\det(\partial ^2 H^{0}/\partial p^2)=
\det(\partial \omega /\partial p) \neq 0$
must be satisfied. We numerically compute the period and plot the frequency
as a function of the action giving the frequency map shown in Figure $3$.
This graph gives the frequencies for the closed orbits in regions
$I,II$ as well as the closed orbits in regions $III,IV$.
The frequency at $p=0$ corresponds to both elliptic equilibrium points in regions
$I$ and $IV$. The frequency falls to zero for a $p$ value
corresponding to both heteroclinic connections as one would expect.
The frequencies to the right of this value of $p$ correspond to
the closed orbits in regions $II$ and $III$ reaching a maximum for the
symmetry trajectory. The nondegeneracy condition is therefore satisfied for all
tori except for the symmetry trajectory.

\section{Aside Calculations}
\label{lastappendix}
We take an aside to check  $(\ref{eqnthree})$ reduces to
 the appropriate expression
if the time dependence of $g_{k}(s)$ is not aperiodic but is 
indeed periodic  with frequency $\omega_{0}$. Without loss of generality 
we assume $\omega_{0}$ to be positive.  In this case we
write the Fourier series expansion of $g_{k}(s)$
\\
\begin{eqnarray*}
g_{k}(s)= \sum_{l\in {\mathbb{Z}}} g_{kl} e^{il\omega _{0} s}
\end{eqnarray*}
\\
and substitute this in $(\ref{eqnthree})$ which results in
\\
\begin{eqnarray*}
f_{k}(t)= f_{k}(0) e^{-i(\tilde{\lambda} \cdot k)t}+ \int ^{t}_{0}
\sum_{l\in {\mathbb{Z}}} g_{kl}e^{il\omega _{0} s}e^{i(\tilde{\lambda} \cdot k)
(s-t)}ds.
\end{eqnarray*}
\\
Now let $\sum _{l\in {\mathbb{Z}}}s_{l} =1$ be any series 
whose sum is one. Then we have
\begin{eqnarray*}
f_{k}(t)&=&f_{k}(0)e^{-i(\tilde{\lambda} \cdot k)t}+\sum_{l\in {\mathbb{Z}}}
g_{kl}\int ^{t}_{0}e^{il\omega _{0} s}e^{i(\tilde{\lambda} \cdot k)
(s-t)}ds\\
&=&f_{k}(0)e^{-i(\tilde{\lambda} \cdot k)t}
\sum _{l\in {\mathbb{Z}}}s_{l}  +\sum_{l\in {\mathbb{Z}}}
g_{kl}\int ^{t}_{0}e^{il\omega _{0} s}e^{i(\tilde{\lambda} \cdot k)
(s-t)}ds\\
&=&\sum _{l\in {\mathbb{Z}}} f_{k}(0)e^{-i(\tilde{\lambda} \cdot k)t}
s_{l}  +\sum_{l\in {\mathbb{Z}}}
g_{kl}\int ^{t}_{0}e^{il\omega _{0} s}e^{i(\tilde{\lambda} \cdot k)
(s-t)}ds\\
&=&\sum _{l\in {\mathbb{Z}}}\left[ 
f_{k}(0)s_{l}e^{-i(\tilde{\lambda} \cdot k)t}+
\frac{g_{kl}e^{-i(\tilde{\lambda} \cdot k)t}}{i(\tilde{\lambda} \cdot k
+\omega _{0}l)}\left( e^{i(\tilde{\lambda} \cdot k+\omega _{0}l)t}-1 
\right) \right]
\end{eqnarray*}
\begin{eqnarray*}
\quad \quad \quad \quad \quad \quad = \sum _{l\in {\mathbb{Z}}}\left[ 
f_{k}(0)s_{l}e^{-i(\tilde{\lambda} \cdot k+\omega _{0}l)t}+
\frac{g_{kl}}{i(\tilde{\lambda} \cdot k
+\omega _{0}l)}
\left(1-e^{-i(\tilde{\lambda} \cdot k+\omega _{0}l)t}\right) \right]
e^{i\omega _{0}lt}
\end{eqnarray*}
\\
\begin{equation}
\label{eqnthreefour}
\quad \quad \quad \quad \quad 
= \sum _{l\in {\mathbb{Z}}}\left[ 
\left(f_{k}(0)s_{l}-
\frac{g_{kl}}{i(\tilde{\lambda} \cdot k
+\omega _{0}l)}\right) e^{-i(\tilde{\lambda} \cdot k+\omega _{0}l)t}
+\frac{g_{kl}}{i(\tilde{\lambda} \cdot k
+\omega _{0}l)}\right] e^{i\omega _{0}lt}
\end{equation}
\\
Note, that except for the time dependence inside  the square bracket,
 $(\ref{eqnthreefour})$ is almost in the form of the Fourier
series for $f_{k}(t)$ 
\\
\begin{equation}
\label{eqnthreefourfour}
f_{k}(t) = \sum _{l\in {\mathbb{Z}}} f_{kl} e^{il\omega _{0} t}.
\end{equation}
\\
Comparing $(\ref{eqnthreefour})$ and $(\ref{eqnthreefourfour})$
we see that we can eliminate the time dependence in the square 
bracket by setting
\\
\begin{eqnarray*}
f_{k}(0)s_{l}= \frac{g_{kl}}{i(\tilde{\lambda} \cdot k
+\omega _{0}l)}.
\end{eqnarray*}
\\
By summing on both sides of the equation, this
is equivalent to 
\\
\begin{eqnarray*}
&\quad & \sum_{l\in {\mathbb{Z}}}f_{k}(0)s_{l}=\sum_{l\in {\mathbb{Z}}}
\frac{g_{kl}}{i(\tilde{\lambda} \cdot k
+\omega _{0}l)}\\
&\quad & f_{k}(0)\sum_{l\in {\mathbb{Z}}}s_{l}=
\sum_{l\in {\mathbb{Z}}}
\frac{g_{kl}}{i(\tilde{\lambda} \cdot k
+\omega _{0}l)}
\end{eqnarray*}
\begin{eqnarray}
&\quad& f_{k}(0)=
\sum_{l\in {\mathbb{Z}}}
\frac{g_{kl}}{i(\tilde{\lambda} \cdot k
+\omega _{0}l)}
\label{eqnthreefourfourf}
\end{eqnarray}
\\
In which case, $(\ref{eqnthreefour})$ reduces to 
\\
\begin{equation}
\label{eqnthreefourfourff}
f_{k}(t)=\sum _{l\in {\mathbb{Z}}} \frac{g_{kl}}{i(\tilde{\lambda} \cdot k
+\omega _{0}l)}
e^{il\omega _{0}t}
\end{equation}
\\
with 
\\
\begin{eqnarray*}
f_{kl}=\frac{g_{kl}}{i(\tilde{\lambda} \cdot k
+\omega _{0}l)}.
\end{eqnarray*}
\\
Note that $(\ref{eqnthreefourfourff})$ reduces to the expression for 
$f_{k}(0)$ obtained in $(\ref{eqnthreefourfourf})$.
\\
\\
We now must obtain an estimate for $|f_{k}(t)|$. Since $G(q,t),F(q,t)\in
 {\cal{A}}_{\rho}$, the Fourier coefficients $f_{k}(t)$ and $g_{k}(t)$ must
be exponentially small with respect to the index $k$. First we analyze
the estimate for $g_{k}(t)$.
\\
\\
We have 
\begin{eqnarray*}
G(q,t) &=& \sum_{q\in {\mathbb{Z}}^{n}} g_{k}(t) e^{ik\cdot q}\\
&=& \sum_{q\in {\mathbb{Z}}^{n}} \sum _{l\in {\mathbb{Z}}}
g_{kl} e^{il\omega_{0}t}e^{ik\cdot q}.
\end{eqnarray*}
\\
We calculate $|g_{kl}|$
\\
\begin{eqnarray*}
g_{kl}= \frac{1}{(2\pi)^{n}}\frac{1}{S}\int ^{\pi}_{-\pi} \cdot \cdot
\cdot \int ^{\pi}_{-\pi} \int ^{S/2}_{-S/2}G(q,s)e^{-ik\cdot q}
e^{-i\omega _{0}ls}dsdq
\end{eqnarray*}
\\
where the integral is along the real axis.
Lift the integral as follows
\\
\begin{eqnarray*}
g_{kl}= \frac{1 }{(2\pi)^{n}}\frac{1}{S}\int ^{\pi}_{-\pi} \cdot \cdot
\cdot \int ^{\pi}_{-\pi} \int ^{S/2}_{-S/2}
G(q_{1}- i\frac{k_{1}}{|k_{1}|}\rho,...,q_{n}-i\frac{k_{n}}{|k_{n}|}\rho,
s-i\frac{l}{|l|}\sigma)
\end{eqnarray*}
\begin{eqnarray*}
\quad \quad \quad  \cdot \Big \{ \prod ^{n}_{j=1}e^{-ik_{j}(q_{j}-i\frac{k_{j}}{|k_{j}|} \rho)} \Big\}
e^{-il\omega_{0}(s-i\frac{l}{|l|}\sigma)}dqds \quad \quad \quad \quad \quad \quad
\end{eqnarray*}
\begin{eqnarray*}
g_{kl}=\frac{e^{-|k|\rho}e^{-|l|\omega_{0}\sigma} }{(2\pi)^{n}S }
\int ^{\pi}_{-\pi} \cdot \cdot
\cdot \int ^{\pi}_{-\pi} \int ^{S/2}_{-S/2}
G(q_{1}- i\frac{k_{1}}{|k_{1}|}\rho,...,q_{n}-i\frac{k_{n}}{|k_{n}|}\rho,
s-i\frac{l}{|l|}\sigma)
\end{eqnarray*}
\\
\begin{eqnarray*}
\cdot e^{-ik\cdot q}
e^{-i\omega _{0}ls}dsdq\quad \quad \quad \quad \quad \quad \quad \quad \quad \quad \quad
\quad \quad \quad 
\end{eqnarray*}
\\
and
\begin{eqnarray*}
|g_{kl}| \leq e^{-|k|\rho}e^{-|l|\omega_{0}\sigma}\| G(q,t) \|_{\rho ,\sigma}.
\end{eqnarray*}
\\
Now, since 
\begin{eqnarray*}
g_{k}(t) = \sum _{l\in {\mathbb{Z}}}g_{kl} e^{il\omega_{0} t},
\end{eqnarray*}
\\
to calculate the estimate $|g_{k}(t)|$ we need to calculate  the following
\\
\begin{eqnarray*}
|e^{il\omega_{0} t}| = \left(e^{il\omega_{0} t}e^{-il\omega_{0} \overline{t}}\right)^{1/2}
\end{eqnarray*}
\\
where $t\in {\mathbb{C}}$ and $\overline{t}$ indicates the complex conjugate.
Then\\
\begin{eqnarray*}
|e^{il\omega_{0} t}| = e^{-\omega_{0}l({\mbox{Im }t})}
\end{eqnarray*}
where ${\mbox{Im }}t$ is the imaginary part of $t$.
Assume $|{\mbox{Im }t}| \leq \sigma - \delta$. Then
\\
\begin{eqnarray*}
-\omega_{0}l({\mbox{Im }t}) \leq |\omega_{0}l({\mbox{Im }t})| \leq |l||\omega_{0}|(\sigma - \delta).
\end{eqnarray*}
\\
We then obtain
\begin{eqnarray*}
|g_{k}(t)| &\leq& \sum _{l\in {\mathbb{Z}}}|g_{kl}||e^{il\omega_{0} t}|\\
&\leq& e^{-|k|\rho}\| G(q,t) \|_{\rho ,\sigma} \sum _{l\in {\mathbb{Z}}}e^{-|l|\omega_{0}\sigma}
e^{ |l|\omega_{0}(\sigma - \delta)}\\
&\leq& e^{-|k|\rho}\| G(q,t) \|_{\rho ,\sigma} \sum _{l\in {\mathbb{Z}}}
e^{ -|l|\omega_{0}\delta}\\
&\leq&2 e^{-|k|\rho}\| G(q,t) \|_{\rho ,\sigma}\sum _{l\in {\mathbb{Z}}^{+}}
e^{-\omega_{0}\delta l}\\
&=& 2 e^{-|k|\rho}\| G(q,t) \|_{\rho ,\sigma}\left( \frac{1}{1-e^{-\omega_{0}\delta}}\right).
\end{eqnarray*}
\\
We proceed similarly to obtain an estimate for $f_{k}(t)$. We know
\\
\begin{eqnarray*}
f_{k}(t)=\sum _{l\in {\mathbb{Z}}} \frac{g_{kl}e^{il\omega_{0}t}}{
i(\tilde{\lambda} \cdot k+ l\omega_{0})}.
\end{eqnarray*}
\\
We obtain from the previous estimates
\\
\begin{eqnarray*}
|f_{k}(t)|\leq e^{-|k|\rho}\| G(q,t) \|_{\rho ,\sigma}\sum _{l\in {\mathbb{Z}}}
\frac{e^{ -|l|\omega_{0}\delta}}{|\tilde{\lambda} \cdot k+ l\omega_{0}|}.
\end{eqnarray*}
\\
The diophantine condition on the frequencies gives an estimate for the denominator 
of this expression. First recall that for $k\in {\mathbb{Z}}^{n}$ we define
$|k|= \sum _{i=1}^{n} |k_{i}|$ and $\| k \| = \sup _{i} |k_{i}|$. Furthermore, 
define the vectors $h=(k,l)$ and $\Delta = (\tilde{\lambda}, \omega_{0})$.
Consider first the case $h=(k,l) \neq 0$.
 The diophantine condition is given by
\\
\begin{eqnarray*}
|\Delta \cdot h | \geq \Gamma \| h \|^{-(n+1)}.
\end{eqnarray*}
\\
For any $K,Y,\delta ' >0$ recall the inequality
\\
\begin{eqnarray*}
K^{Y} \leq \left( \frac{Y}{e\delta '} \right)^{Y}e^{K\delta '}.
\end{eqnarray*}
\\
It follows that
\\
\begin{eqnarray*}
\frac{1}{|\Delta \cdot h |} &\leq& \Gamma ^{-1}\| h \|^{n+1}\\
&\leq& \Gamma ^{-1}| h |^{n+1}\\
&\leq& \Gamma ^{-1}\left( \frac{n+1}{e\delta '} \right)^{n+1}
e^{| h |\delta '}.
\end{eqnarray*}
\\
We choose 
\begin{eqnarray*}
\delta ' = \delta \frac{\omega_{0}}{\omega_{0}+1}< \delta
\end{eqnarray*}
so that
\begin{eqnarray*}
|f_{k}(t)|&\leq& e^{-|k|\rho}\| G(q,t) \|_{\rho ,\sigma}\Gamma^{-1}
\left( \frac{(n+1)(\omega_{0}+1)}{e\omega_{0} \delta}\right)^{n+1}
\sum _{l\in {\mathbb{Z}}}e^{ -|l|\omega_{0}\delta}e^{|h|\frac{\delta \omega_{0}}{\omega_{0} +1}}\\
&=&\| G(q,t) \|_{\rho ,\sigma}\Gamma^{-1}
\left( \frac{(n+1)(\omega_{0}+1)}{e\omega_{0} \delta}\right)^{n+1}e^{-|k|(\rho-
\frac{\delta \omega_{0}}{\omega_{0} +1})}
\sum _{l\in {\mathbb{Z}}}e^{-|l|\omega_{0}\delta \left(\frac{\omega_{0}}{\omega_{0} +1}\right)}\\
&\leq& \| G(q,t) \|_{\rho ,\sigma}\Gamma^{-1}
\left( \frac{(n+1)(\omega_{0}+1)}{e\omega_{0} \delta}\right)^{n+1}
e^{-|k|(\rho-\delta)}
\sum _{l\in {\mathbb{Z}}}e^{-|l|\omega_{0}\delta}\\
&\leq& \| G(q,t) \|_{\rho ,\sigma}\Gamma^{-1}
\left( \frac{(n+1)(\omega_{0}+1)}{e\omega_{0} \delta}\right)^{n+1}
\left( \frac{2}{1-e^{-\omega_{0}\delta}} \right)
e^{-|k|(\rho-\delta)}
\end{eqnarray*} 
For the case $h=(k,l)=0$ we use the assumption $\overline{G} (q,t)=0$,
where $\overline{G} (q,t)$ represents the average of the function over
the variables $q$ and $t$.
In this case , we use the following Lemma to show 
 $f_{0}(t)$ or $f_{00}$ are bounded.
\begin{lem}
\quad \\
Given $G(q,t)$ is periodic with respect to $q$ and $t$ and has 
Fourier coefficients $g_{kl}$,
$g_{00}=0$ iff $\overline{G}(q,t)=0$.
\end{lem}
Proof:\\
\quad\\
We have 
\begin{eqnarray*}
G(q,t) =\sum _{k\in {\mathbb{Z}}^{n}} \sum _{l\in {\mathbb{Z}}}
g_{kl} e^{i\omega_{0}lt}e^{ik\cdot q}
\end{eqnarray*}
and
\begin{eqnarray*}
\overline{G}(q,t)&=& \left(\frac{1}{2\pi} \right)^{n} \frac{1}{S}
\int^{\pi}_{-\pi} \cdot \cdot \cdot \int^{\pi}_{-\pi}\int ^{s/2}_{-s/2}
\sum _{k\in {\mathbb{Z}}^{n}} \sum _{l\in {\mathbb{Z}}}
g_{kl} e^{i\omega_{0}lt}e^{ik\cdot q}dtdq\\
&=& \left(\frac{1}{2\pi} \right)^{n} \frac{1}{S}
\sum _{k\in {\mathbb{Z}}^{n}} \sum _{l\in {\mathbb{Z}}}
\left[g_{kl}\left[\frac{e^{i\omega_{0}lt}}{i\omega_{0}l} \right]^{s/2}_{-s/2}
\left[\frac{e^{ik\cdot q}}{(i)^{n} \prod ^{n}_{j=1}k_{j}} \right]^{\pi}_{-\pi}
\right]\\
&=& g_{00}
\end{eqnarray*}
\squareforqed\\
\quad\\
Therefore since 
\begin{eqnarray*}
f_{kl}= \frac{g_{kl}}{i(\tilde{\lambda}\cdot k + l\omega_{0})}
\end{eqnarray*}
we have $f_{00}=0$ and given
\begin{eqnarray*}
f_{k}(t)= \sum _{l\in {\mathbb{Z}}} \frac{g_{kl}e^{il\omega_{0}t}}{
i(\tilde{\lambda}\cdot k + l\omega_{0})}
\end{eqnarray*}
it follows
\begin{eqnarray*}
|f_{0}(t)| &\leq&\sum _{\tiny{
\begin{array}{c}
l\in {\mathbb{Z}}\\
{l\neq 0}
\end{array}}}
\frac{|g_{0l}||e^{il\omega_{0}t}|}
{|l\omega_{0}|}\\
&\leq& \| G(q,t) \|_{\rho ,\sigma}\gamma^{-1}
\left( \frac{(n+1)(\omega_{0}+1)}{e\omega_{0} \delta}\right)^{n+1}
\left( \frac{2}{1-e^{-\omega_{0}\delta}} \right)
\end{eqnarray*}
\quad \\
Now we want to check that, given $G(q,t)$ is  periodic 
with respect to time, $(\ref{eqneight})$ reduces to the 
appropriate expression for $f_{k}(t)$, $(\ref{eqnthreefourfourff})$.
Assuming $G(q,t)$ is time periodic with frequency $\omega _{0}$ implies 
the Fourier coefficients $f_{k}(t)$ are periodic functions
with
\\
\begin{eqnarray*}
G(q,t)= \sum _{k\in {\mathbb{Z}}^{n}} g_{k}(t)e^{ik\cdot q}
\end{eqnarray*}
\\
and
\\
\begin{eqnarray*}
g_{k}(t)=\sum _{l\in {\mathbb{Z}}} g_{kl}e^{il\omega _{0}t}.
\end{eqnarray*}
\\
The Fourier transform of a periodic function is equals
to a series of the following from
\\
\begin{equation}
\label{eqnnine}
{\cal{G}}_{k}(\omega)= \sum _{l\in{\mathbb{Z}}}g_{kl} \delta
(\omega - l\omega_{0}).
\end{equation}
That is, the Fourier transform consists of equally spaced
delta functions which are weighted according to the Fourier 
coefficients of the function.
\\
\\
Substituting $(\ref{eqnnine})$ in $(\ref{eqneight})$ gives the following
\\
\begin{eqnarray*}
f_{k}(t)&=&\int ^{\infty+ i\beta }_{-\infty+ i\beta } \frac{{\cal{G}}_{k}(\omega)}
{i(\tilde{\lambda} \cdot k +\omega)} e^{i\omega t}d\omega \\
&=& \int ^{\infty+ i\beta }_{-\infty+ i\beta }
\frac{\sum _{l\in{\mathbb{Z}}}g_{kl} \delta
(\omega - l\omega_{0})}{i(\tilde{\lambda} \cdot k +\omega)}
e^{i\omega t}d\omega 
\end{eqnarray*}
\begin{eqnarray}
&=& \sum _{l\in{\mathbb{Z}}}g_{kl}\int ^{\infty+ i\beta }_{-\infty+ i\beta }
\frac{\delta
(\omega - l\omega_{0})}{i(\tilde{\lambda} \cdot k +\omega)}
e^{i\omega t}d\omega.
\label{eqnten}
\end{eqnarray}
\\
We use the following  property of the delta function for a function $h(\omega)$
continuous at $\omega=0$
\\
\begin{eqnarray*}
\int^{\infty+ i\beta }_{-\infty+ i\beta } \delta (\omega)h(\omega)d\omega =
h(0)
\end{eqnarray*}
\\
and $(\ref{eqnten})$ becomes
\\
\begin{eqnarray*}
f_{k}(t)= \sum _{l\in{\mathbb{Z}}}\frac{g_{kl}}{i(\tilde{\lambda} \cdot k +
l\omega_{0})}e^{il\omega_{0}t}.
\end{eqnarray*}
\quad \\
\quad \\
We derive a bound for the sum $\sum _{m\in{\mathbb{Z}}^{n}}e^{-|m|\rho}
e^{-|k-m|(\rho -2\delta)}$. We  examine the cases $n=1$, $n=2$ and 
derive an expression for the general case.
\\
\quad\\
\framebox{$n=1$}
\quad\\
\quad\\
We obtain two  bounds, one for $k\geq 0$
and another for $k<0$. These bounds are equal and thus  the sum
$\sum _{m\in{\mathbb{Z}}^{n}}e^{-|m|\rho}
e^{-|k-m|(\rho -2\delta)}$ is bounded
for any $k$.
\\
\quad\\
Let $k\geq 0$ and divide the range of $m$ in three intervals,
$m\leq 0$, $0<m<k$ and $m\geq k$. We obtain bounds for the 
three intervals
\quad\\
\begin{eqnarray*}
m\leq 0 \quad
\Big \{ \quad \sum _{m\leq 0}e^{-|m|\rho}
e^{-|k-m|(\rho -2\delta)} \leq
e^{-|k|(\rho -2\delta)}
 \sum _{m\leq 0}e^{-|m|\rho},
\end{eqnarray*}
\quad\\
\begin{eqnarray*}
0<m<k \quad \Big \{ \quad
\sum _{0<m<k}
e^{-|m|\rho}
e^{-|k-m|(\rho -2\delta)} &=&
\sum _{0<m<k}
e^{-|k|(\rho -2\delta)}e^{-|m|(2\delta)}\\
\quad\\
 &\leq&
e^{-|k|(\rho -2\delta)}
\sum _{0<m}
e^{-|m|(2\delta)},
\end{eqnarray*}
\quad\\
\begin{eqnarray*}
m\geq k \quad \Big \{ \quad
\sum _{m\geq k}
e^{-|m|\rho}
e^{-|k-m|(\rho -2\delta)}\leq 
e^{-|k|\rho }
\sum _{k-m \leq 0 }
e^{-|k-m|(\rho -2\delta)}.
\end{eqnarray*}
Therefore for $k\geq 0$ 
\begin{eqnarray*}
\sum_{m\in{\mathbb{Z}}}
e^{-|m|\rho}e^{-|k-m|(\rho-2\delta)}
&\leq&
e^{-|k|(\rho -2\delta)}
 \sum _{m\leq 0}e^{-|m|\rho}\\
\quad\\
&+&
e^{-|k|(\rho -2\delta)}
\sum _{0<m}
e^{-|m|(2\delta)}\\
\quad\\
&+&
e^{-|k|\rho }
\sum _{k-m \leq 0 }
e^{-|k-m|(\rho -2\delta)}\\
\quad\\
&\leq&
3e^{-|k|(\rho -2\delta)}\sum _{m\geq 0}
e^{-|m|(2\delta)}.
\end{eqnarray*}
\quad\\
Let $k<0$ and divide the range of $m$ in three intervals,
$m\leq k$, $k<m<0$ and $m\geq 0$. We obtain bounds for the three
intervals.
\quad\\
\begin{eqnarray*}
m\leq k \quad \Big \{ \quad 
\sum _{m\leq k}
e^{-|m|\rho}
e^{-|k-m|(\rho -2\delta)}
 \leq
e^{-|k|\rho}\sum_{k-m\geq 0}
e^{-|k-m|(\rho -2\delta)},
\end{eqnarray*}
\quad\\
\begin{eqnarray*}
k<m<0 \quad \Big \{ \quad
\sum _{k<m<0}
e^{-|m|\rho}
e^{-|k-m|(\rho -2\delta)}
&=&
\sum _{k<m<0}
e^{k(\rho -2\delta)}e^{m(2\delta)}\\
\quad\\
&=&
e^{-|k|(\rho -2\delta)}
\sum _{k<m<0}
e^{-|m|(2\delta)}\\
\quad\\
&\leq&
e^{-|k|(\rho -2\delta)}
\sum _{m<0}
e^{-|m|(2\delta)},
\end{eqnarray*}
\quad\\
\begin{eqnarray*}
m\geq 0 \quad \Big \{ \quad
\sum _{m\geq 0}
e^{-|m|\rho}
e^{-|k-m|(\rho -2\delta)} =
e^{-|k|(\rho -2\delta)}
\sum _{m\geq 0}
e^{-|m|\rho}.
\end{eqnarray*}
Therefore for $k<0$
\begin{eqnarray*}
\sum_{m\in{\mathbb{Z}}}
e^{-|m|\rho}e^{-|k-m|(\rho-2\delta)}
&\leq&
e^{-|k|\rho}\sum_{k-m\geq 0}
e^{-|k-m|(\rho -2\delta)}\\
\quad\\
&+&
e^{-|k|(\rho -2\delta)}
\sum _{m<0}
e^{-|m|(2\delta)}\\
\quad\\
&+&
e^{-|k|(\rho -2\delta)}
\sum _{m\geq 0}
e^{-|m|\rho}\\
\quad\\
&\leq&
3e^{-|k|(\rho -2\delta)}
\sum_{m\geq 0}
e^{-|m|(2\delta)}.
\end{eqnarray*}
Finally for any $k\in {\mathbb{Z}}$ we have the bound
\begin{eqnarray*}
\sum_{m\in{\mathbb{Z}}}
e^{-|m|\rho}e^{-|k-m|(\rho-2\delta)}
&\leq&
3e^{-|k|(\rho -2\delta)}
\sum_{m\geq 0}
e^{-|m|(2\delta)}= 3e^{-|k|(\rho -2\delta)}
\left(
\frac{1}{1-e^{-2\delta}}\right).
\end{eqnarray*}
\quad\\
\quad\\
\framebox{$n=2$}
\quad\\
\quad\\
In this case
we have vectors $k=(k_{1},k_{2})\in {\mathbb{Z}}^{2}$
and $m=(m_{1},m_{2})\in {\mathbb{Z}}^{2}$.
We must keep in mind four sub-cases $(1)$ $k_{1}\geq 0$ and
$k_{2}\geq 0$,  $(2)$ $k_{1}\geq 0$ and $k_{2} < 0$, 
 $(3)$ $k_{1}< 0$ and $k_{2}\geq 0$, $(3)$ $k_{1}< 0$ and 
$k_{2} < 0$. We will find a bound for sub-case
$(1)$ and the rest will follow without loss of generality.
We now find the bound for  $k_{1},k_{2}\geq 0$.
\quad\\
\quad\\
\begin{eqnarray*}
\begin{array}{c}
m_{1}\leq 0\\
m_{2}\leq 0
\end{array}
\quad
\Big \{
\quad 
\sum _{
\tiny{
\begin{array}{c}
m_{1}\leq 0\\
m_{2}\leq 0
\end{array}}}
e^{-|m|\rho}e^{-|k-m|(\rho-2\delta)}
\leq
e^{-|k|(\rho-2\delta)}
\sum _{
\tiny{
\begin{array}{c}
m_{1}\leq 0\\
m_{2}\leq 0
\end{array}}}
e^{-|m|\rho},
\end{eqnarray*}
\begin{eqnarray*}
\begin{array}{c}
m_{1}\leq 0\\
0<m_{2}<k_{2}
\end{array}
\quad
\Big \{
&\quad& 
\sum _{
\tiny{
\begin{array}{c}
m_{1}\leq 0\\
0<m_{2}<k_{2}
\end{array}}}
e^{-|m|\rho}e^{-|k-m|(\rho-2\delta)}
\\
\quad\\
&\leq&
\sum _{
\tiny{
\begin{array}{c}
m_{1}\leq 0\\
0<m_{2}<k_{2}
\end{array}}}
e^{-|m_{1}|\rho}
e^{-|k_{1}|(\rho -2\delta)}
e^{-|k_{2}|(\rho -2\delta)}
e^{-|m_{2}|(2\delta)}
\\
\quad\\
&=&
e^{-|k|(\rho -2\delta)}
\sum _{
\tiny{
\begin{array}{c}
m_{1}\leq 0\\
0<m_{2}<k_{2}
\end{array}}}
e^{-|m_{1}|\rho}
e^{-|m_{2}|(2\delta)}\\
\quad\\
&\leq&
e^{-|k|(\rho -2\delta)}
\sum _{
\tiny{
\begin{array}{c}
0<m_{1}\\
0<m_{2}
\end{array}}}
e^{-|m|(2\delta)},
\end{eqnarray*}
\begin{eqnarray*}
\begin{array}{c}
m_{1}\leq 0\\
m_{2}>k_{2}
\end{array}
\quad
\Big \{
&\quad& 
\sum _{
\tiny{
\begin{array}{c}
m_{1}\leq 0\\
m_{2}>k_{2}
\end{array}}}
e^{-|m|\rho}e^{-|k-m|(\rho-2\delta)}
\\
\quad\\
&\leq&
\sum _{
\tiny{
\begin{array}{c}
m_{1}\leq 0\\
m_{2}>k_{2}
\end{array}}}
e^{-|m_{1}|\rho}e^{-|k_{1}|(\rho -2\delta)}
e^{-|k_{2}|\rho}e^{-|k_{2}-m_{2}|(\rho -2\delta)}
\\
\quad\\
&\leq&
e^{-|k|(\rho -2\delta)}
\sum_{
\tiny{
\begin{array}{c}
m_{1}\leq 0\\
m_{2}>k_{2}
\end{array}}}
e^{-|m_{1}|\rho}
e^{-|k_{2}-m_{2}|(\rho -2\delta)},
\end{eqnarray*}
\begin{eqnarray*}
\begin{array}{c}
0<m_{1}<k_{1}\\
m_{2}\leq 0
\end{array}
\quad
\Big \{
&\quad& 
\sum _{
\tiny{
\begin{array}{c}
0<m_{1}<k_{1}\\
m_{2}\leq 0
\end{array}}}
e^{-|m|\rho}e^{-|k-m|(\rho-2\delta)}
\\
\quad\\
&\leq&
e^{-|k|(\rho -2\delta)}
\sum _{
\tiny{
\begin{array}{c}
0<m_{1}<k_{1}\\
m_{2}\leq 0
\end{array}}}
e^{-|m_{2}|\rho}
e^{-|m_{1}|(2\delta)}\\
\quad\\
&\leq& 
e^{-|k|(\rho -2\delta)}
\sum _{
\tiny{
\begin{array}{c}
m_{2}\leq 0
\end{array}}}
e^{-|m_{2}|\rho},
\end{eqnarray*}
\begin{eqnarray*}
\begin{array}{c}
0<m_{1}<k_{1}\\
0<m_{2}<k_{2}
\end{array}
\quad
\Big \{
&\quad& 
\sum _{
\tiny{
\begin{array}{c}
0<m_{1}<k_{1}\\
0<m_{2}<k_{2}
\end{array}}}
e^{-|m|\rho}e^{-|k-m|(\rho-2\delta)}
\\
\quad\\
&\leq&
\sum _{
\tiny{
\begin{array}{c}
0<m_{1}<k_{1}\\
0<m_{2}<k_{2}
\end{array}}}
e^{-|k_{1}|(\rho -2\delta)}e^{-|m_{1}|(2\delta)}e^{-|k_{2}|(\rho-2\delta)}
e^{-|m_{2}|(2\delta)}\\
\quad\\
&=&
e^{-|k|(\rho -2\delta)}
\sum _{
\tiny{
\begin{array}{c}
0<m_{1}<k_{1}\\
0<m_{2}<k_{2}
\end{array}}}
e^{-|m_{1}|(2\delta)}
e^{-|m_{2}|(2\delta)}\\
\quad\\
&\leq&
e^{-|k|(\rho -2\delta)}
\sum _{
\tiny{
\begin{array}{c}
0<m_{1}\\
0<m_{2}
\end{array}}}
e^{-|m_{1}|(2\delta)}
e^{-|m_{2}|(2\delta)}
,
\end{eqnarray*}
\begin{eqnarray*}
\begin{array}{c}
0<m_{1}<k_{1}\\
m_{2}\geq k_{2}
\end{array}
\quad
\Big \{
&\quad& 
\sum _{
\tiny{
\begin{array}{c}
0<m_{1}<k_{1}\\
m_{2}\geq k_{2}
\end{array}}}
e^{-|m|\rho}e^{-|k-m|(\rho-2\delta)}
\\
\quad\\
&\leq&
\sum _{
\tiny{
\begin{array}{c}
0<m_{1}<k_{1}\\
m_{2}\geq k_{2}
\end{array}}}
e^{-|k_{1}|(\rho -2\delta)}e^{-|m_{1}|(2\delta)}
e^{-|k_{2}|\rho}
e^{-|k_{2}-m_{2}|(\rho -2\delta)}\\
\quad\\
&\leq&
e^{-|k|(\rho -2\delta)}
\sum _{
\tiny{
\begin{array}{c}
0<m_{1}<k_{1}\\
m_{2}\geq k_{2}
\end{array}}}
e^{-|m_{1}|(2\delta)}
e^{-|k_{2}-m_{2}|(\rho -2\delta)},
\end{eqnarray*}
\begin{eqnarray*}
\begin{array}{c}
m_{1}\geq k_{1}\\
m_{2}\leq 0
\end{array}
\quad
\Big \{
&\quad& 
\sum _{
\tiny{
\begin{array}{c}
m_{1}\geq k_{1}\\
m_{2}\leq 0
\end{array}}}
e^{-|m|\rho}e^{-|k-m|(\rho-2\delta)}
\\
\quad\\
&\leq&
\sum _{
\tiny{
\begin{array}{c}
m_{1}\geq k_{1}\\
m_{2}\leq 0
\end{array}}}
e^{-|k_{1}|\rho}
e^{-|m_{2}|\rho}
e^{-|k_{1}-m_{1}|(\rho-2\delta)}
e^{-|k_{2}|(\rho-2\delta)}\\
\quad\\
&\leq&
e^{-|k|(\rho-2\delta)}
\sum _{
\tiny{
\begin{array}{c}
m_{1}\geq k_{1}\\
m_{2}\leq 0
\end{array}}}
e^{-|m_{2}|\rho}
e^{-|k_{1}-m_{1}|(\rho-2\delta)},
\end{eqnarray*}
\begin{eqnarray*}
\begin{array}{c}
m_{1}\geq k_{1}\\
0<m_{2}<k_{2}
\end{array}
\quad
\Big \{
&\quad& 
\sum _{
\tiny{
\begin{array}{c}
m_{1}\geq k_{1}\\
0<m_{2}<k_{2}
\end{array}}}
e^{-|m|\rho}e^{-|k-m|(\rho-2\delta)}
\\
\quad\\
&\leq&
\sum _{
\tiny{
\begin{array}{c}
m_{1}\geq k_{1}\\
0<m_{2}<k_{2}
\end{array}}}
e^{-|k_{1}|\rho}e^{-|k_{2}|(\rho-2\delta)}
e^{-|m_{2}|(2\delta)}
e^{-|k_{1}-m_{1}|(\rho-2\delta)}\\
\quad\\
&\leq&
e^{-|k|(\rho-2\delta)}
\sum _{
\tiny{
\begin{array}{c}
m_{1}\geq k_{1}\\
0<m_{2}<k_{2}
\end{array}}}
e^{-|m_{2}|(2\delta)}
e^{-|k_{1}-m_{1}|(\rho-2\delta)},
\end{eqnarray*}
\begin{eqnarray*}
\begin{array}{c}
m_{1}\geq k_{1}\\
m_{2}\geq k_{2}
\end{array}
\quad
\Big \{
&\quad& 
\sum _{
\tiny{
\begin{array}{c}
m_{1}\geq k_{1}\\
m_{2}\geq k_{2}
\end{array}}}
e^{-|m|\rho}e^{-|k-m|(\rho-2\delta)}
\\
\quad\\
&\leq&
\sum _{
\tiny{
\begin{array}{c}
m_{1}\geq k_{1}\\
m_{2}\geq k_{2}
\end{array}}}
e^{-|k_{1}|\rho}
e^{-|k_{1}-m_{1}|(\rho-2\delta)}
e^{-|k_{2}|\rho}
e^{-|k_{2}-m_{2}|(\rho-2\delta)}\\
\quad\\
&\leq&
e^{-|k|\rho}
\sum _{
\tiny{
\begin{array}{c}
m_{1}\geq k_{1}\\
m_{2}\geq k_{2}
\end{array}}}
e^{-|k-m|(\rho-2\delta)}.
\end{eqnarray*}
Finally we have
\begin{eqnarray*}
\sum _{
\tiny{
\begin{array}{c}
m_{1}\in{\mathbb{Z}}\\
m_{2}\in{\mathbb{Z}}
\end{array}}}
e^{-|m|\rho}e^{-|k-m|(\rho-2\delta)}
&\leq& 
e^{-|k|(\rho-2\delta)}
\sum _{
\tiny{
\begin{array}{c}
m_{1}\leq 0\\
m_{2}\leq 0
\end{array}}}
e^{-|m|\rho}\\
\quad\\
&+&
e^{-|k|(\rho -2\delta)}
\sum _{
\tiny{
\begin{array}{c}
0<m_{1}\\
0<m_{2}
\end{array}}}
e^{-|m|(2\delta)}
\\
\quad\\
&+&
e^{-|k|(\rho -2\delta)}
\sum_{
\tiny{
\begin{array}{c}
m_{1}\leq 0\\
m_{2}>k_{2}
\end{array}}}
e^{-|m_{1}|\rho}
e^{-|k_{2}-m_{2}|(\rho -2\delta)}\\
\quad\\
&+&
e^{-|k|(\rho -2\delta)}
\sum _{
\tiny{
\begin{array}{c}
m_{2}\leq 0
\end{array}}}
e^{-|m_{2}|\rho}\\
\quad\\
&+&
e^{-|k|(\rho -2\delta)}
\sum _{
\tiny{
\begin{array}{c}
0<m_{1}\\
0<m_{2}
\end{array}}}
e^{-|m_{1}|(2\delta)}
e^{-|m_{2}|(2\delta)}\\
\quad\\
&+&
e^{-|k|(\rho -2\delta)}
\sum _{
\tiny{
\begin{array}{c}
0<m_{1}<k_{1}\\
m_{2}\geq k_{2}
\end{array}}}
e^{-|m_{1}|(2\delta)}
e^{-|k_{2}-m_{2}|(\rho -2\delta)}\\
\quad\\
&+&
e^{-|k|(\rho-2\delta)}
\sum _{
\tiny{
\begin{array}{c}
m_{1}\geq k_{1}\\
m_{2}\leq 0
\end{array}}}
e^{-|m_{2}|\rho}
e^{-|k_{1}-m_{1}|(\rho-2\delta)}\\
\quad\\
&+&
e^{-|k|(\rho-2\delta)}
\sum _{
\tiny{
\begin{array}{c}
m_{1}\geq k_{1}\\
0<m_{2}<k_{2}
\end{array}}}
e^{-|m_{2}|(2\delta)}
e^{-|k_{1}-m_{1}|(\rho-2\delta)}\\
\quad\\
&+&
e^{-|k|\rho}
\sum _{
\tiny{
\begin{array}{c}
m_{1}\geq k_{1}\\
m_{2}\geq k_{2}
\end{array}}}
e^{-|k-m|(\rho-2\delta)}\\
\quad\\
&\leq&
9e^{-|k|(\rho-2\delta)}
\sum _{
\tiny{
\begin{array}{c}
0\geq m_{1}\\
0\geq m_{2}
\end{array}}}
e^{-|m_{1}|(2\delta)}
e^{-|m_{2}|(2\delta)}\\
\quad\\
&=&
9e^{-|k|(\rho-2\delta)}
\left(
\frac{1}{1-e^{-2\delta}}
\right)^{2}.
\end{eqnarray*}
Without loss of generality the same bound is obtained
for the cases involving different values of $k_{1}$ and
$k_{2}$.Therefore we have for any $k\in {\mathbb{Z}}^{2}$
\begin{eqnarray*}
\sum_{m\in {\mathbb{Z}}^{2}}
e^{-|m|\rho}e^{-|k-m|(\rho -2\delta)}
\leq
9e^{-|k|(\rho-2\delta)}
\left(
\frac{1}{1-e^{-2\delta}}
\right)^{2}.
\end{eqnarray*}
Following the same procedure we can write a bound for the arbitrary
$n$ case. That is, for any $k\in {\mathbb{Z}}^{n}$
\begin{eqnarray*}
\sum_{m\in {\mathbb{Z}}^{n}}
e^{-|m|\rho}e^{-|k-m|(\rho -2\delta)}
\leq
3^{n}e^{-|k|(\rho-2\delta)}
\left(
\frac{1}{1-e^{-2\delta}}
\right)^{n}.
\end{eqnarray*}

\end{document}